%% file: disser.tex
\documentclass [PhD] {uclathes}
\usepackage{epsfig,epsf}

\input {BPS-macros}                     
\input {nearpp-macros}                  

\input {prelim}                           

\begin {document}

\makeintropages

%
%

\input {chapter0}                         
\input {BPS-2pt}                          
\input {BPS-systematic}                   
\input {BPS-3pt}                          
\input {nearpp-chapter}                   




\end {document}

%% file: BPS-macros.tex
\newcommand{\thalf}{\mbox{\tiny {$\! \half$}}}
\newcommand{\tquarter}{\mbox{\tiny {$\! \quarter$}}}

 \def\cA{{\mathcal A}}
 
 \def\cC{{\mathcal C}}

 \def\cO{{\mathcal O}}

 \def\a{\alpha}
 \def\b{\beta}
 
 \def\d{\delta}

 \def\L{\Lambda}

 \def\adt{\dot \alpha}
 \def\bdt{\dot \beta}

 \def\be{\begin{equation}}\def\ee{\end{equation}}
 \def\bea{\begin{eqnarray}}\def\eea{\end{eqnarray}}

 \def\del{\partial}

 \def\xz{\times}

 \def\nab{\nabla}

 \def\no{\nonumber}
 \def\del{\partial}

 \def\3dt{\dot{3}}
 
\def\qu{{1\over4}}
 \let\la=\label

 \def\nn{\nonumber}
 \def\bd{\begin{document}}
 \def\ed{\end{document}}
 \def\bea{\begin{eqnarray}}\def\barr{\begin{array}}\def\earr{\end{array}}
 \def\eea{\end{eqnarray}}
 \newcommand{\eq}[1]{(\ref{#1})}
 \def\eqs#1#2{(\ref{#1}-\ref{#2})}
\def\diag{{\mathrm diag}}

\newcommand{\ho}[1]{$\, ^{#1}$}
\newcommand{\hoch}[1]{$^{#1}$}

%% file: nearpp-macros.tex
\def\ssl#1{\rlap{\hbox{$\mskip 3 mu /$}}#1} 
\newcommand{\ft}[2]{{\textstyle\frac{#1}{#2}}}                                
\newcommand{\eqn}[1]{(\ref{#1})}
\newcommand{\beq}{\begin{equation}}
\newcommand{\beqa}{\begin{eqnarray}}
\newcommand{\bega}{\begin{array}}
\newcommand{\ea}{\end{array}}
\newcommand{\al}{{\alpha^\prime}}
\newcommand{\ale}{\alpha^\prime_{eff}}
\newcommand{\g}{g_{as}}
\newcommand{\eeq}{\end{equation}}
\newcommand{\eeqa}{\end{eqnarray}}
\newcommand{\p}{\partial}
\newcommand{\N}{{\mathsf N}}
\newcommand{\h}{{\hat h}}
\newcommand{\at}{{\tilde a}}
\newcommand{\M}{{\mathcal M}}
\newcommand{\Or}{{\mathcal O}}
\newcommand{\F}{{\textbf F}}
\newcommand{\bm}{{\bar m}}
\newcommand{\bn}{{\bar n}}
\newcommand{\gss}{\gamma_{\sigma \sigma}}
\newcommand{\gst}{\gamma_{\sigma \tau}}
\newcommand{\intl}{\int_0^l d\sigma \,}
\newcommand{\Yd}{{\dot Y^-}}
\newcommand{\Yp}{{(Y^-)'}}
\newcommand{\si}{\sum_{i=1}^2\,}
\newcommand{\sn}{\sum_{n \ge 0 \,}}
\newcommand{\T}{{\mathcal T}}
\newcommand{\A}{{\mathcal A}}
\newcommand{\U}{{\textbf U}}
\newcommand{\lra}{\leftrightarrow}

\newcommand{\NN}{\mbox{$\mathcal N$}}
\newcommand{\NNN}{\mbox{$\mathcal N$}}
\newcommand{\so}{\mbox{$\Rightarrow$}}
\newcommand{\bone}{\mbox{\textbf 1}}
\newcommand{\tr}{\mbox{tr$\,$}}
\newcommand{\str}{\mbox{str$\,$}}
\newcommand{\sym}{\mbox{sym$\,$}}
\newcommand{\half}{\mbox{$1\over2$}}
\newcommand{\third}{\mbox{$1\over3$}}
\newcommand{\quarter}{\mbox{$1\over4$}}
\newcommand{\eigth}{\mbox{$1\over8$}}
\newcommand{\ihalf}{\mbox{$i\over2$}}
\newcommand{\BB}{\mbox{$\mathcal B$}}
\newcommand{\OO}{\mbox{$\mathcal O$}}
\newcommand{\PP}{\mbox{$\mathcal P$}}
\newcommand{\QQ}{\mbox{$\mathcal Q$}}
\newcommand{\RR}{\mbox{$\mathcal R$}}
\newcommand{\FF}{\mbox{$\mathcal F$}}
\newcommand{\KK}{\mbox{$\mathcal K$}}
\newcommand{\LL}{\mbox{$\mathcal L$}}
\newcommand{\YY}{\mbox{$\mathcal Y$}}
\newcommand{\s}{\sigma}
\newcommand{\e}{\epsilon}
\newcommand{\Ot}{{\tilde \Omega}}
\newcommand{\X}{{\mathcal X}}
\newcommand{\Y}{{\mathcal Y}}

\newcommand{\Aa}{\bega{c}k,l \\ i_k{\neq}i_l\ea}
\newcommand{\Ab}{\bega{c}k{\neq}l \\ i_k{=}i_l\ea}

\newcommand{\XII}[6]{\mbox{
\setlength{\unitlength}{0.7em}
\begin{picture}(9.5,3)
\put(0,1.5){\line(1,0){9}}
\put(0,-0.5){\line(1,0){9}}
\put(1,1.5){\line(1,-1){2}}
\put(3,1.5){\line(-1,-1){2}}
\put(5,1.5){\line(0,-1){2}}
\put(8,1.5){\line(0,-1){2}}
\put(5.5,0){\makebox (2,1){...}}
\put(1,0){\makebox (2,1){\tiny $\bullet$}}
\put(0.5,2){\makebox (1,1){$#1 \phantom{d\!\!\!}$}}
\put(0.5,-2){\makebox (1,1){$#2 \phantom{d\!\!\!}$}}
\put(2.5,2){\makebox (1,1){$#3 \phantom{d\!\!\!}$}}
\put(2.5,-2){\makebox (1,1){$#4 \phantom{d\!\!\!}$}}
\put(4.5,2){\makebox (1,1){$#5 \phantom{d\!\!\!}$}}
\put(4.5,-2){\makebox (1,1){$#5 \phantom{d\!\!\!}$}}
\put(7.5,2){\makebox (1,1){$#6 \phantom{d\!\!\!}$}}
\put(7.5,-2){\makebox (1,1){$#6 \phantom{d\!\!\!}$}}
\end{picture}$\phantom{_{\Big|}}$}}

\newcommand{\XIIm}[8]{\mbox{
\setlength{\unitlength}{0.7em}
\begin{picture}(9.5,3)
\put(0,1.5){\line(1,0){9}}
\put(0,-0.5){\line(1,0){9}}
\put(1,1.5){\line(1,-1){2}}
\put(3,1.5){\line(-1,-1){2}}
\put(5,1.5){\line(0,-1){2}}
\put(8,1.5){\line(0,-1){2}}
\put(5.5,0){\makebox (2,1){...}}
\put(1,0){\makebox (2,1){\tiny $\bullet$}}
\put(0.5,2){\makebox (1,1){$#1 \phantom{d\!\!\!}$}}
\put(0.5,-2){\makebox (1,1){$#2 \phantom{d\!\!\!}$}}
\put(2.5,2){\makebox (1,1){$#3 \phantom{d\!\!\!}$}}
\put(2.5,-2){\makebox (1,1){$#4 \phantom{d\!\!\!}$}}
\put(4.5,2){\makebox (1,1){$#5 \phantom{d\!\!\!}$}}
\put(4.5,-2){\makebox (1,1){$#6 \phantom{d\!\!\!}$}}
\put(7.5,2){\makebox (1,1){$#7 \phantom{d\!\!\!}$}}
\put(7.5,-2){\makebox (1,1){$#8 \phantom{d\!\!\!}$}}
\end{picture}$\phantom{_{\Big|}}$}}

\newcommand{\IIX}[8]{\mbox{
\setlength{\unitlength}{0.7em}
\begin{picture}(9.5,3)
\put(0,1.5){\line(1,0){9}}
\put(0,-0.5){\line(1,0){9}}
\put(1,1.5){\line(0,-1){2}}
\put(3,1.5){\line(0,-1){2}}
\put(6,1.5){\line(1,-1){2}}
\put(8,1.5){\line(-1,-1){2}}
\put(4.0,0){\makebox (2,1){...}}
\put(1,0){\makebox (12,1){\tiny $\bullet$}}
\put(0.5,2){\makebox (1,1){$#1 \phantom{d\!\!\!}$}}
\put(0.5,-2){\makebox (1,1){$#2 \phantom{d\!\!\!}$}}
\put(2.5,2){\makebox (1,1){$#3 \phantom{d\!\!\!}$}}
\put(2.5,-2){\makebox (1,1){$#4 \phantom{d\!\!\!}$}}
\put(5.5,2){\makebox (1,1){$#5 \phantom{d\!\!\!}$}}
\put(5.5,-2){\makebox (1,1){$#6 \phantom{d\!\!\!}$}}
\put(7.5,2){\makebox (1,1){$#7 \phantom{d\!\!\!}$}}
\put(7.5,-2){\makebox (1,1){$#8 \phantom{d\!\!\!}$}}
\end{picture}$\phantom{_{\Big|}}$}}

\newcommand{\III}[6]{\mbox{
\setlength{\unitlength}{0.7em}
\begin{picture}(9.5,3)
\put(0,1.5){\line(1,0){9}}
\put(0,-0.5){\line(1,0){9}}
\put(1,1.5){\line(0,-1){2}}
\put(4.5,1.5){\line(0,-1){2}}
\put(8,1.5){\line(0,-1){2}}
\put(5.0,0){\makebox (2,1){...}}
\put(1.5,0){\makebox (2,1){...}}
\put(0.5,2){\makebox (1,1){$#1 \phantom{d\!\!\!}$}}
\put(0.5,-2){\makebox (1,1){$#2 \phantom{d\!\!\!}$}}
\put(4.0,2){\makebox (1,1){$#3 \phantom{d\!\!\!}$}}
\put(4.0,-2){\makebox (1,1){$#4 \phantom{d\!\!\!}$}}
\put(7.5,2){\makebox (1,1){$#6 \phantom{d\!\!\!}$}}
\put(7.5,-2){\makebox (1,1){$#6 \phantom{d\!\!\!}$}}
\end{picture}$\phantom{_{\Big|}}$}}

\newcommand{\blobII}[6]{\mbox{
\setlength{\unitlength}{0.7em}
\begin{picture}(9.5,3)
\put(0,1.5){\line(1,0){9}}
\put(0,-0.5){\line(1,0){9}}
\put(1.5,1.5){\line(0,-1){2}}
\put(2.5,1.5){\line(0,-1){2}}
\put(5,1.5){\line(0,-1){2}}
\put(8,1.5){\line(0,-1){2}}
\put(5.5,0){\makebox (2,1){...}}
\put(1,0){\makebox (2,1)
{\large $\bullet \!\!\rule[0.04em]{0.7em}{0.4em} \!\!\bullet$}}
\put(0.5,2){\makebox (1,1){$#1 \phantom{d\!\!\!}$}}
\put(0.5,-2){\makebox (1,1){$#2 \phantom{d\!\!\!}$}}
\put(2.5,2){\makebox (1,1){$#3 \phantom{d\!\!\!}$}}
\put(2.5,-2){\makebox (1,1){$#4 \phantom{d\!\!\!}$}}
\put(4.5,2){\makebox (1,1){$#5 \phantom{d\!\!\!}$}}
\put(4.5,-2){\makebox (1,1){$#5 \phantom{d\!\!\!}$}}
\put(7.5,2){\makebox (1,1){$#6 \phantom{d\!\!\!}$}}
\put(7.5,-2){\makebox (1,1){$#6 \phantom{d\!\!\!}$}}
\end{picture}$\phantom{_{\Big|}}$}}

%% file: prelim.tex
\title          {Operators in the $d=4$, $\NN=4$ SYM \\
                and the AdS/CFT correspondence
		}
\author         {Anton Vladimirovich Ryzhov}
\department     {Physics}
\degreeyear     {2003}

\dedication     {\textsl{to my teachers}}

\acknowledgments {\input {acknowledgments}}

\abstract       {\input {abstract}}


%% file: acknowledgments.tex

First of all, 
I would like to express my gratitude to my thesis adviser Eric D'Hoker, 
for sharing with me his insights and physical intuition, 
and teaching me many things about physics --- 
as well as about how to do physics. 
I am also in debt to my other advisers: 
I wish to thank Larry Yaffe, who taught me field theory; 
Nick Dorey, who introduced me to the supersymmetric world; 
and Per Kraus, from whom I learned a lot about gauge theories, 
supergravity, and other things.
It is also my pleasure to thank 
Masaki Shigemori, Paul Heslop, Paul Howe, 
and especially Andrei Parnachev for wonderful collaborations. 
All these people have influenced my growth both as a physicist, 
and as a person.

Finally, I am grateful to my parents, my sister, and my grandmother, 
for their love and support, 
and for giving me the opportunity to carry out my Ph.D. studies.


Chapter \ref{chapter: BPS: 2pt} is a version of 
A.~V.~Ryzhov,
``Quarter BPS operators in N = 4 SYM,''
JHEP \textbf{0111}, 046 (2001).
Here I would like to thank Sergio Ferrara for useful discussions. 
I am especially grateful to Eric D'Hoker, 
for suggesting the problem and continuous 
involvement in this paper. 

Chapter \ref{chapter:systematic} is a version of 
E.~D'Hoker, P.~Heslop, P.~Howe and A.~V.~Ryzhov,
``Systematics of quarter BPS operators in N = 4 SYM,''
JHEP \textbf{0304}, 038 (2003). 

Chapter \ref{chapter: BPS: 3pt} is a version of 
E.~D'Hoker and A.~V.~Ryzhov,
``Three-point functions of quarter BPS operators in N = 4 SYM,''
JHEP \textbf{0202}, 047 (2002).
Here we wish to thank 
Daniel Z. Freedman and Sergio Ferrara 
for useful conversations 
during various stages of this project. 
We also thank the referee for suggesting 
clarifications on a number of issues. 

Chapter \ref{chapter:nearpp} is a version of 
A.~Parnachev and A.~V.~Ryzhov,
``Strings in the near plane wave background and AdS/CFT,''
JHEP \textbf{0210}, 066 (2002). 
Here we would like to thank Per Kraus, David Kutasov and David Sahakyan
for many useful discussions. 
We are also grateful to Dan Freedman and Lubos Motl 
for sharing their insights on their construction of 
leading order SYM operators. 
%

This research was supported in part by NSF Grants 
No. PHY-98-19686 and PHY-01-40151. 

%% file: abstract.tex
In this dissertation we explore various aspects 
of the AdS/CFT correspondence, 
which is a duality between $d=4$, $\NN=4$ SYM, 
and IIB superstring theory on $AdS_5\times S^5$ 
with selfdual RR field strength. 
String quantization on general backgrounds with 
fluxes is very difficult. 
So instead, one uses the duality at the level of 
canonical fields of supergravity and the corresponding 
\half-BPS operators in SYM, since they both 
belong to the shortest multiplets of the superconformal group 
$SU(2,2|4)$.

In addition to \half-BPS operators, 
there are others with non-renormalization properties. 
One such class of operators is the \quarter-BPS operators, 
which are dual to threshold bound states 
of elementary supergravity excitations. 
Their scaling dimension 
is also determined by their internal quantum numbers. 
In Chapter \ref{chapter: BPS: 2pt}, 
we consider scalar composites with the right quantum numbers, 
and construct \quarter-BPS operators 
as the ones with $\OO(g_{\mathrm {Y\!M}}^2)$-protected 
two-point functions. 
Extended superspace methods (Chapter \ref{chapter:systematic}) 
make it simple to identify and remove descendant pieces 
from \quarter-BPS candidates. 
In Chapter \ref{chapter: BPS: 3pt}, 
we compute three-point functions involving \quarter-BPS operators, 
and explain how their non-renormalization 
translates into statements about the dual supergravity quantities. 

But we can go beyond discussing 
supergravity modes and protected SYM operators. 
The GS superstring on $AdS_5\times S^5$ 
can be quantized exactly in the limit where the $AdS_5$ radius 
$R\rightarrow\infty$ and the R-charge $J\sim R^2$. 
String states with finite energy and momentum 
(BMN states) are then dual to 
single trace operators with certain phases inserted (BMN operators). 
BMN operators are another natural generalization of 
\half-BPS operators, to which they reduce in the zero-momentum limit.

The perturbative expansion of scaling dimensions of BMN operators is 
in powers of $gN/J^2$. 
Moreover, one can do perturbation theory around the 
$R\rightarrow\infty$, $J\sim R^2$ limit. 
Both expansions have the same regime of validity 
in string theory and in SYM. 
In Chapter \ref{chapter:nearpp}, 
we calculate the first $1/R^2$ corrections to BMN states and their energies, 
and the $1/J$ corrections for the corresponding BMN operators. 
We find complete agreement between the dual quantities. 

%% file: chapter0.tex
\chapter{Introduction}
\label{chapter: introduction}

The celebrated AdS/CFT correspondence asserts that 
the dual description of \NNN=4 four dimensional 
super Yang Mills is type IIB string theory 
in $AdS_5 \times S^5$ with self-dual
RR five-form field strength 
\cite{AdS-CFT}. 
The radius of curvature of $AdS_5$ and $S^5$ 
scales like $R/l_s \sim (g_{\mathrm {Y\!M}}^2 N)^{1/4}\sim (g N)^{1/4}$. 
The spectrum of string states in this background 
corresponds to the spectrum of 
single trace operators in SYM. 
Part of the reason that the AdS/CFT conjecture 
has not been verified directly, 
is that string quantization in the
presence of RR flux is notoriously difficult.
Type IIB supergravity, which
describes the dynamics of massless string modes, is
only valid for the large values of $R/l_s$, while
on the SYM side one can perform 
reliable computations only for small 't Hooft 
coupling $g N$: 
SYM and SUGRA calculations have 
complementary regimes of validity.

Thus on the one hand, the AdS/CFT conjecture 
still has not been proven directly, 
despite the abundance of circumstantial evidence. 
On the other hand, the AdS/CFT correspondence 
provides a powerful tool for deriving
dynamical information in \NN=4 superconformal YM theory outside 
the regime of weak coupling perturbation theory. 
Comparison of weak SYM coupling 
(small $g_{\mathrm {Y\!M}}^2$, perturbative gauge theory calculations) 
and strong SYM coupling 
(large AdS radius $R$, reliable supergravity calculations) 
behaviors has given rise to a number of
surprising new conjectures 
\cite{LMRS, DFS, G-rKP, Skiba, ZANON, BKRS, extremal:refs, near-extremal:refs}, 
which were later confirmed using 
extended superspace methods 
\cite{Intriligator, HSW, AF}.

\section{Protected operators in SYM and SUGRA}

There are very few things one can calculate exactly in the
four-dimensional superconformal Yang-Mills theories. One class of
such quantities consists of a set of correlation functions of the
Bogomolnyi Prasad Sommerfield (BPS) operators. In $\NN=4$
superconformal Yang-Mills, there are \half-BPS, \quarter-BPS and \eigth-BPS
operators, which are operators that are invariant under 8, 4 and 2
(out of 16) Poincar\'e supercharges respectively. Based on very
general arguments involving only the supersymmetry algebra
\cite{Dobrev85, Ferr, minwalla}, the anomalous dimension of any of
these operators vanishes identically in the full quantum theory.

\subsection{Chiral primaries}

In the AdS/CFT correspondence, the local gauge invariant operators of
$\NN=4$ superconformal Yang-Mills are mapped to the physical
states of the Type IIB superstring on AdS$_5\times$S$^5$. (See
\cite{AdS-CFT} for the original papers and
\cite{ADS-CFT:protected} for reviews.) 
The single trace
\half-BPS operators (also referred to as chiral primary operators or
CPOs) play a special role as they are in one-to-one correspondence
with the short multiplets of supergravity and Kaluza-Klein states
with spins $\leq 2$. Driven by the success of this correspondence,
several authors have derived non-renormalization results for
various correlation functions of these operators. Results on the
perturbative non-renormalization of two- and three-point functions were
derived in \cite{LMRS, ADS:2-and-3, DFS} in components and in superspace in
\cite{ZANON, HSW}; for further references on three-point computations
see \cite{Bastianelli:1999vm}. An argument for the complete
non-renormalization of the three-point function of the supercurrent
multiplet based on anomalies was given in \cite{ADS:2-and-3} and a
superspace version of this was presented in \cite{HSW}. An
argument for the (non-perturbative) non-renormalization of all 
two- and three-point functions of BPS operators based on an extra $U(1)_Y$
symmetry was given in \cite{Intriligator} and this was verified using
analytic superspace methods in \cite{EHW} following on from  the
earlier work of \cite{HW}. Generalizations to $n$-point functions
were obtained for extremal \cite{extremal:refs} and
near-extremal correlators 
\cite{near-extremal:refs, Arut99, ADS:4-pt:superspace}.

Other BPS operators are also important, both from the perspective
of superconformal Yang-Mills theory, and from that of the AdS/CFT
correspondence. The simplest generalization is to multi-trace 1/2
BPS operators \cite{Sk99, AF}, for which non-renormalization results
are the same as for single trace \half-BPS operators; see also
\cite{Fer00}. Indeed, the arguments of \cite{Intriligator} and \cite{EHW}
apply in this case too.

\subsection{Quarter BPS operators}

A more delicate generalization is to the multi-trace scalar
operators obeying a \quarter-BPS shortening rule. A general group
theoretic classification of such operators in free field theory
was amongst the results derived in \cite{AFSZ}. 
These \quarter-BPS operators 
share many non-renormalization properties with \half-BPS operators. 
However, they are much more involved, 
which renders their construction nontrivial 
in the fully interacting theory
\cite{Ryzhov:2001bp}. 
In the full quantum interacting theory, the true \quarter-BPS operators
involve admixtures of classical \quarter-BPS operators of \cite{AFSZ}
with descendants of non-BPS operators that occur in long
supersymmetry multiplets. 
In Chapter \ref{chapter: BPS: 2pt} 
we calculate $\OO(g^2)$ two-point functions 
of local, polynomial, scalar composite operators 
within a given representation 
of the $SU(4)$ $R$-symmetry group. 
By studying these two-point functions, 
we identify the eigenstates of the dilatation operator, 
which turn out to be complicated mixtures of single and
multiple trace operators. 



However, this procedure for constructing the candidate
operators is somewhat ad hoc and difficult to generalize.
In Chapter \ref{chapter:systematic}, we use 
extended superspace methods to study \quarter-BPS operators 
in a systematic fashion. 
In the framework of analytic superspace it turns out that 
they 
are described as tensor superfields carrying
superindices
\cite{D'Hoker:2003vf}. 
%
In the construction, the
operators of the classical theory annihilated by 4 out of 16
supercharges are arranged into two types. The first type consists
of those operators that contain \quarter-BPS operators in the full
quantum theory. The second type consists of descendants of
operators in long unprotected multiplets which develop anomalous
dimensions in the quantum theory. The \quarter-BPS operators of the
quantum theory are defined to be orthogonal to all the descendant
operators with the same classical quantum numbers. It is shown, to
order $g_{\mathrm {Y\!M}}^2$, that these \quarter-BPS operators have protected
dimensions. 
In fact, they are the same as the ones found in 
Chapter \ref{chapter: BPS: 2pt}.


\subsection{Comparison with supergravity}

Using the operators thus constructed, 
in Chapter \ref{chapter: BPS: 3pt} 
we will compute three-point functions involving \half-and
\quarter-BPS operators. 
The combinatorics of the problem is rather involved, and 
we concentrate on 
certain classes of three-point functions; 
some results are valid for general $N$, 
while others are large $N$ approximations
\cite{D'Hoker:2001bq}. 
In all cases studied, these correlators%
\footnote{
Correlation functions with $n \ge 4$
operators are, in general, expected to receive quantum
corrections, just as the multipoint functions of \half-BPS operators
do \cite{df98, AF}. 
See also \cite{ADS-CFT:protected} for further references.
} 
are shown to be non-renormalized to order $g_{\mathrm {Y\!M}}^2$.

In the AdS/CFT correspondence, 
\quarter-BPS chiral primaries 
are dual to threshold bound states 
of elementary supergravity excitations. 
We present a supergravity discussion 
of two- and three-point correlators 
involving these bound states, 
and show agreement of the SYM correlators 
with their large $N$, large
$g_{\mathrm {Y\!M}}^2 N$ limit accessible through the AdS/CFT correspondence.


\section{Near the plane wave limit of AdS}

So far, we have been studying the properties of 
supergravity modes, and the corresponding protected SYM operators, 
appealing to nonrenormalization theorems to compare 
their correlators in the dual descriptions 
\cite{ADS-CFT:protected}. 

An improvement to the \half-BPS chiral primaries 
was constructed by Berenstein, Maldacena and Nastase (BMN) in \cite{BMN}. 
The GS superstring can be quantized exactly in the plane wave background 
\cite{Metsaev,Metsaev:2002re}, 
which can be viewed as a double scaling of the $AdS_5 \times S^5$ 
geometry \cite{BMN,Blau:2002dy}.
Remarkably, the parameter controlling perturbative expansion of scaling
dimensions of such operators is $\lambda' = g N / J^2$, which
can be made small to allow reliable gauge theory computations.
String states 
with finite plane wave light cone energy and momentum
correspond
to single trace operators in the gauge theory with certain phases
inserted
\cite{BMN, Gross:2002su,Santambrogio:2002sb,Kristjansen:2002bb}.
Later, string interactions were studied 
both in the plane wave string theory and in the gauge theory
\cite{Kristjansen:2002bb,Spradlin:2002ar,Spradlin:2002rv,Klebanov:2002mp,Berenstein:2002sa,CFHMMPS,Kiem:2002xn,Huang:2002wf,Chu:2002pd,Lee:2002rm}. 

The plane wave limit is an improvement over 
being able to handle only supergravity states and protected operators. 
But we would still like 
to get closer to the full AdS string theory.
One way to gain insight is to do systematic perturbation 
theory around the plane wave limit, taking $1/R^2$ 
as a small parameter
\cite{Parnachev:2002kk}. 
This approach has been tested in \cite{p1} on the
$AdS_3 \times S^3$ background with NS-NS flux.
%
In chapter \ref{chapter:nearpp}, 
we study the AdS/CFT correspondence for string states which
flow into plane wave states in the Penrose limit.
Leading finite radius corrections to the string spectrum are compared
with scaling dimensions of finite R-charge BMN-like operators.
We find agreement between string and gauge theory results. 
This is a constructive step towards proving the 
AdS/CFT conjecture.%
\footnote{
However, it should be mentioned that 
$1/R^2$ expansion around the BMN limit is still much closer 
to the plane wave geometry, 
than it is to 
doing string theory on the full AdS background. 
}


The AdS/CFT correspondence is in turn just 
an example of the general phenomenon of holography, 
by which a CFT on the (conformal) boundary of some 
manifold is dual to a string theory in the bulk. 
The study of holography remains an open and exciting field, 
which keeps giving theorists valuable insights 
into both field theory phenomena, and into string theory. 
Hopefully, dualities will bring us closer to understanding 
confinement in gauge theories, 
or even to having a nonperturbative formulation of string theory.

%% file: BPS-2pt.tex
\chapter{Quarter BPS Operators in \NN=4 SYM}
\label{chapter: BPS: 2pt}


During the past years, 
there has been a renewed interest in the study of chiral operators 
in the \NN=4 supersymmetric 
Yang-Mills theory in four dimensions. 
Forming short representations 
of the global $SU(2,2|4)$ superconformal symmetry group, 
chiral operators have tightly constrained quantum 
numbers. 
In particular, the scaling dimension of a chiral operator 
is not renormalized.%
\footnote{
\label{footnote:non-renormalized nonBPS}
	The possibility that certain non-chiral operators 
	may have vanishing anomalous dimension
	was raised in \cite{AF}. 
	}

Chiral primary operators have been classified in \cite{AFSZ, DP}. 
They can be \half-BPS, \quarter-BPS, and \eigth-BPS. 
The \half-BPS operators provide 
the simplest example of chiral primaries. 
These are 
scalar composite operators in the $[0,q,0]$ representations 
of the $R$-symmetry group $SU(4) \sim SO(6)$; 
their scaling dimension is $\Delta = q$, see \cite{AFSZ}. 
\half-BPS operators are annihilated by eight out of the sixteen Poincar\'e 
supercharges 
of the theory. 
Similarly, \quarter-BPS primaries belong to 
$[p,q,p]$ representations of the $R$-symmetry group, 
are annihilated by four supercharges, and have protected 
scaling dimension of $\Delta = 2p+q$. 
Finally, \eigth-BPS primaries 
live in $[p,q,p+2k]$ of $SU(4)$, 
are killed by only two supercharges, 
and their $\Delta = 3k+2p+q$. 
Quantum numbers of the descendant operators 
are related to those of their primaries by the 
\NN=4 superconformal algebra.

\half-BPS operators have been much studied. 
Using the conjectured AdS/CFT correspondence \cite{AdS-CFT}, 
it was shown, 
that for gauge groups 
$SU(N)$ with $N$ large, two and three point functions 
of \half-BPS chiral primaries are the same at weak 
and strong coupling \cite{LMRS}.%
\footnote{
	Higher $n$-point functions also 
	agree with supergravity predictions in the large $N$ limit 
	\cite{DHKMV}. 
	}
It was then verified that these SYM correlators get no 
$\OO(g^2)$ corrections, for all $N$ \cite{DFS}. 
Chiral descendant operators share these 
non-renormalization properties with their parent primaries \cite{DFS}. 
$\OO(g^4)$ and  instanton contributions to 
two and three point functions 
of \half-BPS primaries 
turn out to vanish as well 
\cite{HSW, BKRS, AF, ZANON}. 
Non-renormalization of these correlators 
was further established on general grounds 
in \cite{Intriligator, HSW}. 
Besides $SU(N)$ theories and single trace chiral primaries, 
multiple trace operators with the same $SU(2,2|4)$ quantum numbers, 
as well as 
arbitrary gauge groups 
were considered \cite{Skiba}. 
In these cases, two and three point functions 
were also found to receive no $\OO(g^2)$ corrections.

It is natural to ask whether other chiral operators, 
for example \quarter-BPS primaries, 
have protected correlators. 
Here the situation is much less straightforward 
than for $[0,q,0]$ operators. 
In fact, except for 
the simplest operator 
found in \cite{BKRS}, 
no other \quarter-BPS chiral primaries 
were written down%
\footnote{
	\quarter-BPS operators have been studied indirectly 
	through OPEs of \half-BPS chiral primaries, see 
	\cite{BKRS, AF, ADS:4-pt, ADS:4-pt:superspace}.
	} 
in the fully interacting theory. 
The main difficulty is that unlike in the free theory, 
where a kinematical (group theoretical) treatment 
of \cite{AFSZ} is sufficient, for nonzero coupling 
the problem of determining primary operators 
becomes a dynamical question.%
\footnote{
	We would like to thank Sergio Ferrara 
	for bringing this to our attention. 
	}

Apart from the double trace scalar composite operators 
in the $[p,q,p]$ of the $R$-symmetry (flavor) group $SU(4)$ 
(the free theory chiral primaries from the classification of \cite{AFSZ}), 
there are other single and multiple trace scalar composites 
with the same $SU(4)$ quantum numbers 
and the same $\OO(g^0)$ 
scaling dimension. 
Unlike in the \half-BPS case where this 
phenomenon occurs \cite{Skiba}, 
scalar composites in the $[p,q,p]$ generally do not have 
a well defined scaling dimension. 
Thus, one should first find their linear combinations 
which are eigenstates of the dilatation operator, 
which we call pure operators. 
To this end, we calculate two point functions 
of 
local, gauge invariant, polynomial, scalar composite 
operators in a given $[p,q,p]$ representation; 
diagonalize the dilatation operator within 
each representation of $SU(4)$; 
and find that 
some of the pure operators receive no $\OO(g^2)$ corrections to 
their scaling dimension or normalization. 
These operators have the right $SU(4)$ quantum numbers 
and protected $\Delta = 2p+q$, 
and are the only candidates for being 
the \quarter-BPS chiral primaries 
from the classification of \cite{AFSZ}.

Calculating the symmetry factors for Feynman diagrams 
is a formidable combinatorial problem for general 
representation $[p,q,p]$ of $SU(4)$, 
and general $N$ of the gauge group $SU(N)$. 
So to keep the formulas manageable, 
we concentrate on two special cases. 
For low dimensional operators ($2p+q < 8$), 
we perform explicit computations for arbitrary $N$; 
in particular, we recover the simplest 
\quarter-BPS operator 
studied previously in \cite{BKRS}. 
Alternatively, we give a leading plus subleading 
large $N$ argument (valid for general $[p,q,p]$ representations) 
for a class of 
\quarter-BPS chiral primaries, 
which are linear combinations 
of double- and single-trace scalar composite operators.

The plan of this Chapter is as follows. 
First we review some aspects of 
$SU(2,2|4)$ group theory, and 
describe the scalar composite operators 
we will be dealing with. 
Then we set the stage for $\OO(g^2)$ 
calculations of two-point functions, 
and outline the main ingredients 
of these calculations. 
After that, we explicitly compute the simplest 
sets of correlators. 
In the course of these computations, 
it turns out that only one 
type of Feynman diagrams contributes 
to the correlators at order $g^2$, 
and we provide a simple explanation of this fact.%
\footnote{
	The argument we give applies more generally. 
	In particular, it provides 
	an alternative interpretation of the work in 
	\cite{DFS} and \cite{Skiba}. 
	}
We present the full calculation for these two point functions. 
For higher $\Delta$, calculations were done using {\textit {Mathematica}} 
and only the results are shown. 
Several new features come into play, and we 
describe them as we go along. 
Finally, 
we switch gears and do a large $N$ analysis 
of \quarter-BPS operators with arbitrary scaling dimension.


\section{$SU(2,2|4)$ group theory}
\label{section:superconformal group}

Four dimensional  \NN=4 superconformal Yang-Mills theory 
has been studied extensively 
for a long time, and we begin by reviewing 
some well known facts. 

\NN=4 SYM can be formulated in several (equivalent) ways; 
see Appendix  \ref{n=4 susy:section} for some of the descriptions. 
None of them shows all the features of the 
theory explicitly. For example, working with 
six scalars $\phi^I = \phi^I_a t^a$ 
(where 
$a=1, ... , N^2-1$ runs over the gauge group $SU(N)$, 
and $\phi^I_a(x)$, $I=1, ... , 6$ are real scalar fields), 
and grouping the fermions 
as $\lambda^i_a$, $i=1,...,4$, 
makes the full $SU(4)$ $R$-symmetry group manifest, but hides all the 
supersymmetries. On the other hand, formulating 
the theory in terms of \NN=1 superfields shows 
some of the supersymmetry, but the Lagrangian 
looks invariant just under the $SU(3) \times U(1)$
subgroup of the full $SU(4)$. 
In practice, the more supersymmetries we use, the
simpler it is to perform actual calculations.%
\footnote{
	E.g., the order $g^4$ calculations 
	in \cite{BKRS} were done in the \NN=2 
	harmonic superspace formalism. 
	}
For the purposes of $\OO(g^2)$ computations, 
it suffices to use component fields of the \NN=1 superfield 
formulation of the theory, 
with the (Euclidean signature) Lagrangian 
\cite{DFS} 
\begin	{eqnarray}
\label	{lagrangian:for Feynman rules}
\LL &\!\!=\!\!& \tr \left\{ 
\quarter F_{\mu\nu} F^{\mu\nu} 
+ \half \bar \lambda \gamma^\mu D_\mu \lambda 
+ \overline{D_\mu z_j} D^\mu z_j 
+ \half \bar \psi^j \gamma^\mu D_\mu \psi^j 
\right\} 
\nonumber\\ 
&&
+ i \sqrt{2} g f^{abc} \left( 
\bar \lambda_a \bar z_b^j L \psi_c^j - 
\bar \psi_a^j R z_b^j \lambda_c 
\right) 
- \half Y f^{abc} \e_{ijk} \left( 
\bar \psi_a^i z_b^j L \psi_c^k - 
\bar \psi_a^i R \bar z_b^j \psi_c^k 
\right) 
\nonumber\\ 
&&
- \half g^2 (f^{abc} \bar z_b^j z_c^j ) (f^{ade} \bar z_d^k z_e^k ) 
+ \quarter Y^2 f^{abc} f^{ade} \e_{ijk} \e_{ilm} 
z_b^j z_c^k \bar z_d^l \bar z_e^m 
\end	{eqnarray}
($L$ and $R$ are chirality projectors). 
The theory defined by (\ref{lagrangian:for Feynman rules}) 
has \NN=1 supersymmetry. 
We use separate coupling constants $g$ and $Y$ 
to distinguish the terms coming from the gauge and 
superpotential sectors. 
When $Y = g \sqrt{2}$, SUSY is enhanced to \NN=4.

Since the manifest symmetry group is now $SU(3) \times U(1)$, 
we first project onto it the representations of 
the full $SU(4)$. 
This can be done by mapping the quantum numbers as 
\begin{equation}
\label{projection:def}
[p,q,r] \mapsto [p,q]^{ - {1\over2} (p+2q+3r)}
\end{equation}
Under this projection, the fermions in the theory 
are mapped as:
$\lambda_{1,2,3} \mapsto \psi_{1,2,3} \in [1,0]^{-{1\over2}}$, 
$\lambda_4 \mapsto \lambda = [0,0]^{{3\over2}}$, so 
${\mathbf 4} = [1,0,0] \rightarrow [1,0]^{-{1\over2}} \oplus [0,0]^{{3\over2}}$. 
Similarly the scalars are projected 
as 
\begin{equation}
{\mathbf 6} ~=~ 
[0,1,0] ~~\rightarrow~~ 
[1,0]^1 \oplus [0,1]^{-1} ~=~ \{ z_j \} \oplus \{ \bar z^k \} 
\end{equation}
Put more simply, 
this amounts to 
rewriting the real scalars $\phi^I$, and fermions $\lambda^i$ 
as 
$\phi^i = {1\over\sqrt{2}}(z_i + \bar z_i)$, 
$\phi^{i+3} = {1\over i \sqrt{2}}(z_i - \bar z_i)$, 
and 
$\lambda^i = \psi_i$, $\lambda^4 = \lambda$. 
Index $i=1,2,3$ labels the ${\mathbf 3}$ or ${\mathbf {\bar 3}}$
of the $SU(3)$ factor of the manifest symmetry group 
of (\ref{lagrangian:for Feynman rules}).

The $R$-symmetry group 
of the theory is $SU(4) \sim SO(6)$, which 
is a part of the larger superconformal 
$SU(2,2|4)$. 
Unitary representations of \NN=4 SYM 
were classified in \cite{DP}. 
As in any conformal theory, 
operators 
are classified by their scaling dimension $\Delta$. 
Each multiplet 
of $SU(2,2|4)$ 
contains an operator of lowest 
dimension, which is called a primary operator. 
The action of generators of the conformal group%
\footnote{
	See for example \cite{MS}, 
	or one of the big reviews \cite{ADS-CFT:protected}. 
	}
on a primary operator $\Phi(x)$ is given by
\begin{eqnarray}
\label{begin:conformal algebra}
\left[ P_\mu , \Phi(x) \right] &=& i \partial_\mu \Phi(x) \\
\left[ M_{\mu\nu} , \Phi(x) \right] &=& 
\left[ i (x_\mu \partial_\nu - x_\nu \partial_\mu) 
+ \Sigma_{\mu\nu} \right] \Phi(x) \\
\left[ D , \Phi(x) \right] &=& 
i \left( - \Delta + x^\mu \partial_\mu \right) \Phi(x) \\
\left[ K_\mu , \Phi(x) \right] &=& 
\left[ i (x^2 \partial_\mu - 2 x_\mu x^\nu \partial_\nu + 2 x_\mu \Delta) 
- 2 x^\nu \Sigma_{\mu\nu} \right] \Phi(x) 
\label{end:conformal algebra}
\end{eqnarray}
Notice that 
$\left[ M_{\mu\nu} , \Phi(0) \right] = \Sigma_{\mu\nu} \Phi(0)$, 
$\left[ D , \Phi(0) \right] = - i \Delta \Phi(0)$, and 
$\left[ K_\mu , \Phi(0) \right] = 0$. 
Together with the 16 
Poincar\'e supersymmetry generators $Q$ (and $\bar Q$), 
and 16 special conformal fermionic generators $S$ (and $\bar S$),
these close in a superconformal algebra of $SU(2,2|4)$. 
The additional (anti)commutation relations are schematically 
given by 
\begin{eqnarray}
\label{begin:superconformal algebra}
\left[ D , Q \right] &=& - \ihalf Q , \quad 
\left[ D , S \right] ~=~ + \ihalf S , \quad 
\left[ K , Q \right] ~\sim~ S \quad 
\left[ P , S \right] ~\sim~ Q , \\ 
\left[ Q , S \right] &\sim& M + D + R , \quad 
\left[ S , S \right] ~\sim~ K , \quad 
\left[ Q_i , Q_j \right] ~\sim~ P \, \delta_{ij} ~ (i,j=1,...,4) 
\hspace{2em}
\label{end:superconformal algebra}
\end{eqnarray}
where $R$ stands for the quantum numbers of the $R$-symmetry group $SU(4)$. 
The Lagrangian of the theory, 
as well as the action of supersymmetry generators 
on the elementary fields, 
are listed in Appendix \ref{n=4 susy:section}.

Primary operators of the superconformal group 
which are annihilated by at least some of the $Q$-s 
are called chiral primaries. 
Descendants of chiral primaries are then 
chiral operators, in the \NN=4 sense. 
Chirality is a property of the whole $SU(2,2|4)$ 
multiplet; 
just being annihilated by say 8 Poincar\'e SUSY generators
doesn't make an operator \half-BPS. 
Since the supercharges anticommute, 
we can take a 
non-chiral operator and act on it with 
some of the $Q$-s. 
The resulting (non-chiral!) operator 
will be annihilated by the same $Q$-s. 

For a chiral primary field $\Phi$ 
annihilated by a Poincar\'e supercharge $Q$, 
we can write 
$\left[ Q , \Phi(x) \right] =0$ and $\left[ K , \Phi(0) \right] = 0$, 
and so 
$\left[ S , \Phi(0) \right] \sim \left[ [K,Q] , \Phi(0) \right] = 0$ 
as well. 
Hence we can express the conformal dimension $\Delta$ of $\Phi$ 
entirely in terms of its spin $\Sigma$ 
and $SU(4)$ quantum numbers 
$R$ 
\begin{equation}
0 = \left[ [Q,S] , \Phi(0) \right] \sim 
\left[ M+D+R , \Phi(0) \right] 
= \left( \Sigma - i \Delta + R \right) \Phi(0)
\end{equation}
by the superconformal algebra 
(\ref{begin:conformal algebra}-\ref{end:superconformal algebra}). 
Quantum numbers of descendants 
are related to those of their parent primaries 
by 
(\ref{begin:conformal algebra}-\ref{end:superconformal algebra}) 
as well. 
In particular, 
$\Delta$ of any chiral operator 
can not receive quantum corrections.

\section{Gauge invariant scalar composite operators}
\label{section:operators:BPS-paper}

A kinematic (group theoretic) classification of 
BPS operators was given in \cite{AFSZ}. 
Chiral primaries%
\footnote{
	When referring to ``primary'' fields, 
	we often have in mind the entire 
	$SU(4)$ multiplet to which 
	the actual primary belongs. 
	This slight abuse of notation is common in the literature. 
	} 
are Lorentz scalars, 
which are made by taking local 
gauge invariant polynomial combinations of the 
$\phi^I(x)$, $I=1, ... , 6$. 
They fall into one of the three families \cite{DP}. 
The simplest one consists of \half-BPS operators. These 
chiral primaries are annihilated by half of the $Q$-s, 
and live in short multiplets with spins ranging 
from zero to 2. 
\half-BPS chiral primaries 
are totally symmetric traceless rank $q$ tensors 
of the flavor $SO(6)$. 
$SU(4)$ labels of these representations are $[0,q,0]$ 
with the corresponding 
$SO(6)$ Young tableau%
\footnote{
	See for example \cite{Hamermesh} for a general 
	discussion on constructing irreducible tensors 
	of $SO(n)$. 
	}
$\mbox{
\setlength{\unitlength}{0.7em}
\begin{picture}(3.5,1)
\put(-.8,0){\framebox (1,1){}}
\put(0.2,0){\framebox (2,1){\scriptsize $...$}}
\put(2.2,0){\framebox (1,1){}}
\end{picture}}$, 
one row of length $q$. 
Operators with the highest $SU(4)$ weight in the $[0,q,0]$ 
have the form 
$\tr (\phi^1)^q$, modulo the $SO(6)$ traces.%
\footnote{
	For example, the highest weight state in the [2,0,2] 
	is $\tr (\phi^1)^2 - {1\over6} \sum_{I=1}^6 \tr \phi^I \phi^I$. 
	Operators in this representation are usually referred to 
	as ``$\tr X^2$'' in the literature, and are special since 
	their descendants include the $SU(4)$ flavor currents 
	and the stress tensor. 
	}
Because the color group is $SU(N)$ rather than $U(N)$, 
$\tr \phi^I = 0$ 
so $q \ge 2$. 
Conformal dimension of a \half-BPS chiral primary is related to 
its flavor quantum numbers as $\Delta = q$.

\quarter-BPS operators form 
the next simplest family of chiral operators 
in the classification of \cite{AFSZ}. 
Their multiplets have spins from zero to 3. 
The primaries 
belong to $[p,q,p]$ representations, 
and are annihilated by four out of sixteen 
Poincar\'e supercharges. 
There is a restriction $p \ge 2$: 
for $p=0$ the operators are \half-BPS; 
and in the case $p=1$, they vanish 
after we take the $SU(N)$ traces. 
The highest weight state 
of $[p,q,p]$ 
corresponds to the 
\begin{equation}
\label{pqp:representation} 
\mbox{
\setlength{\unitlength}{1em}
\begin{picture}(7.5,2)
\put(0,1){\framebox (1,1){\scriptsize $1$}}
\put(1,1){\framebox (2,1){\scriptsize $...$}}
\put(3,1){\framebox (1,1){\scriptsize $1$}}
\put(4,1){\framebox (1,1){\scriptsize $1$}}
\put(5,1){\framebox (1,1){\scriptsize $...$}}
\put(6,1){\framebox (1,1){\scriptsize $1$}}
\put(0,0){\framebox (1,1){\scriptsize $2$}}
\put(1,0){\framebox (2,1){\scriptsize $...$}}
\put(3,0){\framebox (1,1){\scriptsize $2$}}
\put(1.8,-.7){\scriptsize $p$}
\put(5.4,.3){\scriptsize $q$}
\end{picture}}
\end{equation}
$SO(6)$ Young tableau. 
In the free theory, \quarter-BPS primaries 
corresponding to (\ref{pqp:representation}) 
are of the form 
$\tr (\phi^1)^{p+q} \, \tr (\phi^2)^{p}$ 
(modulo $(\phi^1,\phi^2)$ antisymmetrizations, 
and subtraction of the $SO(6)$ traces). 
However, there are many other ways to 
partition a given Young tableau, 
and each may result in 
a different operator after we take the $SU(N)$ traces. 
A priori, we do not know if any of them 
are pure (i.e. eigenstates of the dilatation operator $D$), 
or are mixtures of operators with different 
scaling dimensions. 
So these operators should be regarded 
just as a basis of gauge invariant, local, 
polynomial, scalar composite operators in the $[p,q,p]$ 
of $SU(4)$. 
By taking linear combinations of these, we will 
construct eigenstates of $D$ in general, 
and \quarter-BPS primaries in particular.


For completeness, let us mention the \eigth-BPS operators, 
which form the last family of chiral operators in the classification 
of \cite{AFSZ}. \eigth-BPS multiplets are also short, with 
spins from zero to 7/2, and the chiral primaries 
are of the form 
$\tr (\phi^1)^{p+k+q} \, \tr (\phi^2)^{p+k} \, \tr (\phi^3)^{k}$ 
(modulo $(\phi^1,\phi^2,\phi^3)$ antisymmetrizations, 
and minus the $SO(6)$ traces), 
in the free theory. 
As before, there is a $k \ge 2$ restriction on the 
quantum numbers: $k \ge 1$ so the operators 
are annihilated by exactly two Poincar\'e supercharges;
while operators with $k=1$ necessarily contain commutators 
after we take the $SU(N)$ traces, as $\tr \phi^I = 0$. 
\eigth-BPS chiral primaries have $SU(4)$ labels $[p,q,p+2k]$, and 
their scaling dimensions have protected values of 
$\Delta = 3 k + 2 p + q$. 
Although 
these operators are also interesting, 
we will not study them here.

When calculating $n$-point functions, it suffices to consider 
one (nonzero) correlator for a given choice of representations; 
all others will be related to it by 
$SU(4)$ Clebsch-Gordon coefficients 
(by the Wigner-Eckart theorem). 
Therefore, we are free to take the most 
convenient representatives of the full $SU(4)$ 
representations, or of the smaller $SU(3) \times U(1)$ 
bits into which a given representation of $SU(4)$ breaks down.%
\footnote{
	All correlators in the 
	resulting $SU(3) \times U(1)$ representations 
	will have identical spatial dependence, 
	since they come from the same $SU(4)$ representation. 
	}
The combinatorics of the problem 
simplifies if we consider operators 
of the form $[(z_1)^{p+q} (z_2)^p]$ 
and their conjugates, 
which is what we will do in this Chapter.

Finally, suppose we have disentangled the mixtures 
of $[p,q,p]$ scalar composite operators 
annihilated by a quarter of the Poincar\'e supercharges,
into linear combinations of operators 
with definite scaling dimension. 
Furthermore, assume we found an operator $\YY$ 
whose scaling dimension is protected. 
Since $\YY$ is a pure operator annihilated by four 
Poincar\'e supercharges, it can be either 
a \quarter-BPS primary; 
or a level two descendant of a \eigth-BPS primary, 
but this case is excluded%
\footnote{
\label{footnote: no eigth-BPS descendants}
	If $\YY$ came from a \eigth-BPS primary, 
	the parent primary would in the $[p',q',p'+2 k]$ 
	representation of $SU(4)$, with $k \ge 2$. 
	On the other hand, to make the scaling dimension 
	and $SU(4)$ Dynkin labels 
	work out right, the only allowed choice is 
	$[p,q,p+2]$, or $k=1$. 
	}
by group theory; 
or a level four descendant of a non-chiral primary. 
If $\YY$ were non-chiral, 
its primary would be 
a scalar composite operator of the form $[z^{2p+q-3} \bar z]$;
and in all examples that we studied in this Chapter, such 
operators do receive $\OO(g^2)$ corrections to their 
scaling dimension.%
\footnote{
	But see footnote \ref{footnote:non-renormalized nonBPS}. 
	}
We conclude that 
a scalar composite operator in the $[p,q,p]$, 
which is annihilated by a quarter of the supercharges 
and has a protected scaling dimension $\Delta = 2p+q$, 
is a \quarter-BPS chiral primary.

\section{Contributing diagrams}
\label{section:contributing diagrams}

The two point functions we will be calculating 
are 
of the 
form 
\begin{equation}
\label{eq:general two-point}
\langle 
\left[ {z_1}^{(p+q)} {z_2}^p \right] \!(x) 
\,
\left[ {\bar z_1}^{(p+q)} {\bar z_2}^p \right] \!(y) \rangle 
\end{equation}
where $[...]$ stands for gauge invariant combinations. 
The free field part of such a correlator 
is given by a power of the free
scalar propagator $[G(x,y)]^{(2p+q)}$, times a combinatorial factor.

\begin{figure}
{\begin{center}
\epsfig{width=5.5in, file=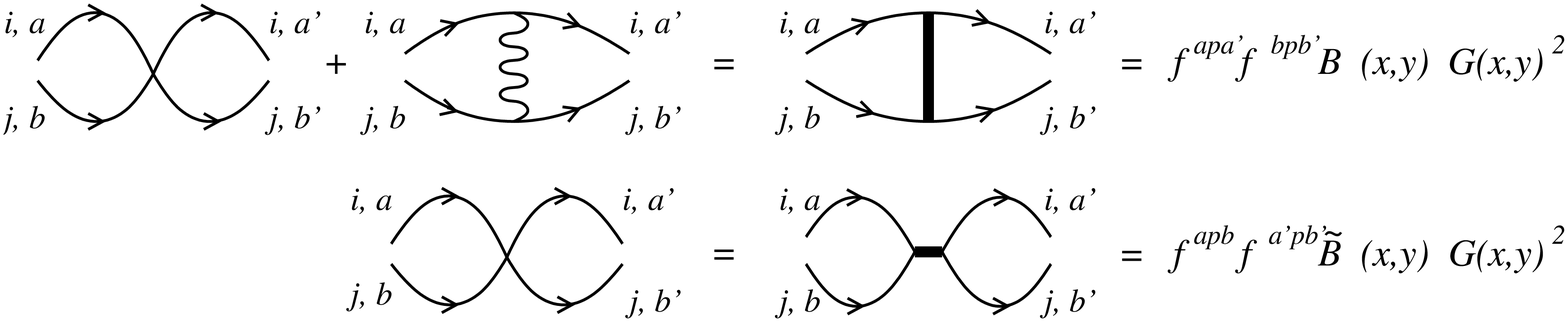, angle=0}
\epsfig{width=5.2in, file=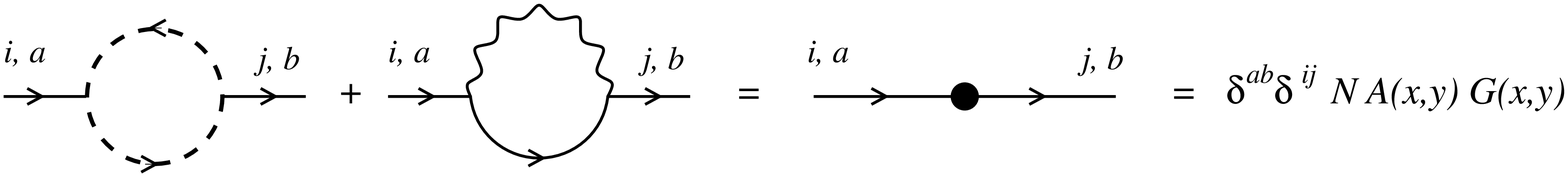, angle=0}
\end{center}}
\caption{%
Structures contributing to two-point functions of scalars at order $g^2$ 
through four-scalar blocks and the propagator. 
Thick lines correspond to exchanges of the gauge boson, 
and of the auxiliary fields $F_i$ and $D$
(in the \NN=1 formulation; 
after integrating out $F_i$ and $D$, 
the $z z \bar z \bar z$ vertex). 
The scalar propagator remains diagonal in both color and 
flavor indices at order $g^2$. 
At order $g^2$, there are corrections to the scalar propagator 
coming from a fermion loop (dashed line) 
and a gauge boson semi-loop (wiggly line). 
Also, blocks involving four scalars get contributions from a single 
gauge boson exchange, and from the four-scalar vertex. 
Gauge fixing and ghost terms in the Lagrangian do not contribute 
to (\ref{eq:general two-point}) at $\OO(g^2)$. 
\label {fig:four-scalar and propagator}
}%
\end {figure}

\begin{figure}[t!]
{\begin{center}
\epsfig{width=0.8in, file=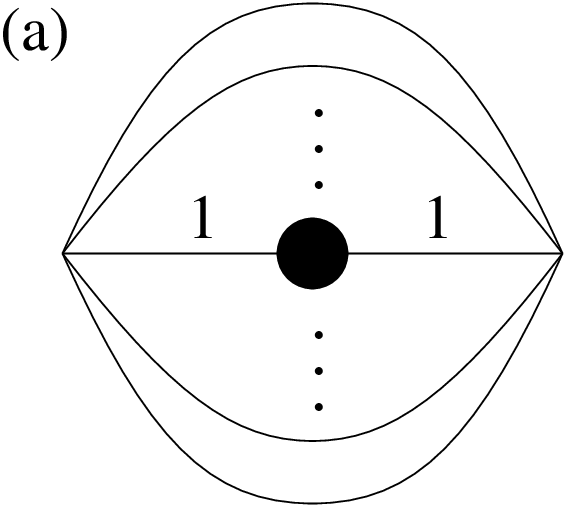, angle=-90} 
\epsfig{width=0.8in, file=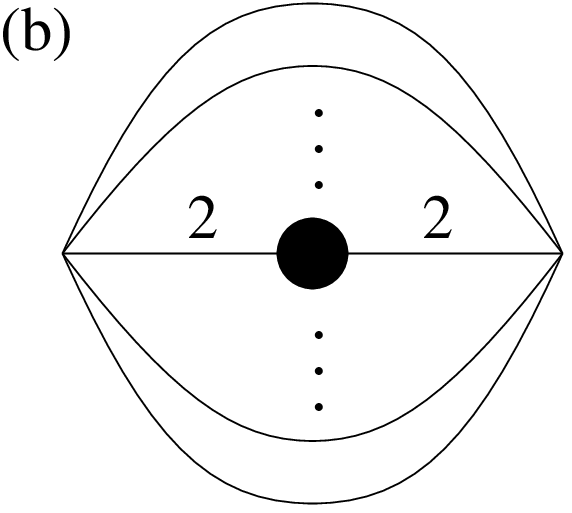, angle=-90} 
\epsfig{width=0.8in, file=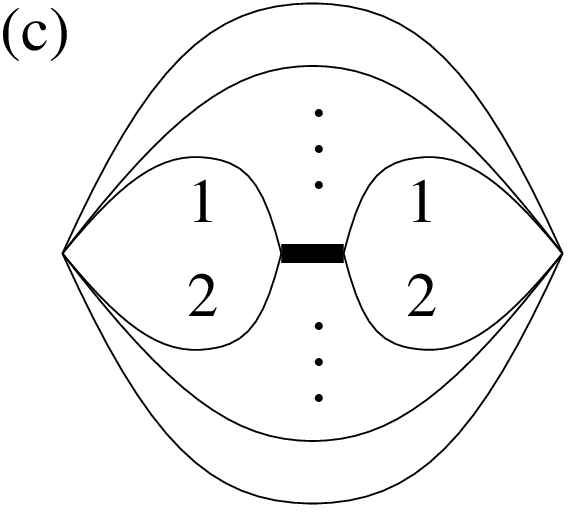, angle=-90}
\epsfig{width=0.8in, file=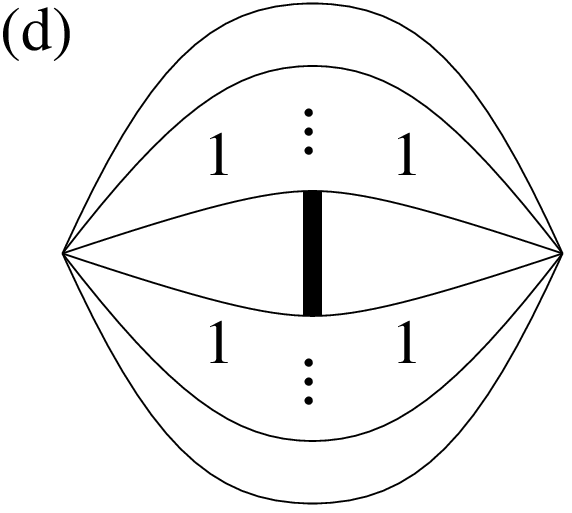, angle=-90} 
\epsfig{width=0.8in, file=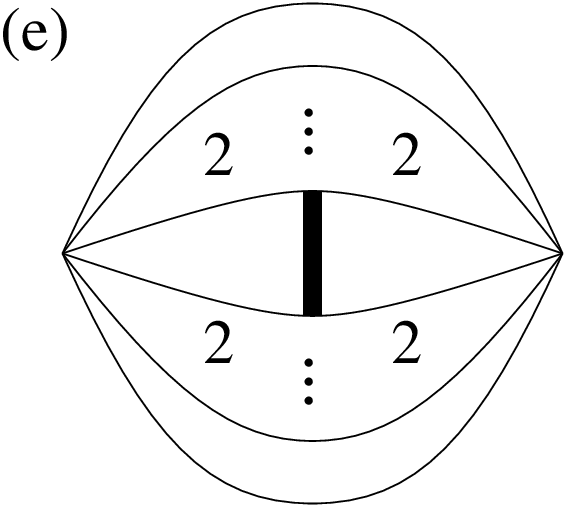, angle=-90} 
\epsfig{width=0.8in, file=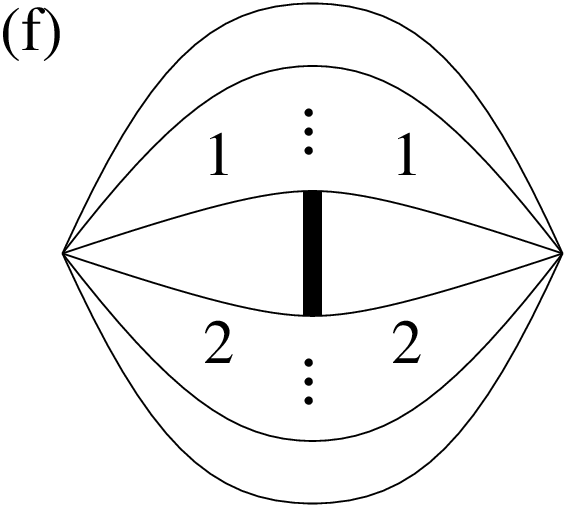, angle=-90} 
\end{center}}
\caption{%
Diagrams contributing to two-point functions of scalars at order $g^2$. 
\label {fig:contributing diagrams}
}%
\end {figure}

From the Lagrangian (\ref{lagrangian:for Feynman rules}) 
we can read off the structures contributing to 
the four-scalar blocks, and the leading
correction to the propagator 
at order $g^2$. 
These are shown in Figure \ref{fig:four-scalar and propagator}, 
where they are categorized according to their 
gauge group (color) index structure 
(we will use the same notation as in \cite{DFS}). 
The scalar propagator remains diagonal in both color and 
flavor indices at order $g^2$. 
Notice that the corrections proportional to $\tilde B$ 
are antisymmetric in $i$ and $j$, hence they are absent when the 
scalars in the four legs have the same flavor. 
Thus we will have to compute contributions of six types%
\footnote{
	If all scalars were of the same flavor, say 1 
	(as is the case for \half-BPS operators considered 
	in \cite{DFS}and \cite{Skiba}), we would 
	only have to consider diagrams of types (a) and (d). 
	}
(see Figure \ref{fig:contributing diagrams}). 

Most Feynman diagrams we come across are easier 
to evaluate in position space, where they factorize 
into products of free propagators and the 
blocks shown in Figure \ref{fig:four-scalar and propagator}, 
and everything except for the combinatorial 
factors out front is almost trivial. 
In momentum space, on the other hand, even the simplest 
$\OO(g^2)$ graphs contain divergent subdiagrams.

The functions $A$ and $B$ will be discussed in detail 
in Section \ref{section:gauge-dependent}.
Coordinate dependence of 
$\tilde B$ is parametrically determined %
by conformal invariance, 
$\tilde B(x,0) = \tilde a \log (x^2 \mu^2) + \tilde b$. 
The coefficients $\tilde a$ and $\tilde b$ can be found  
using, for example, differential regularization \cite{FJMV}, 
or a simpler equivalent prescription:
replace 
$1/x^2 \to 1/(x^2 + \e^2)$ for 
propagators inside 
integrals ($\e \sim \mu^{-1}$ is related to the renormalization scale). 
With this, 
\begin{eqnarray}
\tilde B (x,0) &=& 
\label{b-tilde}
- \quarter Y^2 
\int { (d^4 z) \left[ 4 \pi^2 x^2 \right]^2 \over 
\left[ 4 \pi^2 ((z-x)^2 + \e^2)\right]^2 \left[ 4 \pi^2 (z^2 + \e^2)\right]^2}
\nonumber\\&=& 
- Y^2 
{1 \over 32 \pi^2 } 
\left[ \log (x^2/\e^2) - 1\right] 
\end{eqnarray}
(for \NN=4 SUSY, $Y^2 = 2 g^2$); 
the same result is obtained in dimensional regularization.

\section{The simplest cases} 
\label{section:simplest}

We begin by considering scalar composite operators 
in representations $[p,q,p]$ of the color $SU(4)$, 
which have $2p+q = $ 4 and 5. 

The case of $\Delta = 4 + \OO(g^2)$ has been 
studied before. 
For example, the authors of \cite {BKRS}
argued that there are two operators%
\footnote{
	For the notation and definitions, see Section \ref{2-0-2}. 
	}
$\OO^{[2,0,2]}_1$ and $\OO^{[2,0,2]}_2$, 
which are made of four scalars and 
annihilated by four supercharges. 
$\OO^{[2,0,2]}_1$ is a descendant 
of the Konishi scalar $\left( \sum_{I=1}^6 \tr \phi^I \phi^I \right)$
and therefore is pure (i.e. is an eigenstate of the dilatation operator), 
since the Konishi scalar is pure. 
The other operator, $\OO^{[2,0,2]}_2$, 
contains 
a piece proportional to $\OO^{[2,0,2]}_1$, 
but the rest is
a chiral primary. The method in \cite{BKRS} was to analyze four-point
correlators of certain \half-BPS operators, and to look at 
the possible operators in exchange channels. 
They found that there is a \quarter-BPS operator exchanged
by demonstrating that there is a pole 
corresponding to an operator of 
scaling dimension $\Delta = 4$. 
They determined this operator to be 
$\YY_{[2,0,2]} = \OO^{[2,0,2]}_2 - {4\over N} \OO^{[2,0,2]}_1$. 

Unfortunately, this method does not generalize to 
chiral primaries with scaling dimension $\Delta \ge 6$, 
as we shall see in Section \ref{6 and higher}. 
So instead we explicitly compute two-point functions
of scalar composite operators of a 
given scaling dimension, and find the ones
which do not get corrected. 
This allows us to fix the normalization 
of \quarter-BPS operators as well.

\subsection{Scalar composites with weight [2,0,2]}
\label{2-0-2}

The simplest operators annihilated by 
four out of sixteen Poincar\'e supercharges 
correspond to the highest weight state of the 
${\mathbf {84}}$ = [2,0,2] of $SU(4)$. 
The $SO(6)$ 
Young tableau for representation is 
$
\mbox{
\setlength{\unitlength}{.5em}
\begin{picture}(1.7,1.5)
\put(-.8,0.5){\framebox (1,1){}}
\put(0.2,0.5){\framebox (1,1){}}
\put(-.8,-.5){\framebox (1,1){}}
\put(0.2,-.5){\framebox (1,1){}}
\end{picture}}
$. 
An $SO(6)$ irreducible tensor $T$ 
with this symmetry is made from the corresponding 
$Gl(6)$ irreducible tensor $T^0$ by subtracting 
all possible $SO(6)$ traces: 
\begin{eqnarray}
\label{202:representation} 
T_{\mbox{\setlength{\unitlength}{.7em}
\begin{picture}(2,2.2)
\put(-.3,1){\framebox (1,1){\scriptsize $a$}}
\put(0.7,1){\framebox (1,1){\scriptsize $b$}}
\put(-.3,0){\framebox (1,1){\scriptsize $c$}}
\put(0.7,0){\framebox (1,1){\scriptsize $d$}}
\end{picture}}}
&=& 
T^0_{\mbox{\setlength{\unitlength}{.7em}
\begin{picture}(2,2.2)
\put(-.3,1){\framebox (1,1){\scriptsize $a$}}
\put(0.7,1){\framebox (1,1){\scriptsize $b$}}
\put(-.3,0){\framebox (1,1){\scriptsize $c$}}
\put(0.7,0){\framebox (1,1){\scriptsize $d$}}
\end{picture}}}
- {1\over4} \Big( 
T^0_{\mbox{\setlength{\unitlength}{.7em}
\begin{picture}(2,2.2)
\put(-.3,1){\framebox (1,1){\scriptsize $\bullet$}}
\put(0.7,1){\framebox (1,1){\scriptsize $\bullet$}}
\put(-.3,0){\framebox (1,1){\scriptsize $a$}}
\put(0.7,0){\framebox (1,1){\scriptsize $b$}}
\put(-.3,1){\framebox (2,1){\scriptsize $-$}}
\end{picture}}} \delta_{cd} + 
T^0_{\mbox{\setlength{\unitlength}{.7em}
\begin{picture}(2,2.2)
\put(-.3,1){\framebox (1,1){\scriptsize $\bullet$}}
\put(0.7,1){\framebox (1,1){\scriptsize $\bullet$}}
\put(-.3,0){\framebox (1,1){\scriptsize $c$}}
\put(0.7,0){\framebox (1,1){\scriptsize $d$}}
\put(-.3,1){\framebox (2,1){\scriptsize $-$}}
\end{picture}}} \delta_{ab} + 
T^0_{\mbox{\setlength{\unitlength}{.7em}
\begin{picture}(2,2.2)
\put(-.3,1){\framebox (1,1){\scriptsize $\bullet$}}
\put(0.7,1){\framebox (1,1){\scriptsize $\bullet$}}
\put(-.3,0){\framebox (1,1){\scriptsize $a$}}
\put(0.7,0){\framebox (1,1){\scriptsize $d$}}
\put(-.3,1){\framebox (2,1){\scriptsize $-$}}
\end{picture}}} 
\delta_{bc} + T^0_{\mbox{\setlength{\unitlength}{.7em}
\begin{picture}(2,2.2)
\put(-.3,1){\framebox (1,1){\scriptsize $\bullet$}}
\put(0.7,1){\framebox (1,1){\scriptsize $\bullet$}}
\put(-.3,0){\framebox (1,1){\scriptsize $b$}}
\put(0.7,0){\framebox (1,1){\scriptsize $c$}}
\put(-.3,1){\framebox (2,1){\scriptsize $-$}}
\end{picture}}} \delta_{ad} 
\Big)
\nonumber\\ 
&& \hspace{2.5em} 
+ {1\over20} 
\left( 
\delta_{ab} \delta_{cd} - \delta_{ad} \delta_{bc} 
\right) T^0_{\mbox{\setlength{\unitlength}{.7em}
\begin{picture}(2,2.2)
\put(-.3,1){\framebox (1,1){\scriptsize $\bullet$}}
\put(0.7,1){\framebox (1,1){\scriptsize $\bullet$}}
\put(-.3,0){\framebox (1,1){\scriptsize $\bullet$}}
\put(0.7,0){\framebox (1,1){\scriptsize $\bullet$}}
\put(-.3,1){\framebox (2,1){\scriptsize $-$}}
\put(-.3,0){\framebox (2,1){\scriptsize $-$}}
\end{picture}}} 
\end{eqnarray}
where 
$
\mbox{
\setlength{\unitlength}{.7em}
\begin{picture}(1.7,1)
\put(-.8,0){\framebox (1,1){\scriptsize $\bullet$}}
\put(0.2,0){\framebox (1,1){\scriptsize $\bullet$}}
\put(-.8,0){\framebox (2,1){\scriptsize $-$}}
\end{picture}}
= \sum_{a=1}^6 \!\!\!\! 
\mbox{
\setlength{\unitlength}{.7em}
\begin{picture}(2.5,1)
\put(0,0){\framebox (1,1){\scriptsize $a$}}
\put(1,0){\framebox (1,1){\scriptsize $a$}}
\end{picture}}
$. 
Recall that the $SU(4) \to SU(3)\times U(1)$ projection 
(\ref{projection:def}) is realized on the elementary fields 
as $\phi_a = {1\over\sqrt{2}}(z_a + \bar z_a)$, 
$\phi_{a+3} = {1\over i \sqrt{2}}(z_a - \bar z_a)$, $a=1,2,3$. 
Under this ``3+1 split,'' 
the highest weight state of [2,0,2] becomes 
\begin{eqnarray}
\label{202:highest weight} 
T_{\mbox{\setlength{\unitlength}{.7em}
\begin{picture}(2,2.2)
\put(-.3,1){\framebox (1,1){\scriptsize $1$}}
\put(0.7,1){\framebox (1,1){\scriptsize $1$}}
\put(-.3,0){\framebox (1,1){\scriptsize $2$}}
\put(0.7,0){\framebox (1,1){\scriptsize $2$}}
\end{picture}}}
&=&
{1\over4} \Big(
T_{\mbox{\setlength{\unitlength}{.7em}
\begin{picture}(2,2.2)
\put(-.3,1){\framebox (1,1){\scriptsize $\mathbf 1$}}
\put(0.7,1){\framebox (1,1){\scriptsize $\mathbf 1$}}
\put(-.3,0){\framebox (1,1){\scriptsize $\mathbf 2$}}
\put(0.7,0){\framebox (1,1){\scriptsize $\mathbf 2$}}
\end{picture}}}
+ 2 
T_{\mbox{\setlength{\unitlength}{.7em}
\begin{picture}(2,2.2)
\put(-.3,1){\framebox (1,1){\scriptsize $\mathbf {\bar 1}$}}
\put(0.7,1){\framebox (1,1){\scriptsize $\mathbf 1$}}
\put(-.3,0){\framebox (1,1){\scriptsize $\mathbf 2$}}
\put(0.7,0){\framebox (1,1){\scriptsize $\mathbf 2$}}
\end{picture}}}
+ 2 
T_{\mbox{\setlength{\unitlength}{.7em}
\begin{picture}(2,2.2)
\put(-.3,1){\framebox (1,1){\scriptsize $\mathbf 1$}}
\put(0.7,1){\framebox (1,1){\scriptsize $\mathbf 1$}}
\put(-.3,0){\framebox (1,1){\scriptsize $\mathbf {\bar 2}$}}
\put(0.7,0){\framebox (1,1){\scriptsize $\mathbf 2$}}
\end{picture}}}
+ 2 
T_{\mbox{\setlength{\unitlength}{.7em}
\begin{picture}(2,2.2)
\put(-.3,1){\framebox (1,1){\scriptsize $\mathbf {\bar 1}$}}
\put(0.7,1){\framebox (1,1){\scriptsize $\mathbf 1$}}
\put(-.3,0){\framebox (1,1){\scriptsize $\mathbf {\bar 2}$}}
\put(0.7,0){\framebox (1,1){\scriptsize $\mathbf 2$}}
\end{picture}}}
+ 
T_{\mbox{\setlength{\unitlength}{.7em}
\begin{picture}(2,2.2)
\put(-.3,1){\framebox (1,1){\scriptsize $\mathbf {\bar 1}$}}
\put(0.7,1){\framebox (1,1){\scriptsize $\mathbf {\bar 1}$}}
\put(-.3,0){\framebox (1,1){\scriptsize $\mathbf 2$}}
\put(0.7,0){\framebox (1,1){\scriptsize $\mathbf 2$}}
\end{picture}}}
\Big) + \mbox{c.c.} 
\nonumber\\ 
&=& 
{1\over4} 
T^0_{\mbox{\setlength{\unitlength}{.7em}
\begin{picture}(2,2.2)
\put(-.3,1){\framebox (1,1){\scriptsize $\mathbf 1$}}
\put(0.7,1){\framebox (1,1){\scriptsize $\mathbf 1$}}
\put(-.3,0){\framebox (1,1){\scriptsize $\mathbf 2$}}
\put(0.7,0){\framebox (1,1){\scriptsize $\mathbf 2$}}
\end{picture}}}
+ \mbox{terms with lower $U(1)$ charge}
\end{eqnarray}
where in the left hand side, $1 = \phi^1$; 
and in the right hand side, 
${\mathbf 1} = z_1$, ${\mathbf {\bar 1}} = \bar z_1$, etc. 
We see that after this projection, we don't 
have to worry about subtracting the $SO(6)$ 
traces, if we are only interested in the 
highest $U(1)$ charge operators. 
Henceforth, we will consider operators made by 
applying the Young (anti)symmetrizers corresponding 
to this tableau, to the string of $z_i$-s.%
\footnote{
	Computing two point functions of operators 
	in representations of $SU(3)\times U(1)$ 
	is as good as computing two point functions of the 
	original $SU(4)$ irreducible operators, 
	see Section \ref{section:superconformal group}. 
	In the following, 
	we will neglect the $SO(6)$ traces 
	without a comment. 
	}

We can construct one single trace and one double trace 
operators with the highest [2,0,2] weight. 
When projected onto $SU(3)\times U(1)$, they become 
\begin{eqnarray}
\OO^{[2,0,2]}_1 
&=& \tr z_1 z_1 z_2 z_2 - \tr z_1 z_2 z_1 z_2 
= 
- \half \tr [z_1 , z_2] [z_1 , z_2] 
\\ 
\OO^{[2,0,2]}_2 
&=& 2 \left( \tr z_1 z_1 ~ \tr z_2 z_2 - 
\tr z_1 z_2 ~ \tr z_1 z_2 \right) 
\end{eqnarray}
$\OO^{[2,0,2]}_1$ 
is a descendant; by consecutively applying 
four SUSY transformations,%
\footnote{
	Supersymmetry transformations are listed in 
	Appendix \ref{n=4 susy:section}, see 
	equations (\ref{SUSY-su(3) form:begin}-\ref{SUSY-su(3) form:end}). 
	}
it can be obtained from the Konishi scalar, 
$\KK_0 = \tr z_j \bar z^j$. 
More explicitly, acting with the SUSY generator 
$\bar Q_{\bar \zeta}$ 
gives 
\begin{equation}
\delta_{\bar \zeta} z_j = 0, 
\quad 
\delta_{\bar \zeta} \bar z^j = \sqrt{2} \bar\zeta \bar\psi^j, 
\quad 
\delta_{\bar \zeta} \bar\psi^j = i \e^{jkl} [z_k,z_l] \bar\zeta, 
\end{equation}
and with 
$Q_{\zeta_3}$, 
\begin{equation}
\delta_{\zeta_3} z_j = -\sqrt{2} \zeta_3 \lambda \delta_{3j}, 
\quad 
\delta_{\zeta_3} \lambda = 2 i [z_1,z_2] \zeta_3, 
\end{equation}
thus 
\begin{eqnarray}
(\bar Q_{\bar \zeta})^2 \KK_0 &=& 
\bar Q_{\bar \zeta} \tr z_j \sqrt{2} \bar\zeta \bar\psi^j = 
6 i \sqrt{2} (\bar\zeta \bar\zeta) \tr [z_1,z_2] z_3 
,
\nonumber\\
(Q_{\zeta_3})^2 (\bar Q_{\bar \zeta})^2 \KK_0 &=& 
- 12 i (\bar\zeta \bar\zeta) Q_{\zeta_3} \tr [z_1,z_2] \zeta_3 \lambda = 
24 (\bar\zeta \bar\zeta) (\zeta_3 \zeta_3) \tr [z_1,z_2]^2 
\nonumber\\
&=&
- 48 (\bar\zeta \bar\zeta) (\zeta_3 \zeta_3) \OO^{[2,0,2]}_1 
\label{konishi:descendant}
\end{eqnarray}
The four generators which annihilate $\OO^{[2,0,2]}_1$, 
are the ones we acted with to obtain it from $\KK_0$. 
However, since $\KK_0$ is a non-chiral primary, 
its descendant $\OO^{[2,0,2]}_1$ 
is not \quarter-BPS, 
despite being annihilated by 
a quarter of SUSY generators.

The free field results for two point functions of 
$\OO^{[2,0,2]}_1$ and $\OO^{[2,0,2]}_2$ are 
\begin{eqnarray}
\label{2:0:2}
\left( \matrix{
\langle \OO_1 \bar{\OO}_1 \rangle & 
\langle \OO_1 \bar{\OO}_2 \rangle \cr 
\langle \OO_2 \bar{\OO}_1 \rangle & 
\langle \OO_2 \bar{\OO}_2 \rangle  
} \right)_{\mathrm {free}} 
&=&
{3 (N^2 -1) G^4 \over 16} 
\left( \matrix{
N^2 & 4 N \cr 
4 N & 8 (N^2 - 2)  
} \right)
\end{eqnarray}
while the leading corrections are found to be 
\begin{eqnarray}
\label{2:0:2:corrections}
\left( \matrix{
\langle \OO_1 \bar{\OO}_1 \rangle & 
\langle \OO_1 \bar{\OO}_2 \rangle \cr 
\langle \OO_2 \bar{\OO}_1 \rangle & 
\langle \OO_2 \bar{\OO}_2 \rangle  
} \right)_{g^2} 
&=&
{9 (N^2 - 1) G^4 (\tilde B N) \over 16}
\left( \matrix{
N^2 & 4 N \cr 
4 N & 16 
} \right) 
; 
\end{eqnarray}
here 
$\langle \OO_i \bar{\OO}_j \rangle \equiv 
\langle \OO^{[2,0,2]}_i (x) \bar{\OO}^{[2,0,2]}_j (y) \rangle$, 
and $G \equiv G(x,y)$, $\tilde B \equiv \tilde B(x,y)$. 
Some helpful formulae we used for deriving 
(\ref{2:0:2}-\ref{2:0:2:corrections}) 
are collected in Appendix \ref{suN-identities}.

By looking at (\ref{2:0:2:corrections}), we conclude that 
neither 
$\OO^{[2,0,2]}_1$ nor $\OO^{[2,0,2]}_2$ are chiral. 
However, there is a linear combination of these two operators 
which has protected two point functions at order $g^2$. 
The operator 
\begin{eqnarray}
\YY_{[2,0,2]}(x) &\equiv& \OO^{[2,0,2]}_2(x) - 
{4\over N} \OO^{[2,0,2]}_1(x) 
\label{2:0:2:protected}
\end{eqnarray}
satisfies 
$\langle \YY \bar {\YY} \rangle = 
\langle \YY \bar {\OO}_1 \rangle = 0$, 
so $\YY_{[2,0,2]}$ is orthogonal to the 
descendant of the Konishi operator $\OO^{[2,0,2]}_1$, and 
has protected dimension $\Delta_{\mbox{\scriptsize $\YY$}} = 4$ at order $g^2$. 
Computationally, 
this cancellation is rather intricate: 
all $\langle \OO_i \bar{\OO}_j \rangle$ 
have very different large $N$ behavior.

We can also calculate the two point 
function of the Konishi scalar with itself: 
\begin{eqnarray}
\langle \KK_0(x) \bar{\KK}_0(y) \rangle &=& 
3 (N^2-1) [G(x,y)]^2 
\left\{
1 + 3 \tilde B(x,y) N + \OO(g^4) 
\right\}
\end{eqnarray}
so 
$
\langle \OO^{[2,0,2]}_1(x) \bar{\OO}^{[2,0,2]}_1(y) \rangle = 
\mbox{$1\over16$} N^2 [G(x,y)]^2
\langle \KK_0(x) \bar{\KK}_0(y) \rangle + \OO(g^4)$, which 
is just the free theory result. 
In particular, $\KK_0$ and its descendant $\OO^{[2,0,2]}_1$ 
have the same normalization and 
their scaling dimensions differ by 2, as they must: 
with $\tilde B(x,0)$ given by (\ref{b-tilde}), 
the scaling dimension of $\OO^{[2,0,2]}_1$ is 
$\Delta_1 = 4 + {3 g^2 N \over 16 \pi^2} + \OO(g^4)$, 
and that of $\KK_0$ is 
$\Delta_{\mbox{\scriptsize $\KK$}} 
= 2 + {3 g^2 N \over 16 \pi^2} + \OO(g^4)$, 
in agreement with 
\cite{BKRS}.
The mixture 
$\OO^{[2,0,2]}_2 = \YY_{[2,0,2]} + {4\over N} \OO^{[2,0,2]}_1$
has the two-point function with itself which 
breaks down into two pieces, 
\begin{eqnarray}
\langle \OO^{[2,0,2]}_2(x) \bar{\OO}^{[2,0,2]}_2(y) \rangle 
&=& 
\langle \YY_{[2,0,2]}(x) \bar{\YY}_{[2,0,2]}(y) \rangle +
{16\over N^2} \, 
\langle \OO^{[2,0,2]}_1(x) \bar{\OO}^{[2,0,2]}_1(y) \rangle  
\nonumber\\
&=&
{C_{\mbox{\scriptsize $\YY$}} \over (x-y)^{2 \Delta_{\mbox{\scriptsize $\YY$}}}} + 
{16\over N^2} {C_1 \over (x-y)^{2 \Delta_1}} 
\end{eqnarray}
Logarithmic corrections to 
$\langle \OO^{[2,0,2]}_2(x) \bar{\OO}^{[2,0,2]}_2(y) \rangle$
are due entirely to the second term 
${16\over N^2} \langle \OO^{[2,0,2]}_1(x) \bar{\OO}^{[2,0,2]}_1(y) \rangle$.

$\OO(g^2)$ corrections to all two-point functions just considered 
are proportional to $\tilde B$. 
In other words, only the contributions due to diagrams of 
type (c) in Figure \ref{fig:contributing diagrams} 
survive, and all other corrections cancel.
As we will show in 
Section \ref{section:gauge-dependent}, 
this is a general phenomenon. 
Also, in the large $N$ limit 
the fraction 
$\langle \OO_1 \bar{\OO}_2 \rangle / \langle \OO_1 \bar{\OO}_1 \rangle$ 
is suppressed; 
it vanishes in the limit $N\rightarrow\infty$, $g^2 N$ fixed.%
\footnote{
	The large $N$ limit will be analyzed in more 
	detail in Section \ref{section:large N}. 
	}

One can show that conformal dimension of $\YY_{[2,0,2]}$ is protected 
perturbatively (to order $g^4$) 
and nonperturbatively (for any instanton number) 
as well, see \cite{BKRS}. 
Non-renormalization of scaling dimension of $\YY$ hints at its 
BPS property. 
Following the authors of \cite{BKRS}, 
we suggest that it is 
indeed a \quarter-BPS chiral primary  operator.

\subsection{Scalar composites with weight [2,1,2]}
\label{2-1-2}

The story is similar in the case of the ${\mathbf {300}} = [2,1,2]$ of $SU(4)$.
The only scalar composite operators corresponding to the highest 
weight of this representation are  
(after the projection onto $SU(3)\times U(1)$, 
as discussed in Sections \ref{section:superconformal group} and \ref{2-0-2}) 
\begin{eqnarray}
\OO^{[2,1,2]}_1 &=& \tr z_1 z_1 z_1 z_2 z_2 - \tr z_1 z_1 z_2 z_1 z_2 
= 
- \half \tr [z_1 , z_2] [z_1^2 , z_2] 
\\ 
\OO^{[2,1,2]}_2 &=& \tr z_1 z_1 z_1 ~ \tr z_2 z_2 - 
2 \tr z_1 z_2 ~ \tr z_1 z_2 z_1 + \tr z_1 z_2 z_2 ~ \tr z_1 z_1 
\end{eqnarray}
$\OO^{[2,1,2]}_1$ is a descendant; 
$(Q_{\zeta_3})^2 (\bar Q_{\bar \zeta})^2 
\str z_1 z_i \bar z^i 
\propto \OO^{[2,1,2]}_1$
just like in Equation (\ref{konishi:descendant}).
Born level and order $g^2$ two point functions are 
\begin{eqnarray}
\label{2:1:2-begin}
\left( \matrix{
\langle \OO_1 \bar{\OO}_1 \rangle & 
\langle \OO_1 \bar{\OO}_2 \rangle \cr 
\langle \OO_2 \bar{\OO}_1 \rangle & 
\langle \OO_2 \bar{\OO}_2 \rangle 
} \right) 
&=&
{(N^2 -1) (N^2 -4) G^5 \over 16 N} 
\left\{
\left( \matrix{
N^2 & 6 N \cr 
6 N & 6 (N^2 -3)  
} \right)
\right. \nonumber\\ && \left. \quad\quad\quad\quad\quad\quad
+ ~
4 \tilde B N  
\left( \matrix{
N^2 & 6 N \cr 
6 N & 36 
} \right)
+ \OO(g^4) 
\right\}
\quad\quad
\label{2:1:2-end}
\end{eqnarray}
where 
$\langle \OO_i \bar{\OO}_j \rangle \equiv 
\langle \OO^{[2,1,2]}_i (x) \bar{\OO}^{[2,1,2]}_j (y) \rangle$, 
and $G \equiv G(x,y)$, $\tilde B \equiv \tilde B(x,y)$ as before. 
(Note that corrections proportional to $A$ and $B$ cancel again.) 
There is a linear combination of $\OO^{[2,1,2]}_1$ and $\OO^{[2,1,2]}_2$
\begin{eqnarray}
\label{2:1:2:o-hat}
\YY_{[2,1,2]}(x) &\equiv& \OO^{[2,1,2]}_2(x) - {6\over N} \OO^{[2,1,2]}_1(x) 
\end{eqnarray}
whose two point functions with arbitrary operators 
do not receive perturbative order $g^2$ corrections. 
Again, it seems reasonable to conclude that $\YY_{[2,1,2]}$ is a 
\quarter-BPS operator, 
as it is annihilated by four out of sixteen supercharges, 
has a protected scaling dimension $\Delta_{\mbox{\scriptsize $\YY$}} = 5$ 
(at order $g^2$), 
and contains no descendant pieces, 
$\langle \YY_{[2,1,2]}(x) \bar{\OO}^{[2,1,2]}_1(y) \rangle = 0$.

\section{A Gauge Invariance Argument}
\label{section:gauge-dependent}

In Sections \ref{2-0-2} and \ref{2-1-2}, we explicitly 
calculated the $\OO(g^2)$ corrections to 
two point functions of scalar composite operators. 
We found that corrections proportional to $A$ and $B$ cancel, 
i.e. gauge group combinatorics 
demands that diagrams containing a gauge boson exchange 
do not arise in the correlator. 
Here we give a general derivation of this fact, 
which boils down to gauge invariance of the operators in question, 
and gauge dependence of $A$ and $B$. 

The two point functions we have been considering are of the form 
\begin{equation}
\label	{two-point:D-term}
\langle 
\left[ {z}^m \right] \!(x) 
\,
\left[ {\bar z}^{m} \right] \!(y) \rangle 
\end{equation}
where $\left[ {z}^m \right] \!(x)$ is some gauge-invariant 
homogeneous polynomial (of degree $m$) in the $z_i^a$-s. 
Diagrams involving a gauge boson exchange 
which contribute to the two-point functions of the form 
(\ref{two-point:D-term}), are proportional to either $A(x,y)$ 
or $B(x,y)$, 
see 
Figure \ref{fig:four-scalar and propagator}.
By using nonrenormalization of 
the two point function 
$\langle \tr z_1 z_2 (x) \; \tr \bar z_1 \bar z_2 (y) \rangle$, 
one can immediately see \cite{DFS} that 
$B(x,y) = - 2 A(x,y)$, so 
\begin{eqnarray}
\label	{two-point:D-term:gauge-inv}
\langle 
\left[ {z}^m \right](x) 
\left[ {\bar z}^{m} \right](y) \rangle_{(A+B)} 
&=& c_g A(x,y) [G(x,y)]^m 
\end{eqnarray}
where $c_g$ is some combinatorial coefficient.

Conformal invariance restricts 
$A(x,y) = a \log x^2 \mu^2 + b$. 
The constants $a$ and $b$ turn out to be gauge dependent. 
The gauge fixing parameter $\xi$ enters the expression 
for the scalar propagator as%
\footnote{
	By changing 
	$\xi$, 
	we can vary both the pole piece and the $\OO(\e^0)$ term
	(but we can't make them both zero simultaneously; 
	the $\xi$-independent part is proportional
	to a different integral). 
	Compare this with \cite{Kovacs}, where 
	order $g^2$ corrections to the scalar propagators 
	were found to vanish in super-Feynman gauge 
	of \NN=1 formulation of the theory.
	}
\begin 	{eqnarray}
\mbox{
\setlength{\unitlength}{0.7em}
\begin{picture}(4,1)
\put(0,0.4){\framebox (4,0){\Large $\bullet$}}
\end{picture}}
&=& 
\xi g^2 \mu^{4 - d} 
{1 \over p^4}
\int 
{ (d^d k) \over (2\pi)^d} 
{
[(2 p + k) \cdot k]^2
\over
k^4 (p+k)^2
}
+ (\mbox{$\xi$-independent}) 
\nonumber\\&=& 
\xi g^2 
\half \pi^{2+\e} 
{1 \over (p^2)^{1-\e} \mu^{2 \e}} 
\left[
{1\over\e} + \gamma + \OO(\e)
\right]
+ (\mbox{$\xi$-independent}) \quad\quad
\end 	{eqnarray}
in momentum space 
(in dimensional regularization \cite{Collins}; 
$\e = {d\over2}-2$ and 
$\gamma$ 
is Euler's gamma constant), 
so in position space 
\begin 	{eqnarray}
A(x,0) &=& 
\half \pi^2 g^2 \xi 
\left[ 
\log x^2 \mu^2 + \log 4\pi - \gamma 
\right]
+ (\mbox{$\xi$-independent}) 
\nonumber\\
&\equiv& a \log x^2 \mu^2 + b 
\end 	{eqnarray}
after factoring out the free propagator $G(x,0)$. 
Both 
$a$ and $b$ have pieces linear in the 
gauge fixing parameter $\xi$. 

Since a correlator of gauge invariant 
operators 
must be gauge independent, 
the combinatorial coefficient multiplying 
$A(x,y)$ in equation (\ref{two-point:D-term:gauge-inv}) 
must vanish; 
we necessarily have $c_g=0$. 
This is a general phenomenon, 
illustrated by an explicit calculation of 
Sections \ref{2-0-2} and \ref{2-1-2}: 
gauge dependent contributions 
are proportional to $2 A + B = 0$. 

In the $\OO(g^2)$ calculations of correlators of 
\half-BPS operators \cite{DFS} and \cite{Skiba}, 
there were no other contributions to 
two-point functions except for 
$A$ and $B$. 
Thus, gauge invariance together with \NN=4 SUSY 
(which is needed to make $2 A + B = 0$) guarantees 
that the correlators of \cite{DFS} and \cite{Skiba} 
receive no order $g^2$ corrections.


\section{Operators of dimension 6 and higher}
\label{6 and higher}

At this point, we would like to consider 
operators made of $2p+q \ge 6$ scalar fields. 
According to the classification of \cite{AFSZ}, 
these $[p,q,p]$ operators are the candidates for 
\quarter-BPS chiral primaries. 
However, a new complication arises 
compared to the cases of $2p+q \le 5$ studied in 
Sections \ref{2-0-2} and \ref{2-1-2}. 
Now, there are many ways in which we can make 
gauge invariant combinations of fields, and hence 
many scalar composites have to be taken into account. 
Apart from single and double trace operators 
we have seen so far, 
operators made of three or more traces also
have to be considered. 

This phenomenon has a counterpart in the context 
of \half-BPS operators, see \cite{Skiba}. 
The crucial difference is 
that in our case, 
none of the scalar composites are pure, 
and only some special mixtures have a well 
defined scaling dimension. 
In general, the ``naive'' 
scalar composite operators 
will have nonvanishing two point functions 
with each other, whenever this is allowed by group theory. 
Unlike in the simplest cases of Section \ref{section:simplest}, 
operators containing commutators are not pure, but 
contain pieces which are descendants of different operators. 

To find pure operators, 
in this Section we calculate the 
two point functions of highest weight $[p,q,p]$ scalar composites 
$\OO^{[p,q,p]}_i$, 
and arrange them as%
\footnote{
	The operators we are working with are 
	after the projection onto $SU(3)\times U(1)$, 
	as discussed in Sections \ref{section:superconformal group} 
	and \ref{2-0-2}.
	The $\OO^{[p,q,p]}_i$ are made of only $z$-s and no $\bar z$-s. 
	}
\begin	{equation}
\label{eq:oo-correlator-general}
\langle \OO^{[p,q,p]}_i(x) \bar{\OO}^{[p,q,p]}_j(y) \rangle
\equiv 
[G(x,y)]^{(2p+q)} 
\left[ 
{\mathbf F}_{ij} + \tilde B(x,y) N \, {\mathbf G}_{ij} + \OO(g^4) 
\right] 
.
\end	{equation}
Here, ${\mathbf F}$ the matrix of combinatorial factors at free level; 
and ${\mathbf G}$, of order $g^2$ correction combinatorial factors. 
Note that there can be no corrections proportional to $A$ or $B$, 
as was argued in Section \ref{section:gauge-dependent}. 
Both ${\mathbf F}$ and ${\mathbf G}$ are matrices of pure numbers; 
they are still functions of $N$, but coordinate and $g^2$ 
dependence are all absorbed in 
$\tilde B$ and $[G(x,y)]^{(2p+q)}$. 
Now the problem becomes one of linear algebra: 
starting with a basis of 
$\OO^{[p,q,p]}_i$, we want to find their linear combinations 
$\YY^{[p,q,p]}_j$ that are pure operators. 
The $\YY^{[p,q,p]}_j$ have a well defined renormalized scaling dimension 
$\Delta_j = \Delta^0 + \Delta^1_j + \OO(g^4)$; 
$\Delta^0 = 2 p + q$ for all $\OO^{[p,q,p]}_i$ 
and hence for all $\YY^{[p,q,p]}_j$. 
Such operators can be chosen orthogonal 
at Born level, and so 
\begin	{equation}
\langle \YY^{[p,q,p]}_i \bar{\YY}^{[p,q,p]}_j \rangle 
= {C^{[p,q,p];0}_i \delta_{ij} \over x^{2 \Delta^0}} 
\left[ 
1 + \beta_i - \Delta^1_i \log \mu^2 x^2 + \OO(g^4)
\right] 
\end	{equation}
to order $g^2$. 
Coefficients $\Delta^1_j \sim \beta_j \sim g^2$ 
correspond to corrections of $\YY^{[p,q,p]}_j$'s scaling dimension 
and its normalization; $\beta_j$ 
depends on the 
renormalization scale $\mu$. 
To distinguish the pure operators which do receive 
corrections to their scaling dimension, we will denote them 
by $\tilde{\YY}$, and reserve the notation $\YY$ 
for the ones that 
have $\OO(g^2)$ protected two point functions.

This is a standard problem, 
analogous to finding the normal modes of small oscillations 
of a mechanical system 
(see \cite{Goldstein}, for example). 
We have to diagonalize 
a symmetric matrix ${\mathbf G}$ of corrections 
with respect to the symmetric positive definite matrix 
${\mathbf F}$ of free correlators. 
In other words, we need to find the eigenvalues of 
matrix ${\mathbf F}^{-1} {\mathbf G}$. 
If some of them 
vanish, the corresponding eigenvectors 
are operators whose two point functions 
(with themselves as well as with other operators) 
do not get order $g^2$ corrections. 
We conjecture that 
these are in fact the \quarter-BPS operators we are after.

We will now illustrate how this method works 
with the example of the [3,1,3] representation of $SU(4)$.
%
%
%
The scalar composites corresponding to the highest weight of the 
[3,1,3] = ${\mathbf {960}}$ of $SU(4)$ are 
the following five linearly independent%
\footnote{
	For $N \le 4$, the number of independent 
	gauge invariant operators is smaller.
	}
operators: 
\begin	{eqnarray}
\label{equation: 313 operators:start}
\OO^{[3,1,3]}_0 &\equiv& 
\tr z_1 z_1 z_1 z_2 z_1 z_2 z_2 - \tr z_1 z_1 z_1 z_2 z_2 z_1 z_2
\\
\OO^{[3,1,3]}_1 &\equiv& 
\third \tr z_1 z_1 z_1 z_1 z_2 z_2 z_2 
-\half \tr z_1 z_1 z_1 z_2 z_1 z_2 z_2 
-\half \tr z_1 z_1 z_1 z_2 z_2 z_1 z_2 
\hspace{-2em}\nonumber\\&&
+ 
\third \tr z_1 z_1 z_2 z_1 z_2 z_1 z_2 + 
\third \tr z_1 z_1 z_2 z_2 z_1 z_1 z_2 
\quad\quad
\\
\OO^{[3,1,3]}_2 &\equiv& 
\tr z_1 z_1 z_1 z_1 ~\tr z_2 z_2 z_2 
- 3 \, \tr z_1 z_1 z_1 z_2 ~\tr z_1 z_2 z_2 
\\ && 
+ 
\left( 
2 \, \tr z_1 z_1 z_2 z_2 + \tr z_1 z_2 z_1 z_2 
\right) ~\tr z_1 z_1 z_2 
- \tr z_1 z_2 z_2 z_2 ~\tr z_1 z_1 z_1 
\nonumber\\ 
\OO^{[3,1,3]}_3 &\equiv& 
- \left( 
\tr z_1 z_1 z_2 z_2 z_2 - \tr z_1 z_2 z_1 z_2 z_2 
\right) ~\tr z_1 z_1  
\nonumber\\ && 
+ 
\left( 
\tr z_1 z_1 z_1 z_2 z_2 - \tr z_1 z_1 z_2 z_1 z_2 
\right) ~\tr z_1 z_2 
\\
\OO^{[3,1,3]}_4 &\equiv& 
\tr z_1 z_2 
\left( 
2 \, \tr z_1 z_2 ~\tr z_1 z_2 z_2 - 
\tr z_2 z_2 ~\tr z_1 z_1z_1 
- 3 \tr z_1 z_1 ~\tr z_1 z_2z_2 
\right) 
\hspace{-2em}\nonumber\\&&
+
\tr z_1 z_1 
\left( 
\tr z_2 z_2 ~\tr z_1 z_1 z_2 + 
\tr z_1 z_1 ~\tr z_2 z_2z_2
\right) 
\label{equation: 313 operators:end}
\end	{eqnarray}
The $\OO^{[3,1,3]}_i$ are constructed by applying 
Young symmetrizers to all possible gauge invariant combinations 
of $(z_1)^4 (z_2)^3$.
The symmetrizers correspond to the tableau 
$
\mbox{
\setlength{\unitlength}{0.5em}
\begin{picture}(4,1.5)
\put(-.5,-.5){\framebox (3,2){}}
\put(-.5,-.5){\framebox (2,2){}}
\put(-.5,-.5){\framebox (1,2){}}
\put(-.5,.5){\framebox (4,1){}}
\end{picture}}$
of $SO(6)$,
while gauge 
invariant combinations amount to 
grouping the $z_i$-s into traces. 
Operators resulting from other partitions and symmetrizations 
turn out to be linear combinations of the $\OO^{[3,1,3]}_i$ above.

More explicitly, the operators listed in 
(\ref{equation: 313 operators:start}-\ref{equation: 313 operators:end})
are constructed as 
\begin{eqnarray}
\label{equation:313:partitions:begin}
7 = 7: &&
\OO^{[3,1,3]}_0 ,
\OO^{[3,1,3]}_1 \sim 
\left(\hspace{-1ex}\mbox{
\setlength{\unitlength}{0.7em}
\begin{picture}(4,1.7)
\put(0,0.5){\framebox (1,1){}}
\put(1,0.5){\framebox (1,1){}}
\put(2,0.5){\framebox (1,1){}}
\put(3,0.5){\framebox (1,1){}}
\put(0,-0.5){\framebox (1,1){}}
\put(1,-0.5){\framebox (1,1){}}
\put(2,-0.5){\framebox (1,1){}}
\end{picture}} \; \right) 
\\
7 = 4 + 3: &&
\phantom{\Big|^B}\hspace{-2ex}
\OO^{[3,1,3]}_2 \sim 
\left(\hspace{-1ex}\mbox{
\setlength{\unitlength}{0.7em}
\begin{picture}(4,1.7)
\put(0,0.6){\framebox (1,1){}}
\put(1,0.6){\framebox (1,1){}}
\put(2,0.6){\framebox (1,1){}}
\put(3,0.6){\framebox (1,1){}}
\put(0,-0.8){\framebox (1,1){}}
\put(1,-0.8){\framebox (1,1){}}
\put(2,-0.8){\framebox (1,1){}}
\end{picture}} \; \right) 
\\
7 = 5 + 2: &&
\phantom{\Big|^B}\hspace{-2ex}
\OO^{[3,1,3]}_3 \sim 
\left(\hspace{-1ex}\mbox{
\setlength{\unitlength}{0.7em}
\begin{picture}(4.5,1.7)
\put(0,0.2){\framebox (1,1){}}
\put(1,0.2){\framebox (1,1){}}
\put(0,-0.8){\framebox (1,1){}}
\put(1,-0.8){\framebox (1,1){}}
\put(2,-0.8){\framebox (1,1){}}
\put(2.5,0.6){\framebox (1,1){}}
\put(3.5,0.6){\framebox (1,1){}}
\end{picture}} \; \right) 
\\
7 = 3 + 2 + 2: &&
\phantom{\Big|^B}\hspace{-2ex}
\OO^{[3,1,3]}_4 \sim 
\left(\hspace{-1ex}\mbox{
\setlength{\unitlength}{0.7em}
\begin{picture}(4.5,1.7)
\put(0,0.6){\framebox (1,1){}}
\put(1,0.6){\framebox (1,1){}}
\put(2.5,0.6){\framebox (1,1){}}
\put(3.5,0.6){\framebox (1,1){}}
\put(0,-0.8){\framebox (1,1){}}
\put(1,-0.8){\framebox (1,1){}}
\put(2,-0.8){\framebox (1,1){}}
\end{picture}} \; \right) 
\label{equation:313:partitions:end}
\end{eqnarray}
where each continuous group of boxes stands for a single trace. 
Although it does not appear in this example, 
in general it matters not only how we partition the string 
of letters, but exactly which letters we put in the groups
before we apply the symmetrizers. 
Also, there can be more than one operator corresponding to the 
same partition, provided there are multiple 
Young symmetrisers for a given tableau 
($\OO^{[3,1,3]}_0$ and $\OO^{[3,1,3]}_1$ here).

Here we should mentions that 
operator $\OO^{[3,1,3]}_0$ has some 
special properties. 
It satisfies $(\OO^{[3,1,3]}_0)^\dagger = - (\OO^{[3,1,3]}_0)^*$
while for all other operators 
$(\OO^{[3,1,3]}_i)^\dagger = + (\OO^{[3,1,3]}_i)^*$.
As a result, $\OO^{[3,1,3]}_0$ has zero correlators with
everything else. 
It is also the only pure operator out of all the 
$\OO^{[3,1,3]}_i$ appearing in 
(\ref{equation: 313 operators:start}-\ref{equation: 313 operators:end}).
It is a descendant, and we will not consider it below.

So we calculate explicitly 
the $\half \cdot 4 \cdot (4+1) = 10$ 
two point functions of [3,1,3] operators,  
and arrange them as in Equation (\ref{eq:oo-correlator-general}). 
We calculate the matrix ${\mathbf F}$ of free correlator 
combinatorial factors; and the matrix 
${\mathbf G}$, of order $g^2$ correction combinatorial factors. 
The matrix ${\mathbf F}^{-1} {\mathbf G}$ is $4 \times 4$; it has 
two zero eigenvalues, while the other two satisfy a quadratic equation. 
We find 
\begin{small}
\begin	{equation} \!
{\frac{768}{5 {N^3} \left( {N^2} -1\right) \left( {N^2} -4 \right) }}
\, {\mathbf F} 
= 
\left( 
\matrix{ {\frac{{N^2} + 3}{{N^2}}} & {\frac{12}{N}} & -{\frac{12}{N}} & {\frac{72}
    {{N^2}}} \cr {\frac{12}{N}} & {\frac{36 \left( {N^4} - 8 {N^2} + 18 \right) }{{N^4}}} & 
    -{\frac{108}{{N^2}}} & {\frac{72 \left( 2 {N^2} -9\right) }{{N^3}}} \cr {\frac{-12}
    {N}} & -{\frac{108}{{N^2}}} & {\frac{36 \left( {N^2} + 6 \right) }{5 {N^2}}} & -{\frac{72}
    {N}} \cr {\frac{72}{{N^2}}} & {\frac{72 \left( 2 {N^2} -9 \right) }{{N^3}}} & -{\frac{72}
    {N}} & {\frac{72 \left( {N^2} -3 \right) }{{N^2}}} }
\right)
\end	{equation}
\end{small}
for the matrix of free combinatorial factors, and 
\begin{small}
\begin{eqnarray}
{\frac{128}{25 {N^3} \left( {N^2} -1 \right) \left( {N^2} -4 \right) }}
\, {\mathbf G} 
=
\left( 
\matrix{ {\frac{{N^2} +7 }{{N^2}}} & {\frac{12 \left( {N^2} +3 \right) }{{N^3}}} & -{\frac{72 
      \left( {N^2}+1 \right) }{5 {N^3}}} & {\frac{96}{{N^2}}} \cr {\frac{12 
      \left( {N^2}+3 \right) }{{N^3}}} & {\frac{144}{{N^2}}} & -{\frac{144}{{N^2}}} & {\frac{864}
    {{N^3}}} \cr -{\frac{72 \left( {N^2}+1 \right) }{5 {N^3}}} & -{\frac{144}{{N^2}}} & {
     \frac{144 \left( {N^2}+16 \right) }{25 {N^2}}} & 
-{\frac{288 \left( {N^2} +6\right) }
    {5 {N^3}}} \cr {\frac{96}{{N^2}}} & {\frac{864}{{N^3}}} & 
-{\frac{288 \left( {N^2} + 6 \right) }{5 {N^3}}} & {\frac{576}{{N^2}}} }
\right) 
\hspace{-1em} 
\nonumber\\
\end	{eqnarray}
\end{small}
for the matrix of corrections proportional to $\tilde B(x,y) N$.

The vectors killed by ${\mathbf F}^{-1} {\mathbf G}$ 
work out to be 
\begin	{eqnarray}
\label	{eq:3-1-3:y1}
\YY^{[3,1,3]}_1 &=& 
- {\frac{12 N}{{N^2}-2}} \, \OO^{[3,1,3]}_1 + \OO^{[3,1,3]}_2 
- {\frac{5}{{N^2}-2}} \, \OO^{[3,1,3]}_3
\\
\YY^{[3,1,3]}_2 &=& 
{\frac{96}{{N^2}-4}} \OO^{[3,1,3]}_1 
- {\frac{4 N}{{N^2}-4}} \OO^{[3,1,3]}_2 + 
{\frac{10 N}{{N^2}-4}} \OO^{[3,1,3]}_3 + \OO^{[3,1,3]}_4
\quad
\end{eqnarray}
They correspond to 
zero eigenvalues of ${\mathbf F}^{-1} {\mathbf G}$, 
and so are the candidates for \quarter-BPS 
primaries in the [3,1,3].%
\footnote{
	We chose $\YY^{[3,1,3]}_2$ to be orthogonal 
	to $\YY^{[3,1,3]}_1$, 
	in the sense that 
	$\langle \YY^{[3,1,3]}_2 \bar{\YY}^{[3,1,3]}_1\rangle = 0$. 
	}
The remaining eigenvectors of ${\mathbf F}^{-1} {\mathbf G}$ 
\begin{eqnarray}
\tilde{\YY}^{[3,1,3]}_3 &=& 
\OO^{[3,1,3]}_1 
- {\frac{10}{3 \left( N + {\sqrt{ N^2 + 160 }} \right) }} \OO^{[3,1,3]}_3 
\\
\tilde{\YY}^{[3,1,3]}_4 &=& 
\OO^{[3,1,3]}_3 
- {\frac{3 \left( N - {\sqrt{ N^2 + 160 }} \right) }{10}} \OO^{[3,1,3]}_1 
\label	{eq:3-1-3:y4}
\end	{eqnarray}
correspond to eigenvalues 
$27 + {\frac{3 {\sqrt{160 + {N^2}}}}{N}}$ 
for $\tilde{\YY}^{[3,1,3]}_3$, 
and $27 - {\frac{3 {\sqrt{160 + {N^2}}}}{N}}$ 
for $\tilde{\YY}^{[3,1,3]}_4$. 
Expressions 
(\ref{eq:3-1-3:y1}-\ref{eq:3-1-3:y4})
are exact in $N$, 
and are not just large $N$ approximations. 
As expected, the descendants 
$\tilde{\YY}$ 
are mixtures of operators involving commutators. 

	Note that both the $g^2$ corrections to the scaling 
	dimension of $\tilde{\YY}$ 
	and their expansion 
	coefficients involve radicals (so there is really no way to ``guess'' 
	the pure primaries such operators came from). 
	Also, radiative corrections to all $\Delta$-s are non-negative, 
	since at free field level the $\OO^{[3,1,3]}_i$ are annihilated by 
	a quarter of the supercharges, and hence saturate the BPS 
	bound. 


Other representations are similar and we do not give 
the details here. 
However, it gets quite cumbersome 
as the number of $\OO$'s increases. 
The prescription of Sections \ref{2-0-2} and \ref{2-1-2} 
(find the pure non-BPS primaries, list their descendants, 
then subtract these pieces from the candidate 
\quarter-BPS operator) also becomes difficult to implement 
in the \NN=1 component approach. 
On the other hand, 
harmonic superspace gives us valuable insigths, 
and this more efficient approach is used in Chapter \ref{chapter:systematic}
to analyse these representations.

\section{Large $N$ analysis}
\label{section:large N}

As we have seen, computations get more and more 
cumbersome as one tries to find \quarter-BPS 
operators for bigger representations of the color group; 
even the number of operators one has to consider is a 
nontrivial function of the representation. 
Symmetry factors multiplying 
the Feynman graphs show no immediate pattern, 
and most of the results presented in 
Sections \ref{2-0-2}, \ref{2-1-2}, and \ref{6 and higher} 
had to be calculated using {\textit {Mathematica}}.%
\footnote{
	The calculations took from 
	0.003 hours for the [2,0,2] representation 
	to
	23 hours for [3,1,3], 
	per single $\OO(g^2)$ two point function. 
	We used a Sparc 10 with 2048 M memory 
	and 440 MHz speed. 
	Born level calculations were considerably 
	(about 20 times) faster.  
	} 

The next best thing we can do is 
consider the large $N$ limit. Specifically, we shall 
concentrate on the leading behavior as $N \to \infty$,
plus the first $1/N$ correction. 

\subsection{Operators $\OO_{[p,q,p]}$ and $\KK_{[p,q,p]}$}
\label{section:oo and kk}

Let us take another look at the results of 
Section 
\ref{6 and higher}, 
where we managed to perform $\OO(g^2)$ analysis exactly 
in $N$ rather than in the large $N$ approximation. 
In all cases considered so far, there is a special 
\quarter-BPS chiral primary $\YY^{[2,3,2]}_1$
which is made of only the double trace and single trace operators. 
At large $N$, this operator is a combination 
of only 
a particular double trace operator, 
and the single trace operator, whose contribution 
is $1/N$ suppressed. 
The goal of Section \ref{section:large N} 
is to show that this is in fact what happens for 
general $[p,q,p]$ representations. 
Here, we begin by defining 
these operators.

Recall that the $SO(6)$ Young tableau for the $[p,q,p]$ of $SU(4)$ 
consists of two rows 
(one of length $p+q$, and the other of length $p$). 
Among the possible partitions of the highest weight tableau, 
there are two special ones 
\begin{eqnarray}
\OO_{[p,q,p]} \sim 
\left(
\mbox{
\setlength{\unitlength}{1em}
\begin{picture}(7.5,1.6)
\put(0,.5){\framebox (1,1){\scriptsize $1$}}
\put(1,.5){\framebox (2,1){\scriptsize $...$}}
\put(3,.5){\framebox (1,1){\scriptsize $1$}}
\put(4,.5){\framebox (1,1){\scriptsize $1$}}
\put(5,.5){\framebox (1,1){\scriptsize $...$}}
\put(6,.5){\framebox (1,1){\scriptsize $1$}}
\put(0,-1){\framebox (1,1){\scriptsize $2$}}
\put(1,-1){\framebox (2,1){\scriptsize $...$}}
\put(3,-1){\framebox (1,1){\scriptsize $2$}}
\put(1.8,-1.7){\scriptsize $p$}
\put(5.4,-0.7){\scriptsize $q$}
\end{picture}}
\right) 
, \quad 
\KK_{[p,q,p]} \sim 
\left(
\mbox{
\setlength{\unitlength}{1em}
\begin{picture}(7.5,1.6)
\put(0,.2){\framebox (1,1){\scriptsize $1$}}
\put(1,.2){\framebox (2,1){\scriptsize $...$}}
\put(3,.2){\framebox (1,1){\scriptsize $1$}}
\put(4,.2){\framebox (1,1){\scriptsize $1$}}
\put(5,.2){\framebox (1,1){\scriptsize $...$}}
\put(6,.2){\framebox (1,1){\scriptsize $1$}}
\put(0,-.8){\framebox (1,1){\scriptsize $2$}}
\put(1,-.8){\framebox (2,1){\scriptsize $...$}}
\put(3,-.8){\framebox (1,1){\scriptsize $2$}}
\put(1.8,-1.5){\scriptsize $p$}
\put(5.4,-0.5){\scriptsize $q$}
\end{picture}}
\right) 
\end{eqnarray}
where each continuous group of boxes stands for 
a single trace, as before. 
Explicitly, the corresponding operators are 
\begin{eqnarray}
\label{def:OO}
\OO_{[p,q,p]} &=& 
\sum_{k=0}^p {(-1)^k p! \over k! (p-k)!} ~
\tr \left( {z_1}^{p+q-k} {z_2}^{k} \right)_s ~ 
\tr \left( {z_1}^{k} {z_2}^{p-k} \right)_s 
\\
\label{def:KK}
\KK_{[p,q,p]} &=& 
\sum_{k=0}^p {(-1)^k p! \over k! (p-k)!} ~
\tr \left[ \left( {z_1}^{p+q-k} {z_2}^{k} \right)_s 
\left( {z_1}^{k} {z_2}^{p-k} \right)_s \right] 
\end{eqnarray}
(after projecting onto $SU(4) \to SU(3) \times U(1)$ 
and keeping only 
the highest $U(1)$-charge pieces, 
as discussed in 
Sections \ref{section:superconformal group} and \ref{2-0-2}). 
Made of only $z_1$ and $z_2$, 
both types of operators 
are annihilated by four out of the sixteen 
Poincar\'e supersymmetry generators: 
using the SUSY 
transformations 
spelled out in Appendix \ref{n=4 susy:section}, 
we find 
$\bar Q_{\bar \zeta} z_j = 0$, 
$Q_{\zeta_3} z_j = - \sqrt{2} (\lambda \zeta_3) \delta_{3j}$, 
so 
\begin{eqnarray}
\bar Q_{\bar \zeta} \OO_{[p,q,p]} = 
Q_{\zeta_3} \OO_{[p,q,p]} = 
\bar Q_{\bar \zeta} \KK_{[p,q,p]} = 
Q_{\zeta_3} \KK_{[p,q,p]} = 
0 
. 
\end{eqnarray}

It is clear why $\KK_{[p,q,p]}$ is special: it is 
the only single trace $[p,q,p]$ operator 
which can be constructed out of these fields. 
On the other hand, $\OO_{[p,q,p]}$ is 
``the most natural'' double trace composite operator 
in this representation. 
We also recognize it as the 
free theory chiral primary 
from the classification of \cite{AFSZ}.

As we have seen, 
neither the single trace $\KK_{[p,q,p]}$ nor the double trace $\OO_{[p,q,p]}$ 
are eigenstates of the dilation operator, for general $N$. 
Below we calculate correlators 
$\langle \OO \bar{\OO} \rangle$, 
$\langle \OO \bar{\KK} \rangle$, 
$\langle \KK \bar{\OO} \rangle$, and 
$\langle \KK \bar{\KK} \rangle$, 
in the large $N$ 
limit, and determine the pure operators 
and their scaling dimension 
in this approximation.

\subsection{General correlators 
$\langle {\OO}_{[p,q,p]} \bar{\OO}_{[p,q,p]} \rangle$ 
to order $g^2$}
\label{section:OO}

Let us first consider the 
$\langle {\OO}_{[p,q,p]}(x) \bar{\OO}_{[p,q,p]}(y) \rangle$ 
correlators. 
The free contribution is just a power of the free scalar propagator 
$G(x,y) = [4\pi (x-y)^2]^{-1}$, times a combinatorial factor: 
\begin{eqnarray}
\label{def:RR}
\langle {\OO}_{[p,q,p]}(x) \bar{\OO}_{[p,q,p]}(y) \rangle |_{\mathrm {free}} 
&=&
\sum_{k,l=0}^p 
{(-1)^k p! \over k! (p-k)!} ~ {(-1)^l p! \over l! (p-l)!} ~
(\RR_{k,l}^{p+q,p}) |_{\mathrm {free}} 
\\&=& \nonumber 
[G(x,y)]^{(2p+q)} 
\sum_{k,l=0}^p 
{(-1)^k p! \over k! (p-k)!} ~ {(-1)^l p! \over l! (p-l)!} ~
\FF_{k,l}^{p+q,p}
\end{eqnarray}
where 
\begin{eqnarray}
\RR_{k,l}^{p+q,p} &=& 
\langle 
\left[ 
\str (z_1)^{(p+q-k)} (z_2)^{k} 
\right] (x) 
\left[ 
\str (z_1)^{k} (z_2)^{(p-k)} 
\right] (x) 
\nonumber\\&&~\,
\left[ 
\str (z_1)^{(p+q-l)} (z_2)^{l} 
\right] (y) 
\left[ 
\str (z_1)^{l} (z_2)^{(p-l)} 
\right] (y) 
\rangle 
,\\
\label{def:FF}
\FF_{k,l}^{p+q,p} &=& 
\sum_{\sigma , \rho } 
\left[ \str t^{a_{k+1}} ... t^{a_{p+q}} t^{b_1} ... t^{b_k} \right] 
\left[ \str t^{a_1} ... t^{a_k} t^{b_{k+1}} ... t^{b_p} \right] 
\nonumber \\ && \quad 
\left[ \str t^{a_{\sigma(l+1)}} ... t^{a_{\sigma(p+q)}} 
t^{b_{\rho(1)}} ... t^{b_{\rho(l)}} \right] 
\left[ \str t^{a_{\sigma(1)}} ... t^{a_{\sigma(l)}} 
t^{b_{\rho(l+1)}} ... t^{b_{\rho(p)}} \right] 
\nonumber\\ 
\end{eqnarray}
($\sigma$ and $\rho$ sample over groups of permutations 
$S_{p+q}$ and $S_p$ on $p+q$ and $p$ letters, respectively). 

Like in the \half-BPS case, 
the leading contribution%
\footnote{
	See Appendix \ref{suN-identities} for useful $SU(N)$ identities. 
	}
to 
$\FF_{k,l}^{p+q,p} \sim (N/2)^{(2p+q)}$ 
comes 
from terms in which generators appear in reverse 
order for $z$-s and $\bar z$-s. 
To estimate the large $N$ behavior 
we can use equation (\ref{merging traces}) to ``merge traces,'' 
$(\tr t^{d_1} ... t^{d_s} t^c) (\tr t^c t^{d_s} ... t^{d_1}) 
\sim \half \tr t^{d_1} ... t^{d_s} t^{d_s} ... t^{d_1} 
\sim (N/2)^{s+1}$. 
In order to find the numerical factor out front 
(which does not scale with $N$ but depends on $p$ and $q$), 
we should determine exactly 
which terms have this structure. 

The generators can appear in opposite order in two pairs of 
traces in (\ref{def:FF}) under the following circumstances. 
First, it can happen 
when $k=l$ and the 
traces are merged as 1 with 3 
and 2 with 4. The factors which arise are: 
$[1/p!]^2$ from symmetrizations in the 2-nd and 4-th traces; 
$[1/(p+q)!]^2$ from symmetrizations in the 2-nd and 4-th traces; 
$p!$ because for any ordering in the 1-st trace there is an identical 
one in the 3-d trace; 
$(p+q)!$ for the same reason for the 2-nd and 4-th trace; 
$k! (p-k)!$ since any permutation of just $t^a$-s or just $t^b$-s 
in the 1-st trace can be ``undone'' 
by $\sigma$-s and $\rho$-s in the 3-d trace; 
and similarly $k! (p+q-k)!$ for the 2-nd and 4-th trace; 
$p (p+q)$ because of trace cyclicity.  
There is also an overall factor from the definition (\ref{def:OO}). 
Second, if $q=0$, 
we can 
merge traces the other way: 1 with 4 and 2 with 3; in this case 
$k=p-l$ and all other factors are the same. 
Thus, the leading contributions%
\footnote{
	The error we are committing is of order $\OO(N^{-2})$. 
	}
add up to 
\begin{eqnarray}
\label{o-o:leading}
\langle {\OO}_{[p,q,p]}(x) \bar{\OO}_{[p,q,p]}(y) \rangle |_{\mathrm {free}} 
\sim \left( \half N 
G(x,y) \right)^{(2p+q)} 
\hspace{-3.5em} 
\hspace{-16em}&&
\nonumber\\&&~~~~\times
\sum_{k=0}^p 
(-)^{k} 
\left[ {p! \over k! (p-k)!} \right]^2
{ k! (p-k)! k! (p+q-k)! \over (p-1)! (p+q-1)! } 
\left[ (-)^k + \delta_{q,0} (-)^{p-k} \right]
\nonumber\\&&=
\left( \half N G(x,y) \right)^{(2p+q)} 
\left[ 1 + \delta_{q,0} (-)^{p} \right] 
{ p (p+q) (p+q+1) \over (q+1) } 
\end{eqnarray}
The reproduces the leading order correlators 
in the low dimensional cases considered in Sections 
\ref{2-0-2}, \ref{2-1-2}, and \ref{6 and higher}.
Also note that if $q=0$ and $p$ 
is odd, both operators 
$\OO_{[p,q,p]}$ and $\KK_{[p,q,p]}$ 
vanish identically, in agreement with (\ref{o-o:leading}).

Now consider the 
corrections to this result. 
Diagrams contributing 
to two point functions of scalar composite operators 
at $\OO(g^2)$ 
fall into two categories, see Figure \ref{fig:contributing diagrams}. 
On the one hand, there are 
Feynman graphs involving a gauge boson exchange 
(these are proportional to $A$ or $B$). 
On the other hand, we also have 
gauge independent ones (proportional to $\tilde B$) 
coming entirely from the 
$z z \bar z \bar z$-vertex. 
These two types of corrections have different combinatorial 
(index) structure, and we shall handle them separately.

\subsubsection{Gauge dependent contributions: Combinatorial Argument}
\label{section:OO-combinatorial}

In Section \ref{section:gauge-dependent}, we argued that 
two point functions of gauge-invariant operators can not 
contain pieces proportional to the gauge dependent functions
$A$ and $B$. Here we show this explicitly for operators
${\OO}_{[p,q,p]}$ and ${\KK}_{[p,q,p]}$. This is the only 
part of Section \ref{section:large N} which is exact in $N$,
and is not just a large $N$ approximation.

The simplest order $g^2$ contribution 
to $\langle {\OO}_{[p,q,p]} \bar{\OO}_{[p,q,p]} \rangle$ 
comes from corrections to the scalar propagator 
(diagrams of type (a) and (b) in 
Figure \ref{fig:contributing diagrams}). 
It has the same index structure as the 
free field result, and so is the 
same up to a factor $(p+q) N A$ for (a)- 
and $p N A$ for (b)-type diagrams. 
These factors simply count the number of $z_1$-s
and $z_2$-s.

Next consider the other diagrams where the 
correction comes from blocks with the same flavor 
in the four legs, ones of type (d) and (e). 
Each term in the $k$, $l$ sum in (\ref{def:RR}) 
receives corrections of the form 
\begin{eqnarray}
\label{e:basic}
&&
(\half)(\half)(-1)(2) B 
\sum_{\sigma,\rho} \sum_{i \ne j = 1}^{p+q} 
\left[ \tr t^{b_{k+1}} ... t^{b_p} t^{a_1} ... t^{a_k} \right] 
\left[ \tr t^{a_{k+1}} ... t^{a_{p+q}} t^{b_1} ... t^{b_k} \right] 
\nonumber \\ && \quad 
\left[ t^{b_{\rho(l+1)}} ... t^{b_{\rho(p)}} 
t^{a_{\sigma(1)}} ... 
\left[ t^{a_{\sigma(i)}} , t^c \right] 
... 
\left[ t^{a_{\sigma(j)}} , t^c \right] 
... t^{a_{\sigma(p+q)}} 
t^{b_{\rho(1)}} ... t^{b_{\rho(l)}} \right] 
\end{eqnarray}
where all traces have to be symmetrized, 
and the second line also contains two traces.%
\footnote{
	We will omit the $[G(x,y)]^{(2p+q)}$ factor 
	which is common to all diagrams. 
	} 
The 
prefactor multiplying the sum 
comes about in the following manner: 
a factor of (\half), since the sum is on $i \ne j$ rather than 
on $i < j$; similarly the other (\half) arises because using
$\{ \sigma(i) , \sigma(j) \}$ and $\{ \sigma(j) , \sigma(i) \}$ 
counts the same pair twice; a factor of 2 has to be taken into 
account as the two cross-symmetric pairs give the same contribution;
and finally $(-1)$ is there from two factors of $i$ which are 
needed to convert $f$-s to commutators. 

The structure in the sum (\ref{e:basic}) 
consists of four kinds of terms. 
The two commutators can be both in the third trace, 
or both in the fourth trace, 
or one in either trace. 
When both commutators are 
in the same trace, we can play the same game as for \half-BPS 
operators. 
Fix $i$ and do the sum on $j$ first; this assembles 
$
( ... [t^{a_{\sigma(1)}} , t^c] ... ) + ... + 
( ... [t^{a_{\sigma(l)}} , t^c] ... ) 
= ( ... [t^{a_{\sigma(1)}} ... t^{a_{\sigma(l)}} , t^c] ... )
$
for example. 
Then, use trace cyclicity in the form 
$\tr [A,B]C = \tr A[B,C]$
to move one of the traces over 
to the other commutator and $t^{b_\rho}$-s; 
here we pick up a minus sign which cancels the $(-1)$ in 
(\ref{e:basic}). 
As $[[ t^a , t^c],t^c] = N t^a$, the first bit is easy --- 
just like in the \half-BPS case, it is proportional to 
$(+\half B N)$ times the free-field combinatorial factor; 
when we do the sum on $i$ we also get a factor of $(p+q)$ here. 
As $B+2A=0$ (by \NN=4 SUSY), this part cancels 
the diagrams of type (a). 
The leftovers, together with the terms with 
commutators in different traces, add up to 
\begin{eqnarray}
\label{leftover} 
&&
\half B 
\left[ \tr t^{b_{k+1}} ... t^{b_p} t^{a_1} ... t^{a_k} \right] 
\left[ \tr t^{a_{k+1}} ... t^{a_{p+q}} t^{b_1} ... t^{b_k} \right] 
\sum_{\sigma,\rho} 
\nonumber \\ && \quad 
\left[ \tr t^{b_{\rho(l+1)}} ... t^{b_{\rho(p)}} 
t^{a_{\sigma(1)}} ... t^{a_{\sigma(l)}} \right] 
\left[ \tr 
\left[ t^{a_{\sigma(l+1)}} ... t^{a_{\sigma(p+q)}} , t^c \right] 
\left[ t^{b_{\rho(1)}} ... t^{b_{\rho(l)}}  , t^c \right] 
\right] 
\nonumber \\ && 
+ 
\left[ \tr 
\left[ t^{b_{\rho(l+1)}} ... t^{b_{\rho(p)}}  , t^c \right] 
\left[ t^{a_{\sigma(1)}} ... t^{a_{\sigma(l)}} , t^c \right] 
\right] 
\left[ \tr 
t^{a_{\sigma(l+1)}} ... t^{a_{\sigma(p+q)}} 
t^{b_{\rho(1)}} ... t^{b_{\rho(l)}} 
\right] 
\nonumber \\ && 
- 2
\left[ \tr 
t^{b_{\rho(l+1)}} ... t^{b_{\rho(p)}} 
\left[ t^{a_{\sigma(1)}} ... t^{a_{\sigma(l)}} , t^c \right] 
\right] 
\left[ \tr 
\left[ t^{a_{\sigma(l+1)}} ... t^{a_{\sigma(p+q)}} , t^c \right] 
t^{b_{\rho(1)}} ... t^{b_{\rho(l)}} 
\right] 
\hspace{3em}
\end{eqnarray}

Similarly, 
there are diagrams of type (e) in Figure \ref{fig:contributing diagrams}, 
where all of the flavors are ``2'' in all four legs. 
Here, we have a term proportional to the free-field result (the only
difference being an overall factor of $p$ rather than $p+q$), 
which cancels contributions from diagrams of type (b). 
The leftover term 
is the same as what we have just computed. 
This removes the factor of \half~ from 
(\ref{leftover}). 

Finally consider the diagrams of type (f). 
Now we have both 
flavors ``1'' and ``2'' in the four-scalar blocks, while the 
index structure is the same as that of (d) and (e). 
The discussion goes through as above, 
but with a few small modifications. 
First, 
the prefactor is just $(-1) B$ as now the indices $i$ and $j$
run over different flavors (so there is no 
``$i \ne j$ overcounting,'' 
no ``\{$\sigma(i), \sigma(j)\}$ overcounting,''
and no factor of 2 from crossing symmetry). 
Second, 
we do not pick up a minus sign when transferring commutators
under the traces (before, both commutators with $t^c$ were on, say, 
$t^a$-s, whereas now one is on $t^a$ and one on $t^b$). 
Therefore, the result of 
adding the diagrams of type (f) is to precisely cancel 
the whole leftover structure of (twice that given in) equation 
(\ref{leftover}). 

Thus we have explicitly reproduced the general result of 
Section \ref{section:gauge-dependent},
but with a lot more work: $A$ and $B$ contributions to two-point
functions of 
gauge-invariant scalar composite 
operators combine as $2 A + B$, which vanishes by 
\NN=4 SUSY. 

\subsubsection{Contributions proportional to $\tilde B$}
\label{section:OO-F-term}

So far we have established that adding all 
gauge dependent Feynman graphs, i.e. 
diagrams of types (a), (b), (d), (e), and (f) 
shown in Figure \ref{fig:contributing diagrams}, 
gives a vanishing $\OO(g^2)$ contribution to the two-point 
functions (\ref{def:RR}), once we impose 
\NN=4 SUSY. 
Just as in the cases of low dimensional operators considered 
in Sections \ref{section:simplest} and \ref{6 and higher}, the 
$\OO(g^2)$ corrections to the two-point function 
of $[p,q,p]$ operators 
come exclusively from diagrams of type (c), 
and are proportional to $\tilde B$. 
To find the combinatorial factor multiplying $\tilde B$, 
we need to perform a calculation similar to the one 
in Section \ref{section:OO-combinatorial}.

Here, ``contracted with $f$-s'' are $z$-s with $z$-s and 
$\bar z$-s with $\bar z$-s (unlike in diagrams of types 
(d), (e) and (f), where it was one $z$ and one $\bar z$), 
and it is more convenient to label the generators 
slightly differently. For example, the free-field 
result (\ref{def:FF}) can be rewritten as 
\begin{eqnarray}
\label{def:FF-other}
\FF_{k,l}^{p+q,p} &=& 
\sum_{\sigma , \rho } 
\left[ \str 
t^{a_{\sigma(k+1)}} ... t^{a_{\sigma(p+q)}} 
t^{b_{1}} ... t^{b_{k}} 
\right] 
\left[ \str 
t^{a_{\sigma(1)}} ... t^{a_{\sigma(k)}} 
t^{b_{k+1}} ... t^{b_{p}} 
\right] 
\nonumber \\ && \quad 
\left[ \str 
t^{a_{l+1}} ... t^{a_{p+q}} 
t^{b_{\rho(1)}} ... t^{b_{\rho(l)}} 
\right] 
\left[ \str 
t^{a_{1}} ... t^{a_{l}} 
t^{b_{\rho(l+1)}} ... t^{b_{\rho(p)}} 
\right] 
\end{eqnarray}
For the Born level combinatorial factor 
it makes no difference, 
but in calculating the order $g^2$ diagrams of type (c) 
we ``$f$-contract'' 
$i$-th $z_1$ with $\rho(j)$-th $z_2$, and 
$\sigma(i)$-th $\bar z_1$ with $j$-th $\bar z_2$. 
This will exhaust all pairs without overcounting 
(because $i$ and $j$ are again on different flavors), 
so the prefactor will be just $(-1) \tilde B(x,y)$. Apart from 
this prefactor (and a factor of $[G(x,y)]^{(2p+q)}$), 
the (c)-type correction reads 
(sums on $\sigma$ and $\rho$ implied) 
\begin{eqnarray}
\label{h2 contribution}
&&
\!\!\!\!\!\!\!
\sum
\left[ \tr 
t^{a_{\sigma(k+1)}} ... t^{a_{\sigma(p+q)}} 
t^{b_{1}} ... t^{b_{j-1}} [t^{a_{i}}, t^c] t^{b_{j+1}} ... t^{b_{k}} 
\right] 
\left[ \tr 
t^{a_{\sigma(1)}} ... t^{a_{\sigma(k)}} 
t^{b_{k+1}} ... t^{b_{p}} 
\right] 
\nonumber \\ && \quad 
\left[ \tr 
t^{a_{l+1}} ... t^{a_{p+q}} 
t^{b_{\rho(1)}} ... t^{b_{\rho(l)}} 
\right] 
\left[ \tr 
t^{a_{1}} ... t^{a_{i-1}} [t^{b_{j}}, t^c] t^{a_{i+1}} ... t^{a_{l}} 
t^{b_{\rho(l+1)}} ... t^{b_{\rho(p)}} 
\right] 
\nonumber\\&+&
\!\!\!\!\!\!\!
\sum
\left[ \tr 
t^{a_{\sigma(k+1)}} ... t^{a_{\sigma(p+q)}} 
t^{b_{1}} ... t^{b_{j-1}} [t^{a_{i}}, t^c] t^{b_{j+1}} ... t^{b_{k}} 
\right] 
\left[ \tr 
t^{a_{\sigma(1)}} ... t^{a_{\sigma(k)}} 
t^{b_{k+1}} ... t^{b_{p}} 
\right] 
\nonumber \\ && \quad 
\left[ \tr 
t^{a_{l+1}} ... t^{a_{i-1}} [t^{b_{j}}, t^c] t^{a_{i+1}} ... t^{a_{p+q}} 
t^{b_{\rho(1)}} ... t^{b_{\rho(l)}} 
\right] 
\left[ \tr 
t^{a_{1}} ... t^{a_{l}} 
t^{b_{\rho(l+1)}} ... t^{b_{\rho(p)}} 
\right] 
\nonumber\\&+&
\!\!\!\!\!\!\!
\sum
\left[ \tr 
t^{a_{\sigma(k+1)}} ... t^{a_{\sigma(p+q)}} 
t^{b_{1}} ... t^{b_{k}} 
\right] 
\left[ \tr 
t^{a_{\sigma(1)}} ... t^{a_{\sigma(k)}} 
t^{b_{k+1}} ... t^{b_{j-1}} [t^{a_{i}}, t^c] t^{b_{j+1}} ... t^{b_{p}} 
\right] 
\nonumber \\ && \quad 
\left[ \tr 
t^{a_{l+1}} ... t^{a_{p+q}} 
t^{b_{\rho(1)}} ... t^{b_{\rho(l)}} 
\right] 
\left[ \tr 
t^{a_{1}} ... t^{a_{i-1}} [t^{b_{j}}, t^c] t^{a_{i+1}} ... t^{a_{l}} 
t^{b_{\rho(l+1)}} ... t^{b_{\rho(p)}} 
\right] 
\nonumber\\&+&
\!\!\!\!\!\!\!
\sum
\left[ \tr 
t^{a_{\sigma(k+1)}} ... t^{a_{\sigma(p+q)}} 
t^{b_{1}} ... t^{b_{k}} 
\right] 
\left[ \tr 
t^{a_{\sigma(1)}} ... t^{a_{\sigma(k)}} 
t^{b_{k+1}} ... t^{b_{j-1}} [t^{a_{i}}, t^c] t^{b_{j+1}} ... t^{b_{p}} 
\right] 
\nonumber \\ && \quad 
\left[ \tr 
t^{a_{l+1}} ... t^{a_{i-1}} [t^{b_{j}}, t^c] t^{a_{i+1}} ... t^{a_{p+q}} 
t^{b_{\rho(1)}} ... t^{b_{\rho(l)}} 
\right] 
\left[ \tr 
t^{a_{1}} ... t^{a_{l}} 
t^{b_{\rho(l+1)}} ... t^{b_{\rho(p)}} 
\right] 
\nonumber\\
\end{eqnarray}
with all traces symmetrized again; 
the sums on $\{i,j\}$ are: 
in the first line, from $\{1,1\}$ to $\{l,k\}$,  
in the second line, from $\{l+1,1\}$ to $\{p+q,k\}$, 
in the third line, from $\{1,k+1\}$ to $\{l,p\}$, 
and in the last line, from $\{l+1,k+1\}$ to $\{p+q,p\}$.  

In the large $N$ limit, such terms can scale as 
$\half (N/2)^{(2p+q-1)}$, at best. 
(Because of the commutators with 
$t^c$, we have to merge traces three times: 
they don't eat each other up in pairs as they did before). 
After including the factor of $\tilde B$, 
together with (\ref{o-o:leading}) this means that 
\begin{eqnarray}
\langle {\OO}_{[p,q,p]} (x) \bar{\OO}_{[p,q,p]} (y) \rangle 
&\!\!=\!\!& 
\left( {N\over2} G(x,y) \right)^{(2p+q)} 
\\&&\hspace{-4em} \times 
\left( 
\left[ 1 + \delta_{q,0} (-)^{p} \right] 
{ p (p+q) (p+q+1) \over (q+1) } 
+ 
\tilde B(x,y) N \times \OO(N^{-2}) 
\right) 
\nonumber
\end{eqnarray}
to order $g^2$. 
We might be tempted to stop here. 
By observing that since to working precision, 
the two point function of ${\OO}_{[p,q,p]}$ with itself 
does not get $\OO(g^2)$ corrections, 
we could try to conclude it is chiral, and 
in particular has a protected $\Delta = 2 p + q$. 
However, as the explicit examples of 
Sections \ref{section:simplest} and \ref{6 and higher} 
show, ${\OO}_{[p,q,p]}$ may not be a pure operator, 
in which case it doesn't make sense to talk 
about its scaling dimension.

\subsection{General correlators 
$\langle {\KK}_{[p,q,p]} \bar{\OO}_{[p,q,p]} \rangle$}

The analysis exactly parallels that of the previous section. 
Again, 
\begin{eqnarray}
\label{def:PP}
\langle {\KK}_{[p,q,p]}(x) \bar{\OO}_{[p,q,p]}(y) \rangle 
&=&
\sum_{k,l=0}^p 
{(-1)^k p! \over k! (p-k)!} ~ {(-1)^l p! \over l! (p-l)!} ~
\PP_{k,l}^{p+q,p}
\end{eqnarray}
with
\begin{eqnarray}
\PP_{k,l}^{p+q,p} &=& 
\langle 
\tr \left[ 
\left( (z_1)^{(p+q-l)} (z_2)^{l} \right)_s 
\left( (z_1)^{l} (z_2)^{(p-l)} \right)_s 
\right] (x) 
\nonumber\\&&~\,
\left[ 
\str (z_1)^{(p+q-k)} (z_2)^{k} 
\right] (y) 
\left[ 
\str (z_1)^{k} (z_2)^{(p-k)} 
\right] (y) 
\rangle 
\end{eqnarray}

The leading large $N$ contributions to the free correlators 
now come from terms which contain the combinatorial factor 
similar to 
\begin{eqnarray}
\label{k-o:leading}
&&\hspace{-2em}
(\tr t^{a_1} \! ... t^{a_k} t^{b_{k+1}} \! ... t^{b_p}
t^{a_{k+1}} \! ... t^{a_{p+q}} t^{b_{1}} \! ... t^{b_k}) 
~ 
(\tr t^{b_{p}} \! ... t^{b_{k+1}} t^{a_k} \! ... t^{a_1})
(\tr t^{b_{k}} \! ... t^{b_1} t^{a_{p+q}} \! ... t^{a_{k+1}})
\nonumber\\&&\hspace{4em} \sim
(\half)^2 N (N/2)^{(2p+q-2)} = \half (N/2)^{(2p+q-1)}
\end{eqnarray}
(the two halves and $-2$ in the exponent 
are because we have to merge traces twice, 
and the factor of $N=\tr \bone$ is there as usual). 
To get the other numerical factors we have to carefully analyze
which terms in the sums and symmetrizations scale with $N$ this way. 
So far, we do not need them.

As before, individual terms in the $k$, $l$ sum get corrections 
similar to (\ref{e:basic}): 
\begin{eqnarray}
&&
(\half)(\half)(-1)(2) B 
\sum_{\sigma,\rho} \sum_{i \ne j = 1}^{p+q} 
 \tr 
\left[ 
\left( t^{b_{k+1}} ... t^{b_p} t^{a_1} ... t^{a_k} \right) 
\left( t^{a_{k+1}} ... t^{a_{p+q}} t^{b_1} ... t^{b_k} \right)
\right]  
\nonumber \\ && \quad 
\left[ t^{b_{\rho(l+1)}} ... t^{b_{\rho(p)}} 
t^{a_{\sigma(1)}} ... 
\left[ t^{a_{\sigma(i)}} , t^c \right] 
... 
\left[ t^{a_{\sigma(j)}} , t^c \right] 
... t^{a_{\sigma(p+q)}} 
t^{b_{\rho(1)}} ... t^{b_{\rho(l)}} \right] 
\end{eqnarray}
with proper symmetrizations (and omitted factor of $[G(x,y)]^{(2p+q)}$). 
The only difference is that now 
there are three traces (rather than four). 
The discussion of gauge dependent diagrams 
(all types other than (c), see Figure \ref{fig:contributing diagrams}) 
goes through verbatim, since we were only manipulating 
the second set of traces. 
As before, when we impose \NN=4 SUSY they cancel, 
and order $g^2$ corrections to the two-point functions (\ref{def:PP}) 
are due to diagrams of type (c) only.  

Diagrams of type (c) are only slightly different
from those contributing to the 
$\langle \OO \bar{\OO} \rangle$ correlator; they add up to 
$(-1) \tilde B(x,y) [G(x,y)]^{(2p+q)}$ 
times 
\begin{eqnarray}
\label{h2 contribution for OK}
&&
\!\!\!\!\!\!\!
\sum
\tr \!\! \left[ 
\left( 
t^{a_{\sigma(k+1)}} ... t^{a_{\sigma(p+q)}} 
t^{b_{1}} ... t^{b_{j-1}} [t^{a_{i}}, t^c] t^{b_{j+1}} ... t^{b_{k}} 
\right) 
\left( 
t^{a_{\sigma(1)}} ... t^{a_{\sigma(k)}} 
t^{b_{k+1}} ... t^{b_{p}} 
\right) 
\right] 
\nonumber \\ && \quad 
\left[ \tr 
t^{a_{l+1}} ... t^{a_{p+q}} 
t^{b_{\rho(1)}} ... t^{b_{\rho(l)}} 
\right] 
\left[ \tr 
t^{a_{1}} ... t^{a_{i-1}} [t^{b_{j}}, t^c] t^{a_{i+1}} ... t^{a_{l}} 
t^{b_{\rho(l+1)}} ... t^{b_{\rho(p)}} 
\right] 
\nonumber\\&+&
\!\!\!\!\!\!\!
\sum
\tr \!\! \left[ 
\left( 
t^{a_{\sigma(k+1)}} ... t^{a_{\sigma(p+q)}} 
t^{b_{1}} ... t^{b_{j-1}} [t^{a_{i}}, t^c] t^{b_{j+1}} ... t^{b_{k}} 
\right) 
\left( 
t^{a_{\sigma(1)}} ... t^{a_{\sigma(k)}} 
t^{b_{k+1}} ... t^{b_{p}} 
\right) 
\right] 
\nonumber \\ && \quad 
\left[ \tr 
t^{a_{l+1}} ... t^{a_{i-1}} [t^{b_{j}}, t^c] t^{a_{i+1}} ... t^{a_{p+q}} 
t^{b_{\rho(1)}} ... t^{b_{\rho(l)}} 
\right] 
\left[ \tr 
t^{a_{1}} ... t^{a_{l}} 
t^{b_{\rho(l+1)}} ... t^{b_{\rho(p)}} 
\right] 
\nonumber\\&+&
\!\!\!\!\!\!\!
\sum
\tr \!\! \left[ 
\left( 
t^{a_{\sigma(k+1)}} ... t^{a_{\sigma(p+q)}} 
t^{b_{1}} ... t^{b_{k}} 
\right) 
\left( 
t^{a_{\sigma(1)}} ... t^{a_{\sigma(k)}} 
t^{b_{k+1}} ... t^{b_{j-1}} [t^{a_{i}}, t^c] t^{b_{j+1}} ... t^{b_{p}} 
\right) 
\right] 
\nonumber \\ && \quad 
\left[ \tr 
t^{a_{l+1}} ... t^{a_{p+q}} 
t^{b_{\rho(1)}} ... t^{b_{\rho(l)}} 
\right] 
\left[ \tr 
t^{a_{1}} ... t^{a_{i-1}} [t^{b_{j}}, t^c] t^{a_{i+1}} ... t^{a_{l}} 
t^{b_{\rho(l+1)}} ... t^{b_{\rho(p)}} 
\right] 
\nonumber\\&+&
\!\!\!\!\!\!\!
\sum
\tr \!\! \left[ 
\left( 
t^{a_{\sigma(k+1)}} ... t^{a_{\sigma(p+q)}} 
t^{b_{1}} ... t^{b_{k}} 
\right) 
\left( 
t^{a_{\sigma(1)}} ... t^{a_{\sigma(k)}} 
t^{b_{k+1}} ... t^{b_{j-1}} [t^{a_{i}}, t^c] t^{b_{j+1}} ... t^{b_{p}} 
\right) 
\right] 
\nonumber \\ && \quad 
\left[ \tr 
t^{a_{l+1}} ... t^{a_{i-1}} [t^{b_{j}}, t^c] t^{a_{i+1}} ... t^{a_{p+q}} 
t^{b_{\rho(1)}} ... t^{b_{\rho(l)}} 
\right] 
\left[ \tr 
t^{a_{1}} ... t^{a_{l}} 
t^{b_{\rho(l+1)}} ... t^{b_{\rho(p)}} 
\right] 
\nonumber\\
\end{eqnarray}
with proper symmetrizations; 
the sums on $\{i,j\}$ are: 
in the first line, from $\{1,1\}$ to $\{l,k\}$,  
in the second line, from $\{l+1,1\}$ to $\{p+q,k\}$, 
in the third line, from $\{1,k+1\}$ to $\{l,p\}$, 
and in the last line, from $\{l+1,k+1\}$ to $\{p+q,p\}$.  

We can get correlators 
$\langle {\OO}_{[p,q,p]}(x) \bar{\KK}_{[p,q,p]}(y) \rangle$
by just complex conjugating the sum 
(\ref{h2 contribution for OK})
times the same prefactor; 
both the free propagator $G(x,y)$ and the function $\tilde B(x,y)$
are real and depend only on $(x-y)^2$, so exchanging the arguments 
$x \lra y$ and conjugating doesn't change anything.

Large $N$ dependence of (\ref{h2 contribution for OK})
can be again estimated by merging traces. This time, we have to 
merge traces only twice (there are three traces total), 
so it scales a power of $N$ higher than a similar 
$\langle {\OO}(x) \bar{\OO}(y) \rangle |_{g^2}$
correction (where traces had to be merged three times). 
Hence, 
$\langle {\OO}_{[p,q,p]}(x) \bar{\KK}_{[p,q,p]}(y) \rangle |_{g^2} 
\sim (N/2)^{(2p+q)}$.

\subsection{General correlators 
$\langle {\KK}_{[p,q,p]} \bar{\KK}_{[p,q,p]} \rangle$}
\label{section:KK}

The analysis is similar as for 
$\langle {\KK}_{[p,q,p]}(x) \bar{\OO}_{[p,q,p]}(y) \rangle$. 
The only surviving contribution to 
$\langle {\KK}_{[p,q,p]} \bar{\KK}_{[p,q,p]} \rangle|_{g^2}$
is due to diagrams of type (c) again, and 
equals 
\begin{eqnarray}
\label{h2 contribution for KK}
&&
\!\!\!\!\!\!\!
\sum
\tr \!\! \left[ 
\left( 
t^{a_{\sigma(k+1)}} ... t^{a_{\sigma(p+q)}} 
t^{b_{1}} ... t^{b_{j-1}} [t^{a_{i}}, t^c] t^{b_{j+1}} ... t^{b_{k}} 
\right) 
\left( 
t^{a_{\sigma(1)}} ... t^{a_{\sigma(k)}} 
t^{b_{k+1}} ... t^{b_{p}} 
\right) 
\right] 
\nonumber \\ && \quad 
\tr \!\! \left[ 
\left( 
t^{a_{l+1}} ... t^{a_{p+q}} 
t^{b_{\rho(1)}} ... t^{b_{\rho(l)}} 
\right) 
\left( 
t^{a_{1}} ... t^{a_{i-1}} [t^{b_{j}}, t^c] t^{a_{i+1}} ... t^{a_{l}} 
t^{b_{\rho(l+1)}} ... t^{b_{\rho(p)}} 
\right) 
\right] 
\nonumber\\&+&
\!\!\!\!\!\!\!
\sum
\tr \!\! \left[ 
\left( 
t^{a_{\sigma(k+1)}} ... t^{a_{\sigma(p+q)}} 
t^{b_{1}} ... t^{b_{j-1}} [t^{a_{i}}, t^c] t^{b_{j+1}} ... t^{b_{k}} 
\right) 
\left( 
t^{a_{\sigma(1)}} ... t^{a_{\sigma(k)}} 
t^{b_{k+1}} ... t^{b_{p}} 
\right) 
\right] 
\nonumber \\ && \quad 
\tr \!\! \left[ 
\left( 
t^{a_{l+1}} ... t^{a_{i-1}} [t^{b_{j}}, t^c] t^{a_{i+1}} ... t^{a_{p+q}} 
t^{b_{\rho(1)}} ... t^{b_{\rho(l)}} 
\right) 
\left( 
t^{a_{1}} ... t^{a_{l}} 
t^{b_{\rho(l+1)}} ... t^{b_{\rho(p)}} 
\right) 
\right] 
\nonumber\\&+&
\!\!\!\!\!\!\!
\sum
\tr \!\! \left[ 
\left( 
t^{a_{\sigma(k+1)}} ... t^{a_{\sigma(p+q)}} 
t^{b_{1}} ... t^{b_{k}} 
\right) 
\left( 
t^{a_{\sigma(1)}} ... t^{a_{\sigma(k)}} 
t^{b_{k+1}} ... t^{b_{j-1}} [t^{a_{i}}, t^c] t^{b_{j+1}} ... t^{b_{p}} 
\right) 
\right] 
\nonumber \\ && \quad 
\tr \!\! \left[ 
\left( 
t^{a_{l+1}} ... t^{a_{p+q}} 
t^{b_{\rho(1)}} ... t^{b_{\rho(l)}} 
\right) 
\left( 
t^{a_{1}} ... t^{a_{i-1}} [t^{b_{j}}, t^c] t^{a_{i+1}} ... t^{a_{l}} 
t^{b_{\rho(l+1)}} ... t^{b_{\rho(p)}} 
\right) 
\right] 
\nonumber\\&+&
\!\!\!\!\!\!\!
\sum
\tr \!\! \left[ 
\left( 
t^{a_{\sigma(k+1)}} ... t^{a_{\sigma(p+q)}} 
t^{b_{1}} ... t^{b_{k}} 
\right) 
\left( 
t^{a_{\sigma(1)}} ... t^{a_{\sigma(k)}} 
t^{b_{k+1}} ... t^{b_{j-1}} [t^{a_{i}}, t^c] t^{b_{j+1}} ... t^{b_{p}} 
\right) 
\right] 
\nonumber \\ && \quad 
\tr \!\! \left[ 
\left( 
t^{a_{l+1}} ... t^{a_{i-1}} [t^{b_{j}}, t^c] t^{a_{i+1}} ... t^{a_{p+q}} 
t^{b_{\rho(1)}} ... t^{b_{\rho(l)}} 
\right) 
\left( 
t^{a_{1}} ... t^{a_{l}} 
t^{b_{\rho(l+1)}} ... t^{b_{\rho(p)}} 
\right) 
\right] 
\nonumber\\
\end{eqnarray}
up to a factor of $(-1) \tilde B(x,y) [G(x,y)]^{(2p+q)}$, 
and proper symmetrizations. 
The sums on $\{i,j\}$ are as before: 
in the first line, from $\{1,1\}$ to $\{l,k\}$,  
in the second line, from $\{l+1,1\}$ to $\{p+q,k\}$, 
in the third line, from $\{1,k+1\}$ to $\{l,p\}$, 
and in the last line, from $\{l+1,k+1\}$ to $\{p+q,p\}$).

At Born level, 
$\langle {\KK}_{[p,q,p]} \bar{\KK}_{[p,q,p]} \rangle|_{\mathrm {free}} 
\sim (\half N)^{(2p+q)}$ at large $N$:
we have to merge traces once, and there are 
$2p+q$ generators involved.

\subsection{Quarter BPS operators}
\label{section:consistency}

Given the leading large $N$ dependence of a 
$\langle \OO \bar{\KK} \rangle |_{\mathrm {free}}$
correlator, it's easy to determine the 
leading large $N$ dependence of the corresponding 
$\langle \KK \bar{\KK} \rangle |_{\mathrm {free}}$. 
Indeed, suppose that a particular term e.g.
\begin{eqnarray} 
\label{term:o-k}
(\tr t^a ... t^a t^b ... t^b)(\tr t^a ... t^a t^b ... t^b)
\tr (t^a ... t^a t^b ... t^b)(t^a ... t^a t^b ... t^b)
\sim\nonumber\\
(2/N)
(\tr t^a ... t^a t^b ... t^b)(\tr t^c t^c t^a ... t^a t^b ... t^b)
\tr (t^a ... t^a t^b ... t^b)(t^a ... t^a t^b ... t^b)
\end{eqnarray}
contributes to $\langle \OO \bar{\KK} \rangle$. 
We can insert contents of the first and second 
traces into the third trace 
(using $2 (\tr A t^r) (\tr B t^r) \sim \tr A B$)
to reduce this expression to a single trace. 
On the other hand, the term in the $\langle \KK \bar{\KK} \rangle$
with the same order of generators can be written as 
\begin{eqnarray} 
\label{term:k-k}
\tr (t^a ... t^a t^b ... t^b)(t^a ... t^a t^b ... t^b)
\tr (t^a ... t^a t^b ... t^b)(t^a ... t^a t^b ... t^b)
\sim\nonumber\\
2
(\tr t^a ... t^a t^b ... t^b t^c)(\tr t^c t^a ... t^a t^b ... t^b)
\tr (t^a ... t^a t^b ... t^b)(t^a ... t^a t^b ... t^b)
\end{eqnarray}
and we can insert the first and second traces into the third trace 
again in the same locations. 
This term gives a leading contribution provided all 
generators collapse 
after consecutively applying $2 t^r t^r \sim N \bone$
without having to commute generators past one another. 
In this case, the term in (\ref{term:o-k}) also gives a leading 
contribution. 
The only other terms 
which 
have the same large $N$ behavior are the ones 
that differ from it by cyclic permutations within 
the first and second traces. 
This gives an overall factor of 
$p(p+q)$. Comparing (\ref{term:o-k}) and (\ref{term:k-k}) then 
shows that to leading order in $N$, 
the difference is a factor of 
$\beta \equiv p(p+q)/N$. 

For large $N$, the analysis of order $g^2$ corrections 
is analogous to the case of free field contributions. 
The structure of terms 
in (\ref{h2 contribution for OK}) and (\ref{h2 contribution for KK})
is similar, 
and leading contributions come from terms with 
the same order of generators (modulo cyclic permutations 
for $\langle \OO \bar{\KK} \rangle$ corrections); 
the difference is the multiplicative factor 
$\beta$, the same for all such terms. 

Bringing this together with the results of 
Sections \ref{section:OO}-\ref{section:KK}, we can write down the 
large $N$ leading order 
two point functions as 
\begin{eqnarray}
\label{pqp:largeN}
\left( \matrix{
\langle \KK \bar{\KK} \rangle & 
\langle \KK \bar{\OO} \rangle \cr 
\langle \OO \bar{\KK} \rangle & 
\langle \OO \bar{\OO} \rangle 
} \right) 
= 
\alpha
(G N)^{2p+q}
\left\{
\left( \matrix{
1 & \beta \cr 
\beta & * 
} \right)
+ 
\tilde \alpha 
(\tilde B N) 
\left( \matrix{
1 & \beta \cr 
\beta & \OO(N^{-2}) 
} \right)
\right\}
\end{eqnarray}
where $\alpha$, $*$, 
and $\tilde \alpha$ are some constants of 
order $\OO(N^0)$, 
and $\beta = p(p+q)/N$. 
As before, $G \equiv G(x,y)$; 
$\tilde B \equiv \tilde B(x,y)$; 
and 
$\langle \OO \bar{\OO} \rangle \equiv 
\langle \OO_{[p,q,p]}(x) \bar{\OO}_{[p,q,p]}(y) \rangle$, etc. 
Each (order one) coefficient 
above is valid 
to $\OO(N^{-2})$, and of course (\ref{pqp:largeN}) is 
perturbative in the coupling constant --- 
we have neglected $\OO(g^4)$ terms. 

Diagonalizing the matrix of corrections 
with respect to the matrix of free correlators 
as in Section \ref{6 and higher}, we find that 
\begin{eqnarray}
\label{K-hat:large N}
\tilde{\YY}_{[p,q,p]} &=& 
\KK_{[p,q,p]}   + \OO(N^{-2}) 
\\
\label{O-hat:large N}
\YY_{[p,q,p]} &=& 
\OO_{[p,q,p]} - {p(p+q)\over N} \KK_{[p,q,p]}  + \OO(N^{-2}) 
\end{eqnarray}
are pure operators, 
and as such have well defined scaling dimension. 
At this order, 
the scaling dimension of $\tilde{\YY}_{[p,q,p]}$ receives 
an $\OO(g^2N)$ correction proportional to the coefficient $\tilde \alpha$ 
in (\ref{pqp:largeN}), while 
the scaling dimension 
$\Delta_{\mbox{\scriptsize $\YY$ }} = 2 p + q$ 
of $\YY_{[p,q,p]}$ 
is protected. 
Finally, the normalization of $\YY_{[p,q,p]}$ does not 
get any $g^2$ corrections, and is given by the 
Born level expression 
\begin{eqnarray}
\label{normalization:general}
\langle \YY_{[p,q,p]}(x) \bar{\YY}_{[p,q,p]}(y) \rangle 
&\!\!=\!\!& 
\left[ 1 + \delta_{q,0} (-)^{p} \right]
{ p (p+q) (p+q+1) \over (q+1) } 
\left[ {N \over 8 \pi (x-y)^2} \right]^{2p+q}
\nonumber\\
&& \quad 
\times \left( 1 + \, \OO(N^{-2}; g^4) \right) 
\end{eqnarray}
This is found from formulae (\ref{o-o:leading}), 
(\ref{pqp:largeN}), and (\ref{O-hat:large N}). 
The exact expressions of 
Sections \ref{section:simplest} and \ref{6 and higher} 
agree with (\ref{O-hat:large N}) and (\ref{normalization:general})
in the large $N$ limit. 
We conclude that to working precision, 
$\YY_{[p,q,p]}$ is a \quarter-BPS chiral primary 
operator. 

\section{Conclusion}

In this Chapter we studied local, polynomial, gauge invariant 
scalar composite operators
in $[p,q,p]$ representations of $SU(4)$ in 
$d=4$, \NN=4 SYM. 
We found that certain such operators 
have protected 
two-point functions at order $g^2$, with each other 
and 
with other operators. 
We presented ample evidence that these $\OO(g^2)$ protected operators 
are \quarter-BPS chiral primaries 
in the fully interacting theory. 

These operators are 
not just the double trace operators from 
the classification of \cite{AFSZ}, 
but mixtures of all 
gauge invariant local composite operators 
made of the same scalars:
single trace operators, 
other double trace operators, 
triple trace operators, etc. 
As our exact in $N$, explicit 
construction of \quarter-BPS 
primaries of low scaling dimension 
($\Delta = 2p+q < 8$) shows, 
the $N$ dependence of the coefficients 
in these linear combinations is quite complicated. 

Apart from 
operators
of low dimensions for arbitrary $N$, 
we considered the large $N$ behavior 
of two point functions for all $[p,q,p]$. 
A leading and subleading analysis 
reveals that for every $[p,q,p]$, 
there is a \quarter-BPS operator 
which is a 
certain linear combinations of 
double and single trace operators. 
We give closed form expressions 
for the operators involved in this 
linear combination and the coefficients with 
which they enter, 
all valid to next-to-leading order in $N$.


\section{Appendix}

\subsection{\NN = 4 SUSY in various forms}
\label{n=4 susy:section}

In the \NN=1 component notation, the classical Lagrangian 
(see \cite{Sohnius}, p. 158) takes the form
(in geometric notation, i.e. with ${1\over g^2}$ 
multiplying the whole action)
\begin{eqnarray}
\LL &=& \mbox{$1\over g^2$ } \tr 
\left\{ 
- \mbox{$1\over 4$} F_{\mu\nu} F^{\mu\nu} 
+ \ihalf \bar \lambda \gamma^\mu D_\mu \lambda 
+ \half D^2
\right. \\ \nonumber &&\quad~~ \left. 
+ \half D_\mu A_j D^\mu A_j 
+ \half D_\mu B_j D^\mu B_j 
+ \ihalf \bar \psi_j \gamma^\mu D_\mu \psi_j
+ \half F_j F_j 
+ \half G_j G_j 
\right. \\ \nonumber &&\quad~~ \left. 
- i [ A_j , B_j ] D 
- i \bar \psi_j [ \lambda , A_j ] 
- i \bar \psi_j \gamma_5 [ \lambda , B_j ] 
\right. \\ \nonumber &&\quad~~ \left. 
- \ihalf \e_{jkl} 
\left( \right.\!\!  
\bar \psi_j [\psi_k , A_l ] - \bar \psi_j \gamma_5 [\psi_k , B_l ] 
\right. \\ \nonumber &&\quad\quad\quad\quad\quad~ \left. 
+ [ A_j , A_k ] F_l - [ B_j , B_k ] F_l 
+ 2 [ A_j , B_k ] G_l 
\left.\!\! \right) 
\right\}
\end{eqnarray}
(with Lorentz signature); 
there should be no confusion between the auxiliary field $D$ of 
the vector multiplet and the covariant derivative 
$D_\mu = \partial_\mu + i A_\mu$. 

This Lagrangian can also be rewritten in a manifestly 
$SU(4)$-invariant form. 
Combining the three chiral spinors and 
the gaugino into 
\begin{eqnarray}
\lambda_4 = \lambda ; \quad\quad
\lambda_j = \psi_j , \quad j = 1,2,3 
\end{eqnarray}
and making $4\times4$ antisymmetric matrices of scalars and pseudoscalars by
\begin{eqnarray}
A_{j k} = - \e_{jkl} A_l ; &\quad& A_{4 j} = - A_{j 4} = A_j ; \\
B_{j k} = \phantom{-} \e_{jkl} B_l ; &\quad& B_{4 j} = - B_{j 4} = B_j 
\end{eqnarray}
the Lagrangian becomes (sums on indices $j,k,l$ now run from 1 to 4)
\begin{eqnarray}
\LL &=& \mbox{$1\over g^2$ } \tr 
\left\{ 
- \mbox{$1\over 4$} F_{\mu\nu} F^{\mu\nu} 
+ \ihalf \bar \lambda_j \gamma^\mu D_\mu \lambda_j 
+ \half D_\mu A_{jk} D^\mu A_{jk} 
+ \half D_\mu B_{jk} D^\mu B_{jk} 
\right.\nonumber  \\ \nonumber &&\quad~~ \left. 
+ \ihalf \bar \lambda_j [ \lambda_k , A_{jk} ] 
+ \ihalf \bar \lambda_j \gamma_5 [ \lambda_k , B_{jk} ] 
+ \mbox{$1\over32$} [ A_{jk} , B_{lm} ] [ A_{jk} , B_{lm} ] 
\right. \\ &&\quad~~ \left. 
+ \mbox{$1\over64$} [ A_{jk} , A_{lm} ] [ A_{jk} , A_{lm} ] 
+ \mbox{$1\over64$} [ B_{jk} , B_{lm} ] [ B_{jk} , B_{lm} ] 
\right\}
\end{eqnarray}
after integrating out the auxiliary fields $D$, $F_j$, and $G_j$. 
The $A_{jk}$ and $B_{jk}$ are self-dual and antiself-dual tensors of
$O(4)$: 
\begin{equation}
A_{jk} = \half \e_{jklm} A_{lm} ; 
\quad 
B_{jk} = - \half \e_{jklm} B_{lm} 
\end{equation}
Alternatively, the fields $A_i$ and $B_i$ form a {\textbf 6} of the $R$-symmetry 
group $SU(4) \sim SO(6)$: we can group them as 
$\phi^i = A_i$, $\phi^{i+3} = B_i$, $i=1,2,3$.

This form of the Lagrangian is only manifestly $O(4)$ symmetric, however. 
If we define a complex matrix of scalars 
\begin{equation}
M_{jk} \equiv \half \left( A_{lm} + i B_{jk} \right) 
\end{equation}
subject to a reality condition 
\begin{equation}
\bar M^{jk} \equiv (M_{jk})^\dagger = \half \e^{jklm} M_{lm} 
\end{equation}
the \NN=4 Lagrangian 
\begin{eqnarray}
\LL &=& \mbox{$1\over g^2$ } \tr 
\left\{ 
- \mbox{$1\over 4$} F_{\mu\nu} F^{\mu\nu} 
+ i \lambda_j \sigma^\mu D_\mu \bar \lambda^j 
+ \half D_\mu M_{jk} \bar D^\mu \bar M^{jk} 
\right. \\ \nonumber &&\quad~~ \left. 
+ i \lambda_j [ \lambda_k , \bar M^{jk} ] 
+ i \bar \lambda^j [ \bar \lambda^k , M_{jk} ] 
+ \mbox{$1\over 4$} [ M_{jk} , M_{lm} ] [ \bar M^{jk} , \bar M^{lm} ] 
\right\}
\end{eqnarray}
is then manifestly $SU(4)$ covariant, as are the SUSY transformation laws 
\begin{eqnarray}
\delta A_\mu &=& 
i \zeta_j \sigma_\mu \bar \lambda^j - 
i \lambda_j \sigma_\mu \bar \zeta^j \\
\delta M_{jk} &=& 
\zeta_j \lambda_k - \zeta_k \lambda_j + 
\e_{jklm} \bar \zeta^l \bar \lambda^m \\
\delta \lambda_j &=& 
- \ihalf \sigma^{\mu\nu} F_{\mu\nu} \zeta_j 
+ 2 i \sigma^\mu D_\mu M_{jk} \bar \zeta^k 
+ 2 i [ M_{jk}, \bar M^{kl} ] \zeta_l 
\end{eqnarray}
(notice that now $\lambda_j$ and $\bar \lambda^j$ are Weyl spinors; there
should be no confusion: when spinors are multiplied by $2\times2$ 
$\sigma$ matrices they are Weyl, and when the $4\times4$ $\gamma$ matrices
are used, they are Dirac).

We can also rewrite this Lagrangian in terms of three unconstrained
(unlike the $M_{jk}$) complex scalar fields 
\begin{equation}
     z_j = \mbox{$1\over\sqrt{2}$} \left( A_j + i B_j \right) , \quad
\bar z^j = \mbox{$1\over\sqrt{2}$} \left( A_j - i B_j \right) 
\end{equation}
and the original fermions $\psi_j$ and $\lambda$:
\begin{eqnarray}
\label{L su3 form}
\LL &=& \mbox{$1\over g^2$ } \tr 
\left\{ 
- \mbox{$1\over 4$} F_{\mu\nu} F^{\mu\nu} 
+ i \lambda \sigma^\mu D_\mu \bar \lambda 
+ i \psi_j \sigma^\mu D_\mu \bar \psi^j 
+ D_\mu z_j \bar D^\mu \bar z^j 
\right. \\ \nonumber &&\quad~~ \left. 
+ \mbox{$i\over\sqrt2$} \lambda [ \psi_j , \bar z^j ] 
- \mbox{$i\over\sqrt2$} \psi_j [ \lambda , \bar z^j ] 
- \mbox{$i\over\sqrt2$} \e^{jkl} \psi_j [ \psi_k , z_l ] 
\right. \\ \nonumber &&\quad~~ \left. 
+ \mbox{$i\over\sqrt2$} \bar \lambda [ \bar \psi^j , z_j ] 
- \mbox{$i\over\sqrt2$} \bar \psi^j [ \bar \lambda , z_j ] 
- \mbox{$i\over\sqrt2$} \e_{jkl} \bar \psi^j [ \bar \psi^k , \bar z^l ] 
\right. \\ \nonumber &&\quad~~ \left. 
+ [ z_j , z_k ] [ \bar z^j , \bar z^k ] 
- \half [ z_j , \bar z^j ] [ z_k , \bar z^k ] 
\right\}
\end{eqnarray}
and the SUSY transformations now are 
\begin{eqnarray}
\label{SUSY-su(3) form:begin}
\delta A_\mu &=& 
i \zeta_j \sigma_\mu \bar \psi^j - 
i \psi_j \sigma_\mu \bar \zeta^j + 
i \zeta \sigma_\mu \bar \lambda - 
i \lambda \sigma_\mu \bar \zeta \\
\delta z_j &=& 
\sqrt{2} \left(
\zeta \psi_j - \zeta_j \lambda - 
\e_{jkl} \bar \zeta^k \bar \psi^l 
\right) \\
\delta \lambda &=& 
- \ihalf \sigma^{\mu\nu} F_{\mu\nu} \zeta 
+ i \sqrt{2} \sigma^\mu D_\mu z_j \bar \zeta^j 
+ i \e^{jkl} [ z_j, z_k ] \zeta_l 
- i [ z_j, \bar z^j ] \zeta \\ 
\delta \psi_j &=& 
- \ihalf \sigma^{\mu\nu} F_{\mu\nu} \zeta_j 
+ i \sqrt{2} \e_{jkl} \sigma^\mu \bar D_\mu \bar z^k \bar \zeta^l 
- i \sqrt{2} \sigma^\mu D_\mu z_j \bar \zeta 
\nonumber \\ && \quad~~
+ i \left( 
[ z_k , \bar z^k ] \zeta_j 
- 2 [ z_j , \bar z^k ] \zeta_k 
- \e_{jkl} [ \bar z^k , \bar z^l ] \zeta 
\right)
\\
\noalign{\noindent and their conjugates}
\delta \bar z^j &=& 
\sqrt{2} \left(
\bar \zeta \bar \psi^j - \bar \zeta^j \bar \lambda - 
\e^{jkl} \zeta_k \psi_l 
\right) \\
\delta \bar \lambda &=& 
+ \ihalf \bar \sigma^{\mu\nu} F_{\mu\nu} \bar \zeta 
- i \sqrt{2} \bar \sigma^\mu \bar D_\mu \bar z^j \zeta_j 
- i \e_{jkl} [ \bar z^j, \bar z^k ] \bar \zeta^l 
+ i [ \bar z^j, z_j ] \bar \zeta \\ 
\delta \bar \psi^j &=& 
+ \ihalf \bar \sigma^{\mu\nu} F_{\mu\nu} \bar \zeta^j 
- i \sqrt{2} \e^{jkl} \bar \sigma^\mu D_\mu z_k \zeta_l 
+ i \sqrt{2} \bar \sigma^\mu \bar D_\mu \bar z^j \zeta 
\nonumber \\ && \quad~~
- i \left( 
[ \bar z^k , z_k ] \bar \zeta^j 
- 2 [ \bar z^j , z_k ] \bar \zeta^k 
- \e^{jkl} [ z_k , z_l ] \bar \zeta 
\right)
\label{SUSY-su(3) form:end}
\end{eqnarray}
This way of writing the Lagrangian and SUSY transformations 
hides the full $SU(4)$ $R$-symmetry of the 
theory; now, only the $SU(3) \times U(1)$ subgroup of it is manifest.

\subsection{Miscellaneous identities for $SU(N)$}
\label{suN-identities}

We can use the following property of generators of 
$SU(N)$ (for $N\ge3$) in the fundamental representation:
\begin{equation}
\{ t^a, t^b \} = {1\over N} \delta^{ab} + d^{abc} t^c
\end{equation}
Together with $[ t^a, t^b ] = i f^{abc} t^c$ (valid in any
representation), we find 
\begin{equation}
t^a t^b = {1\over 2 N} \delta^{ab} \bone + 
{1\over 2} \left( d^{abc} + i f^{abc} \right) t^c
\end{equation}

Let 
\begin{equation}
g^{a_1 ... a_k} \equiv \tr t^{a_1} ... t^{a_k}
,\quad 
g^{(0)} = \tr \bone = N
.
\end{equation}
Then with the standard normalization 
$\tr t^a t^b = {1\over 2} \delta^{ab}$ 
for $SU(N)$ generators in the fundamental, we can 
in principle recursively determine the 
trace of any string of generators in terms of 
$\delta^{ab}$, $d^{abc}$, and $f^{abc}$: 
\begin{eqnarray}
&&
g^{a} = \tr t^{a} = 0 ,~
g^{a b} = \half \delta^{ab} ,~
g^{a b c} =  \quarter \left( d^{abc} + i f^{abc} \right) ,\quad\mbox{and}\\
&&
g^{a_1 ... a_k} = \mbox{$1\over 2 N$} \, \delta^{a_1 a_2} g^{a_3 ... a_k} 
+ 2\, g^{a_1 a_2 c} g^{c a_3 ... a_k} 
\end{eqnarray}
Now we can set up a recursion relation for
\begin{eqnarray}
\label{eq:recursion p}
P_k \equiv g^{a_1 ... a_k} g^{a_1 ... a_k} 
\quad
\mbox{and}
\quad
\tilde P_k \equiv g^{a_1 ... a_k} g^{a_k ... a_1} 
\end{eqnarray}
(with sums on repeated $a_1 , ... ,  a_k$ implied). 
Using 
$t^a t^a = {N^2 - 1 \over 2 N} \bone$, we find 
\begin{eqnarray}
P_k = {N^2 - 1 \over 4 N^2} P_{k-2} 
+ {4\over N^2 - 1} P_3 P_{k-1} \\
\noalign{\noindent and similarly}
\tilde P_k = {N^2 - 1 \over 4 N^2} \tilde P_{k-2} 
+ {4\over N^2 - 1} \tilde P_3 \tilde P_{k-1} 
\end{eqnarray}
The values of $P_2$, $\tilde P_2$, 
$P_3$ and $\tilde P_3$ have to be computed explicitly;
they are 
\begin{equation}
\label{p3 and p3-tilde}
P_2 = \tilde P_2 = {N^2 - 1 \over 4} , 
\quad 
P_3 = - {N^2 - 1 \over 4 N} , 
\quad \mbox{and}\quad 
\tilde P_3 = {(N^2 - 1)(N^2 - 2) \over 8 N} 
\end{equation}
For large $N$, the leading behavior is given by 
\begin{eqnarray}
\label{leading order traces}
&&
g^{a_1 ... a_k} \sim 
\left( 2^{k-3} \right) 
g^{a_1 a_2 c_3} g^{c_3 a_3 c_4} ... 
g^{c_{k-2} a_{k-2} c_{k-1} } g^{c_{k-1} a_{k-1} a_k} 
\\ \noalign{\noindent and} 
&&
P_{2 k + 1} \sim - {N k \over 4^k} , \quad 
P_{2 k} \sim {N^2 \over 4^k} ; \quad 
\tilde P_{k} \sim {N^k \over 2^k} 
\end{eqnarray}
Dependence on $N$ is%
\footnote{
	Note that the way the recursion formulae 
	(\ref{eq:recursion p})  
	work out 
	together with the initial values (\ref{p3 and p3-tilde}),
	leading order 
	large $N$ results are accurate to order $\OO(N^{-2})$ 
	and not to $\OO(N^{-1})$, as one could have thought naively. 
	}
very different for $P_k$ and $\tilde P_k$; 
in fact, taking generators in reverse order in the second trace
(such as in $\tilde P_k$) grows the fastest with $k$, and taking
them in the same order (as in $P_k$), the slowest.

Here are a few more identities we may have a need for in calculating
two-point functions. First, the normalizations of $SU(N)$ generators 
in an arbitrary representation is defined in terms of a constant $C(r)$ as 
\begin{equation}
\tr\! {}_r \; T^a_r T^b_r = C(r) \delta^{ab}
\end{equation}
and there is a quadratic Casimir, 
\begin{equation}
T^c_r T^c_r = C_2(r) \; \bone 
\end{equation}
For the adjoint and fundamental representations, 
$C_2(\mbox{adj}) = N$, $C_2(\mbox{fund}) = (\mbox{$N^2 - 1 \over 2N$})$, 
$C(\mbox{adj}) = N$, $C(\mbox{fund}) = \half$. 
Then, for example, 
\begin{eqnarray}
T^a T^a T^b T^b &=& \left[C_2(r) \right]^2 \; \bone \\
T^a T^b T^b T^a &=& \left[C_2(r) \right]^2 \; \bone \\
\left[T^a , T^b \right] \left[T^a , T^b \right] &=& 
- N C_2(r) \; \bone \\
T^a T^b T^a T^b &=& 
C_2(r) \left( C_2(r) - \mbox{$N\over2$} \right) \; \bone  
\end{eqnarray}
(we have omitted the label ``$r$'' on the generators, e.g. $T^a = T^a_r$).  
Longer expressions are just a little more complicated but not by much: 
\begin{eqnarray}
\tr 
T^b \left[ T^a , T^b \right] T^c \left[ T^a , T^c \right] 
&=& \quarter N^2 (N^2 - 1) C(r) \\
\tr 
T^a T^b T^c T^a T^c T^b 
&=& (N^2 - 1) C(r) (C_2(r) - \mbox{$N\over2$})^2 \\
\tr 
T^a T^b T^c T^a T^b T^c 
&=& (N^2 - 1) C(r) (C_2(r) - \mbox{$N\over2$}) (C_2(r) - N) 
\quad\quad
\end{eqnarray}
In particular, the last expression vanishes in the adjoint representation.

Using the fact that $U(N)_C = Gl(N,C)$, 
any $N \times N$ matrix $A$ can be decomposed 
into generators (in the fundamental) 
of $SU(N)$ plus the unit matrix: 
\begin{equation}
\label{u(n):completeness}
A = 
\left( 2 \tr A t^c \right) t^c 
+ 
\left( \mbox{$1\over N$} \tr A \right) \bone 
\end{equation}
Then, for example, 
we can write down the ``trace merging formula''
\begin{equation}
\label{merging traces}
2 \left( \tr A t^c \right) \left( \tr B t^c \right) 
= 
\tr A B - 
\mbox{$1\over N$} \left( \tr A \right) \left( \tr B \right) 
\end{equation}
and we can arrive at an even simpler recursion relations for 
$\tilde P_k$: 
\begin{eqnarray}
\tilde P_{k+1} &=& 
\left( \tr t^{a_1} ... t^{a_k} t^c \right) 
\left( \tr t^{a_k} ... t^{a_1} t^c \right) 
\nonumber\\&=& 
\half \left( \tr t^{a_1} ... t^{a_k} t^{a_k} ... t^{a_1} \right) 
- 
\mbox{$1\over 2N$} 
\left( \tr t^{a_1} ... t^{a_k} \right) 
\left( \tr t^{a_k} ... t^{a_1} \right) 
\nonumber\\&=& 
\mbox{$N\over 2$} (\mbox{$N^2 - 1 \over 2N$})^k
- \mbox{$1\over 2N$} \tilde P_k 
\end{eqnarray}
with $\tilde P_1 = 0$. 
(Naturally, this gives the same values for $\tilde P_k$ as before.)

Another useful relation satisfied by the generators of $SU(N)$ 
in the fundamental, is 
\begin{equation}
\label{double-line}
(t^a)_{ij} (t^a)_{kl} = \half \left( 
\delta_{il} \delta_{jk} 
- \mbox{$1\over N$}
\delta_{ij} \delta_{kl} 
\right) 
.
\end{equation}
Using this identity, one can easily reproduce 
the trace merging formula (\ref{merging traces}), 
as well as the expressions (\ref{p3 and p3-tilde}) 
for $P_2$, $\tilde P_2$, $P_3$ and $\tilde P_3$.

%% file: BPS-systematic.tex
\chapter{Systematics of Quarter BPS operators in $\NN =4$ SYM}
\label{chapter:systematic}


In this Chapter we use the machinery of (4,1,1) harmonic superspace
to describe 1/4 BPS operators. The use of extended supersymmetry
dramatically simplifies the counting and construction of scalar
composite operators in the $[q,p,q]$ representations of the
R-symmetry group $SU(4)$.

The construction of all 1/4 BPS operators in the fully interacting
quantum theory is carried out as follows. First, in the classical
interacting theory, a basis is produced of all the scalar
operators in the representations of the R-symmetry group $SU(4)$
suitable for 1/4 BPS operators with Dynkin labels $[q,p,q]$,
$q \ge 1$ and with classical dimension $p+2q$. In the classical
interacting theory there are natural candidates for 1/4 BPS
operators as well as other operators which can be identified as
descendants. The basis may be regrouped into operators of these
two types. Harmonic superspace techniques reduce this construction
down to elementary group theory. The (4,1,1) superspace notation
also makes distinguishing candidate 1/4 BPS operators from
descendant operators simple.

After subtracting off all the descendant
pieces we will recover protected operators. 
These 1/4 BPS primaries are the same protected operators 
as the ones found in Chapter \ref{chapter: BPS: 2pt}, 
but the method described here 
is more elegant and efficient.

\section{$\NN =4$ SYM in harmonic superspace}

The leading component fields of the 1/4 BPS multiplets in $\NN =4$
SCFT are given by scalar fields which transform under the internal
symmetry group $SU(4)$  in representations with Dynkin labels of
the form  $[q,p,q]$. The complete supermultiplets can be very
simply described in harmonic superspace, and we briefly recall how
this construction works.

For $\NN$ extended supersymmetry in four dimensions, 
$(\NN ,{\mathrm p},{\mathrm q})$ 
harmonic superspace is obtained from ordinary
Minkowski superspace $M$ by the adjunction of a compact manifold
of the form $K \equiv H \backslash SU(\NN )$ where $H=S(U({\mathrm p})\xz
U(\NN -({\mathrm p}+{\mathrm q}))\xz U({\mathrm q}))$. This construction
allows one to construct p projections of the $\NN$ supercovariant
derivatives $D_{\a i}$ and q projections of their conjugates $\bar
D_{\adt}^i$ which mutually anticommute.  We can therefore define
generalized chiral or G-analytic superfields (G for Grassmann) in
such superspaces which are annihilated by these derivatives. Now
the superfields in harmonic superspace will also depend on the
coordinates of $K$, and as this space is a complex manifold, they
can be analytic in the usual sense (H-analytic) in their
dependence on these coordinates.  As the internal manifold is
compact H-analytic fields will have finite harmonic expansions.
The Lorentz scalar superfields which are both G-analytic and
H-analytic,  which we shall refer to as analytic, are the fields
we are interested in. They can be shown to carry short irreducible
unitary representations of the superconformal group (provided that
they transform under irreducible representations of $H$). 
For original papers and detailed accounts of harmonic and analytic
superspaces, see for example 
\cite{gikos, harthowe, superspace, DP99}.

The \NN=4 extended formulation is an on-shell formulation, 
so it is not appropriate for loop computations. 
This is just as well, 
since will use it only for classifying operators, 
which is a question of kinematics, and not dynamics.

\subsection{(4,1,1) Harmonic Superspace}

For the 1/4 BPS operators in $\NN=4$ the most appropriate harmonic
superspace has $(\NN,{\mathrm p},{\mathrm q})=(4,1,1)$. Since the analytic
fields in this space will be annihilated by one $D$ and one $\bar
D$ it follows that they will only depend at most on $3/4$ of the
odd coordinates of $M$.  Instead of working directly on the coset
defined by the isotropy group $H=S(U(1)\xz U(2)\xz U(1))$ we shall
follow the standard practice of working on the group $SU(4)$ which
amounts to the same thing provided that all the fields have their
dependence on $H$ fixed. We denote an element of $SU(4)$ by
$u_I{}^i$ and its inverse by $(u^{-1})_i{}^I$. The group $H$ is
taken to act on the capital index $I$ which we decompose as
$I=(1,r,4), \ r\in\{2,3\}$, while $SU(4)$ acts on the small
indices $i,j,\ldots$. Using $u$ and its inverse we can convert
$SU(4)$ indices into $H$ indices and vice versa. Thus we can
define
\beqa
D_{\a I} &\equiv& u_I{}^i D_{\a i} = (D_{\a 1},D_{\a r}, D_{\a 4})
\nonumber \\
\bar D_{\adt}^I &\equiv& \bar D_{\adt}^i (u^{-1})_i{}^I = 
(\bar D_{\adt}^1, \bar D_{\adt}^r, \bar D_{\adt}^4)
\eeqa
Clearly, we have $\{D_{\a 1},\bar D_{\bdt}^4\}=0$. To
differentiate in the coset space directions we use the
right-invariant vector fields on $SU(4)$ which we denote by
$D_I{}^J$,  and which satisfy $\bar D^I{}_J=-D_J{}^I$ and
$D_I{}^I=0$.  These derivatives obey the Lie algebra relations of
$su(4)$ and act on $u_K{}^k$ by
\be D_I{}^J u_K{}^k= \d_K{}^J u_I{}^k -\qu\d_I{}^J u_K{}^k \ee
The basic differential operators on $SU(4)$ can be divided into
three sets: the derivatives $(D_1{}^1,D_{r}{}^s,D_4{}^4)$
correspond to the isotropy group, 
while the derivatives $(D_1{}^r,
D_1{}^4, D_r{}^4)$ can be thought of as essentially the components
of the $\bar\del$ operator on $K$ and the derivatives $(D_r{}^1,
D_4{}^1,  D_4{}^r)$ are the complex conjugates of these. Note that
the derivatives $(D_1{}^r, D_1{}^4, D_r{}^4)$ commute with $D_{\a
1}$ and $\bar D_{\adt}^4$. G-analytic fields are annihilated by
$D_{\a 1}$ and $\bar D_{\adt}^4$, H-analytic fields are
annihilated by $(D_1{}^r, D_1{}^4, D_r{}^4)$ and analytic fields
are annihilated by both of these sets of operators.

The $\NN=4$ Yang-Mills theory is described in Minkowski superspace
by a scalar superfield $W_{ij}=-W_{ji}$ which transforms under the
six-dimensional representation of $SU(4)$ and also under the
adjoint representation of the gauge group which we take to be
$SU(N)$. It is real in the sense that $\bar W^{ij}=\half \e^{ijkl}
W_{kl}$. This superfield satisfies the constraints
\bea \nab_{\a i} W_{jk}&=&\e_{ijkl} \L_{\a}{}^l
\no \\
\bar \nab_{\adt}^i W_{jk}&=& 2\d^i_{[j}\bar\L_{\adt k]} \eea
where $\L$ is a superfield whose leading component is the spinor
field of the multiplet.and where $\nab_{\a i}$ is a spinorial
derivative which is covariant with respect to the gauge group.
Using the superspace Bianchi identities one can show that
the only other independent spacetime component of $W$ is the
spacetime Yang-Mills field strength and that all of the component
fields satisfy their equations of motion.

In $(4,1,1)$ superspace we can define the superfield $W_{1r}\equiv
u_1{}^i u_r{}^j W_{ij}$. Using the properties outlined above one
can easily show that
\be
\nab_{\a 1} W_{1r}=\bar\nab_{\adt}^4 W_{1r}=0
\label{eq:delta W = 0}
\ee
and that the derivatives  $(D_1{}^r, D_1{}^4, D_r{}^4)$  all
annihilate $W_{1r}$, so that $W_{1r}$ is a covariantly analytic
field on $(4,1,1)$ harmonic superspace.  However, if we consider
gauge-invariant products of $W$'s, i.e. traces or multi-traces,
the resulting objects will be analytic superfields; they will be
annihilated by $D_{\a 1}$ and $\bar D_{\adt}^4$ rather than the
gauge-covariant versions. These are the superfields which we shall
use to construct the 1/4 BPS states. To make the formulas less
cluttered we shall abbreviate%
\footnote{
    So the index $r$ takes the values $r= 2, 3$ for $W_{1r}$,
    and $r = 1,2$ for $W_r$.
    }
$W_{1r}$ to $W_{r-1}$ and define
$W^r\equiv \e^{rs} W_s$.

\subsection{Quarter BPS Operators}

These superfields are easy to describe. The superfield
corresponding to the representation $[q,p,q]$ contains $p+2q$
powers of $W$ in the representation $p$ of $SU(2)$ (this is the
$SU(2)$ in the isotropy group), i.e. it has $p$ symmetrized
$SU(2)$ indices. If $q=0$ the single trace operators are the
chiral primaries which are 1/2 BPS. These operators we will refer
to as CPOs and denote by $A_p$,
\be A_{r_1\ldots r_p}\equiv \tr (W_{(r_1}\ldots W_{r_p)}) \ee
The lowest CPO is the stress-tensor multiplet $T_{rs}=A_{rs}$. We
can obtain further 1/2 BPS operators by taking products of CPOs
and  symmetrizing on all of the $SU(2)$ indices. The 1/4 BPS
operators (for $q > 0$) fall into two classes. There are operators
that can be constructed as products of the CPOs with at least one
pair of contracted indices, for example \be T_{rs} T^{rs} ,\quad
A_{rst} A^{st} ,\quad A_{rst} A^t{}_{uvw} ,\quad \mbox{and so on.}
\ee These operators have no commutators in their definition, and
so are the candidate 1/4 BPS operators, up to subtleties which we
shall come to in due course. Operators in the other class have at
least one single-trace factor in which the indices of two or more
pairs of $W$'s are contracted, as in \be \tr W^2 W^2 ,\quad
A_{rst} \; \tr W^2 W^2 ,\quad \tr W_r W_s W^2 W^2 ,\quad
\mbox{etc.} \ee where
 \be
W^2\equiv W_r W^r=\e^{rs} W_r W_s =\half \, \e^{rs} [W_r,W_s]
 \ee
These operators are descendants; the superspace Bianchi identities
imply that
\be \e^{\a\b}\nab_{\a i} \nab_{\b j} \bar W^{kl}= 2\d_j{}^{[k}
[W_{im}, \bar W^{l]m}] \ee
From this formula and its conjugate one can see that $W^2$ can be
written as
 \be
 W^2= (\nab_1)^2 \bar W^{14}=-(\bar\nab^4)^2 W_{14}
 \label{eq:delta = W2}
 \ee
where $(\nab_1)^2\equiv \half \e^{\a\b}\nab_{\a 1}\nab _{\b 1}$,
$(\bar\nab^4)^2\equiv - \half \e^{\adt\bdt}
\bar\nab_{\adt}^4\bar\nab_{\bdt}^4$, and where $W_{14}=u_1{}^i
u_4{}^j W_{ij}$. Given a product of $W_r$'s containing a factor of
$W^2$, therefore, the latter can be written in terms of
derivatives as above and the derivatives can be taken to act on
the whole expression with $W^2$ replaced by either $W_{14}$ or its
conjugate. This follows by G-analyticity. Indeed, if there are two
factors of $W^2$ in an operator then all four derivatives can be
brought outside. This is because
 \be
 \nab_{\a 1} W_{14}=0\qquad {\mathrm {and}}\qquad
 \bar\nab_{\adt}^4 \bar W^{14}=0
 \label{eq:delta4 = 0}
 \ee
In fact, the descendant 1/4 BPS operators always have at least two
factors of $W^2$ so that they can be written explicitly as
derivatives of long operators by these means. However, we are not
quite finished yet because the ancestor operators will not be
H-analytic on $(4,1,1)$ harmonic superspace as they stand. This
can be remedied by noting that
\be
(\nab_1)^2(\bar \nab^4)^2 \bar W^{1r}=[W^2,W^r]
\label{eq:delta^4 W}
\ee
with the aid of which we can write, for example, $\tr(W^2 W^2)$ as
\bea
 \label{eq:konishi descendant}
 \tr(W^2W^2)&=&-{1\over3}(\nab_1)^2(\bar \nab^4)^2 \left( \tr(W_{14}\bar
W^{14})+\tr(W_{1r}\bar W^{1r})\right)\nn\\
&=&-{1\over12} (\nab_1)^2(\bar \nab^4)^2 \tr(W_{ij}\bar
W^{ij})\nn\\
&=&-{1\over12} (D_1)^2(\bar D^4)^2 \tr(W_{ij}\bar W^{ij}) \eea
with $D_{\a 1}D_{\b 1}=\e_{\a\b}(D_1)^2$, and $\bar D_{\adt}^
4\bar D_{\bdt}^4=-\e_{\adt\bdt}(\bar D^4)^2$. Hence we see that
$\tr(W^2 W^2)$ is a descendant of the Konishi operator
$K\equiv\tr(W_{ij}\bar W^{ij})$. In general, one can use
(\ref{eq:delta W = 0}), (\ref{eq:delta = W2}),
(\ref{eq:delta4 = 0}) and (\ref{eq:delta^4 W}) to find
\bea
 \label{eq:konishi-like general descendant}
{1\over 2} (\nab_1)^2(\bar \nab^4)^2
\; W_{ij} A \bar W^{ij}
=
W_r A [W^2, W^r] + [W^2, W^r] A W_r - 2 W^2 A W^2
\eea
provided $A$ involves only the $W_r$.
And since we are dealing with gauge invariant operators,
we can replace the $\nab$ by $D$.
We note for future use that each of the descendants that we
consider below can be written as an ancestor superfield acted on
by the differential operator $(D_1)^2(\bar D^4)^2$.

On the other hand, the CPOs themselves cannot be obtained by
differentiation from other operators and so the candidate 1/4 BPS
operators cannot be (entirely) descendants. An operator
annihilated by $D_1$ and $\bar D^4$ can be either a
$(D_1)^2(\bar D^4)^2$ descendant of a long primary; or a
$(D_1)^2$ or $(\bar D^4)^2$ descendant of a 1/8 BPS primary;
or a 1/4 BPS primary. In Chapter \ref{chapter: BPS: 2pt} we saw that a
$[q,p,q]$ scalar composite operator can not be a descendant of a
1/8 BPS primary. Therefore, we argue that after subtracting off
all the descendant pieces from candidate BPS operators, we should
be left with a 1/4 BPS primary; it simply can not be anything
else!

\subsection{Examples of systematic description}
\label{section:systematic description}

We begin by outlining a few rules that determine which tensor
structures are permitted.
First we observe that, since contractions are made using the
antisymmetric tensor $\e^{rs}$ while the tensors $A_{rs ... t}$
are symmetric, contractions within the same $A$ give zero,
\begin{eqnarray}
A^r{}_{r s ... t} = 0 ,
\end{eqnarray}
so we can only contract indices in different $A$'s.

Next, consider $T^2$. Since $T_{rs}$ transforms under the
3-dimensional representation of $SU(2)$ it follows that the
product of two $T$s will decompose into the five and
one-dimensional representations. The former corresponds to
symmetrization on all four indices, i.e. $T_{(rs}T_{uv)}$, while
for a single contraction we have
 \be
 \label{eq:T-T:contract:general}
 T_{r t} T_s{}^t=-T_{s t} T_r{}^t=\half\e_{rs}T_{uv}T^{uv}
 \ee
Similarly, the product of three $T$s contains only the seven-and
three-dimensional representations, so that, for example
 \be
 \label{eq:T-T:contract}
 T_r{}^s T_s{}^t T_t{}^r=0
 \ee

One can look at contractions of other $A$'s in a similar fashion.
For example, $(A_3)^2$ contains only the seven- and
three-dimensional representations of $SU(2)$ corresponding to
tensors obtained by symmetrising on six or two indices with zero
or two contractions respectively. So
 \be
 \label{eq:A3-A3:contract}
 A^{rst} A_{rst}=A_{(rs}{}^v A_{tu)v}=0
 \ee
while
 \be
 A_{r}{}^{tu}A_{s tu}=A_{s}{}^{tu}A_{r tu}
 \ee
or, equivalently,
\begin{eqnarray}
\label{eq:A3-A3:contract:general} 3 A_{rst} A^t{}_{uv} &=& \e_{r
u} ( A_{s t w} A^{t w}{}_v ) + \e_{r v} ( A_{s t w} A^{t w}{}_u )
+ \e_{s v} ( A_{r t w} A^{t w}{}_u ) + \e_{s u} ( A_{r t w} A^{t
w}{}_v ) . 
\nonumber\\
\end{eqnarray}

These equations  generalise in a straightforward manner. Whenever
we contract an odd number of indices in two $A$s of the same
length and symmetrize on the
remaining indices, the resulting tensor vanishes.%
\footnote{
    In general there will be more than one nonvanishing structure.
    For instance, both
    $A_{(rs}{}^{vw} A_{tu)vw}$ and
    $A^{rstu} A_{rstu}$
    are independent nonvanishing tensors.
} If $A$s of different length are contracted (as in $A_{rs}{}^t
T_{tu}$), there is no such restriction.

We shall now discuss some explicit examples. We shall use the
convention that uncontracted $SU(2)$ indices are understood to be
totally symmetrized.

\subsubsection*{The Representation $[1,p,1]$}

Such operators have to have $(p+2)$ $W_r$'s and only one
contraction. There are no single trace operators in this class
because
 \be
 \tr(W_{r_1}\ldots W_{r_p}W^2)=0
 \ee
Hence these operators can only be constructed by contracting CPOs.
They are all protected. This can also be seen from representation
theory because there are no long representations which contain
these representations \cite{Dolan:2002zh}. This result also shows
that any single-trace factor in an operator must have at least two
contractions.

Here we list the lowest dimensional examples of $[1,p,1]$
representations (for $p \le 5$). For [1,1,1] and [1,2,1] we can
not construct any nonvanishing tensors of this form. For [1,3,1],
there is one possible operator,
 \bea
 \label{eq:ops:131}
 \cO&=& A_{rst} T^{t}{}_{u}
 \eea
Similarly, for [1,4,1] the only possible operator is
 \bea
 \label{eq:ops:141}
 \cO&=& A_{rstu} T^{u}{}_{v}
 \eea
Higher representations offer more choices, and already in the
[1,5,1] we find
 \bea
 \label{eq:ops:151}
 \cO_1&=& A_{rstuv} T^{v}{}_{w}
\no \\
 \cO_2&=& A_{rstv} A^{v}{}_{uw}
\no \\
 \cO_3&=& T_{rs} T_{tv} A^{v}{}_{uw}
 \eea
All of these operators have protected two-point functions, as we
will explicitly verify in Section \ref{section:1p1:details}.

\subsubsection*{The Representation [2,0,2]}

This operator is realized as an $SU(2)$ scalar in $(4,1,1)$
harmonic superspace. There are just two possibilities
 \bea
 \label{eq:ops:202}
 \cO_1&=& T_{rs} T^{rs}
\no \\
 \cO_2 &=& \tr(W^2 W^2)
 \eea
Using the rules outlined in the beginning of Section
\ref{section:systematic description}, we see that $\cO_1$ is the
only multiple trace operator one can construct with two pairs of
contracted indices, and there is also no other choice for the
single trace operator but $\cO_2$. $\cO_1$ is a candidate 1/4 BPS
operator while $\cO_2$ is a descendant; as seen in
(\ref{eq:konishi descendant}), 
it is a descendant of the Konishi operator.

\subsubsection*{The Representation [2,1,2]}

This has 5 fields and forms an $SU(2)$ doublet. There are again
only two possibilities
 \bea
 \cO_1&=& A_{rst} T^{st}
\no \\
 \cO_2&=& \tr(W_r W^2 W^2)
 \eea
This case is completely parallel to the [2,0,2] representation.

The operator $\cO_2$ can be written in the form
 \be
 \cO_2=-{1\over16} (D_1)^2(\bar D^4)^2 \,\tr(W_r W_{ij} \bar W^{ij})
 \ee
Note that the ancestor here is defined on (4,1,1) harmonic
superspace; it is not G-analytic but it is H-analytic. One can
easily remove the harmonic variables to obtain the corresponding
superfield on ordinary superspace. In this case it is
$\tr(W_{ij}W_{kl}\bar W^{kl})$.

\subsubsection*{The Representation [2,2,2]}

This has 6 fields and transforms as a triplet under $SU(2)$.
Multiple trace operators are constructed in the following way. We
can partition the set of six fields as 6 = 4 + 2, 6 = 3 + 3, or 6
= 2 + 2 + 2. Two pairs of indices are contracted, and two
remaining indices are symmetrized. The possibilities are
 \bea
 \label{eq:222:fund}
 \cO_1&=& A_{rstu} T^{tu}
\no \\
 \cO_2&=& A_{r}{}^{tu} A_{stu}
\no \\
 \cO_3&=& T_{rs} T_{tu} T^{tu}
\no \\
 \cO_4&=& \tr(W_r W_s W^2 W^2)
\no \\
 \cO_5&=& \tr(W_r W^2 W_s W^2)
\no \\
 \cO_6&=& \tr (W^2 W^2) T_{rs}
 \eea
The first three are candidate 1/4 BPS operators while the last
three are descendants. For the partitions 6 = 4 + 2 and 6 = 3 + 3,
these are the only choices because contractions within the same
$A$ give zero. For the partition 6 = 2 + 2 + 2, equation
(\ref{eq:T-T:contract}) relates any other triple-trace [2,2,2]
operator to $\cO_3$.

The operator $\cO_6$ is a descendant of the product of the 
Konishi operator and the supercurrent, while $\cO_4$ and $\cO_5$ are
descendants of the operators
 \bea
 \cA_1&=&\tr(W_r W_s W_{ij}\bar W^{ij})
 \no\\
 \cA_2&=&\tr(W_r W_{ij} W_s\bar W^{ij})
 \eea
A short calculation yields
 \bea
 \cO_4&=&{1\over40}(D_1)^2(\bar D^4)^2(\cA_2-3 \cA_1)
 \no\\
 \cO_5&=&{1\over20}(D_1)^2(\bar D^4)^2(\cA_1- 2 \cA_2)
 \eea
In terms of ordinary superfields, both $\cA_1$ and $\cA_2$ are in
the $[0,2,0]$ representation of $SU(4)$.

\subsubsection*{The Representation [2,3,2]}

The possibilities are
 \bea
 \label{ops:232}
 \cO_1&=& A_{rstuv} T^{uv}
\no \\
 \cO_2&=& A_{rs}{}^{uv} A_{tuv}
\no \\
 \cO_3&=& A_{rst} T_{uv} T^{uv}
\no \\
 \cO_4&=& A_r{}^{uv} (T^2)_{stuv}
\no \\
 \cO_5&=& \tr(W_r W_s W_t W^2 W^2)
\no \\
 \cO_6&=& \tr(W_r W_s W^2 W_t W^2)
\no \\
 \cO_7&=& T_{rs}\tr(W_t W^2 W^2)
\no \\
 \cO_8&=& A_{rst}\tr(W^2W^2)
 \eea
The first four are candidate 1/4 BPS operators while the second
four are descendants.

The last of these is a descendant of a product of $A_3$
and the Konishi operator, while the ancestor of $\cO_7$ is a
product of $T$ and $\tr(W_r W_{ij}\bar W^{ij})$. For the other two
descendants we have
 \bea
 \cO_5&=&{1\over24}(D_1)^2(\bar D^4)^2(\cA_2-2\cA_1)
 \no\\
 \cO_6&=&{1\over24}(D_1)^2(\bar D^4)^2(\cA_1-2\cA_2)
 \eea
where
 \bea
 \cA_1&=&\tr(W_rW_sW_t W_{ij}\bar W^{ij})
 \no\\
 \cA_2&=&\tr(W_rW_s W_{ij}W_t\bar W^{ij})
 \eea

\subsubsection*{The Representation [3,0,3]}

There is only one possibility: \bea \cO &=& \tr(W^2 W^2 W^2) \eea
This operator is a descendant. There are no candidate 1/4 BPS
operators in this case. As we saw in (\ref{eq:T-T:contract}) and
(\ref{eq:A3-A3:contract}), the operators $A_{rst} A^{rst}$ and
$T_{rs} T_t{}^r T^{st}$ vanish identically. Explicitly, this
operator can be written as
 \be
 \cO=-{1\over8}(D_1)^2(\bar D^4)^2\tr(W^2 W_{ij}\bar W^{ij})
 \ee

\subsubsection*{The Representation [3,1,3]}

This example again has seven fields but the representation of
$SU(2)$ is the doublet. The operators are
 \bea
 \label{ops:313}
 \cO_1&=& A_r{}^{stu} A_{stu}
\no \\
 \cO_2&=& (T^2)_{rstu} A^{stu}
\no \\
 \cO_3&=& \tr(W_r W^2 W^2 W^2)
\no \\
 \cO_4&=& \tr(W_r W^2 W_s W^2 W^s - W_r W_s W^2 W^s W^2)
\no \\
 \cO_5&=& T_r{}^s \tr(W_s W^2 W^2)
 \eea
so there are 3 descendants in this case. We have symmetrized
$\cO_4$ so that $\cO_4^\dagger = + (\cO_4)^*$. This symmetry
amounts to charge conjugation on the fields $X$ in the adjoint
representation of the gauge group and the ${\mathbf 6}$ of $SU(4)$.
Its effect on the $\NN=1$ superfield formulation is to map $z_j
\to z_j ^t$.

The last operator is again a descendant of a product of operators
that we have discussed previously. For the other two we have
 \bea
 \cO_3&=&{1\over30}(D_1)^2(\bar D^4)^2(\cA_2-5\cA_1)
 \no\\
 \cO_4&=&{1\over6} (D_1)^2(\bar D^4)^2(\cA_1+\cA_2)
 \eea
where
 \bea
 \cA_1&=&\tr(W_r W^2 W_{ij}\bar W^{ij})
 \no\\
 \cA_2&=&\tr(W_r W_s W_{ij} W^s\bar W^{ij})
 \eea

\subsection{Multiplicity of quarter BPS operators}

The (classical) quarter BPS operators in the $SU(4)$
representation are built from single trace quarter (and half) BPS
operators. These have the form \be
 \tr (W_{(r_1}\dots W_{r_q)}
(W^2)^p) \ee for operators in the $[p,q,p]$ $SU(4)$
representation, but the order of the $2p+q$ operators inside the
trace is arbitrary.

To find the number of different single trace operators in this
representation, $N_{pq}$, consider the reducible operator
\be X_Q:=\tr(W_{r_1}\dots W_{r_Q}), \ee
where the $SU(2)$ indices are no longer taken to be symmetrised.
This is in a reducible representation of $SU(2)$, and contains all
single trace scalar composite operators of dimension $Q$.
So one obtains the number of
operators in each representation by
expanding this operator as a sum of irreducible representations.
For example to find all single trace
operators of dimension 4 consider $X_4$. This has 6 components
(given by (1111), (1112), (1122), (1212), (2221), (2222)
where $(r_1 r_2 r_3 r_4)$
is short hand for $\tr(W_{r_1}\dots W_{r_4})$.) In
terms of irreducible $SU(2)$ representations it splits as $6=5+1$.
In terms of $SU(4)$ representations the 5 corresponds to $[0,4,0]$
and the 1 corresponds to $[2,0,2]$, and so we find that there is
only one operator in each of these two representations.

More generally, to split $X_Q$ into irreducibles, consider the
components of $X_Q$. Let $c(Q,p)$ denote the number of components
of $X_Q$ with $p$ 1's and $Q-p$ 2's, i.e. the number of ways to
arrange a total of $Q$ objects with $p$ of one type and $Q-p$ of
another type up to circular permutations.

Then $X_Q$ splits into the following irreducible representations:
\be \sum_{p=0}^{\lfloor Q/2 \rfloor} \left(c(Q,p) - c(Q,p-1)
\right) \ [p,q,p] \ee where $q=Q-2p$ and where $\lfloor x \rfloor$
denotes the largest integer less than or equal to $x$. So the
number of single trace operators in the $[p,q,p]$ representation
is \be N_{pq}=c(Q,p)-c(Q,p-1).\la{mult} \ee In general the formula
for $c(Q,p)$ is quite complicated, but in certain cases it
simplifies. For example
\be c(Q,0)=1
,\quad
c(Q,1)=1
,\quad
c(Q,2)=\lfloor {Q /2} \rfloor
,
\ee
and if $Q$ and $p$ are co-prime
then \be
 N_{pq}= {1 \over Q} {Q \choose p}.
\ee

As an example, consider dimension 6 operators: $X_6$ has 14
components and $c(6,p)$ is given by:
\be c(6,0)=c(6,1)=1
,\quad
c(6,2)=3
,\quad
c(6,3)=4.
\ee
Then~\eq{mult} gives
\be N_{06}=1
,\quad
N_{14}=0
,\quad
N_{22} = 2
,\quad
N_{30}=1
,
\ee
reproducing the
correct numbers of single-trace operators discussed above (in
particular there are two operators in the $[2,2,2]$ representation
and 1 in the $[3,0,3]$.)

Since multiple trace operators can be obtained by multiplying
together single trace operators, to find the number of multi-trace
operators in a given representation one just has to consider all
possible ways of obtaining the representation in question from
tensor products of other representations and use the formula for
single-trace operators.

\subsection{Relationship between $\NN=4$ and $\NN=1$ superfields}

The map between quarter BPS operators in the $\NN=1$ formalism and
those in $(4,1,1)$ analytic superspace is straightforward. In the
$\NN=1$ formalism the quarter BPS operators are given by
\be \left[ (z^{2c})^p (z_d)^q \right] \ee
where $[\dots ]$ denote gauge invariant combinations,
$(X_a)^p\equiv X_{(a_1}\dots X_{a_p)}$ and $z^{2c}\equiv z_a z_b
\e^{abc}$. Here the $a,b,\dots = (1,2,3)$ are $SU(3)$ indices.
These operators have highest weight state given by
\be \left[ (z_{1} z_{2}-z_{2} z_{1} )^p (z_1)^q \right]. \ee
In $(4,1,1)$ harmonic superspace on the other hand, this object is
given by
\be \left[ (W^2)^p  (W_r)^q  \right]. \ee
If we relabel the $SU(2)$ indices $r,s,\dots =1,2$ then this
operator has highest weight state
\be \left[ (W_1 W_2-W_2 W_1)^p  (W_1)^q  \right]. \ee
The correspondence between the $N=1$ operators and the harmonic
superspace operators is now clear, one simply replaces $W$ with
$z$ to obtain the highest weight states of each.

\section{Explicit Computations}
\label{section: systematic: explicit}

In this section we will explicitly calculate two point functions
of the above operators. We will work with the lowest components of
superfields, the $z_i^a$ and $\bar z_i^a$. (Here, $i=1, ... , 3$,
and $a$ labels the adjoint representation of the gauge group
$SU(N)$.) We list operators in a given irrep of the R-symmetry
group $SU(4)$, 
and for the descendant operators write out the corresponding
Konishi-like long operator they come from. Then we look at Born
level and order $g^2$ contributions to the two point functions of
the highest weight state operators.

In each representation we will have the descendant operators $L_i$
and ``candidate 1/4-BPS'' operators $\OO$. We will compute the
order $g^0$ two point functions $\langle O L_i^\dagger
\rangle_{\mathrm {Born}}$ and $\langle L_i L_j^\dagger \rangle_{\mathrm
{Born}}$. Then we will consider operators 
\bea \tilde{\OO} \equiv
\OO - \langle O L_i^\dagger \rangle_{\mathrm {Born}} \biggl (\langle L
L^\dagger \rangle_{\mathrm {Born}}^{-1} \biggr )^{ij} L_j \eea 
By construction, they are orthogonal to all the $L_i$ at Born level,
$\langle \tilde{\OO} L_i^\dagger \rangle_{\mathrm {Born}} = 0$. Then we
will show that these operators $\tilde{\OO}$ have protected two
point functions at order $g^2$, $\langle \tilde{\OO} L_i^\dagger
\rangle_{g^2} = 0$ and $\langle \tilde{\OO}' \tilde{\OO}^\dagger
\rangle_{g^2} = 0$ for all such operators $\tilde{\OO}, \tilde
{\OO}'$. The claim is that these operators $\tilde{\OO}$ are
1/4-BPS.

The basis of operators we will choose is slightly different from
the one used in Chapter \ref{chapter: BPS: 2pt}. The operators introduced in
the preceding section are more natural and intuitive.

\subsubsection*{The representation $[1,p,1]$}
\label{section:1p1:details}

$ \bullet $ There is only one operator in the representation
$[1,3,1]$ whose highest $SU(4)$ weight state is \bea \OO &\equiv&
\tr z_1 z_1 \; \tr z_1 z_1 z_2 - \tr z_1 z_2 \; \tr z_1 z_1 z_1
\eea while acting on $\OO$ once with an $SU(4)$ ladder operator
gives \bea \OO' &\equiv& 2 \, \tr z_1 z_1 \; \tr z_1 z_2 z_2 - \tr
z_1 z_2 \; \tr z_1 z_1 z_2 - \tr z_2 z_2 \; \tr z_1 z_1 z_1 \eea
This operator has the same weight as the [2,1,2] operators (but is
of course orthogonal to them). The Born and order $g^2$ overlaps
are \bea \label{eq:131:born} \langle \OO' \OO'{}^\dagger
\rangle_{\mathrm {Born}} &=& {15 \over 32} N (N^2 - 1) (N^2 - 4) , \quad
\langle \OO' \OO'{}^\dagger \rangle_{g^2} = 0 .
\end{eqnarray}
So indeed it is a 1/4-BPS operator.

$\bullet$ There is only one operator in the representation
$[1,4,1]$.  The highest $SU(4)$ weight state operator is \bea \OO
&\equiv& \tr z_1 z_1 \; \tr z_1 z_1 z_1 z_2 - \tr z_1 z_2 \; \tr
z_1 z_1 z_1 z_1 \eea while acting on $\OO$ once with an $SU(4)$
ladder operator gives \bea \label{eq:141:ops} \OO' &\equiv& 2 \,
\tr z_1 z_1 \; \tr z_1 z_1 z_2 z_2 + \tr z_1 z_1 \; \tr z_1 z_2
z_1 z_2 \nonumber\\&& - 2 \, \tr z_1 z_2 \; \tr z_1 z_1 z_1 z_2 -
\tr z_2 z_2 \; \tr z_1 z_1 z_1 z_1 \eea This operator has the same
weight as the [2,2,2] operators (but is of course orthogonal to
them). The Born and order $g^2$ overlaps are \bea
\label{eq:141:born} \langle \OO' \OO'{}^\dagger \rangle_{\mathrm {Born}}
&=& {3 \over 8} (N^2 - 1) (N^2 - 4) (N^2 - 9) , \quad \langle \OO'
\OO'{}^\dagger \rangle_{g^2} = 0 . \eea So indeed it is a 1/4-BPS
operator.

$\bullet$ Finally, there are 3 operators in the representation
$[1,5,1]$. Their highest $SU(4)$ weight state operators are \bea
\OO_1 &\equiv& \tr z_1 z_1 \; \tr z_1 z_1 z_1 z_1 z_2 - \tr z_1
z_2 \; \tr z_1 z_1 z_1 z_1 z_1
\\
\OO_2 &\equiv& \tr z_1 z_1 z_1 \; \tr z_1 z_1 z_1 z_2 - \tr z_1
z_1 z_2 \; \tr z_1 z_1 z_1 z_1
\\
\OO_3 &\equiv& \tr z_1 z_1 \left( \tr z_1 z_1 \; \tr z_1 z_1 z_2 -
\tr z_1 z_2 \; \tr z_1 z_1 z_1 \right) \eea while acting on $\OO$
once with an $SU(4)$ ladder operator gives \bea \OO_1' &\equiv& 2
\, \tr z_1 z_1 \; \tr z_1 z_1 z_1 z_2 z_2 + 2 \, \tr z_1 z_1 \;
\tr z_1 z_1 z_2 z_1 z_2 \nonumber\\&& - 3 \, \tr z_1 z_2 \; \tr
z_1 z_1 z_1 z_1 z_2 - \tr z_2 z_2 \; \tr z_1 z_1 z_1 z_1 z_1
\\
\OO_2' &\equiv& 2 \, \tr z_1 z_1 z_1 \; \tr z_1 z_1 z_2 z_2 + \tr
z_1 z_1 z_1 \; \tr z_1 z_2 z_1 z_2 \nonumber\\&& - \tr z_1 z_1 z_2
\; \tr z_1 z_1 z_1 z_2 - 2 \, \tr z_1 z_2 z_2 \; \tr z_1 z_1 z_1
z_1
\\
\OO_3' &\equiv& 2 \, \tr z_1 z_1 \; \tr z_1 z_1 \; \tr z_1 z_2 z_2
+ \tr z_1 z_1 \; \tr z_1 z_2 \; \tr z_1 z_1 z_2 \nonumber\\&& - 2
\, \tr z_1 z_2 \; \tr z_1 z_2 \; \tr z_1 z_1 z_1 - \tr z_2 z_2 \;
\tr z_1 z_1 \; \tr z_1 z_1 z_1 \eea These operators have the same
weight as the [2,3,2] operators (but are of course orthogonal to
them). The Born and order $g^2$ overlaps are \bea
\label{eq:151:born} \langle \OO_i' \OO_j'{}^\dagger \rangle_{\mathrm
{Born}} &=& \mbox{$ {35 (N^2 - 1) (N^2 - 4) \over 128 N} $}
\nonumber\\&& \hspace{-3em}\times \left(\matrix{ N^4 - 10 N^2 + 72
& -11 N^2 + 36     & 6 N (N^2 - 2)      \cr
                  & N^4 - 4 N^2 + 18 & - 2 N ( 2 N^2 + 3) \cr
                  &                & 2 N^2 (N^2 + 5)
}\right) , \quad
\\
\langle \OO_i' \OO_j'{}^\dagger \rangle_{g^2} &=& 0 . \eea So
indeed they all are 1/4-BPS operators.

\subsubsection*{The Representation [2,0,2]}
\label{section:202:details}

The operators corresponding to (\ref{eq:ops:202}) are \bea
\label{ops:202} \OO_1 &=& 2 \left( \tr z_1 z_1 ~ \tr z_2 z_2 - \tr
z_1 z_2 ~ \tr z_1 z_2 \right)
\no \\
\OO_2 &=& \tr z_1 z_1 z_2 z_2 - \tr z_1 z_2 z_1 z_2 \eea 
The single trace operator $\OO_2$ 
is a descendant of the Konishi scalar 
(see equation \ref{konishi:descendant}), 
\begin{eqnarray}
\OO_2 \sim (Q^2 \bar Q^2 ) \; \tr z_j \bar z^j .
\label{eq:konishi:202} \eea
On the other hand, the operator orthogonal to $\OO_2$
at Born level
\begin{eqnarray}
\tilde {\OO}_1 = \OO_1 - {4 \over N} \OO_2
\label{eq:bps:202}
\eea
stays orthogonal to $\OO_2$,
$\langle \tilde {\OO}_1(x) \bar {\OO}_2(y) \rangle = 0$;
and its two-point function is protected at order $g^2$, 
$\langle \tilde {\OO}_1(x) \bar {\tilde{\OO}}_1(y) \rangle = \langle
\tilde {\OO}_1(x) \bar {\tilde{\OO}}_1(y) \rangle_{\mathrm {Born}}$.

\subsubsection*{The Representation [2,1,2]}
\label{section:212:details}

In this representation we again have only two operators \bea \OO_1
&=& \tr z_1 z_1 z_1 ~ \tr z_2 z_2 - 2 \tr z_1 z_2 ~ \tr z_1 z_2
z_1 + \tr z_1 z_2 z_2 ~ \tr z_1 z_1
\no \\
\OO_2 &=& \tr z_1 z_1 z_1 z_2 z_2 - \tr z_1 z_1 z_2 z_1 z_2 \eea
and the single trace operator is again a descendant, \bea \OO_2
\sim (Q^2 \bar Q^2 )
\; \tr \left[ z_1 z_j \bar z^j + z_1 \bar z^j z_j \right]
\label{eq:konishi:212}
\eea
The operator orthogonal to $\OO_2$ at Born level
\bea \tilde {\OO}_1 = \OO_1 - {6 \over N} \OO_2
\label{eq:bps:212}
\eea
satisfies $\langle \tilde {\OO}_1(x) \bar {\OO}_2(y) \rangle =
0$, $\langle \tilde {\OO}_1(x) \bar {\tilde{\OO}}_1(y) \rangle =
\langle \tilde {\OO}_1(x) \bar {\tilde{\OO}}_1(y) \rangle_{\mathrm {Born}}$
at order $g^2$.

\subsubsection*{The Representation [2,2,2]}
\label{section:222:details}

Here we have a total of six operators. 
The lowest components of superfields
(\ref{eq:222:fund}) are
\begin  {eqnarray}
\label{eq:222:ops} \OO_1 &\equiv& 3 \, \tr z_1 z_1 z_1 z_1 \; \tr
z_2 z_2 - 6 \, \tr z_1 z_1 z_1 z_2 \; \tr z_1 z_2 
\nonumber\\&& \hspace{10em}
+ \left( 2 \,
\tr z_1 z_1 z_2 z_2 + \tr z_1 z_2 z_1 z_2 \right) \, \tr z_1 z_1
\hspace{-2em}
\no \\
\OO_2 &\equiv& \tr z_1 z_1 z_1  ~\tr z_1 z_2 z_2 - \tr z_1 z_1 z_2
~\tr z_1 z_1 z_2
\no \\
\OO_3 &\equiv& \tr z_1 z_1 \left( \tr z_1 z_1 ~\tr z_2 z_2 - \tr
z_1 z_2 ~\tr z_1 z_2 \right)
\no \\
\OO_4 &\equiv& \tr z_1 z_1 z_1 z_1 z_2 z_2 - 2 \, \tr z_1 z_1 z_1
z_2 z_1 z_2 + \tr z_1 z_1 z_2 z_1 z_1 z_2
\no \\
\OO_5 &\equiv& \tr z_1 z_1 z_1 z_2 z_1 z_2 - \tr z_1 z_1 z_2 z_1
z_1 z_2
\no \\
\OO_6 &\equiv&
\tr z_1 z_1 \left( \tr z_1 z_1 z_2 z_2 - \tr z_1 z_2 z_1 z_2 \right)
\end    {eqnarray}
(and we didn't bother to keep the same normalization factors for
all of them --- just whatever looks better). The descendants arise
from the Konishi-like long primary operators as \bea \OO_4 &\sim&
(Q^2 \bar Q^2 )
\; \tr \left[ z_1 z_1 z_j \bar z^j + z_1 z_1 \bar z^j z_j \right]
\no \\
\OO_5 &\sim& (Q^2 \bar Q^2 )
\; \tr \left[ z_1 z_j z_1 \bar z^j \right]
\no \\
\OO_6 &\sim& (Q^2 \bar Q^2 )
\; \left[ \tr z_1 z_1 \right] \; \left[ \tr z_j \bar z^j \right]
\label{eq:konishi:222} \eea
Note that another operator exists in a
long multiplet, whose descendant coincides with $\OO_6$,
\bea
\OO_6 \sim (Q^2 \bar Q^2 )
\; \left[ \tr z_1 z_j \right] \; \left[ \tr z_1 \bar z^j \right]
\eea This may be established by observing
that the difference operator, \bea \label{eq:bianchi:protected} 3
\left[ \tr z_1 z_j \right] \; \left[ \tr z_1 \bar z^j \right] -
\left[ \tr z_1 z_1 \right] \; \left[ \tr z_j \bar z^j \right]
\end{eqnarray}
is semi-short\footnote{This is the non-renormalised 20' operator 
discussed in~\cite{AF, 20', Bianchi:2002rw}}.
A similar phenomenon occurs for higher representations,
and we will not mention it explicitly.

The linear combinations orthogonal to these operators
at Born level can be taken
as \bea \label{eq:bps:222} \tilde {\OO}_1 &=& \OO_1 - {24 \over N}
\OO_4 - {48 (2 N^2 - 3) \over N (3 N^2 - 2)} \OO_5 + {40 \over 3
N^2 - 2} \OO_6
\no \\
\tilde {\OO}_2 &=& \OO_2 - {4 \over N} \OO_4 - {3 (7 N^2 - 8) \over
N (3 N^2 - 2)} \OO_5 + {5 \over 3 N^2 - 2} \OO_6
\no \\
\tilde {\OO}_3 &=& \OO_3 \hspace{3.6em} - {20 \over 3 N^2 - 2} \OO_5
- {10 \over 3 N^2 - 2} \OO_6 \eea The matrix of two point
functions in this basis is \bea \label{eq:2-pt:222} \left(\matrix{
\langle \tilde{\OO}_i \tilde{\OO}_j^\dagger \rangle_{\mathrm {Born}} & 0 \cr
0 & \langle L_i L_j^\dagger \rangle_{\mathrm {Born}} }\right) +
\left(\matrix{ 0 & 0 \cr 0 & \langle L_i L_j^\dagger \rangle_{g^2}
}\right) \eea where the (symmetric) blocks are 
\begin{small}
\bea
\label{eq:222:components} \langle L_i L_j^\dagger \rangle_{\mathrm
{Born}} = \mbox{$ {(N^2 - 1) \over 64} $} \left(\matrix{ 7 N^4 + 20
N^2 + 8 & -4 N^4 -10 N^2 +4 & 2 N (13 N^2 - 2) \cr
                   & 3 N^4 + 2              & - 2 N (6 N^2 + 1) \cr
                   &                        & 2 N^2 (3 N^2 + 13)
}\right) 
\hspace{-1em}
\nonumber\\
\eea 
\end{small}
and 
\begin{small}
\bea \langle L_i L_j^\dagger \rangle_{g^2} =
\mbox{$ {3 \tilde B N (N^2 - 1) \over 32} $} \left(\matrix{ 25 N^4
+ 148 N^2 + 8 & 15 N^4 - 66 N^2 + 4    & 4 N (27 N^2 + 17) \cr
                     & 10 N^4 + 22 N^2 + 2 & - 2 N (28 N^2 + 13) \cr
                     &                        & 2 N^2 (9 N^2 + 89)
}\right)
\hspace{-2.3em}
\no \\
\eea 
\end{small}
and \bea \langle \tilde{\OO}_1 \tilde{\OO}_1^\dagger \rangle_{\mathrm
{Born}} & = & {360 ~\cC_N \over N^2 (3 N^2 - 2)} \times
 (N^6 - 11 N^4 + 70 N^2 - 48)
\no \\
\langle \tilde{\OO}_1 \tilde{\OO}_2^\dagger \rangle_{\mathrm {Born}} & = &
{-120 ~\cC_N \over N^2 (3 N^2 - 2)} \times
 (5 N^4 - 36 N^2 + 24)
\no \\
\langle \tilde{\OO}_1 \tilde{\OO}_3^\dagger \rangle_{\mathrm {Born}} & = &
{240 ~\cC_N \over N (3 N^2 - 2)} \times
 (N^2 - 2) (2 N^2 - 3)
\no \\
\langle \tilde{\OO}_2 \tilde{\OO}_2^\dagger \rangle_{\mathrm {Born}} & = &
{5 ~\cC_N \over N^2 (3 N^2 - 2)} \times (3 N^6 - 41 N^4 + 160 N^2
- 96)
\no \\
\langle \tilde{\OO}_2 \tilde{\OO}_3^\dagger \rangle_{\mathrm {Born}} & = &
{-20 ~\cC_N \over N (3 N^2 - 2)} \times
  (13 N^2 - 12)
\no \\
\langle \tilde{\OO}_3 \tilde{\OO}_3^\dagger \rangle_{\mathrm {Born}} & = &
{60 ~\cC_N \over  (3 N^2 - 2)} \times
 (N^2 + 1) (N^2 - 2)
.
\eea
Here and below we shall use the abbreviation,
\bea
\label{eq:constant:cN:defined} \cC_N  \equiv
(N^2 -1 ) (N^2 -4) / 64
.
\eea

As seen from (\ref{eq:2-pt:222}), the operators defined in
(\ref{eq:bps:222}) have protected two-point functions $\langle
\tilde{\OO}_i \tilde{\OO}_j^\dagger \rangle$ at order $g^2$. This
shows that we can argue that $\tilde {\OO}_1$, $\tilde {\OO}_2$,
$\tilde {\OO}_3$ are 1/4-BPS. Anomalous
scaling dimensions of long operators ($L_i = \OO_4, \OO_5, \OO_6$)
match those of their Konishi-like primaries computed in
\cite{Bianchi:2002rw}.

\subsubsection*{A Better basis for protected [2,2,2] operators}

It may seem odd that the operators mixing of the operators in the
representations $[2,0,2]$ and $[2,1,2]$ are in terms of
coefficients that are merely inverse powers of $N$, while the
mixing coefficients for the operators we identified in the
representation $[2,2,2]$ have more complicated denominators. This
distinction would also be surprising from the perspective of
AdS/CFT, since the more complicated denominators would suggest
that an infinite series of corrections in the string coupling $g_s
= \lambda /N$ would appear for given `t Hooft coupling $\lambda$.
As a matter of fact, the mixing coefficients depend upon the bases
chosen for both the $\OO_1$, $\OO_2$ and $\OO_3$ operators as well
as the pure descendants. In a different basis, the coefficients
are all proportional to inverse powers of $N$. For the
representation $[2,2,2]$, these new operators are found easily,
and we have \bea
\tilde {\OO}_1' &=& \tilde {\OO}_1 + {4 \over N} \tilde {\OO}_3 = \OO_1
+ {4 \over N} \OO_3 - {24 \over N} \OO_4 - {32 \over N} \OO_5
\no \\
\tilde {\OO}_2' &=& \tilde {\OO}_2 + {1 \over 2 N} \tilde {\OO}_3 =
\OO_2 - {4 \over N} \OO_4 - {7 \over N} \OO_5
\no \\
\tilde {\OO}_3' &=& \tilde {\OO}_3 + {2 \over 3 N} \tilde {\OO}_1 - {4
\over N} \tilde {\OO}_2 = \OO_3  + {2 \over 3 N} \OO_1 - {4 \over N}
\OO_2 - {10 \over 3 N} \OO_6 \label{eq:bps:222:better} \eea 
In this new basis, the matrix of 2-pt functions now reads 
\begin{small}
\bea
\label{eq:222:components:better} \langle \tilde{\OO}_i'
\tilde{\OO}_j'{}^\dagger \rangle_{\mathrm {Born}} {=}
{5 \cC_N \over N^2} \left(\matrix{ 24 (N^4 + 3 N^2 + 32) & -16
(N^2 + 9)       & 16 N (4 N^2 - 1) \cr
                      & (N^2 - 3) (N^2 + 9) & - 2 N (N^2 - 9)  \cr
                      &                     & 4 (N^4 + 17 N^2 - 24)
}\right)
\hspace{-2.2em}
\no \\
\eea
\end{small}

\subsubsection*{The Representation [2,3,2]}
\label{section:232:details}

Here we have 8 operators. 
The operators corresponding to the basis
(\ref{ops:232}) are 
\begin{small}
\bea \OO_1 &\equiv& 2 \, \tr z_1 z_1 z_1 z_1
z_1 \; \tr z_2 z_2 - 4 \, \tr z_1 z_1 z_1 z_1 z_2 \; \tr z_1 z_2
\no \\ && \hskip .5in + \left( \tr z_1 z_1 z_1 z_2 z_2 + \tr z_1
z_1 z_2 z_1 z_2 \right) \tr z_1 z_1
\no \\
\OO_2 &\equiv& 3 \, \tr z_1 z_1 z_1 z_1 \; \tr z_1 z_2 z_2 - 6 \,
\tr z_1 z_1 z_1 z_2 \; \tr z_1 z_1 z_2 \no \\ && \hskip .5in +
\left( 2 \, \tr z_1 z_1 z_2 z_2 + \tr z_1 z_2 z_1 z_2 \right) \tr
z_1 z_1 z_1 \quad\quad
\no \\
\OO_3 &\equiv& \tr z_1 z_1 z_1 \left( \tr z_1 z_1 \; \tr z_2 z_2 -
\tr z_1 z_2 \; \tr z_1 z_2 \right)
\no \\
\OO_4 &\equiv& \tr z_1 z_1 z_1 \; \tr z_1 z_1 \; \tr z_2 z_2 -2 \,
\tr z_1 z_1 z_2 \; \tr z_1 z_1 \; \tr z_1 z_2
\nonumber\\&& \hskip .5in
+ \tr z_1 z_2 z_2 \; \tr z_1 z_1 \; \tr z_1 z_1
\hspace{-0.5em}
\no \\
\OO_5 &\equiv& \tr z_1 z_1 z_1 z_1 z_1 z_2 z_2 - 2 \, \tr z_1 z_1
z_1 z_1 z_2 z_1 z_2 + \tr z_1 z_1 z_1 z_2 z_1 z_1 z_2
\no \\
\OO_6 &\equiv& \tr z_1 z_1 z_1 z_1 z_2 z_1 z_2 - \tr z_1 z_1 z_1
z_2 z_1 z_1 z_2
\no \\
\OO_7 &\equiv& \left( \tr z_1 z_1 z_1 z_2 z_2 - \tr z_1 z_1 z_2
z_1 z_2 \right) \; \tr z_1 z_1
\no \\
\OO_8 &\equiv& \left( \tr z_1 z_1 z_2 z_2 - \tr z_1 z_2 z_1 z_2
\right) \; \tr z_1 z_1 z_1 \label{2-3-2:operators} 
\eea 
\end{small}
Out of
these, four are descendants, 
\begin{small}
\bea \OO_5 &\sim& (Q^2 \bar Q^2 )
\; \tr \left[ 2 z_1 z_1 z_1 z_j \bar z^j + 2 z_1 z_1 z_1 \bar z^j
z_j - z_1 z_1 z_j z_1 \bar z^j - z_1 z_1 \bar z^j z_1 z_j \right]
\no \\
\OO_6 &\sim& (Q^2 \bar Q^2 )
\; \tr \left[ z_1 z_1 z_1 z_j \bar z^j + z_1 z_1 z_1 \bar z^j z_j
- 2 z_1 z_1 z_j z_1 \bar z^j - 2 z_1 z_1 \bar z^j z_1 z_j \right]
\no \\
\OO_7 &\sim& (Q^2 \bar Q^2 )
\left[ \tr z_1 z_1 z_j \; \tr z_1 \bar z^j + \tr z_1 z_1 \bar z^j
\; \tr z_1 z_j - 12 \, \tr z_1 z_1 z_1 \; \tr z_j \bar z^j \right]
\no \\
\OO_8 &\sim& (Q^2 \bar Q^2 )
\left[ \tr z_1 z_1 z_1 \; \tr z_j \bar z^j \right]
\label{eq:konishi:232} \eea 
\end{small}
while the combinations orthogonal to
them at Born level
can be taken as 
\begin{small}
\bea \label{eq:bps:232} \tilde {\OO}_1 &=&
\OO_1 - {20 \over N} \OO_5 - {30 (N^2 - 2) \over N^3} \OO_6 + {15
(N^2 - 2) \over N^4} \OO_7 + {10 (N^2 + 2) \over N^4} \OO_8
\no \\
\tilde {\OO}_2 &=& \OO_2 - {30 \over N} \OO_5 - {30 (2 N^2 - 3)
\over N^3} \OO_6 + {15 (2 N^2 - 3) \over N^4} \OO_7 + {10 (N^2 +
3) \over N^4} \OO_8
\no \\
\tilde {\OO}_3 &=& \OO_3 \hspace{3.6em} - {12 \over N^2} \OO_6 - {3
(N^2 - 2) \over N^3} \OO_7 - {2 (N^2 + 2) \over N^3} \OO_8
\no \\
\tilde {\OO}_4 &=& \OO_4 \hspace{3.6em} - {18 \over N^2} \OO_6 - {(7
N^2 - 9) \over N^3} \OO_7 - {2 (2 N^2 + 9) \over 3 N^3} \OO_8 \eea
\end{small}

The matrix of two point functions in this basis is \bea
\label{eq:2-pt:232} \left(\matrix{ \langle \tilde{\OO}_i
\tilde{\OO}_j^\dagger \rangle_{\mathrm {Born}} & 0 \cr 0 & \langle L_i
L_j^\dagger \rangle_{\mathrm {Born}} }\right) + \left(\matrix{ 0 & 0 \cr
0 & \langle L_i L_j^\dagger \rangle_{g^2} }\right) \eea so indeed
at order $g^2$ the operators defined in (\ref{eq:bps:232}) have
protected correlators. This shows that we can argue that $\tilde
{\OO}_1$, $\tilde {\OO}_2$, $\tilde {\OO}_3$, $\tilde {\OO}_4$ are the
1/4-BPS primaries we are after. The (symmetric) blocks in equation
(\ref{eq:2-pt:232}) are 
\begin{small}
\bea \label{eq:232:components:begin}
\langle L_i L_j^\dagger \rangle_{\mathrm {Born}} = \mbox{$ \half N \cC_N
$} \left(\matrix{ 6 N^2 + 45 & -3 N^2 - 9 & 18 N         & 36 N
\cr
               & 2 N^2 - 3    & -4N          & - 15 N \cr
               &              &  4 N^2 + 24 & 36     \cr
               &              &              & 9 N^2 + 54
}\right) 
\eea 
\end{small}
and 
\begin{small}
\bea \langle L_i L_j^\dagger \rangle_{g^2} & = &
\mbox{$ 12 \tilde B N \cC_N $} \no \\ &&\times \left(\matrix{ 9 N
(2 N^2 + 27) & - 9 N (N^2 + 8) & 54 (N^2 + 2)   & 27 (5 N^2 + 6)
\cr
                 & N (5 N^2 + 17)  & -6 (3 N^2 + 2)  & -6 (10 N^2 + 3) \cr
                 &                  & 4 N (2 N^2 + 23) & 174 N         \cr
                 &                  &                  & 9 N (3 N^2 + 35)
}\right)
\hspace{-0.5em}
\no \\
\label{eq:232:components} 
\eea 
\end{small}
The Born level overlaps of
protected operators are given by ugly and not particularly
illuminating expressions. Here we list them for the sake of
completeness: \bea \label{eq:232:protected} \langle \tilde{\OO}_1
\tilde{\OO}_1^\dagger \rangle_{\mathrm {Born}} &=&
 30 ~\cC_N N^{-5} (N^8 - 10 N^6 + 117 N^4 - 720 N^2 + 420)
\no \\
\langle \tilde{\OO}_1 \tilde{\OO}_2^\dagger \rangle_{\mathrm {Born}} &=& -90
~\cC_N N^{-5} (4 N^6 - 69 N^4 + 395 N^2 - 210)
\no \\
\langle \tilde{\OO}_1 \tilde{\OO}_3^\dagger \rangle_{\mathrm {Born}} &=&
 90 ~ \cC_N N^{-4} (N^2 - 2) (N^4 - 7 N^2 + 14)
\no \\
\langle \tilde{\OO}_1 \tilde{\OO}_4^\dagger \rangle_{\mathrm {Born}} &=&
 30 ~ \cC_N N^{-4} (N^2 - 2) (N^2 - 9) (3 N^2 - 7)
\no \\
\langle \tilde{\OO}_2 \tilde{\OO}_2^\dagger \rangle_{\mathrm {Born}} &=&
 45 ~ \cC_N
 N^{-5} (N^8 - 29 N^6 + 328 N^4 - 1290 N^2 + 630)
\no \\
\langle \tilde{\OO}_2 \tilde{\OO}_3^\dagger \rangle_{\mathrm {Born}} &=&
-630 ~\cC_N  N^{-4} (N^2 - 1) (N^2 - 6)
\no \\
\langle \tilde{\OO}_2 \tilde{\OO}_4^\dagger \rangle_{\mathrm {Born}} &=&
 30 ~ \cC_N
 N^{-4} (N^2 - 1) (N^2 - 9) (2 N^2 - 21)
\no \\
\langle \tilde{\OO}_3 \tilde{\OO}_3^\dagger \rangle_{\mathrm {Born}} &=&
 9 ~ \cC_N N^{-3} (N^6 - N^4 - 16 N^2 + 56)
\no \\
\langle \tilde{\OO}_3 \tilde{\OO}_4^\dagger \rangle_{\mathrm {Born}} &=&
 6 ~ \cC_N N^{-3} (N^2 - 9) (N^4 + 3 N^2 - 14)
\no \\
\langle \tilde{\OO}_4 \tilde{\OO}_4^\dagger \rangle_{\mathrm {Born}} &=&
 2 ~\cC_N N^{-3} (N^2 - 9) (7 N^4 + 16 N^2 - 63)
\eea and the constant $\cC_N = (N^2 - 1) (N^2 - 4) / 64$ was
defined in (\ref{eq:constant:cN:defined}).

\subsubsection*{The Representation [3,1,3]}
\label{section:313:details}

Here there are 5 operators. 
The 
operators corresponding to the basis (\ref{ops:313}) are 
\bea
\OO_1 &\equiv& \tr z_1 z_1 z_1 z_1 ~\tr z_2 z_2 z_2 - 3 \, \tr z_1
z_1 z_1 z_2 ~\tr z_1 z_2 z_2 \no \\ && \hskip .5in + \left( 2 \,
\tr z_1 z_1 z_2 z_2 + \tr z_1 z_2 z_1 z_2 \right) \tr z_1 z_1 z_2
- \tr z_1 z_2 z_2 z_2 ~\tr z_1 z_1 z_1
\nonumber\\
\OO_2 &\equiv& \tr z_1 z_2 \left( 2 \, \tr z_1 z_2 ~\tr z_1 z_2
z_2 - \tr z_2 z_2 ~\tr z_1 z_1z_1 - 3 \, \tr z_1 z_1 ~\tr z_1
z_2z_2 \right) \no \\ && \hskip .5in + \tr z_1 z_1 \left( \tr z_2
z_2 ~\tr z_1 z_1 z_2 + \tr z_1 z_1 ~\tr z_2 z_2z_2 \right)
\no \\
\OO_3 &\equiv& \tr z_1 z_1 z_1 z_2 z_1 z_2 z_2 - \tr z_1 z_1 z_1
z_2 z_2 z_1 z_2
\no \\
\OO_4 &\equiv& 2 \, \tr z_1 z_1 z_1 z_1 z_2 z_2 z_2 - 3 \, \tr z_1
z_1 z_1 z_2 z_1 z_2 z_2 - 3 \, \tr z_1 z_1 z_1 z_2 z_2 z_1 z_2 \no
\\ && \hskip .5in + 2 \, \tr z_1 z_1 z_2 z_1 z_2 z_1 z_2 + 2 \,
\tr z_1 z_1 z_2 z_2 z_1 z_1 z_2
\no \\
\OO_5 &\equiv& - \left( \tr z_1 z_1 z_2 z_2 z_2 - \tr z_1 z_2 z_1
z_2 z_2 \right) \tr z_1 z_1 \no \\ && \hskip .5in + \left( \tr z_1
z_1 z_1 z_2 z_2 - \tr z_1 z_1 z_2 z_1 z_2 \right) \tr z_1 z_2
\end    {eqnarray}
Out of these three are descendants, \bea \OO_3 &\sim& (Q^2 \bar
Q^2 )
\; \tr \left[ z_1 z_1 z_2 z_j \bar z^j + z_1 z_1 z_2 \bar z^j z_j
- z_1 z_1 z_j \bar z^j z_2 - z_1 z_1 \bar z^j z_j z_2 \right]
\no \\
\OO_4 &\sim& (Q^2 \bar Q^2 )
\; \tr \left[ z_1 z_1 z_2 z_j \bar z^j + z_1 z_1 z_2 \bar z^j z_j
+ z_1 z_1 z_j \bar z^j z_2 + z_1 z_1 \bar z^j z_j z_2
\right.\nonumber\\&&\left.\hspace{6em} - 2 z_1 z_2 z_1 z_j \bar
z^j - 2 z_1 z_2 z_1 \bar z^j z_j
\right] \\
\OO_5 &\sim& (Q^2 \bar Q^2 )  \; \left[ 2 \, \tr z_1 z_1 z_j \;
\tr z_2 \bar z^j + 2 \, \tr z_1 z_1 \bar z^j \; \tr z_2 z_j - \tr
z_1 z_2 \bar z^j \; \tr z_1 z_j
\right.\nonumber\\&&\left.\hspace{4em} - \tr z_1 z_2 z_j \; \tr
z_1 \bar z^j - \tr z_2 z_1 \bar z^j \; \tr z_1 z_j - \tr z_2 z_1
z_j \; \tr z_1 \bar z^j \right] \quad\quad \label{eq:konishi:313}
\no \eea while the combinations orthogonal to them at Born level
can be taken as
\bea \label{eq:bps:313} \tilde {\OO}_1 &=& \OO_1 - {2 N \over N^2 -
2} \OO_4 - {5 \over N^2 - 2} \OO_5
\no \\
\tilde {\OO}_2 &=& \OO_2 + {8 \over N^2 - 2} \OO_4 + {10 N \over N^2
- 2} \OO_5 \eea 
(which are the operators found in Chapter \ref{chapter: BPS: 2pt}).
The matrix of two point functions in this basis is
\bea \label{eq:2-pt:313} \left(\matrix{ \langle \tilde{\OO}_i
\tilde{\OO}_j^\dagger \rangle_{\mathrm {Born}} & 0 \cr 0 & \langle L_i
L_j^\dagger \rangle_{\mathrm {Born}} }\right) + \left(\matrix{ 0 & 0 \cr
0 & \langle L_i L_j^\dagger \rangle_{g^2} }\right) \eea so indeed
at order $g^2$ the operators defined in (\ref{eq:bps:313}) have
protected correlators. This shows that we can argue that $\tilde
{\OO}_1$, $\tilde {\OO}_2$, are the 1/4-BPS primaries we are after.
The (symmetric) blocks in (\ref{eq:2-pt:313}) are \bea
\label{eq:313:components} \langle L_i L_j^\dagger \rangle_{\mathrm
{Born}} &=& \mbox{$
 \cC_N
$} \left(\matrix{ N (N^2 - 9) & 0              & 0        \cr
              & 15 N (N^2 + 3) & - 30 N^2 \cr
              &                & 3 N (N^2 + 6)
}\right)
\no \\
\langle L_i L_j^\dagger \rangle_{g^2} &=& \mbox{$ 6 \tilde B N
\cC_N $} \left(\matrix{ 5 N (N^4 - 9) & 0              & 0
\cr
                & 75 N (N^2 + 7) & 180 (N^2 + 1) \cr
                &                & 12 N (N^2 + 16)
}\right)
\no \\
\langle \tilde{\OO}_i \tilde{\OO}_j^\dagger \rangle_{\mathrm {Born}} &=& {15
(N^2 - 9) \over  N^2 (N^2 - 2)} ~ \cC_N \left(\matrix{ (N^2 - 1)
(N^2 - 4) & 4 N (N^2 - 1) \cr
                    & 2 N^2 (N^2 - 6)
}\right) \eea

As mentioned before, the operator $\OO_3$ 
has zero correlators with
everything else. The reason is that $\cO_3^\dagger = - (\cO_3)^*$, 
while for all other operators $\cO_i^\dagger = + (\cO_i)^*$.

\subsubsection*{Completeness of the Construction}

An important point which remains to be addressed is whether the
construction of the 1/4 BPS operators given above is exhaustive.
The fact that it is follows from $SU(4)$ group theory in the
following manner. Given a 1/4 BPS representation of $SU(4)$ of the
type $[q,p,q]$, one begins by listing all possible monomial scalar
composite operators built out of $(p+q)$ $z_1$'s
and $q$ $z_2$'s.%
\footnote{
    The remaining scalar field $z_3$ never enters
    in the highest weight of a 1/4 BPS representation, though it will
    also be needed when describing 1/8 BPS operators.
} These monomials form a basis for the linear space of scalar
composite operators of the form $[ (z_1)^{(p+q)} \, (z_2)^q ]$.
They can occur in representations $[0,p+2q,0]$, $[1,p+2q-2,1]$,
... , $[q,p,q]$. Then we have to show that the number of such
monomials matches the total number of operators we constructed in
these representations.

Let us illustrate how this works with an example. Consider the
[2,2,2] representation. The complete set of scalar composite
operators we can build out of 4 $z_1$'s and 2 $z_2$'s is
\begin  {eqnarray}
\label{eq:222:monomials} 6: && \tr z_1 z_1 z_1 z_1 z_2 z_2 , ~~
\tr z_1 z_1 z_1 z_2 z_1 z_2 , ~~ \tr z_1 z_1 z_2 z_1 z_1 z_2
\no \\
4+2: && \tr z_1 z_1 z_1 z_1 \; \tr z_2 z_2 , ~~ \tr z_1 z_1 z_1
z_2 \; \tr z_1 z_2 , ~~ \tr z_1 z_1 z_2 z_2 \; \tr z_1 z_1 ,
\no \\&& 
\tr z_1 z_2 z_1 z_2 \; \tr z_1 z_1
\no \\
3+3: && \tr z_1 z_1 z_1 \; \tr z_1 z_2 z_2 , ~~ \tr z_1 z_1 z_2 \;
\tr z_1 z_1 z_2
\no \\
2+2+2: && \tr z_1 z_1 \; \tr z_1 z_1 \; \tr z_2 z_2 , ~~ \tr z_1
z_1 \; \tr z_1 z_2 \; \tr z_1 z_2
\end    {eqnarray}
or the total of 11 operators. By taking linear combinations of
these, we can construct:

$\bullet$ 4 
tensors in the [0,6,0] corresponding
to the partitions of 6 in (\ref{eq:222:monomials});

$\bullet$ 1 tensor in the representation [1,4,1], given in
(\ref{eq:141:ops});

$\bullet$ 6 tensors in the representation [2,2,2], listed in
(\ref{eq:222:ops}).

\noindent
Thus there are no other scalar composite operators of the form $[
(z_1)^4 \, (z_2)^2 ]$.

In the same fashion, we can go through all other representations
we have considered in this Chapter and verify that we didn't leave
out any operators.

%% file: BPS-3pt.tex
\chapter{Three-Point Functions of Quarter BPS Operators in \NN=4 SYM}
\label{chapter: BPS: 3pt}


In Chapters \ref{chapter: BPS: 2pt} and \ref{chapter:systematic}, 
we constructed \quarter-BPS chiral primaries in the fully interacting 
\NN=4 SYM theory. 
In general, these operators are linear combinations of all 
local, polynomial, gauge invariant, scalar composite operators 
in the $[p,q,p]$ representations of the $R$-symmetry group $SU(4)$. 
The coefficients with which they all combine into 
operators with a well defined scaling dimension are quite involved. 
The \quarter-BPS chiral primaries, like the 
\half-BPS operators extensively studied in the literature, 
have protected 
two-point functions 
(at least at order $g^2$).

Here we investigate the (non-) renormalization 
properties of three-point correlators involving \quarter-BPS operators 
along with \half-BPS operators. 
Given the elaborate combinatorics of the problem, 
we concentrate on the following special cases. 
First, we discuss several group theoretic simplifications 
of the combinatorial factors multiplying the Feynman 
graphs that contribute to three-point functions of 
chiral primaries. 
Based on $SU(4)$ group theory and conformal invariance only, 
we argue that certain classes of such correlators 
are protected at order $g^2$, for 
all $N$. 
In particular, 
this allows us to compute $\OO(g^2)$ corrections 
to correlators of the form 
$\langle \OO_{\thalf} \OO_{\thalf} \OO_{\mathrm {BPS}} 
\rangle$, 
where $\OO_{\thalf}$ are two \half-BPS operators, and 
$\OO_{\mathrm {BPS}}$ is an arbitrary (\half-, \quarter-, or \eigth-BPS) 
chiral primary. 
Next, 
we look at the three-point functions 
$\langle \OO_{\thalf} \OO_{\tquarter} \OO_{\tquarter} 
\rangle$ and  
$\langle \OO_{\tquarter} 
\OO_{\tquarter} \OO_{\tquarter} 
\rangle$, 
also for general $N$, 
where $\OO_{\tquarter}$ are the $\Delta \le 7$ 
\quarter-BPS primaries found in 
Chapters \ref{chapter: BPS: 2pt} and \ref{chapter:systematic}. 
Then, we carry out a large $N$ analysis 
of 
$\langle \OO_{\thalf} \OO_{\tquarter} \OO_{\tquarter} 
\rangle$ 
correlators involving the special \quarter-BPS operators 
(mixtures of single and double trace scalar composite operators), 
for arbitrary $\Delta$. 
Also in the large $N$ limit, 
we identify the corresponding objects 
in the supergravity description, 
and compute the correlators on the AdS side of the correspondence. 

Finally, we make some speculations. 
Based on the broad range of special cases, 
we conjecture that two- and three-point functions 
of \half- and \quarter-operators receive no quantum corrections, 
for arbitrary $N$. 
Additionally, a set of group theoretic considerations 
discussed here 
extends straightforwardly from three-point functions 
to extremal correlators. 
Therefore, we suggest 
that extremal correlators involving 
\half- and \quarter-operators 
are protected as well.

\section{Contributing diagrams}
\label{section:contributing diagrams:3pt}


The two-point functions of the 
scalar composites discussed in 
Chapters \ref{chapter: BPS: 2pt} and \ref{chapter:systematic}, 
have the schematic form 
\begin{equation}
\label{eq:2-pt generic}
\langle 
\left[ {z_1}^{(p+q)} {z_2}^p \right](x) 
\left[ {\bar z_1}^{(p+q)} {\bar z_2}^p \right](y) \rangle 
\end{equation}
where $[...]$ are some gauge invariant combinations. 
The free field part of (\ref{eq:2-pt generic}) 
is given by a power of the free
correlator $[G(x,y)]^{(2p+q)}$, times a combinatoric factor. 
From the Lagrangian 
(\ref{lagrangian:for Feynman rules}) 
we can read off the structures
for the leading
correction to the propagator, and the four-scalar blocks, 
see Figure 
\ref{fig:four-scalar and propagator}.

\begin{figure}[t!]
{\begin{center}
\epsfig{width=2.4in, file=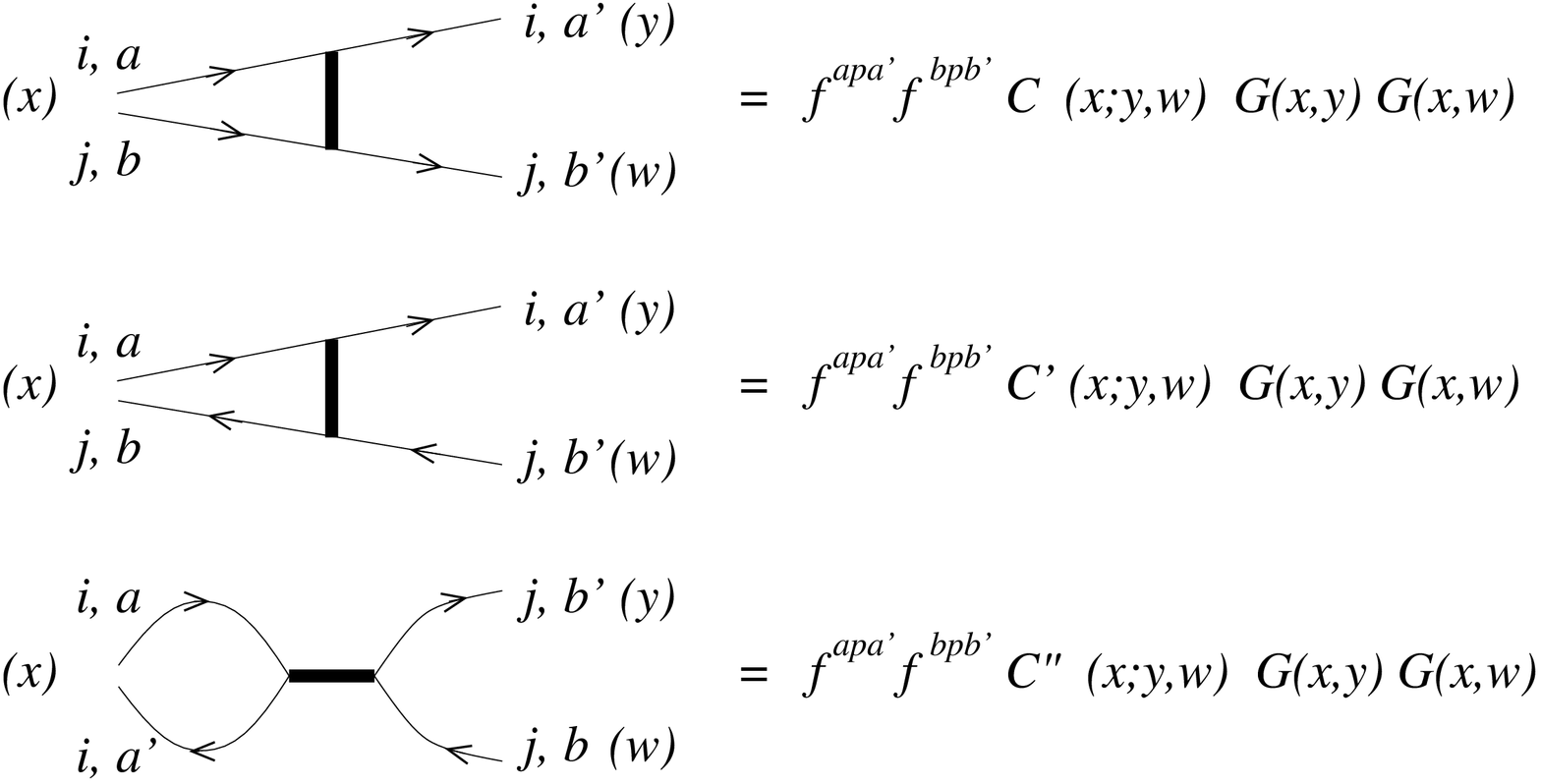, angle=-90}
\epsfig{width=1.5in, file=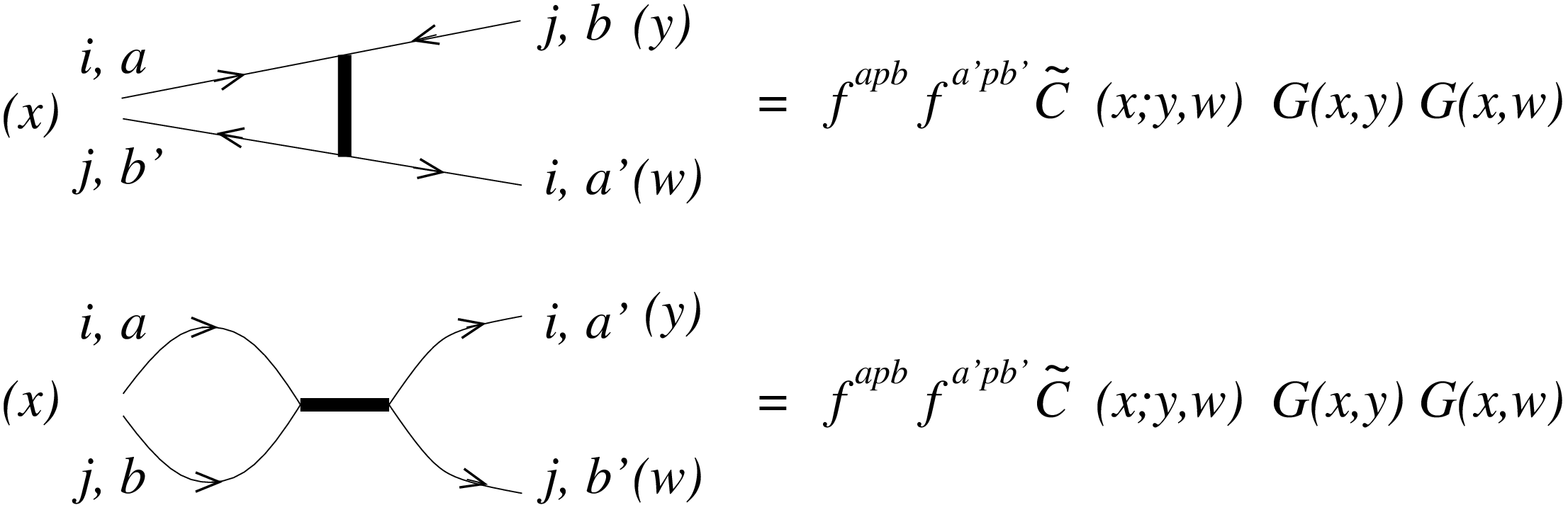, angle=-90}\quad${}\!\!$
\end{center}}
\caption{%
Building blocks for $g^2$ corrections to three-point functions. 
The three space-time points are $x$ (with two legs attached) and $y$ and $w$ 
(single leg each). 
\label {fig:three-pt-blocks}
}%
\end {figure}

Three-point functions to be considered in this Chapter 
are of the form 
\begin{equation}
\label{eq:3-pt generic}
\langle 
\left[ z^{k+l} \right](x) 
\left[ \bar z^{k+m} \right](y) 
\left[ \bar z^l z^m \right](w) 
\rangle 
\end{equation}
The free result is just 
the product of appropriate powers of 
free correlators $[G(x,y)]^k [G(x,w)]^l [G(y,w)]^m$. 
The same structures that contribute to the two-point functions 
at order $g^2$ 
(see Figure 
\ref{fig:four-scalar and propagator}), 
also contribute to the three-point functions (\ref{eq:3-pt generic}). 
Apart from these, 
there are new building blocks, shown in Figure \ref{fig:three-pt-blocks}. 
They have the same index structure, 
but are now functions of three 
space-time coordinates rather than two. 

Notice that the $F$-term corrections proportional to $\tilde B(x,y)$ 
in Figure 
\ref{fig:four-scalar and propagator}, 
and the last graph (proportional to $\tilde C(x;y,w)$)
in Figure \ref{fig:three-pt-blocks}, 
are antisymmetric in $i$ and $j$, hence they are absent when the 
scalars in the four legs have pairwise the same flavor. 
For the same reason, 
these corrections are also absent when 
the operator at point $x$ 
is symmetric in all of its flavor indices. 
In particular, this is the case when the operator 
at $x$ is \half-BPS.

\section{Restrictions from \NN=4 SUSY and gauge invariance}
\label{section:gauge-dependent:3pt}

The form of quantum corrections to two and three-point functions 
is known \cite{DFS}. 
Space-time coordinate dependence 
of the Feynman diagrams contributing to these 
correlators at order $\OO(g^2)$ 
is constrained, since all exchanged fields are massless. 
We know the parametric form of 
the functions $A(x,y)$, $B(x,y)$, $\tilde B(x,y)$; and 
$C(x;y,w)$, $C'(x;y,w)$, $C''(x;y,w)$, and $\tilde C(x;y,w)$, 
without having to perform integrals explicitly. 
Functions which depend on two space-time points, 
are of the form $A(x_1, x_2) = a \log x_{12}^2 \mu^2 + b$ 
with $x_{ij} \equiv x_i-x_j$; 
three-point contributions look like 
$C(x_1;x_2,x_3) = a' \log x_{12}^2 x_{13}^2 \mu^4 - a'' \log x_{23}^2 + b'$ 
(making use of the $x_2 \lra x_3$ symmetry of these building blocks).

\NN=4 SUSY tells us more. 
From non-renormalization of two and three-point functions 
of operators in the stress tensor multiplet, one can see 
\cite{DFS} that 
$B(x,y) = - 2 A(x,y)$, 
and 
$C'(0;x,y) + \tilde C(0;x,y) = -C(0;x,y)$; 
the authors of \cite{DFS} chose to combine these and call it just $C'$.%
\footnote{
	One way to see this is to consider the protected 
	correlators of [0,2,0] scalar composite operators 
	$\langle \tr z_1 z_2 (x) \; \tr \bar z_1 \bar z_2 (y) \rangle$, 
	and 
	$\langle [z^2](x) [\bar z^2](y) [z \bar z](w) \rangle$
	and
	$\langle [z \bar z](x) [z \bar z](y) [z \bar z](w) \rangle$. 
	}
The coefficients $a'$, 
$a''$ and $b'$ are determined%
\footnote{
	This follows from 
	$C(x;y,w) + C(y;x,w) + C(w;x,y) + A(x,y) + A(y,w) + A(x,w)  =0$. 
	}
in terms of $a$ and $b$: 
\begin{eqnarray}
A (x,0) ~=~ 
-\half B (x,0) 
&=& a \log {x^2 \mu^2} + b
\nonumber\\
- C (0;x,y) &=& a \log {x^2 y^2 \mu^2\over (x-y)^2} + b
\end{eqnarray}
Therefore, the net contribution 
to the three-point function (\ref{eq:3-pt generic}) 
of the $\OO(g^2)$ diagrams involving a gauge boson exchange 
(diagrams $A$, $B$, and $C$), 
is 
\begin{eqnarray}
\label{three-point:D-term}
\langle 
[z^k](x_1) [\bar z^l](x_2) [z^m \bar z^n](x_3) 
\rangle |_{(A+B+C)} 
\hspace{-15em}
\nonumber\\
&=& 
a 
(
c_g^{12} \log x_{12}^2 \mu^2 + 
c_g^{13} \log x_{13}^2 \mu^2 + 
c_g^{23} \log x_{23}^2 \mu^2 
)
+
b c_g^{123}
\end{eqnarray}
where and $c_g^{ij}$ and $c_g^{123}$ are 
some combinatorial coefficients.

Now we use gauge invariance of the theory. 
On the one hand, 
we observe that 
the coefficients $a$ and $b$ are gauge dependent 
as we saw in 
Chapter \ref{chapter: BPS: 2pt}, Section \ref{section:gauge-dependent}, 
\begin 	{eqnarray}
A(x,0) &=& 
\half \pi^2 g^2 \xi 
\left[ 
\log x^2 \mu^2 + \log 4\pi - \gamma 
\right]
+ (\mbox{$\xi$-independent}) 
\end 	{eqnarray}
where $\xi$ is the gauge fixing parameter. 
On the other hand, 
a correlator of gauge invariant operators 
can not depend on $\xi$. 
Therefore, the combinatorial coefficients 
multiplying $a$ and $b$ in equation (\ref{three-point:D-term}) 
must vanish, 
$c_g^{ij} = c_g^{123} = 0$. 
Hence, the $D$-term diagrams 
proportional to $A$, $B$, and $C$ all cancel; 
their net contribution 
to the three-point functions (\ref{eq:3-pt generic}) is zero. 

So just like in the case of two-point functions, 
we only have to consider the $F$-term graphs. 
They are proportional to $\tilde B$ and $\tilde C$, 
the only gauge independent diagrams around 
($C'=-(C+\tilde C)$ and $C''=C-\tilde C$ do not have to 
be treated separately as they are linear combinations 
of the other ones).


\section{Position dependence of $\tilde B$ and $\tilde C$}
\label{section:c-tilde}

Having shown that $D$-term corrections to three-point functions 
(\ref{eq:3-pt generic}) are absent, it remains to 
consider the $F$-term interactions. 
In this Section we derive a relation between 
functions $\tilde B$ and $\tilde C$, 
which will play a key role in the analysis of 
three-point functions of \quarter-BPS chiral primaries, 
see Section \ref{section:cat diagrams}.

Space-time position dependence of 
$\tilde B$ and $\tilde C$ 
(shown Figures 
\ref{fig:four-scalar and propagator} 
and 
\ref{fig:three-pt-blocks}) 
is parametrically determined to be 
$\tilde B(x,0) = \tilde a \log (x^2 \mu^2) + \tilde b$ and 
$\tilde C(0;x,y) = \tilde a' \log (x^2 y^2 \mu^4) 
- \tilde a'' \log ((x-y)^2 \mu^2) + \tilde c$. 
Furthermore, the leading divergent behavior can be read off 
from the integrals unambiguously, and so from the limit 
$\tilde C(0;x,y \to x)$ we infer $\tilde a' = \half \tilde a$.

To evaluate the remaining coefficients 
$\tilde a$, $\tilde a''$, and $\tilde b$, 
replace 
$1/x^2 \to 1/(x^2 + \e^2)$ for scalar propagators inside 
integrals.%
\footnote{
	This is the fastest way to calculate the 
	integral for $\tilde C$, 
	but one can obtain the same results 
	using dimensional regularization. 
	}
With this prescription 
\begin{eqnarray}
\label{b-tilde:again}
\tilde B (x,0) &\!\!=\!\!& 
- \quarter Y^2  
\int { (d^4 z) \left[ 4 \pi^2 x^2 \right]^2 \over 
\left[ 4 \pi^2 ((z-x)^2 + \e^2)\right]^2 \left[ 4 \pi^2 (z^2 + \e^2)\right]^2}
\nonumber\\&\!\!=\!\!& 
-Y^2  
{1 \over 32 \pi^2 } 
\left[ \log (x^2/\e^2) - 1\right] 
\\
\noalign{\noindent
is the regularized two-point function, 
while the three-point function becomes 
}
\label{c-tilde}
\tilde C (x;y,0) &\!\!=\!\!& 
- \quarter Y^2  
\int { (d^4 z) \left[ 4 \pi^2 x^2 \right] \left[ 4 \pi^2 (x-y)^2 \right] \over 
\left[ 4 \pi^2 ((z-x)^2 + \e^2)\right]^2 
\left[ 4 \pi^2 ((z-y)^2 + \e^2)\right]
\left[ 4 \pi^2 (z^2 + \e^2)\right] }
\nonumber\\&\!\!=\!\!& 
-Y^2  
{1 \over 64 \pi^2 } 
\left[ \log {x^2 (x-y)^2 \over y^2 \e^2}\right] 
\end{eqnarray}
(The numerators inside the integrals come about 
because of the powers of free scalar propagator 
in the definitions of $\tilde B$ and $\tilde C$, 
see Figures 
\ref{fig:four-scalar and propagator} 
and 
\ref{fig:three-pt-blocks}.) 
Hence, 
\begin{eqnarray}
\label{eq:cat diagrams}
\tilde C(x;y,0) + 
\tilde C(y;x,0) - 
\tilde B(x,y) = - Y^2 \times {1 \over 32 \pi^2 } 
\end{eqnarray}
is a nonzero constant 
(for \NN=4 SUSY, $Y^2 = 2 g^2$).
The value of this constant 
does not depend on the regulator $\e$. 
Also note that with the ``point splitting regularization'' 
one would get the incorrect result of vanishing constant 
in (\ref{eq:cat diagrams}).

\section{Structure of the three-point functions}
\label{section:3-pt structure}

With the results of Section 
\ref{section:gauge-dependent:3pt}
at hand, we can write down the form of a general 
three-point function of scalar composite operators 
(\ref{eq:3-pt generic}) to order $g^2$: 
\begin{eqnarray}
\label{eq:3-pt g^2}
\langle 
\left[ z^{k+l} \right] (x) 
\left[ \bar z^{k+m} \right] (y) 
\left[ \bar z^l z^m \right] (w) 
\rangle 
&\!\!=\!\!& 
G(x,y)^k 
G(x,w)^l 
G(w,y)^m 
\nonumber\\&&\hspace{-5.55em} \times
\Big( \Big. 
\alpha_{\mathrm {free}} 
+ \tilde \beta_{xy} \tilde B(x,y) 
+ \tilde \beta_{xw} \tilde B(x,w) 
+ \tilde \beta_{yw} \tilde B(y,w) 
\nonumber\\&&\hspace{-2em}
+ \tilde \gamma_{x} \tilde C(x;y,w) 
+ \tilde \gamma_{y} \tilde C(y;x,w) 
+ \tilde \gamma_{w} \tilde C(w;x,y) 
\nonumber\\&&\hspace{-2em}
+ \OO(g^4) 
\Big. \Big) 
\end{eqnarray}
where $\alpha_{\mathrm {free}}$, $\tilde \beta$-s and $\tilde \gamma$-s 
are some combinatorial coefficients. 
Using the expressions 
(\ref{b-tilde:again}) 
and (\ref{c-tilde}) 
from Section \ref{section:c-tilde}, 
we can determine the $\OO(g^2)$ position dependence of 
(\ref{eq:3-pt g^2}) completely 
--- if we know these combinatorial coefficients. 
Together with conformal invariance, 
and the $SU(4)$ symmetry properties 
of the operators in (\ref{eq:3-pt g^2}), 
we can often go a long way to figuring out 
which of the combinatorial coefficients must 
vanish, without doing any actual calculations.

\subsection{Space-time coordinate dependence}

Like in the case of two-point functions, 
conformal invariance restricts position 
dependence of three-point correlators of
pure operators (i.e. ones which have a well 
defined scaling dimension). 
Consider three (gauge invariant Lorentz scalar) 
operators $\OO_1$, $\OO_2$, and $\OO_3$, 
of dimensions 
$\Delta_i = k_i + \delta_i$, 
inserted at corresponding points $x_i$. 
Let $k_i$ be integers; and $\delta_i$, 
the order $g^2$ corrections to the scaling dimensions 
(which may or may not be zero). 
The three-point function 
$\langle \OO_1 \OO_2 \OO_3 \rangle$
is completely determined up to 
a multiplicative constant $C_{123} = C^0_{123} + C^1_{123}$ 
(where again $C^0_{123}$ is the free field result 
and $C^1_{123} \sim g^2$), 
\begin{eqnarray}
\label{eq:3-pt constraint}
\langle
\OO_1 (x_1) \OO_2 (x_2) \OO_3 (x_3) 
\rangle
&=& 
{
C_{123} 
\over 
x_{12}^{\Delta_1 + \Delta_2 - \Delta_3}
x_{13}^{\Delta_1 + \Delta_3 - \Delta_2}
x_{23}^{\Delta_2 + \Delta_3 - \Delta_1}
}
\nonumber\\
&=& 
\langle
\OO_1 \OO_2 \OO_3 
\rangle_{\mathrm {free}} 
\Big( \Big. 
1 
+ C^1_{123} / C^0_{123} 
\nonumber\\&&\quad 
- \delta_1 \log {x_{12}^2 x_{13}^2 \over x_{23}^2 \e^2}
- \delta_2 \log {x_{21}^2 x_{23}^2 \over x_{13}^2 \e^2}
- \delta_3 \log {x_{31}^2 x_{32}^2 \over x_{12}^2 \e^2}
\nonumber\\&&\quad 
+ \OO(g^4) 
\Big. \Big) 
\end{eqnarray}
with $x_{ij} = x_i - x_j$ as usual.

Suppose that all three operators have 
protected scaling dimensions, $\delta_i = 0$. 
Then (\ref{eq:3-pt constraint}) reduces to 
\begin{eqnarray}
\label{eq:3-pt constraint:all protected}
\langle
\OO_1 (x_1) \OO_2 (x_2) \OO_3 (x_3) 
\rangle
= 
\langle
\OO_1 \OO_2 \OO_3 
\rangle_{\mathrm {free}} 
\Big( \Big. 
1 
+ C^1_{123} / C^0_{123} 
\Big. \Big) 
\end{eqnarray}
and no logs arise. In this case, the 
combinatorial factors in (\ref{eq:3-pt g^2}) 
satisfy 
\begin{eqnarray}
\label{eq:coeffs - all three:beg}
\tilde \gamma_x 
&=& 
- \left( 
\tilde \beta_{xy} + \tilde \beta_{xw} 
\right) 
\\
\tilde \gamma_y 
&=& 
- \left( 
\tilde \beta_{xy} + \tilde \beta_{yw} 
\right) 
\\
\tilde \gamma_w 
&=& 
- \left( 
\tilde \beta_{xw} + \tilde \beta_{yw} 
\right) 
\label{eq:coeffs - all three:end}
\end{eqnarray}
and we need to calculate only three coefficients 
(the $\tilde \beta$-s, for example), 
to find all the $\OO(g^2)$ corrections 
to this correlator. 
In fact, the only allowed correction 
is the constant 
$
\tilde \beta \equiv 
\tilde \beta_{xy} + \tilde \beta_{xw} + \tilde \beta_{yw}$
times $(- Y^2 / 32 \pi^2)$. 
To show that a three-point function of BPS operators is protected, 
we have to demonstrate that $\tilde \beta = 0$.

\subsection{Group theory simplifications}
\label{section:simplifications}

There are several simplifications which set some of the 
combinatorial coefficients in (\ref{eq:3-pt g^2}) to zero. 
These considerations are based on the underlying 
$SU(4) \sim SO(6)$ symmetry of the theory only, 
and are applicable for general $N$. 
We will leave aside the trivial case when the 
correlator is forced to vanish by group theory, 
and assume that the Born level coefficient 
in (\ref{eq:3-pt g^2}) $\alpha_{\mathrm {free}} \ne 0$.

The simplest BPS operators are \half-BPS chiral primaries, 
gauge invariant scalar composites in $[0,q,0]$ 
representations of $SO(6)$. 
These are 
totally symmetric tensors of $SO(6)$, so if 
for example the operator $\OO_x$ is \half-BPS, 
the coefficients 
$\tilde \beta_{xy} = \tilde \beta_{xw} = \tilde \gamma_x = 0$
since the diagrams they multiply are antisymmetric 
in the flavor indices of $\OO_x$.%
\footnote{
	If we chose $\OO_w$ as such a \half-BPS operator, 
	we would not be able to conclude that $\tilde \gamma_w = 0$
	just from the symmetries of $\OO_w$: 
	the fourth diagram of Figure \ref{fig:three-pt-blocks}, 
	is not antisymmetric in flavor indices 
	at the vertex where the operator is made 
	of both $z$-s and $\bar z$-s. 
	However, using equation (\ref{eq:coeffs - all three:end})
	we find $\tilde \gamma_w = \tilde \beta_{xw} + \tilde \beta_{yw}= 0$
	since $\tilde \beta_{xw} = \tilde \beta_{yw}= 0$ 
	when all three operators have protected scaling dimensions. 
	}
Similarly, if both $\OO_x$ and $\OO_y$ are \half-BPS and 
$\OO_w$ is any BPS operator, we have 
$\tilde \beta_{xy} = \tilde \beta_{xw} = \tilde \gamma_x = 
\tilde \beta_{yw} = \tilde \gamma_y = 0$, 
and hence 
$\tilde \beta = 
\tilde \beta_{xy} + \tilde \beta_{xw} + \tilde \beta_{yw} = 0$; 
there are no $\OO(g^2)$ corrections in this case. 
In particular, this reproduces the result 
of \cite{DFS} when all three $\OO_{x,y,w}$ are \half-BPS chiral primaries.

\begin{figure}[t!]
{\begin{center}
\epsfig{height=1.45in, file=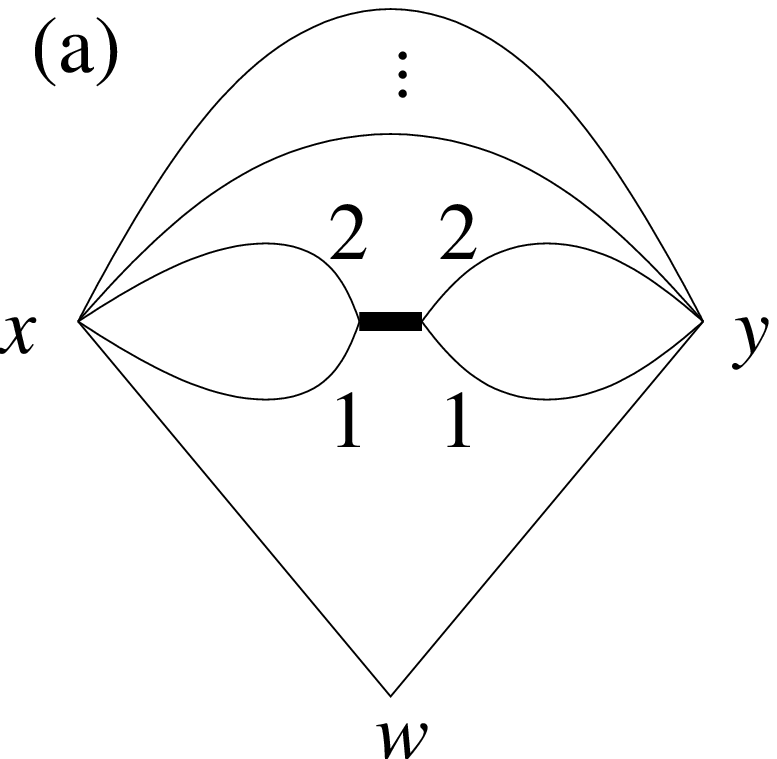, angle=0}
\hspace{1.6em}
\epsfig{height=1.45in, file=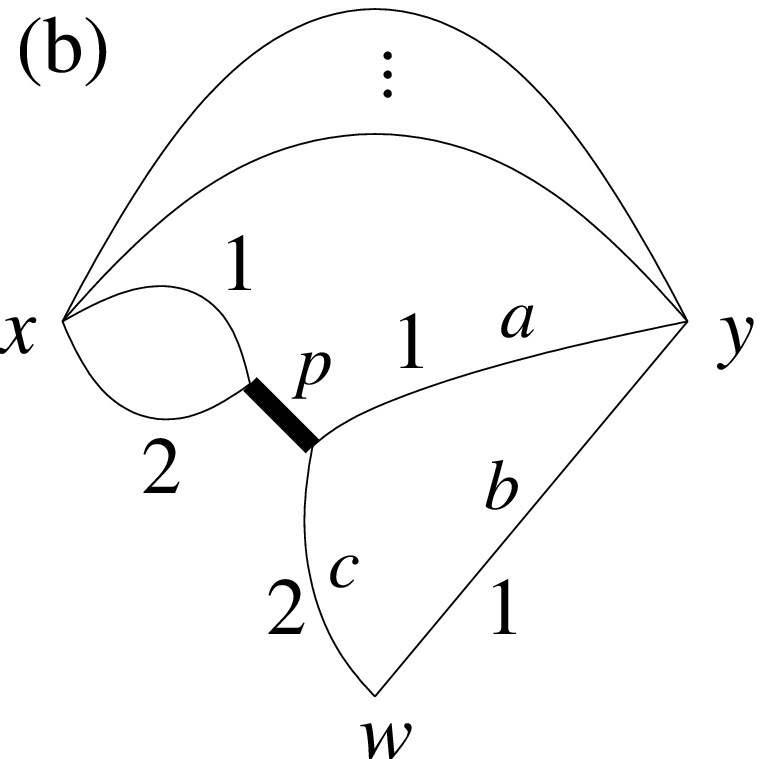, angle=0}
\hspace{1.6em}
\epsfig{height=1.45in, file=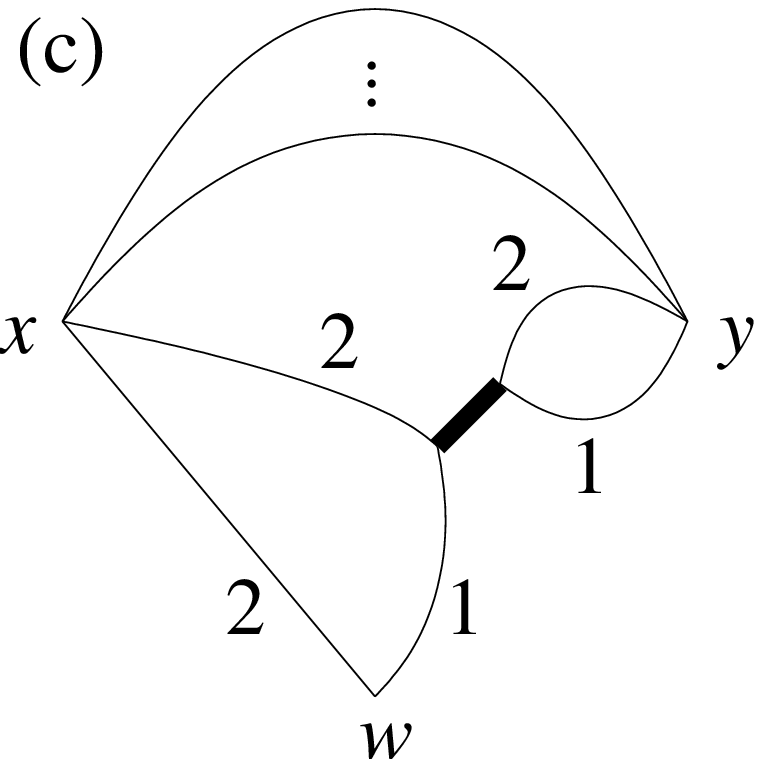, angle=0}
\end{center}}
\caption{%
$F$-term contributions to 
$\langle \OO_{\mathrm {BPS}}(x) \, \OO'_{\mathrm {BPS}}(y) \, \OO_{\thalf}(w)
\rangle_{g^2}$ 
in the case when the correlator can be ``partitioned into two flavors:'' 
(a) proportional to $\tilde B(x,y)$; 
(b) proportional to $\tilde C(x;y,w)$; 
(c) proportional to $\tilde C(y;x,w)$. 
\label {fig:3-point-z1-z2}
}%
\end {figure}

The ``standard'' way to choose $SU(4)$ weights of operators 
in a three point function is to take 
the highest weight state $\lambda$ in its representation; 
another, 
the lowest weight state $-\lambda'$ in its; 
and the third operator to have weight $\lambda'-\lambda$.
But in some cases, the combinatorics simplifies if we chose the 
weights differently, as will be illustrated below. 

Suppose that $\OO_w$ is \half-BPS, and furthermore we can 
``partition the correlator into two flavors,''
i.e. choose the operators such that 
$\OO_w = [\bar z_1^m z_2^n]$ 
while 
$\OO_x = [z_1^k z_2^l]$ 
and $\OO_y = [\bar z_1^{(k-m)} \bar z_2^{(l-n)}]$. 
Consider a diagram proportional to $\tilde C(x;y,w)$, 
see Figure \ref{fig:3-point-z1-z2}(b). 
The sum of all such diagrams 
is symmetric in the color indices $a$ and $b$ at $y$ 
(since the scalars at $y$ used in this diagram have the same flavors, 
$z_1^a$ and $z_1^b$), 
and symmetric in colors at $w$ (since $\OO_w$ is \half-BPS, 
it is symmetric in the indices $b$ and $c$). 
But it must be antisymmetric in the color indices $a$ and $c$ 
of the $z_1^a$ and $z_2^c$ leaving the interaction vertex 
(as this diagram comes with a factor of $f^{apc}$, 
see Figure 
\ref{fig:four-scalar and propagator}). 
We conclude that all such diagrams cancel, and so $\tilde \gamma_x = 0$. 
In the same fashion, we conclude that $\tilde \gamma_y = 0$ as well, 
and together with 
$\tilde \beta_{yw} = \tilde \beta_{xw} = 0$
(as $\OO_w$ is \half-BPS), we find that $\beta = 0$ 
when $\OO_x$ and $\OO_y$ are any operators with 
protected scaling dimensions.

\begin{figure}[t!]
{\begin{center}
\epsfig{width=5.5in, file=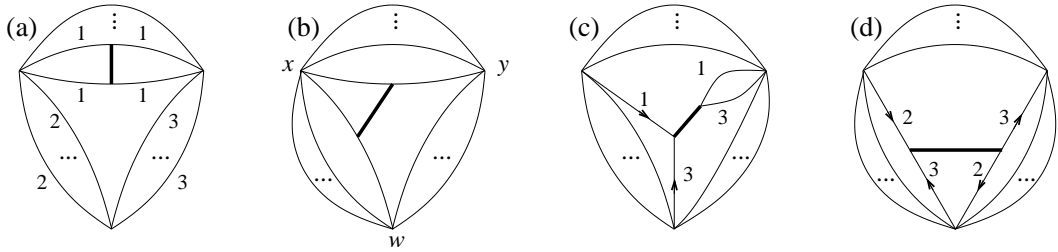, angle=0}
\end{center}}
\caption{%
Order $g^2$ corrections to correlators of the form 
(\ref{disjoint-flavors}): 
(a) and (b) includes a gauge boson exchange;
(c) and (d) $F$-terms.
Self energy contributions (not shown) also 
include a gauge boson exchange. 
\label {fig:disjoint-flavors}
}%
\end {figure}

There is another type of three-point functions 
of BPS chiral primaries which receive no 
$\OO(g^2)$ corrections by similar considerations. 
Consider a correlator 
(\ref{eq:3-pt g^2}) such that $\OO_x$ 
is made of $z_1$ and $z_2$; 
$\OO_y$ made of $\bar z_1$ and $\bar z_3$; 
and $\OO_w$, made of $\bar z_2$ and $z_3$, 
i.e. a correlator of the form  
\begin	{eqnarray}
\langle 
\label{disjoint-flavors}
[z_1^m z_2^n] (x) \; [\bar z_1^m \bar z_3^k] (y) \; [z_3^k \bar z_2^n] (w) 
\rangle 
\end	{eqnarray}
This correlator is ``partitioned into three disjoint flavors.'' 
Order $g^2$ contributions to this three-point function are shown in 
Figure \ref{fig:disjoint-flavors}. 
There are no corrections proportional to any $\tilde B$-s 
since all lines within any rainbow carry the same flavor, 
so immediately 
$\tilde \beta_{xy} = \tilde \beta_{yw} = \tilde \beta_{xw} = 0$ 
and hence
there are no $\OO(g^2)$ corrections here, as well.

\begin{figure}[t!]
{\begin{center}
\epsfig{width=5.5in, file=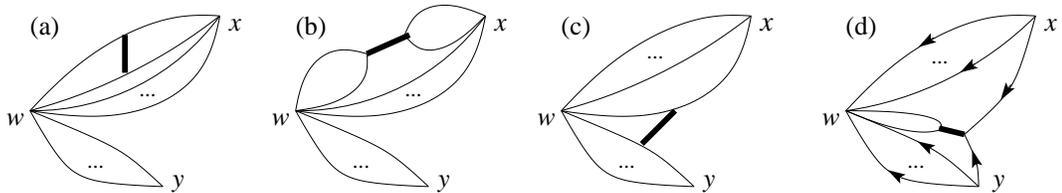, angle=0}
\end{center}}
\caption{%
Order $g^2$ corrections to extremal correlators:
(a) and (b) within a single peddle;
(c) and (d) between the two peddles. 
Self energy contributions (not shown) and diagrams (a) and (c) 
are gauge dependent, while (b) and (d) diagrams arise from 
the $F$-terms. 
\label {fig:extremal}
}%
\end {figure}

Finally, extremal three-point functions 
can be analyzed in a simple way. 
Here, the scaling dimension of one of the operators 
is equal to the sum of scaling dimensions of the other two.%
\footnote{
	In general, 
	$(n+1)$-point functions 
	$\langle \OO_0 (x_0) \OO_1(x_1) ... \OO_n(x_n) \rangle$ 
	are called extremal if one of the scaling dimensions 
	is the sum of all the others, 
	$\Delta_0 = \Delta_1 + ... + \Delta_n$.
	}
Suppose that $\Delta_x + \Delta_y = \Delta_w$ in 
(\ref{eq:3-pt g^2}). 
At Born level, there are no $G(x,w)$ propagators, 
and so there are no corrections proportional 
to $\tilde B(x,y)$, $\tilde C(x;y,w)$, or $\tilde C(y;x,w)$, 
see Figure \ref{fig:extremal}. 
Together with the constraints 
(\ref{eq:coeffs - all three:beg}-\ref{eq:coeffs - all three:end}), 
this determines $\tilde \beta = 0$ 
when the three (Lorentz scalar) operators inserted $x$, $y$, and $w$ 
are arbitrary operators with protected scaling dimensions.

Another remark about extremal correlators is in order. 
As it is easy to see, one of the above group theory 
simplifications 
generalizes straightforwardly to 
extremal correlators of chiral primaries. 
Namely, if all operators except for one are \half-BPS 
(and the remaining one is an arbitrary chiral primary), 
extremal correlators 
receive no order $g^2$ corrections.

\section{Three-point functions of BPS operators}

We are now ready to discuss correlators of 
three BPS chiral primaries. 
The simplest correlators 
$\langle 
\OO_{\thalf} \OO_{\thalf} \OO_{\thalf} 
\rangle$ 
(where each $\OO_{\thalf}$ stands for a \half-BPS operator), 
were considered by the authors of 
\cite{DFS}, who found that three-point functions
of \half-BPS operators do not get corrected at order $g^2$, 
for any $N$. 
These are a special case of 
correlators of the form 
$\langle \OO_{\thalf} \OO_{\thalf} \OO_{\mathrm {BPS}} 
\rangle$, 
which we discussed in Section \ref{section:simplifications} 
(here $\OO_{\mathrm {BPS}}$ is an arbitrary BPS operator). 
Such three-point functions receive no $\OO(g^2)$ corrections 
by group theory reasoning.

\subsection
{Correlators $\langle 
\OO_{\tquarter} \OO_{\tquarter} \OO_{\thalf} 
\rangle$}

Not all three-point functions of chiral primaries can be simplified 
using the results of Section \ref{section:simplifications}, 
so occasionally we will have to actually 
compute some of the combinatorial coefficients. 
In this Section we will look at correlators 
of two \quarter-BPS operators with one \half-BPS operator. 

\subsubsection{$\langle \OO_{[p,q,p]}(x) \bar{\OO}_{[p,q,p]}(y) (\tr \bar z t z) (w) \rangle$}
\label{section:cat diagrams}

The simplest 
$\langle 
\OO_{\tquarter} \OO_{\tquarter} \OO_{\thalf} 
\rangle$ 
three-point functions 
are of the form 
\begin{eqnarray}
\label{o-o-x-square}
\langle \OO (x) \bar{\OO}' (y) (\tr X^2) (w) \rangle 
\end{eqnarray}
where the \half-BPS primary $\tr X^2$ is a 
scalar composite operator in the [0,2,0] of $SU(4)$. 
Group theory restricts the quantum numbers of 
operators which can have nontrivial three point functions. 
Tensoring $[p,q,p] \otimes [0,2,0]$ 
using Young diagrams of $SO(6)$ gives 
\begin{eqnarray}
\label{pqp-times-020}
\mbox{
\setlength{\unitlength}{0.6em}
\begin{picture}(7.5,2)
\put(0,1){\framebox (4,1){\scriptsize $p$}}
\put(4,1){\framebox (3,1){\scriptsize $q$}}
\put(0,0){\framebox (4,1){\scriptsize $p$}}
\end{picture}}
\otimes
\mbox{
\setlength{\unitlength}{0.6em}
\begin{picture}(2,2)
\put(-0.5,1){\framebox (1,1){$~$}}
\put(0.5,1){\framebox (1,1){$~$}}
\end{picture}}
&=& 
\mbox{
\setlength{\unitlength}{0.6em}
\begin{picture}(9.5,2)
\put(0,1){\framebox (4,1){$~$}}
\put(4,1){\framebox (3,1){$~$}}
\put(0,0){\framebox (4,1){$~$}}
\put(7,1){\framebox (1,1){$~$}}
\put(8,1){\framebox (1,1){$~$}}
\end{picture}}
\oplus
\mbox{
\setlength{\unitlength}{0.6em}
\begin{picture}(8.5,2)
\put(0,1){\framebox (4,1){$~$}}
\put(4,1){\framebox (3,1){$~$}}
\put(0,0){\framebox (4,1){$~$}}
\put(7,1){\framebox (1,1){$~$}}
\put(4,0){\framebox (1,1){$~$}}
\end{picture}}
\oplus
\mbox{
\setlength{\unitlength}{0.6em}
\begin{picture}(7.5,2)
\put(0,1){\framebox (4,1){$~$}}
\put(4,1){\framebox (3,1){$~$}}
\put(0,0){\framebox (4,1){$~$}}
\put(4,0){\framebox (1,1){$~$}}
\put(5,0){\framebox (1,1){$~$}}
\end{picture}}
\nonumber\\
&\oplus&
\mbox{
\setlength{\unitlength}{0.6em}
\begin{picture}(9.5,2)
\put(0,1){\framebox (4,1){$~$}}
\put(4,1){\framebox (3,1){$~$}}
\put(0,0){\framebox (4,1){$~$}}
\put(7,1){\framebox (1,1){X}}
\put(8,1){\framebox (1,1){}}
\put(6,1){\framebox (1,1){X}}
\end{picture}}
\oplus
\mbox{
\setlength{\unitlength}{0.6em}
\begin{picture}(8.5,2)
\put(0,1){\framebox (4,1){$~$}}
\put(4,1){\framebox (3,1){$~$}}
\put(0,0){\framebox (4,1){$~$}}
\put(7,1){\framebox (1,1){X}}
\put(4,0){\framebox (1,1){}}
\put(6,1){\framebox (1,1){X}}
\end{picture}}
\oplus
\mbox{
\setlength{\unitlength}{0.6em}
\begin{picture}(11.5,2)
\put(0,1){\framebox (4,1){$~$}}
\put(4,1){\framebox (3,1){$~$}}
\put(0,0){\framebox (4,1){$~$}}
\put(7,1){\framebox (1,1){X}}
\put(8,1){\framebox (1,1){}}
\put(3,0){\framebox (1,1){X}}
\end{picture}}
\nonumber\\
&\oplus&
\mbox{
\setlength{\unitlength}{0.6em}
\begin{picture}(9.5,2)
\put(0,1){\framebox (4,1){$~$}}
\put(4,1){\framebox (3,1){$~$}}
\put(0,0){\framebox (4,1){$~$}}
\put(7,1){\framebox (1,1){X}}
\put(8,1){\framebox (1,1){X}}
\put(5,1){\framebox (1,1){X}}
\put(6,1){\framebox (1,1){X}}
\end{picture}}
\oplus
\mbox{
\setlength{\unitlength}{0.6em}
\begin{picture}(8.5,2)
\put(0,1){\framebox (4,1){$~$}}
\put(4,1){\framebox (3,1){$~$}}
\put(0,0){\framebox (4,1){$~$}}
\put(7,1){\framebox (1,1){X}}
\put(4,0){\framebox (1,1){X}}
\put(3,0){\framebox (1,1){X}}
\put(6,1){\framebox (1,1){X}}
\end{picture}}
\oplus
\mbox{
\setlength{\unitlength}{0.6em}
\begin{picture}(7.5,2)
\put(0,1){\framebox (4,1){$~$}}
\put(4,1){\framebox (3,1){$~$}}
\put(0,0){\framebox (4,1){$~$}}
\put(5,0){\framebox (1,1){X}}
\put(4,0){\framebox (1,1){X}}
\put(3,0){\framebox (1,1){X}}
\put(2,0){\framebox (1,1){X}}
\end{picture}}
\nonumber\\
&\oplus& 
...
\end{eqnarray}
where in the first row there are no contractions (i.e. $SO(6)$ traces), 
only symmetrizations and antisymmetrizations; in the 
second row, one contraction; 
and in the third row, two contractions;
the ``...'' stands for tableaux with more than two rows.%
\footnote{
	Tensoring representations in the manner of 
	equation (\ref{pqp-times-020}) gets messy 
	for larger representations. 
	Another 
	method 
	(of Berenstein-Zelevinsky triangles) 
	is discussed in Appendix \ref{BZ-triangles}. 
	} 
In terms of Dynkin labels, equation (\ref{pqp-times-020})  reads 
\begin{eqnarray}
\label{pqp-times-020:dynkin}
[p,q,p] \otimes [0,2,0] &=& 
[p,q+2,p] \oplus [p+1,q,p+1] \oplus [p+2,q-2,p+2]
\nonumber\\
&\oplus&
[p,q,p] \oplus [p+1,q-2,p+1] \oplus [p-1,q+2,p-1]
\nonumber\\
&\oplus&
[p,q-2,p] \oplus [p-1,q,p-1] \oplus [p-2,q+2,p-2]
\nonumber\\
&\oplus&
...
\end{eqnarray}
Now, the ``...'' stands for representations with 
$[r,s,r+2k]$ Dynkin labels with $k \ne 0$. 
Thus the only three-point functions 
of the form (\ref{o-o-x-square}) 
which can possibly have a nonzero value 
are the extremal correlators 
\begin{eqnarray}
\label{eq:extremal X^2}
\langle\OO_{[p,q,p]}(x)\bar{\OO}_{[p,q-2,p]}(y) (\tr\bar z_1^2)(w) \rangle 
\\
\langle\OO_{[p,q,p]}(x)\bar{\OO}_{[p-2,q+2,p-2]}(y)(\tr\bar z_2^2)(w) \rangle 
\\
\langle\OO_{[p,q,p]}(x)\bar{\OO}_{[p-1,q,p-1]}(y)(\tr\bar z_1\bar z_2)(w) 
\rangle 
\end{eqnarray}
which correspond to those diagrams in (\ref{pqp-times-020}) 
with zero or maximal number of contractions; 
and non-extremal correlators 
\begin{eqnarray}
\label{eq1:non-extremal X^2}
\langle \OO_{[p,q,p]}(x) \bar{\OO}_{[p,q,p]}(y) (\tr \bar z t z)(w) \rangle 
\\
\langle\OO_{[p,q,p]}(x)\bar{\OO}_{[p-1,q+2,p-1]}(y)(\tr\bar z_2 z_1)(w)\rangle 
\label{eq2:non-extremal X^2}
\end{eqnarray}
where $t$ is a diagonal $SU(3)$ generator. 
All other correlators of the form (\ref{o-o-x-square})
either vanish because the tensor product of 
irreps $[p,q,p]$ and $[0,2,0]$ does not contain $[r,s,r]$, 
or are related to the ones 
in (\ref{eq:extremal X^2}-\ref{eq2:non-extremal X^2}).

\begin{figure}[t!]
{\begin{center}
\epsfig{width=1.4in, file=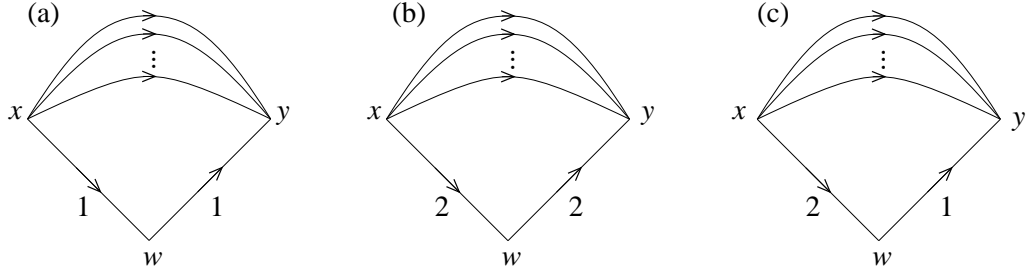, angle=-90}
\end{center}}
\caption{%
Nonvanishing Born diagrams for 
non-extremal 3-point functions 
$\langle \OO_{[p,q,p]} \bar{\OO}_{[p,q,p]} (\tr \bar z t z) \rangle$ 
(a-b); 
$\langle \OO_{[p,q,p]} \bar{\OO}_{[p-1,q+2,p-1]} (\tr \bar z_2 z_1) 
\rangle$ 
(c). 
\label {fig:born-3-point-x-square}
}%
\end {figure}

Extremal three-point functions were discussed 
in Section \ref{section:simplifications}, and were found 
to be protected at order $g^2$. 
The only correlators of the form 
$\langle \OO \bar {\OO}' \tr X^2 \rangle$
we need to consider are those given by 
(\ref{eq1:non-extremal X^2}) and (\ref{eq2:non-extremal X^2}). 
However, 
the three-point functions 
of Figure \ref{fig:born-3-point-x-square}(c), 
must in fact vanish: 
$\tr \bar z_2 z_1 = \half \bar z_2^a z_1^a$ 
is diagonal in color indices, and hence 
the combinatorial factors for the Born graph 
of 
$\langle \OO_{[p,q,p]}(x) \bar{\OO}_{[p-1,q+2,p-1]}(y) 
(\tr \bar z_2 z_1)(w) \rangle$ 
are 
proportional to the ones for the two-point function 
$\langle \OO_{[p,q,p]} \bar{\OO}_{[p-1,q+2,p-1]} \rangle = 0$.
The same thing happens at order $g^2$, etc.%
\footnote{
	Explicitly, in Section \ref{section:simplifications} 
	we saw that the $\OO(g^2)$ part of this 
	three-point function vanishes. 
	} 
So correlators (\ref{eq2:non-extremal X^2}), 
although allowed by 
(\ref{pqp-times-020:dynkin}), 
are in fact forbidden by a combination of 
$SU(N)$ and $SU(4)$ group theory.

\begin{figure}[t!]
{\begin{center}
\epsfig{width=0.85in, file=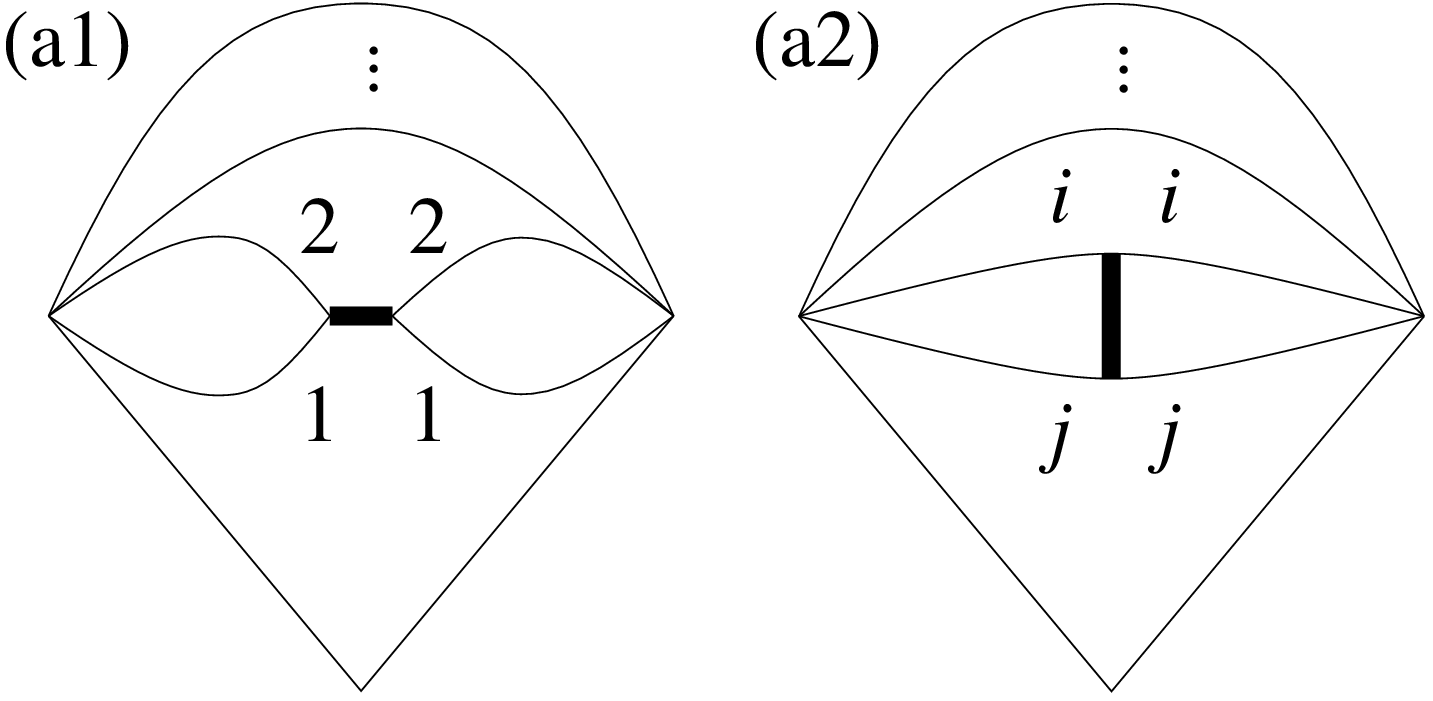, angle=-90}
\quad
\epsfig{width=0.85in, file=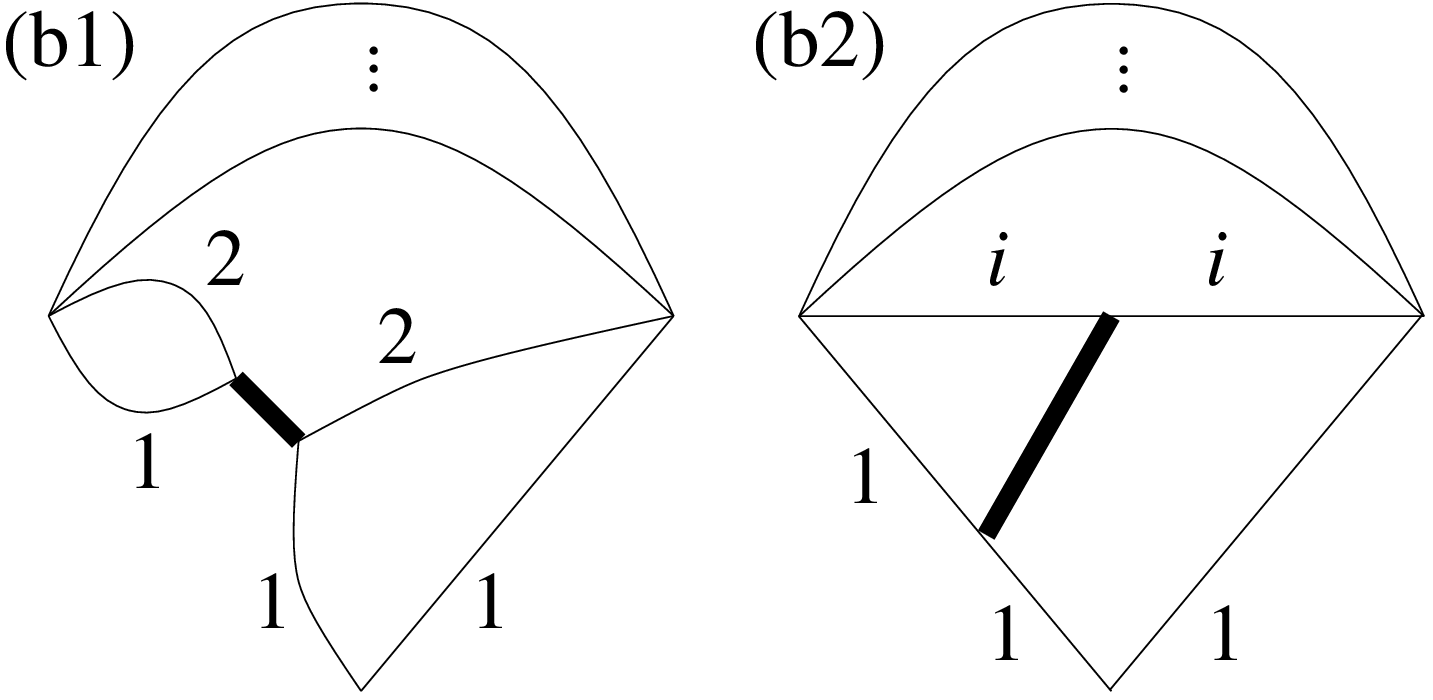, angle=-90}
\quad
\epsfig{width=0.85in, file=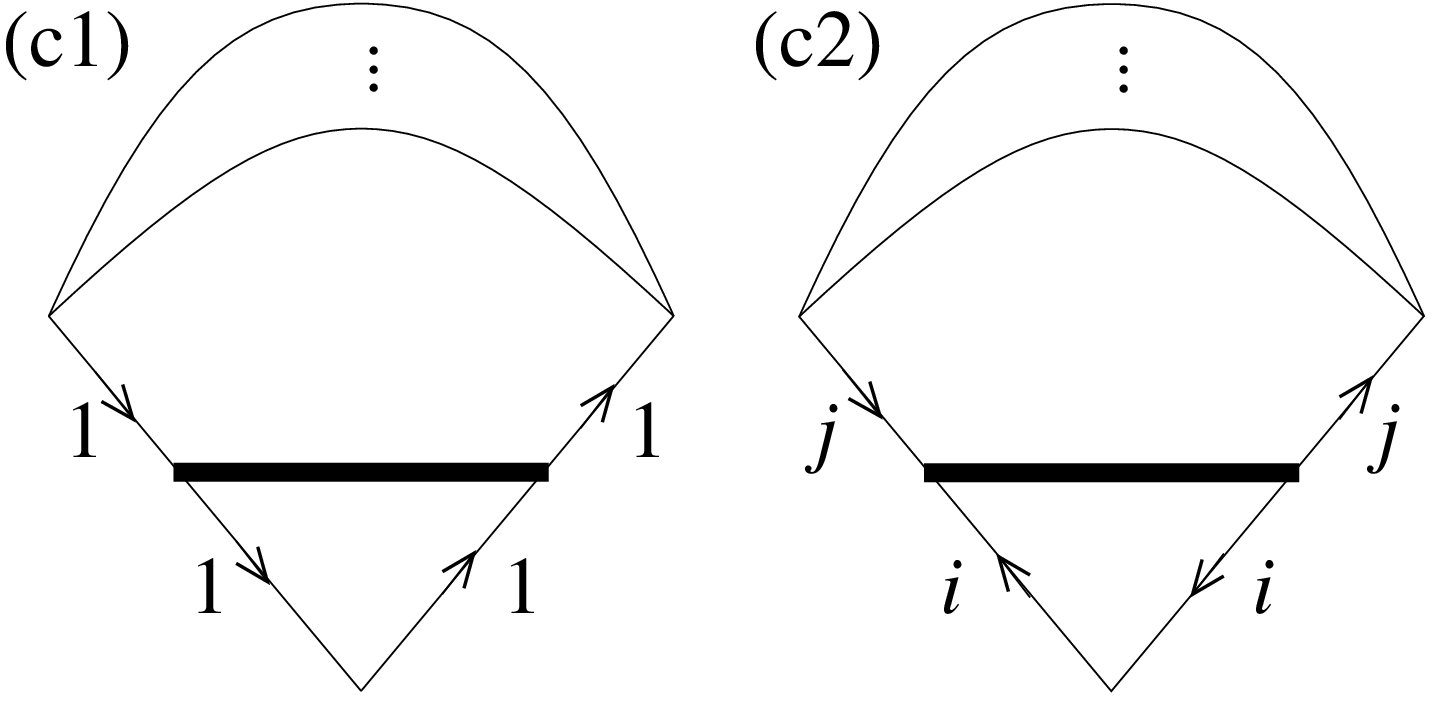, angle=-90}
\end{center}}
\caption{%
Order $g^2$ corrections to 
$\langle \OO_{[p,q,p]}(x) \bar{\OO}_{[p,q,p]}(y) (\tr \bar z t z) (w) \rangle$ 
with $t_{22} = 0$: 
(a) within the rainbow ($i,j=1,2$ in the second diagram); 
(b) from the rainbow to $X^2$ 
(there are similar ones with the other leg of $X^2$ uncorrected). 
\label {fig:3-point-x-square}
}%
\end {figure}

Correlators 
$\langle \OO_{[p,q,p]}(x) \bar{\OO}_{[p,q,p]}(y) (\tr \bar z t z)(w)\rangle$ 
are the only ones that remain to be considered. 
The contributing 
Born level diagrams are shown in 
Figure \ref{fig:born-3-point-x-square}(a,b), 
and the $\OO(g^2)$ graphs 
appear 
in Figure \ref{fig:3-point-x-square}
(corrections to the scalar propagator are not shown, 
but are also present). 
Repeating the 
arguments of \cite{DFS} from the 
\half-BPS calculations, we see that the combinatorial structure 
of this three-point function 
$\langle \OO \bar{\OO} (\tr \bar z t z) \rangle$
is the same as that of the 
two-point function $\langle \OO \bar{\OO} \rangle$. 
At Born level, we find that 
\begin{eqnarray}
\label{eq:o-o-tr-ztz Born}
\langle \OO_{[p,q,p]}(x) \bar{\OO}_{[p,q,p]}(y) (\tr \bar z t z) (w) \rangle 
|_{\mathrm {free}} && \nonumber\\
&&
\hspace{-14em}
= 
\half [(p+q) t_{11} + p t_{22} ] 
[{G(x,w) G(y,w) \over G(x,y)}]
\langle \OO_{[p,q,p]}(x) \bar{\OO}_{[p,q,p]}(y) \rangle 
|_{\mathrm {free}} 
\quad\quad 
\end{eqnarray}
At order $g^2$, the contributions 
proportional to $\tilde B$ and $\tilde C$
(diagrams (a1) and (b1) in Figure \ref{fig:3-point-x-square})
have the same index structure, 
which in turn is identical to that of the two-point functions 
$\langle \OO_{[p,q,p]}(x) \bar{\OO}_{[p,q,p]}(y) \rangle$. 
Because $\tr \bar z_1 z_1$ is diagonal in color indices, 
its only effect on the combinatorics 
is to distinguish the pair of indices which go 
to $\OO_w$ rather than stretch directly between 
$\OO_x$ and $\OO_y$.

There is a curious relation between the functions
$\tilde B(x,y)$ and $\tilde C(x;y,w)$, 
which can be graphically expressed as%
\footnote{
	The fact that $\tilde C(x;y,w) + \tilde C(y;x,w) - \tilde B(x,y)$ 
	is just a constant was established in 
	Section \ref{section:c-tilde} by an explicit calculation. 
	The value of this constant was also found there. 
	}
\begin{equation}
\label{cat:diagrams}
\epsfig{width=5.3in, file=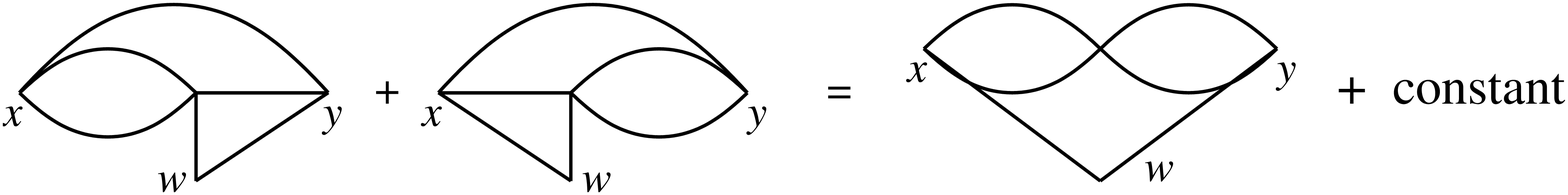, angle=0}
\end{equation}
This is a consequence of conformal invariance and 
nonrenormalization of the scaling dimension of $\tr X^2$. 
To see how this comes about, let $\OO$ be an 
arbitrary (not necessarily BPS) 
scalar operators of same scaling dimension $\Delta$; 
then 
\begin{eqnarray}
\label{eq:o-o-conformal}
\langle \OO(x) \bar{\OO}(y) \rangle &=& 
{C \over (x-y)^{2 \Delta} }
\\
\langle \OO(x) \bar{\OO}(y) \, \tr X^2(w) \rangle &=& 
{\tilde C \over (x-y)^{2 \Delta} } 
{G(x,w) G(y,w) \over G(x,y) }
\label{eq:o-o-X^2-conformal}
\end{eqnarray}
in $d=4$; 
$\tr X^2 = \tr \bar z t z$ as in Equation (\ref{eq:o-o-tr-ztz Born}). 
In other words, coordinate dependence 
(modulo the ratio of free scalar correlators)
is the same, 
and the difference is just a constant factor. 
Assume now that $\OO$ is constructed of only $z_i$-s 
(then $\bar{\OO}$ is made of only $\bar z_j$-s).  
Then, the only $\OO(g^2)$ contributions to the two-point function 
$\langle \OO(x) \bar{\OO}(y) \rangle$ are proportional to $\tilde B(x,y)$. 
Similarly, the $\OO(g^2)$ corrections to the correlator 
$\langle \OO(x) \bar{\OO}(y) \, \tr X^2(w) \rangle$ 
are proportional to 
$\tilde B(x,y)$ and to $[\tilde C(x;y,w) + \tilde C(y;x,w)]$. 
Index structure of these building blocks 
is the same (as discussed after equation \ref{eq:o-o-tr-ztz Born}), so 
\begin{eqnarray}
\label{eq:c-c-b-1}
{ \langle \OO(x) \bar{\OO}(y) \rangle|_{g^2 ~~} 
\over \langle \OO(x) \bar{\OO}(y) \rangle|_{\mathrm {free}}} &=& 
\zeta \; \tilde B(x,y) 
\quad\quad\\
{\langle \OO(x) \bar{\OO}(y) \, \tr X^2(w) \rangle|_{g^2 ~~} 
\over \langle \OO(x) \bar{\OO}(y) \, \tr X^2(w) \rangle|_{\mathrm {free}}} &=& 
\half \zeta' \left[ \tilde C(x;w,y) + \tilde C(y;w,x) + \tilde B(x,y) \right] 
\quad\quad 
\label{eq:c-c-b-2}
\end{eqnarray}
The difference between $\tilde B(x,y)$ contributions to 
$\langle \OO(x) \bar{\OO}(y) \, \tr X^2(w) \rangle$ 
and to $\langle \OO(x) \bar{\OO}(y) \rangle$ 
is that the pair of indices which go to $w$ rather than stretch between 
$x$ and $y$ directly, is distinguished. Hence, 
$\zeta' = \zeta$ times some combinatorial factor.%
\footnote{
	In particular, if 
	$\Delta = \Delta_0$ is not corrected, then neither 
	$\langle \OO(x) \bar{\OO}(y) \, \tr X^2(w) \rangle$ 
	nor 
	$\langle \OO(x) \bar{\OO}(y) \rangle$
	get any one loop corrections. 
	}
As was discussed in Section \ref{section:c-tilde}, 
$\tilde B$ and $\tilde C$ 
have the form 
$\tilde B(x,y) = \tilde a \log {(x-y)^2 \over \e^2} + \tilde b$, 
$\tilde C(x;y,w) 
= \tilde a' \log {(x-y)^2 (x-w)^2 \over \e^4} 
- \tilde a'' \log {(y-w)^2 \over \e^2} + \tilde b'$. 
By comparing 
(\ref{eq:o-o-X^2-conformal}) and (\ref{eq:o-o-conformal}), 
we see that 
expression (\ref{eq:c-c-b-2}) 
must have the same coordinate dependence as (\ref{eq:c-c-b-1}). 
This restricts 
$\tilde a' = \tilde a''  = \half \tilde a$, 
which reproduces the ``winking cat'' identity (\ref{cat:diagrams}).

Finally, we can relate 
$\langle \OO(x) \bar{\OO}(y) \, \tr X^2(w) \rangle$ 
to 
$\langle \OO(x) \bar{\OO}(y) \rangle$ 
by a Ward identity. 
As shown in \cite{OSBORN}, the ratio 
\begin{equation}
{\langle \OO(x) \bar{\OO}(y) T_{\mu\nu}(0) \rangle \over
\langle \OO(x) \bar{\OO}(y) \rangle} = 
{2 \Delta \over 3 \pi^2} 
{t_{\mu\nu}(\gamma) (x-y)^4 \over x^4 y^4}
\end{equation}
depends on the scaling dimension 
$\Delta$ of the operator $\OO$
(here, $\gamma = {x \over x^2} - {y \over y^2}$ and 
$t_{\mu\nu}(\gamma) =
{\gamma_\mu \gamma_\nu \over \gamma^2} - \quarter \eta_{\mu\nu}$). 
Since the energy momentum tensor 
$T_{\mu \nu}$ is in the same $\NN=4$ multiplet with 
$\tr X^2$, there is also nothing 
peculiar about the fact that $\tilde C_{12} / C_{12}$ 
can in general receive $\OO(g^2)$ correction. 
This ratio also depends on $\Delta$.

\subsubsection{General 
$\langle 
\OO_{\tquarter} \OO_{\tquarter} \OO_{\thalf} 
\rangle$ 
correlators}
\label{section:quarter-quarter-half}

Three-point functions of two \quarter-BPS operator and 
one \half-BPS operator are similar to the ones described 
in Section \ref{section:cat diagrams}. 
It suffices to consider a single three-point function 
(such that the 
Clebsch-Gordon coefficient%
\footnote{
	By Wigner-Eckart theorem, for any three representations 
	we only need to calculate one (nonvanishing) correlator
	of any representatives from these irreps. 
	} for 
these three vectors in the given 
irreps of $SU(4)$ 
is nonzero) for each set of three representations.
Without loss of generality, we can choose 
a $[p,q,p]$ scalar composite $\OO(x)$ to be made of only $z$-s; 
a $[r,s,r]$ scalar composite $\OO'(y)$ to be made of only $\bar z$-s; 
and a $[0,k,0]$ scalar composite $\tr X^{\alpha_1+\alpha_2}$ at $w$ 
of the form 
\begin{eqnarray}
t_{i_1 ... i_{\alpha_1} ; j_1 ... j_{\alpha_2}} 
\str z_{i_1} ... z_{i_{\alpha_1}} \bar z_{j_1} ... \bar z_{j_{\alpha_2}} 
\end{eqnarray}
where $\alpha_1+\alpha_2=k$, and 
$t_{i_1 ... i_{\alpha_1} ; j_1 ... j_{\alpha_2}}$ 
is the appropriate irreducible $SU(3)$ tensor 
(like in \cite{DFS}). 
The correlators we are after are 
\begin{eqnarray}
\label{o-o-X^k}
\langle \OO (x) \OO' (y) (\tr X^{\alpha_1+\alpha_2}) (w) \rangle 
\end{eqnarray}
Position dependence of (\ref{o-o-X^k}) is 
\begin{eqnarray}
\label{eq:o-o-X^k position}
[G(x,y)^{(2p+r)+(2r+s)-k} 
G(x,w)^{k+(2p+r)-(2r+s)} 
G(w,y)^{k-(2p+r)+(2r+s)}]^{1/2}
\end{eqnarray}
at Born level. 
The contributing free diagrams are similar to the ones 
shown in Figure \ref{fig:born-3-point-x-square}; 
and $\OO(g^2)$ diagrams, to those of 
Figure \ref{fig:3-point-x-square}, 
but now there can be a 
different number of lines stretching 
between $x$ and $w$ and between $w$ and $y$. 
Apart from the 
factor (\ref{eq:o-o-X^k position}), 
the general 
$\langle 
\OO_{\tquarter} \OO_{\tquarter} \OO_{\thalf} 
\rangle$ 
correlator (\ref{o-o-X^k}) is given by 
\begin{eqnarray}
\label{eq1:o-o-X^k corrections}
\alpha_{\mathrm {free}} + 
\tilde \beta_{xy} \tilde B (x,y) + 
\tilde \gamma_x \tilde C (x;y,w) + 
\tilde \gamma_y \tilde C (y;x,w) 
+ \OO(g^4) 
.
\end{eqnarray}
According to the discussion of Section \ref{section:3-pt structure}, 
the remaining combinatorial coefficients vanish, 
$\beta_{xw} = \beta_{yw} = \gamma_w = 0$. 
Moreover, 
$\tilde \gamma_x = \tilde \gamma_y = - \tilde \beta_{xy}$ 
as follows from equations 
(\ref{eq:coeffs - all three:beg}-\ref{eq:coeffs - all three:end}), 
so 
(\ref{eq1:o-o-X^k corrections}) 
reads 
\begin{eqnarray}
\label{eq2:o-o-X^k corrections}
\alpha_{\mathrm {free}} - 
\tilde \beta_{xy} 
\left( 
\tilde C (x;y,w) + \tilde C (y;x,w) - \tilde B (x,y) 
\right) 
+ \OO(g^4) 
\end{eqnarray}
Hence, we only need to verify that $\tilde \beta_{xy} = 0$.%
	\footnote{
	The expression multiplying $\tilde \beta_{xy}$ 
	in (\ref{eq2:o-o-X^k corrections}) 
	is a nonzero, renormalization scale independent 
	constant. In Section \ref{section:c-tilde}, 
	its value was computed to be $-\left( Y^2 / 32 \pi^2 \right)$. 
	}

The simplifications we can use to deduce that 
$\tilde \beta_{xy} = 0$ without doing calculations, 
are discussed in Section \ref{section:simplifications}. 
Extremal three-point functions are always 
easy to identify, and 
with the BPS primaries in representations 
$[p,q,p]$, $[r,s,r]$, and $[0,k,0]$, 
the restrictions on the scaling dimension translate into 
\begin{equation}
\label{restriction:extremal}
2r+s = 2p+q + k, 
\quad 
2p+q = 2r+s + k, 
\quad\mbox{or}\quad 
2p+q + 2r+s = k, 
\end{equation}
depending on which scaling dimension is the 
sum of the other two.

The ``three flavor partition'' boils down to being able to 
choose a single flavor (at Born level) for the lines between 
the two \quarter-BPS operators, when the third operator is \half-BPS. 
This is possible whenever 
\begin{equation}
\label{restriction:disjoint-flavors}
2 r + s \le k+q 
\quad\mbox{and}\quad 
2 p + q \le k+s. 
\end{equation}

Alternatively, if we can choose 
the \half-BPS operator $\OO_w$ 
to be made of only $\bar z_1$-s and $z_2$-s; 
and the \quarter-BPS operators as 
$\OO_x$ of $z_1$-s and $z_2$-s, 
$\OO_y$ of $\bar z_1$-s and $\bar z_2$-s, 
we get the ``two flavor partition'' 
of Section \ref{section:simplifications}. 
This can happen if 
\begin{equation}
\label{restriction:two-flavors}
k \le q + s. 
\end{equation}

In all three cases 
(\ref{restriction:extremal}), 
(\ref{restriction:disjoint-flavors}), 
(\ref{restriction:two-flavors}) 
there are no $\OO(g^2)$ corrections, 
as established in Section \ref{section:simplifications} 
using only $SU(4)$ group theory and conformal invariance 
arguments. 
However, there are allowed three-point functions 
of the form 
$\langle 
\OO_{\tquarter} \OO_{\tquarter} \OO_{\thalf} 
\rangle$ 
where we can not choose irrep representatives 
in such a nice way.

Throughout the rest of this Section, we will concentrate 
on the \quarter-BPS operators with $\Delta \le 7$, 
constructed in 
Chapters \ref{chapter: BPS: 2pt} and \ref{chapter:systematic}. 
In particular, we will consider scalar composite 
operators in $SU(4)$ representations of the form $[p,q,p]$, 
with $2p+q \le 7$. 
These are [2,0,2], [2,1,2], [2,2,2], [3,1,3], and [2,3,2]. 
We will take \half-BPS (single trace) operators as whichever ones are 
allowed by group theory. 
Of the triple products of the form 
$[p,q,p] \otimes [r,s,r] \otimes [0,k,0]$ 
containing the singlet, 
most satisfy at least one of the 
simplifying constraints 
(\ref{restriction:extremal}), 
(\ref{restriction:disjoint-flavors}), 
or (\ref{restriction:two-flavors}).%
\footnote{
	We omit the tedious details here. 
	In order to find the allowed triple products, 
	we used the method of BZ triangles, 
	see Appendix \ref{BZ-triangles}. 
	Then we just went through the list 
	and checked if any of the conditions 
	(\ref{restriction:extremal}-\ref{restriction:two-flavors}) 
	applied. 
	}
The exceptions are  
$[2,0,2] \otimes [2,0,2] \otimes [0,2,0]$
and 
$[3,1,3] \otimes [3,1,3] \otimes [0,4,0]$. 

Correlators 
$
\langle\OO_{[p,q,p]}(x)\bar{\OO}_{[p,q,p]}(y) \, \tr X^2 (w)\rangle 
$
were considered in Section \ref{section:cat diagrams}, 
so the only three-point function 
we 
have to calculate is 
$\langle 
\OO_{[3,1,3]} \bar{\OO}_{[3,1,3]} \, \tr X^{2+2} 
\rangle$. 
Explicitly, we can choose 
the [3,1,3] scalar composite operators%
\footnote{
	$\OO_{1,...,4}$ were discussed in 
        Chapter \ref{chapter: BPS: 2pt}. 
	They are written out explicitly in 
	Section \ref{6 and higher}. 
	} 
as 
\begin{eqnarray}
\label{eq:flavor choice:313-313-040:begin}
&&
\OO_x = \sum_{j=1}^{4} C_x^j \OO_j
,\quad
\OO_y = \sum_{j=1}^{4} C_y^j \bar{\OO}_j
\quad\mbox{with}\quad
\OO_j \sim [z_1^4 z_2^2 z_3] , \\ 
&& 
\OO_w \sim [z_2^2 \bar z_2^2] - \mbox{$SO(6)$ traces}. 
\label{eq:flavor choice:313-313-040:end}
\end{eqnarray}
The free combinatorial factor for this three-point function is then 
\begin{eqnarray}
\alpha_{\mathrm {free}} &=& 
{(N^2-1) (N^2-2) \over 41472 N^2} 
      (189540 C_x^2 C_y^2 - 4860 C_x^2 C_y^1 N - 
\nonumber\\&& \hspace{3em} 
        4860 C_x^1 C_y^2 N - 
	131220 C_x^4 C_y^2 N - 131220 C_x^2 C_y^4 N + 
\nonumber\\&& \hspace{3em} 
        360 C_x^1 C_y^1 N^2 + 
        13500 C_x^4 C_y^1 N^2 - 79380 C_x^2 C_y^2 N^2 - 
\nonumber\\&& \hspace{3em} 
        22680 C_x^3 C_y^2 N^2 - 
	22680 C_x^2 C_y^3 N^2 + 5184 C_x^3 C_y^3 N^2 + 
\nonumber\\&& \hspace{3em} 
        13500 C_x^1 C_y^4 N^2 - 
        30780 C_x^4 C_y^4 N^2 + 2700 C_x^2 C_y^1 N^3 - 
\nonumber\\&& \hspace{3em} 
        270 C_x^3 C_y^1 N^3 + 
	2700 C_x^1 C_y^2 N^3 + 43740 C_x^4 C_y^2 N^3 - 
\nonumber\\&& \hspace{3em} 
        270 C_x^1 C_y^3 N^3 - 
        9720 C_x^4 C_y^3 N^3 + 43740 C_x^2 C_y^4 N^3 - 
\nonumber\\&& \hspace{3em} 
        9720 C_x^3 C_y^4 N^3 - 
	115 C_x^1 C_y^1 N^4 - 2760 C_x^4 C_y^1 N^4 + 
\nonumber\\&& \hspace{3em} 
        13500 C_x^2 C_y^2 N^4 + 
        4410 C_x^3 C_y^2 N^4 + 4410 C_x^2 C_y^3 N^4 - 
\nonumber\\&& \hspace{3em} 
        1332 C_x^3 C_y^3 N^4 - 
	2760 C_x^1 C_y^4 N^4 + 13680 C_x^4 C_y^4 N^4 - 
\nonumber\\&& \hspace{3em} 
        450 C_x^2 C_y^1 N^5 + 
        240 C_x^3 C_y^1 N^5 - 450 C_x^1 C_y^2 N^5 - 
\nonumber\\&& \hspace{3em} 
        4500 C_x^4 C_y^2 N^5 + 
	240 C_x^1 C_y^3 N^5 + 2340 C_x^4 C_y^3 N^5 - 
\nonumber\\&& \hspace{3em} 
        4500 C_x^2 C_y^4 N^5 + 
        2340 C_x^3 C_y^4 N^5 - 15 C_x^1 C_y^1 N^6 - 
\nonumber\\&& \hspace{3em} 
        990 C_x^2 C_y^2 N^6 - 126 C_x^3 C_y^3 N^6 - 
        1980 C_x^4 C_y^4 N^6)
\end{eqnarray}
so 
$\langle \OO_x \OO_y \OO_w \rangle \ne 0$ 
in general (and when $\OO_x$ and $\OO_y$ are \quarter-BPS, in particular). 
We have also explicitly calculated%
\footnote{
	This calculation 
	was done using {\textit {Mathematica}} 
	and took about 200 hours.
	The choice of flavors 
	(\ref{eq:flavor choice:313-313-040:begin}-\ref{eq:flavor choice:313-313-040:end})
	was optimal from the computational efficiency point of view. 
	}
the $\OO(g^2)$ combinatorial factor 
in (\ref{eq2:o-o-X^k corrections}): 
\begin{eqnarray}
\tilde \beta_{xy} &=& 
{(N^2-1) (N^2-4) \over 13824} 
     (-10800 C_x^2 C_y^1 + 4320 C_x^3 C_y^1 - 
\nonumber\\&& \hspace{2em} 
       10800 C_x^1 C_y^2 - 
	259200 C_x^4 C_y^2 + 4320 C_x^1 C_y^3 + 
\nonumber\\&& \hspace{2em} 
       103680 C_x^4 C_y^3 - 
       259200 C_x^2 C_y^4 + 103680 C_x^3 C_y^4 - 
\nonumber\\&& \hspace{2em} 
       2025 C_x^1 C_y^1 N - 
	27000 C_x^4 C_y^1 N - 32400 C_x^2 C_y^2 N + 
\nonumber\\&& \hspace{2em} 
       38880 C_x^3 C_y^2 N + 
       38880 C_x^2 C_y^3 N - 25920 C_x^3 C_y^3 N - 
\nonumber\\&& \hspace{2em} 
       27000 C_x^1 C_y^4 N - 
	129600 C_x^4 C_y^4 N - 600 C_x^2 C_y^1 N^2 + 
\nonumber\\&& \hspace{2em} 
       2940 C_x^3 C_y^1 N^2 - 
       600 C_x^1 C_y^2 N^2 + 50400 C_x^4 C_y^2 N^2 + 
\nonumber\\&& \hspace{2em} 
       2940 C_x^1 C_y^3 N^2 - 
	7200 C_x^4 C_y^3 N^2 + 50400 C_x^2 C_y^4 N^2 - 
\nonumber\\&& \hspace{2em} 
       7200 C_x^3 C_y^4 N^2 + 
       175 C_x^1 C_y^1 N^3 + 5400 C_x^4 C_y^1 N^3 + 
\nonumber\\&& \hspace{2em} 
       7200 C_x^2 C_y^2 N^3 - 
	7920 C_x^3 C_y^2 N^3 - 7920 C_x^2 C_y^3 N^3 + 
\nonumber\\&& \hspace{2em} 
       3888 C_x^3 C_y^3 N^3 + 
       5400 C_x^1 C_y^4 N^3 + 28800 C_x^4 C_y^4 N^3 + 
\nonumber\\&& \hspace{2em} 
       600 C_x^2 C_y^1 N^4 - 
	780 C_x^3 C_y^1 N^4 + 600 C_x^1 C_y^2 N^4 - 
\nonumber\\&& \hspace{2em} 
       780 C_x^1 C_y^3 N^4 - 
       2880 C_x^4 C_y^3 N^4 - 2880 C_x^3 C_y^4 N^4 + 
\nonumber\\&& \hspace{2em} 
       50 C_x^1 C_y^1 N^5 + 288 C_x^3 C_y^3 N^5)
\end{eqnarray}
If we choose the coefficients 
$( C_x^1, C_x^2, C_x^3, C_x^4 )$ 
and 
$( C_y^1, C_y^2, C_y^3, C_y^4 )$ 
independently from the set 
$
\{
(
- {\frac{12 N}{{N^2}-2}} , 1 ,  - {\frac{5}{{N^2}-2}} , 0
)
, 
(
{\frac{96}{{N^2}-4}} , - {\frac{4 N}{{N^2}-4}} , 
{\frac{10 N}{{N^2}-4}} , 1
)
\}
$, 
we recover $\tilde \beta_{xy} = 0$. 
This corresponds to taking 
$\OO_x$ and $\OO_y$ as the \quarter-BPS chiral primaries 
found in 
Chapters \ref{chapter: BPS: 2pt} and \ref{chapter:systematic}, 
so there are no 
$\OO(g^2)$ corrections in this case either.

\subsection{Three-point functions of \quarter-BPS operators}
\label{section: all BPS}

When all three operators are \quarter-BPS, 
the arguments get more tedious. 
We will chose $2l+k \le 2p+q \le 2r+s$. 
The simplifications 
discussed in Section \ref{section:simplifications} 
applicable to correlators 
$\langle \OO_{[p,q,p]} \OO_{[r,s,r]} \OO_{[l,k,l]} \rangle$
are: 
the extremality condition 
\begin{equation}
\label{restriction:extremal QQQ}
2r+s = 2p+q + 2l+k; 
\end{equation}
and the ``partition into three disjoint flavors'' condition,%
\footnote{
	Which just says that the number 
	of scalars exchanged between each pair of $\OO$-s 
	is no larger than the length of the first column in 
	the corresponding Young tableaux. 
	There are three more inequalities, 
	but they are satisfied trivially 
	since we took $2l+k \le 2p+q \le 2r+s$. 
}
\begin{eqnarray}
\label{restriction1:disjoint QQQ}
\left\{\matrix{
2r+s &\le& 2l+k + q \cr
2r+s &\le& 2p+q + k \cr
2p+q &\le& 2l+k + s 
}\right.
\label{restriction2:disjoint QQQ}
\end{eqnarray}
(all three inequalities have to be satisfied simultaneously). 
For example, 
(\ref{restriction1:disjoint QQQ}) 
are true 
when all the \quarter-BPS operators are in the $84 = [2,0,2]$ of $SU(4)$; 
take%
\footnote{
	As shown in Appendix \ref{partition-into-two}, 
	such operators are in fact in the 84 of $SU(4)$. 
	}
\begin	{eqnarray}
\label{202:different wts-1}
{\YY} (x) = 
\left\{ 
(\tr z_1^2) (\tr z_2^2) - (\tr z_1 z_2) (\tr z_1 z_2) 
\right\}
+ {1\over N} 
\left\{ 
\tr [z_1 , z_2]^2 
\right\}
,
\\
{\YY} (y) = 
\left\{ 
(\tr \bar z_1^2) (\tr \bar z_3^2) - 
(\tr \bar z_1 \bar z_3) (\tr \bar z_1 \bar z_3) 
\right\}
+ {1\over N} 
\left\{ 
\tr [\bar z_1 , \bar z_3]^2 
\right\}
,
\\
{\YY} (w) = 
\left\{ 
(\tr z_3^2) (\tr \bar z_2^2) - (\tr z_3 \bar z_2) (\tr z_3 \bar z_2) 
\right\}
+ {1\over N} 
\left\{ 
\tr [z_3 , \bar z_2]^2 
\right\}
.
\label{202:different wts-2}
\end	{eqnarray}
The Born amplitude 
does not vanish%
\footnote{
	For $N \le 2$ 
	there are no \quarter-BPS operators in the 84 of $SU(4)$.
	}
for $N > 2$, 
so we can't blame the lack of corrections 
on group theory, 
\begin	{eqnarray}
\label{84-84-84-born}
\langle 
{\YY}(x) \; {\YY}(y) \; {\YY}(w) 
\rangle_{\mathrm {free}}
&\propto& 
(N^2-1)(N^2-4)(2 N^2-15)
\end	{eqnarray}
and since 
$\langle {\YY}(x) \; {\YY}(y) \; {\YY}(w) \rangle$
is of the form (\ref{disjoint-flavors}), 
it receives $\OO(g^2)$ corrections.

Of the allowed 
$\langle \OO_{\tquarter}(x) \OO_{\tquarter}(y) 
\OO_{\tquarter}(w) \rangle$
three-point functions 
where each $\OO_{\tquarter}$ is a scalar composite in a 
$[p,q,p]$ of $SU(4)$ with $2p+q \le 7$, 
ten more satisfy 
(\ref{restriction:extremal QQQ}) or 
(\ref{restriction1:disjoint QQQ}).
\footnote{
	We used the method of BZ triangles 
	(see Appendix \ref{BZ-triangles}) 
	to find the allowed triple products. 
	}
For the remaining five correlators 
\begin{eqnarray}
\label{quarter-all: cases}
&& 
\langle \YY_{[2,0,2]} (x) \YY_{[2,0,2]} (y) \YY_{[2,2,2]} (w) \rangle 
\nonumber\\&& 
\langle \YY_{[2,0,2]} (x) \YY_{[2,1,2]} (y) \YY_{[2,3,2]} (w) \rangle 
\nonumber\\&& 
\langle \YY_{[2,0,2]} (x) \YY_{[2,1,2]} (y) \YY_{[3,1,3]} (w) \rangle 
\nonumber\\&& 
\langle \YY_{[2,0,2]} (x) \YY_{[3,1,3]} (y) \YY_{[2,3,2]} (w) \rangle 
\nonumber\\&& 
\langle \YY_{[2,0,2]} (x) \YY_{[3,1,3]} (y) \YY_{[3,1,3]} (w) \rangle 
\end{eqnarray}
we have to verify that there are no contributions 
proportional to any of the functions 
$\tilde B(x,y)$, $\tilde B(x,w)$, or $\tilde B(y,w)$. 
In fact, with $2l+k \le 2p+q \le 2r+s$, 
\begin{eqnarray}
\label{restriction3:disjoint QQQ}
2l+k \le 2r+s + q \quad\mbox{and}\quad 
2p+q \le 2r+s + k 
\label{restriction4:disjoint QQQ}
\end{eqnarray}
are automatically satisfied, 
so we can always choose the operators as%
\footnote{
	This choice of flavors was motivated by 
	computational efficiency; our {\textit {Mathematica}}
	calculations took more computer time with 
	other flavor breakdowns. 
	}
\begin{eqnarray}
\label{QQQ:operators1}
\YY_{[l,k,l]} (x) &\sim& [\bar z_1^a \bar z_2^b z_3^e] \\
\YY_{[p,q,p]} (y) &\sim& [\bar z_1^c \bar z_2^d \bar z_3^e] \\
\YY_{[r,s,r]} (w) &\sim& [z_1^{r+s} z_2^r] 
\label{QQQ:operators2}
\end{eqnarray}
where $e \equiv \half [(2l+k) + (2p+q) - (2r+s)] 
\le l+k, p+q$; 
and integers $a$, $b$, $c$, $d$ partition 
$r+s = a+c$, $r = b+d$. 
Then 
$\tilde \beta_{xy}=0$ 
since the operators exchanged between $\YY_x$ and $\YY_y$ 
all have the same flavor, and we only 
need to calculate $\tilde \beta_{xw}$ and $\tilde \beta_{yw}$. 
Details of these calculations are given 
in Appendix \ref{tedious details}, 
and here we just quote the result: 
as in the cases considered so far, 
$\tilde \beta_{xy} = \tilde \beta_{xw} = \tilde \beta_{yw} = 0$, 
and none of the three-point functions (\ref{quarter-all: cases}) 
receive any $\OO(g^2)$ corrections.

\section{
$\langle \OO_{\tquarter} \OO_{\tquarter} \OO_{\thalf} \rangle$
correlators in the large $N$ limit}
\label{section:large N:3pt}

Like the two-point functions studied in 
Chapters \ref{chapter: BPS: 2pt} and \ref{chapter:systematic}, 
$\langle \OO_x \OO_y \OO_w \rangle$
calculations get progressively more cumbersome as the 
representations of the $\OO$-s become larger. 
In this Section we will calculate correlators of 
two \quarter-BPS operators with one \half-BPS operator, 
in the large $N$ limit. 
The situation when all three operators are 
\quarter-BPS is even less tractable, 
and we avoid it here.

\subsection{Large $N$ operators}

We will use the \quarter-BPS operators found in 
Chapters \ref{chapter: BPS: 2pt} and \ref{chapter:systematic}. 
Schematically, the special double and single trace operators 
can be written as 
\begin{eqnarray}
\label{def:O and K}
\OO_{[p,q,p]} \sim 
\left(
\mbox{
\setlength{\unitlength}{1em}
\begin{picture}(7.5,2.2)
\put(0,.5){\framebox (1,1){}}
\put(1,.5){\framebox (2,1){\scriptsize $...$}}
\put(3,.5){\framebox (1,1){}}
\put(4,.5){\framebox (1,1){}}
\put(5,.5){\framebox (1,1){\scriptsize $...$}}
\put(6,.5){\framebox (1,1){}}
\put(0,-1){\framebox (1,1){}}
\put(1,-1){\framebox (2,1){\scriptsize $...$}}
\put(3,-1){\framebox (1,1){}}
\put(1.8,-1.7){\scriptsize $p$}
\put(5.4,-0.7){\scriptsize $q$}
\end{picture}}
\right) 
, \quad 
\KK_{[p,q,p]} \sim 
\left(
\mbox{
\setlength{\unitlength}{1em}
\begin{picture}(7.5,2.2)
\put(0,.2){\framebox (1,1){}}
\put(1,.2){\framebox (2,1){\scriptsize $...$}}
\put(3,.2){\framebox (1,1){}}
\put(4,.2){\framebox (1,1){}}
\put(5,.2){\framebox (1,1){\scriptsize $...$}}
\put(6,.2){\framebox (1,1){}}
\put(0,-.8){\framebox (1,1){}}
\put(1,-.8){\framebox (2,1){\scriptsize $...$}}
\put(3,-.8){\framebox (1,1){}}
\put(1.8,-1.5){\scriptsize $p$}
\put(5.4,-0.5){\scriptsize $q$}
\end{picture}}
\right) 
\end{eqnarray}
(each continuous group of boxes stands for an $SU(N)$ trace); 
explicit formulae for highest $SU(4)$ weight operators of this form 
are given in 
Chapter \ref{chapter: BPS: 2pt}, Section \ref{section:oo and kk}. 
In the large $N$ limit the linear combinations 
\begin{eqnarray}
\label{K-hat:large N:again}
\tilde{\YY}_{[p,q,p]} &=& 
\KK_{[p,q,p]}   + \OO(N^{-2}) 
\\
\label{O-hat:large N:again}
\YY_{[p,q,p]} &=& 
\OO_{[p,q,p]} - {p(p+q)\over N} \KK_{[p,q,p]}  + \OO(N^{-2}) 
\end{eqnarray}
are eigenstates of the dilatations operator. 
$\YY_{[p,q,p]}$ 
have protected normalization and 
scaling dimension ($\Delta_{\mbox{\tiny $\YY$}} = 2 p + q$) 
at order $g^2$, 
and were argued to be \quarter-BPS.

We did not specify the $SU(4)$ weights of operators 
$\OO_{[p,q,p]}$ and $\KK_{[p,q,p]}$
in (\ref{def:O and K}). 
The choice of weights will depend on 
the representations in the triple product 
$[p,q,p] \otimes [r,s,r] \otimes [0,k,0]$ in the following way. 
Assume $p \le r$; then it is convenient to choose%
\footnote{
	With this choice of flavors, the manipulations 
	of the following Sections are a simple generalization 
	of the arguments of Section \ref{section:large N} of 
	Chapter \ref{chapter: BPS: 2pt}. 
	}
\begin{eqnarray}
\label{QQH1:large N}
\YY_{[p,q,p]} (x) &\sim& [\bar z_1^l z_2^n z_3^p] \\
\YY_{[r,s,r]} (y) &\sim& [\bar z_1^m \bar z_2^n \bar z_3^p] \\
\OO_{[0,k,0]} (w) &\sim& [z_1^k] 
\label{QQH2:large N}
\end{eqnarray}
with 
$l \equiv \half [(2p+q)+k-(2r+s)]$, 
$m \equiv k-l$ and $n \equiv p+q-l$. 
We will also assume that none of the simplifications 
(\ref{restriction:extremal}-\ref{restriction:two-flavors}) 
apply, since those cases were already discussed in 
Section \ref{section:quarter-quarter-half}.

\subsection{
$\langle \KK \KK \OO_{\thalf} \rangle_{\mathrm {free}}$, 
$\langle \KK \OO \OO_{\thalf} \rangle_{\mathrm {free}}$, 
$\langle \OO \KK \OO_{\thalf} \rangle_{\mathrm {free}}$, 
and 
$\langle \OO \OO \OO_{\thalf} \rangle_{\mathrm {free}}$ 
}
\label{section: Born level - large N}

We can estimate the leading large $N$ behavior of 
the combinatorial factors 
$\alpha_{\mathrm {free}}$ and $\tilde \beta_{xy}$ 
using the ``trace merging formula'' 
\begin{equation}
\label{merging traces:3pt}
2 \left( \tr A t^c \right) \left( \tr t^c B \right) 
= 
\tr A B - 
\mbox{$1\over N$} \left( \tr A \right) \left( \tr B \right) 
\end{equation}
where $A$ and $B$ are arbitrary $N \times N$ matrices 
and $t^c$ are $SU(N)$ generators in the fundamental 
(sums on repeated indices are implied). 
With (\ref{merging traces:3pt}) and the expression for the 
quadratic Casimir%
\footnote{
	For the adjoint and fundamental 
	representations, 
	$C_2(\mbox{adj}) = N$, 
	$C_2(\mbox{fund}) = (\mbox{$N^2 - 1 \over 2N$})$, 
	$C(\mbox{adj}) = N$, $C(\mbox{fund}) = \half$. 
	See Section \ref{suN-identities} for more details. 
	}
\begin{equation}
\label{casimir}
T^c_r T^c_r = C_2(r) \; \bone 
\end{equation}
of $SU(N)$, we find for example that for $k \ge 2$, 
\begin{eqnarray}
\label{leading traces}
(\tr t^{a_1} t^{a_2} ... t^{a_k}) 
(\tr t^{a_k} ... t^{a_2} t^{a_1}) 
&=& 
\left(
{N \over 2} 
\right)^k
\left[
1 + \OO(1/N^2) 
\right]
\end{eqnarray}
To have this large $N$ behavior, generators in 
the two traces should appear in opposite order. 
When the generators are taken in any other order 
(except cyclic permutations inside the traces), 
such products are suppressed 
at least by $1/N^2$.

Calculations proceed along the same lines as in 
Chapter \ref{chapter: BPS: 2pt}. 
We begin by considering correlators of the form 
$\langle \OO \OO \OO_{\thalf} \rangle$ with 
the two $\OO$-s in the same representation $[p,q,p]$. 
The leading contribution to $\alpha_{\mathrm {free}}$ 
comes from terms like 
\begin{eqnarray}
\label{OOH:leading-born:equal}
&& 
( \tr t^{a_1} \! ... t^{a_l} t^{b_1} \! ... t^{b_n} )
( \tr t^{c_1} \! ... t^{c_p} ) 
~ 
( \tr t^{d_1} \! ... t^{d_l} t^{b_n} \! ... t^{b_1} )
( \tr t^{c_p} \! ... t^{c_1} ) 
~ 
( \tr t^{a_l} \! ... t^{a_1} t^{d_l} \! ... t^{d_1} ) 
\nonumber\\
&&\hspace{4em}
\sim 
\left( {1\over2} \right)^2 
N \left( {N\over2} \right)^{2l-2 + n} 
\times 
\left( {N\over2} \right)^p 
= \left( {1\over2} \right) \left( {N\over2} \right)^{2l + n + p - 1}
\hspace{3.5em}
\end{eqnarray}
The factor of $(\half)^2$ comes about because 
we merge traces twice; the exponent $2l + n - 2$ counts 
how many generators collapse using 
$t^c t^c \sim \half N \bone$; 
the extra factor of $N$ is due to $\tr \bone = N$; 
and finally $(N/2)^p$ is from contracting the 
traces of equal length containing the $t^{c_i}$-s. 
All remaining calculations of this and next Sections are 
analogous, and we won't spell things out as much.

If the representations of $[p,q,p]$ and $[r,s,r]$ are 
different, a similar situation occurs when for example 
$p = r+s$, i.e. $\OO_{[p,q,p]}$ and $\OO_{[r,s,r]}$ contain 
traces of equal length. Then we merge traces twice, 
and one set of traces collapses completely 
as in (\ref{leading traces}). 
Otherwise, we have to merge traces three times, so 
the leading contributions to 
$\langle \OO \OO \OO_{\thalf} \rangle_{\mathrm {free}}$ 
are of the form 
\begin{eqnarray}
\label{OOH:leading-born}
&& \hspace{-2em}
( \tr t^{a_1} \! ... t^{a_l} t^{b_1} \! ... t^{b_n} )
( \tr t^{c_1} \! ... t^{c_p} ) 
~ 
( \tr t^{d_1} \! ... t^{d_m} t^{b_n} \! ... )
( \tr \! ... t^{b_1} t^{c_p} \! ... t^{c_1} ) 
~ 
( \tr t^{a_l} \! ... t^{a_1} t^{d_m} \! ... t^{d_1} ) 
\hspace{-.5em}
\nonumber\\
&&\hspace{2.5em}
\sim 
\left( {1\over2} \right) 
\left( {N\over2} \right)^{p+l+m+n-1} 
~\quad 
\mbox{\parbox{10em}{if $p = r$ or $p+q = r+s$ \par
or $p = r+s$ or $r = p+q$}} 
\nonumber\\&&\hspace{2.5em}
\sim 
\left( {1\over2} \right)^3 
\left( {N\over2} \right)^{p+l+m+n-3} 
\quad 
\mbox{otherwise}
\end{eqnarray}

For the other three types of correlators, 
no pair of traces ever collapses completely, 
so the answers are more uniform. We find 
that the large $N$ behavior of 
$\langle \KK \KK \OO_{\thalf} \rangle_{\mathrm {free}}$ 
is defined by terms like 
\begin{eqnarray}
\label{KKH:leading-born}
( \tr t^{a_1} ... t^{a_l} t^{b_1} ... t^{b_n} t^{c_1} ... t^{c_p} ) 
~ 
( \tr t^{d_1} ... t^{d_m} t^{b_n} ... t^{b_1} t^{c_p} ... t^{c_1} ) 
~ 
( \tr t^{a_l} ... t^{a_1} t^{d_m} ... t^{d_1} ) 
\hspace{-28em}&&\nonumber\\
&&\sim 
\left( {1\over2} \right)^2 
\left( {N\over2} \right)^{p+l+m+n-2} N 
= \left( {1\over2} \right) 
\left( {N\over2} \right)^{p+l+m+n-1} 
\hspace{4em}
\end{eqnarray}
as we merge traces twice. 
Similarly, 
$\langle \OO \KK \OO_{\thalf} \rangle_{\mathrm {free}}$ 
scales as the terms 
\begin{eqnarray}
\label{OKH:leading-born}
( \tr t^{a_1} ... t^{a_l} t^{b_1} ... t^{b_n} )
( \tr t^{c_1} ... t^{c_p} ) 
~ 
( \tr t^{d_1} ... t^{d_m} t^{b_n} ... t^{b_1} t^{c_p} ... t^{c_1} ) 
~ 
( \tr t^{a_l} ... t^{a_1} t^{d_m} ... t^{d_1} ) 
\hspace{-30em}&&\nonumber\\
&&\sim 
\left( {1\over2} \right)^2 
\left( {N\over2} \right)^{p+l+m+n-2} 
\hspace{16em}
\end{eqnarray}
since traces have to be merged three times now. 
The three-point functions 
$\langle \KK \OO \OO_{\thalf} \rangle_{\mathrm {free}}$ 
also have the leading large $N$ dependence
(\ref{OKH:leading-born}).

\subsection{
$\langle \KK \KK \OO_{\thalf} \rangle_{g^2}$, 
$\langle \KK \OO \OO_{\thalf} \rangle_{g^2}$, 
$\langle \OO \KK \OO_{\thalf} \rangle_{g^2}$, 
and 
$\langle \OO \OO \OO_{\thalf} \rangle_{g^2}$ 
}
\label{section: g^2 - large N}

Here there are no special cases to consider. 
We have to merge traces 
twice for $\langle \KK \KK \OO_{\thalf} \rangle_{g^2}$, 
three times for 
$\langle \KK \OO \OO_{\thalf} \rangle_{g^2}$ or 
$\langle \OO \KK \OO_{\thalf} \rangle_{g^2}$, and 
four times for 
$\langle \OO \OO \OO_{\thalf} \rangle_{g^2}$. 
The leading behavior of the $\tilde \beta_{xy}$ 
combinatorial coefficient 
for the three-point functions 
$\langle \KK \KK \OO_{\thalf} \rangle_{g^2}$ 
is the same as for terms of the form 
\begin{eqnarray}
\label{KKH:leading-g^2}
&& \hspace{-2em}
( \tr t^{a_1} \!\! ... t^{a_l} t^{b_1} \!\! ... t^{b_{n-1}} [t^{c_1},t^s] 
t^{c_1} \!\! ... t^{c_p} ) 
~ 
( \tr t^{d_1} \!\! ... t^{d_m} t^{c_p} \!\! ... t^{c_2} [t^{b_n},t^s] 
t^{b_n} \!\! ... t^{b_1} ) 
~ 
( \tr t^{a_l} \!\! ... t^{a_1} t^{d_m} \!\! ... t^{d_1} ) 
\nonumber\\
&&\hspace{4em}
\sim 
\left( {1\over2} \right)^2 
\left( {N\over2} \right)^{l+m-2} 
\hspace{-1.5em}
( \tr t^{b_1} ... t^{b_{n-1}} [t^{c_1},t^s] 
t^{c_1} ... t^{c_p} 
t^{c_p} ... t^{c_2} [t^{b_n},t^s] 
t^{b_n} ... t^{b_1} )
\hspace{-1em}
\nonumber\\&&\hspace{4em}
\sim
\left( {1\over2} \right)^2 
\left( {N\over2} \right)^{p+l+m+n-4} 
( \tr [t^{c_1},t^s] t^{c_1} [t^{b_n},t^s] t^{b_n} )
\nonumber\\&&\hspace{4em}
\sim
\left( {1\over2} \right) 
\left( {N\over2} \right)^{p+l+m+n} 
\end{eqnarray}
which give the leading large $N$ contributions to it. 
In the same fashion, the most significant terms 
in the correlators 
$\langle \OO \KK \OO_{\thalf} \rangle_{g^2}$ 
are 
\begin{eqnarray}
\label{OKH:leading-g^2}
&& \hspace{-2em}
( \tr t^{a_1} \!\!\! ... t^{a_l} t^{b_1} \!\!\! ... t^{b_{n-1}} [t^{c_1},t^s] ) 
( \tr t^{c_1} \!\!\! ... t^{c_p} ) 
~ 
( \tr t^{d_1} \!\!\! ... t^{d_m} t^{c_p} \!\!\! ... t^{c_2} [t^{b_n},t^s] 
t^{b_n} \!\!\! ... t^{b_1} ) 
~ 
( \tr t^{a_l} \!\!\! ... t^{a_1} t^{d_m} \!\!\! ... t^{d_1} ) 
\nonumber\\
&&\hspace{4em}
\sim 
\left( {1\over2} \right)^2 
\left( {N\over2} \right)^{p+l+m+n-1} 
\sim 
\langle \KK \OO \OO_{\thalf} \rangle_{g^2} 
\hspace{11em}
\end{eqnarray}
while 
$\langle \OO \OO \OO_{\thalf} \rangle_{g^2}$ 
gets its leading $N$ behavior from terms like 
\begin{eqnarray}
\label{OOH:leading-g^2}
&& \hspace{-1em}
( \tr t^{a_1} ... t^{a_l} t^{b_1} ... t^{b_{n-1}} [t^{c_1},t^s] ) 
( \tr t^{c_1} ... t^{c_p} ) 
~ 
( \tr t^{d_1} ... t^{d_m} t^{c_p} ... ) ( \tr ... t^{c_2} [t^{b_n},t^s] 
t^{b_n} ... t^{b_1} ) 
\nonumber\\ && \times 
( \tr t^{a_l} ... t^{a_1} t^{d_m} ... t^{d_1} ) 
\nonumber\\
&&\hspace{4em}\sim 
\left( {1\over2} \right)^3 
\left( {N\over2} \right)^{p+l+m+n-2} 
\hspace{15em}
\end{eqnarray}

\subsection{
$\langle \OO_{\tquarter} \OO_{\tquarter} \OO_{\thalf} \rangle$
correlators are protected
}

With just a little more work, we can find the ratios 
of the $\OO(g^2)$ corrections to the three-point functions 
$\langle \OO \OO \OO_{\thalf} \rangle_{g^2}$, 
$\langle \OO \KK \OO_{\thalf} \rangle_{g^2}$, 
$\langle \KK \OO \OO_{\thalf} \rangle_{g^2}$, 
and 
$\langle \KK \KK \OO_{\thalf} \rangle_{g^2}$. 
The argument proceeds along the same lines as in 
Chapter \ref{chapter: BPS: 2pt}. 
Given a term with generators in a particular order, 
contributing to $\langle \KK \KK \OO_{\thalf} \rangle_{g^2}$, 
such as the one shown in (\ref{KKH:leading-g^2}), 
we know that a term with the same order of generators 
also gives a leading contribution to 
$\langle \OO \KK \OO_{\thalf} \rangle_{g^2}$ 
as in 
(\ref{OKH:leading-g^2}). 
However, cyclic permutations 
within the two traces (of length $p$ and $p+q$) of $\OO$, 
also contribute to 
$\langle \OO \KK \OO_{\thalf} \rangle_{g^2}$ 
in the same amount as the term (\ref{OKH:leading-g^2}). 
Therefore, 
\begin{equation}
\langle \OO \KK \OO_{\thalf} \rangle_{g^2}
/
\langle \KK \KK \OO_{\thalf} \rangle_{g^2} = 
{p(p+q) \over N} + \OO(N^{-3}) 
\equiv \beta 
\end{equation}  
In the same fashion 
\begin{eqnarray}
\langle \KK \OO \OO_{\thalf} \rangle_{g^2}
/
\langle \KK \KK \OO_{\thalf} \rangle_{g^2} &=& 
{r(r+s) \over N} + \OO(N^{-3}) 
\equiv \beta' 
,\\
\langle \OO \OO \OO_{\thalf} \rangle_{g^2}
/
\langle \KK \OO \OO_{\thalf} \rangle_{g^2} &=& 
{p(p+q) \over N} + \OO(N^{-3}). 
\end{eqnarray}

Next consider the Born level correlators 
of Section \ref{section: Born level - large N}. 
When a pair of traces collapses 
completely 
(see equations \ref{OOH:leading-born}-\ref{OKH:leading-born}), 
we get 
\begin{eqnarray}
\label{QQH:free - collapse completely}
\langle \YY_{[p,q,p]} \YY_{[r,s,r]} 
\OO_{\thalf} \rangle_{\mathrm {free}} 
\sim 
\langle \OO \OO \OO_{\thalf} \rangle_{\mathrm {free}} \sim 
N^{p+l+m+n-1} 
\end{eqnarray}
Otherwise, the contributions add up to 
\begin{equation}
\label{QQH:free - do not collapse completely}
\langle \OO \OO \OO_{\thalf} \rangle_{\mathrm {free}} - 
\beta \langle \KK \OO \OO_{\thalf} \rangle_{\mathrm {free}} - 
\beta' \langle \OO \KK \OO_{\thalf} \rangle_{\mathrm {free}} + 
\beta \beta' \langle \KK \KK \OO_{\thalf} \rangle_{\mathrm {free}} 
\sim 
N^{p+l+m+n-3} 
\end{equation}
The terms in (\ref{QQH:free - do not collapse completely})
are all of the same order 
and do not cancel. 
The factors of $\beta$ and $\beta'$ discussed above are still present, 
but there are other complications. 
First, the string of $t^c$-s can be inserted 
anywhere in the third trace in (\ref{OOH:leading-born}), 
and cyclic permutations of the $t^b$-s in the same trace 
give terms of the same order in $N$. 
This results in an extra factor of $(r-p)^2$.
Second, different terms in the 
sum over antisymmetrizations 
(as in equation (\ref{def:OO}), for example)
contribute differently. 
The combinatorics is more involved, 
and we do not discuss this case in detail.

Bringing everything together, we see that 
the order $g^2$ corrections to the 
three-point function of the BPS operators 
in the large $N$ limit add up to 
\begin{eqnarray}
\label{QQH:leading-g^2}
\langle \YY_{[p,q,p]} \YY_{[r,s,r]} 
\OO_{\thalf} \rangle_{g^2} 
&\propto& 
\left( {1\over2} \right) 
\left( {N\over2} \right)^{p+l+m+n-1} 
\tilde B(x,y) N 
\nonumber\\&&\hspace{3em}\times
\left(
\matrix{
-\beta \cr 
1 
}\right)^t
\left( 
\matrix{
1 + \OO(N^{-2}) & \beta' \cr
\beta & \beta \beta' 
}
\right)
\left( 
\matrix{
-\beta' \cr
1 
}
\right)
\nonumber\\&=& 
\left( {1\over2} \right) 
\left( {N\over2} \right)^{p+l+m+n-1} 
\tilde B(x,y) N 
\times
\OO(N^{-4}) 
\quad\quad 
\end{eqnarray}
A comparison of (\ref{QQH:leading-g^2}) with 
(\ref{QQH:free - collapse completely}) 
or (\ref{QQH:free - do not collapse completely}) 
shows that order $g^2$ corrections to 
three-point functions 
of one \half-BPS operator with two \quarter-BPS operators 
vanish in the large $N$ limit, 
within working precision.

\section{Supergravity considerations}
\label{section:SUGRA}

In the AdS/CFT correspondence, there is a duality mapping 
single trace 
\half-BPS primary operators 
$\tr X^k$ 
of the SYM theory 
onto canonical supergravity fields, \cite{AdS-CFT}. 
Given a set of such \half-BPS primary operators, 
one can compute their two- and three-point functions 
in SYM. 
Two-point functions define the normalization of 
operators, and three-point functions probe the 
interactions between them. 
Independently, both the normalization 
of the SUGRA fields 
as well as their couplings, can be read off from the 
supergravity action (or supergravity equations of motion), \cite{LMRS}. 
So as a check of the AdS/CFT correspondence, 
one can compare the unambiguously defined 
three- and higher $n$-point functions of 
normalized \half-BPS operators in SYM, 
with the correlators of the corresponding 
elementary excitations in supergravity, 
\cite{LMRS, AF, ADS:2-and-3, ADS:4-pt}.

We would like to proceed, in the same spirit, 
with the 
\quarter-BPS chiral primaries of 
the \NN=4 Super Yang Mills. 
We argued that these two- and three-point functions are 
independent of the SYM coupling constant 
(at least to order $g^2$), so it is reasonable 
to expect these correlators to agree with their dual AdS description. 
However, multiple trace operators 
do not correspond to any of the fields appearing in the 
supergravity action, so the discussion 
will be different than 
in the case of 
the previously studied \half-BPS primary operators $\tr X^k$.

\subsection{OPE definition of \quarter-BPS chiral primaries}
\label{section:OPE}

One of the ways to see \quarter-BPS chiral primaries 
is to consider higher $n$-point correlators of \half-BPS operators. 
For example, four-point functions of [0,2,0] operators 
reveal a pole corresponding to the exchange of 
a 
[2,0,2] operator 
with a protected dimension $\Delta=4$, \cite{BKRS}. 
In general, 
the \quarter-BPS primaries 
$\YY_{[p,q,p]}$ 
show up in 
\begin{equation}
\label{double trace:4pt of CPOs}
\langle \tr X^{(p+q)}(x) \tr X^p(x+\e) ~ 
\tr X^{(p+q)}(y) 
~ \tr X^p(w) 
\rangle
\end{equation} 
in the limit $\e\to0$, 
as the $[p,q,p]$ operators with 
the threshold value of scaling dimension 
$\Delta = 2p+q = {\mathrm {dim}} [\tr X^{(p+q)}] + {\mathrm {dim}} [\tr X^p]$. 
In other words, \quarter-BPS chiral primaries 
can be defined by the OPE-s 
of \half-BPS operators as 
\begin{equation}
\label{def:double trace - OPE}
\PP^{\Delta = 2p+q}_{[p,q,p]}
\left[ 
\lim_{\e\to0}
\tr X^{(p+q)}(x) \; \tr X^p(x+\e) 
\right] 
= 
\sum_i c_i \YY^{[p,q,p]}_i(x) 
\end{equation} 
Here, $\PP^{\Delta}_{[p,q,p]}$ projects onto 
the $[p,q,p]$ representation of $SU(4)$, 
and eliminates 
operators with scaling dimension other than $\Delta$ 
(e.g. the non-chiral descendants with the same 
$SU(4)$ quantum numbers). 
Singular terms normally subtracted from 
an OPE such as (\ref{def:double trace - OPE}), are 
automatically removed by applying $\PP^{\Delta = 2p+q}_{[p,q,p]}$. 

On the other hand, 
one can see by calculating three-point correlators that 
all \quarter-BPS primary operators $\YY^{[p,q,p]}_i$ 
are present in the OPE (\ref{def:double trace - OPE}). 
It appears that for general $N$, 
there is no canonical definition 
of the special $\YY_{[p,q,p]}$ 
that is a linear combination of 
the single and double-trace scalar composite $[p,q,p]$ operators only. 
However, this $\YY_{[p,q,p]}$ 
dominates in the $N \to \infty$ limit. 
For large $N$, 
all other terms in the right hand side of (\ref{def:double trace - OPE}) 
are suppressed by at least a factor of $1/N$, and 
the predominantly double-trace \quarter-BPS chiral primary operator 
\begin{equation}
\YY_{[p,q,p]} = 
\left(
\mbox{
\setlength{\unitlength}{1em}
\begin{picture}(7.5,1.6)
\put(0,.5){\framebox (1,1){}}
\put(1,.5){\framebox (2,1){\scriptsize $...$}}
\put(3,.5){\framebox (1,1){}}
\put(4,.5){\framebox (1,1){}}
\put(5,.5){\framebox (1,1){\scriptsize $...$}}
\put(6,.5){\framebox (1,1){}}
\put(0,-1){\framebox (1,1){}}
\put(1,-1){\framebox (2,1){\scriptsize $...$}}
\put(3,-1){\framebox (1,1){}}
\put(1.8,-1.7){\scriptsize $p$}
\put(5.4,-0.7){\scriptsize $q$}
\end{picture}}
\right) 
+ \OO(1/N)
\phantom{{}_\Big|}
\end{equation}
is uniquely defined by the OPE of \half-BPS primaries. 

When translated into the SUGRA language, 
the definition (\ref{def:double trace - OPE}) 
implies that 
\quarter-BPS primary operators of SYM 
should be thought of as 
threshold bound states of 
elementary SUGRA excitations. 
A threshold bound state is a state whose mass 
is precisely equal to the sum of the masses of all its 
constituents, and thus occurs at the lower end of the spectrum. 
Any bound state of BPS states which is itself BPS 
is automatically a threshold bound state. 
A familiar example is provided by an assembly of 
like sign charged Prasad-Sommerfield magnetic monopoles, 
whose classical static solution forms a 
threshold bound state of monopole constituents.

\subsection{
$\langle
\OO_{\tquarter} \OO_{\tquarter} 
\rangle$ 
and 
$\langle
\OO_{\thalf} \OO'_{\thalf} \OO_{\tquarter} 
\rangle$ 
correlators
}

We are now going to 
illustrate the consistency of this dictionary. 
Specifically, 
we will look at two- and three-point functions 
involving \half- and \quarter-BPS operators 
in the large $N$ limit, 
in \NN=4 SYM. 
Then we will compare these correlators with 
their dual supergravity description.

The normalization of \half- and \quarter-BPS operators 
comes from their two-point functions, whose leading  
large $N$ behavior was found in Chapter \ref{chapter: BPS: 2pt} 
(also see \cite{LMRS}):  
\begin{eqnarray}
\label{normalization:large N - half} 
\langle
\tr X^q (x) 
~ 
\tr X^q (y) 
\rangle
&\sim& 
{N^{q} \over (x-y)^{2 q} }
\\
\langle
\YY_{[p,q,p]} (x) 
\; 
\bar{\YY}^{[p,q,p]} (y) 
\rangle
&\sim& 
{N^{(2p+q)} \over (x-y)^{2(2p+q)} }
\label{normalization:large N - quarter} 
\end{eqnarray}
times some $N$-independent factors which we omit. 

The simplest three point functions 
involving \half- and \quarter-BPS operators 
are of the form 
$
\langle
\OO_{\thalf} (x) \OO'_{\thalf} (y) \OO_{\tquarter} (w) 
\rangle
$. 
If the $SU(N)$ traces collapse completely 
(in which case 
$\langle \OO_{\thalf} \OO'_{\thalf} \OO_{\tquarter} \rangle$ 
are extremal), 
the normalized three point-functions are then 
\begin{eqnarray}
\label{normalized:large N:HHQ-extremal} 
{1\over \sqrt{N^{(2p+q)+(p+q)+p}}}
\langle
\tr X^{(p+q)} (x) 
~ 
\tr X^{p} (y) 
~
\YY_{[p,q,p]} (w) 
\rangle 
\sim 
1
\end{eqnarray}
from a field theory calculation; 
the space-time coordinate dependence is fixed by 
conformal invariance, so we will not exhibit it anymore. 
If the traces do not collapse completely, the correlator 
is suppressed by $1/N^2$ 
(see the discussion around equations 
(\ref{OOH:leading-born:equal}-\ref{OOH:leading-born}) 
of Section 
\ref{section:large N:3pt}),
and 
\begin{eqnarray}
\label{normalized:large N:HHQ} 
{1\over \sqrt{N^{(2p+q)+(k+l)+k}}}
\langle
\tr X^{(k+l)} (x) 
~ 
\tr X^{k} (y) 
~
\YY_{[p,q,p]} (w) 
\rangle 
\sim 
{1\over N^2}
\end{eqnarray} 
whenever $k \ne p$ of $l \ne q$. 
All this matches nicely 
with the corresponding supergravity diagrams: 
\begin{eqnarray}
\label {AdS:half-half-quarter}
&&\epsfig{height=1.4in, file=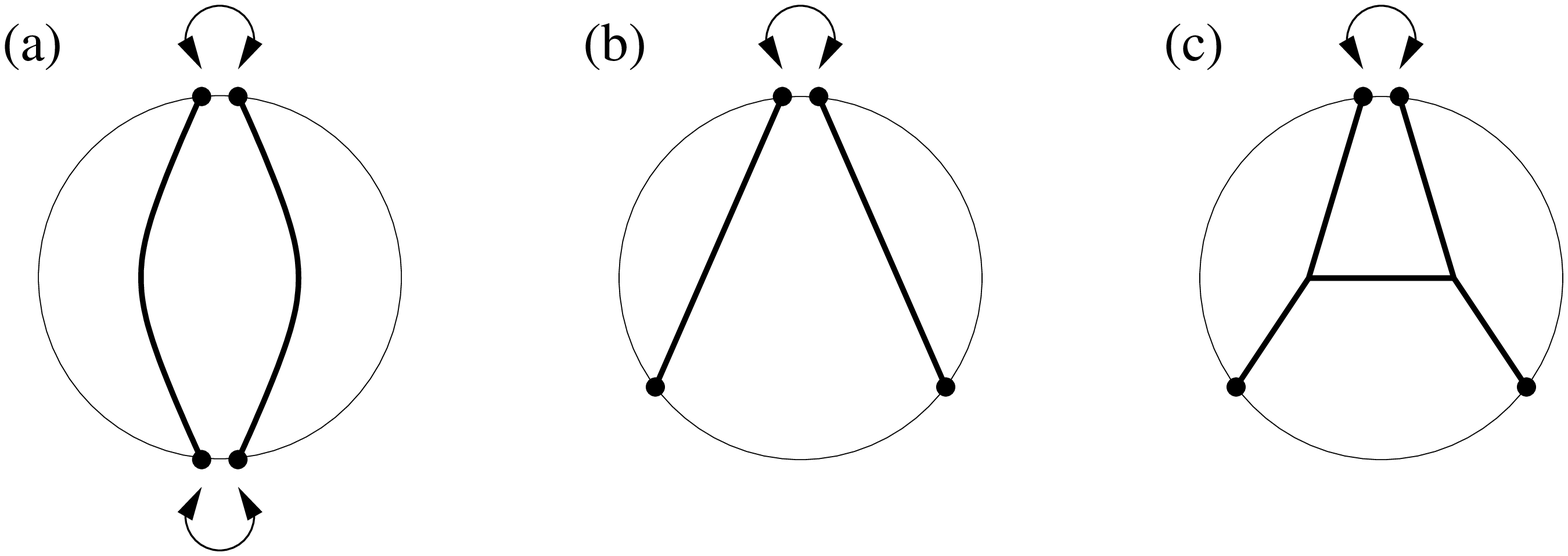, angle=0}
\hspace{-2em}
\\
&&\mbox{
Leading AdS diagrams for 
(a) equation (\ref{normalization:large N - quarter}); 
(b) equation (\ref{normalized:large N:HHQ-extremal}); 
}\hspace{0.6em}
\nonumber
\\
&&\mbox{
(c) equation (\ref{normalized:large N:HHQ}). 
Each cubic bulk interaction vertex goes like $1/N$. 
}
\nonumber
\end{eqnarray}
We denoted \half-BPS primaries by ``$\bullet$''; 
and the predominantly double trace \quarter-BPS 
primaries which arise from bringing two \half-BPS operators 
together by 
``$\hspace{0.2em}
\begin{picture}(1,1)
\put(0,-2){\epsfig{height=0.17in, file=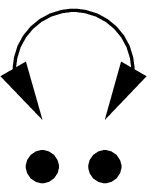, angle=0}}
\end{picture}
{\phantom{\Big|}}\hspace{0.85em}$''.

There are also AdS diagrams with 
quartic interactions in the bulk, 
which have the same large $N$ dependence as 
(\ref{AdS:half-half-quarter}c); 
we will not show these.

\subsection{
$\langle
\OO_{\tquarter} \OO'_{\tquarter} \OO_{\thalf} 
\rangle$ 
correlators
}

Other three point functions involving 
\quarter-BPS as well as \half-BPS operators 
can be analyzed similarly. 
Whenever traces of the SYM operators 
do not collapse completely, 
the supergravity counterparts of such correlators 
have extra bulk interaction vertices. 
The leading dependence 
of such correlators is then suppressed 
by the corresponding power of $1/N$. 
For example, correlators of the form 
$
\langle
\OO_{\tquarter} (x) \OO'_{\tquarter} (y) \OO_{\thalf} (w) 
\rangle
$, 
discussed in Section 
\ref{section:large N:3pt},
behave like 
\begin{equation}
\label{OOH:leading-born:normalized}
{1\over \sqrt{N^{(2p+q)+(2r+s)+k}}}
\langle 
\YY_{[p,q,p]} \YY_{[r,s,r]} \tr X^k \rangle 
\sim 
\left\{
\matrix{
1/N 
& 
\mbox{(a) if one pair of traces}
\cr &
\mbox{collapses completely}\hfill 
\cr 
1/N^3 
& 
\mbox{(b) otherwise}\hfill
}
\right.
\end{equation}
From the AdS point of view, 
this difference is explained by the following diagrams 
\begin{eqnarray}
\label {AdS:quarter-quarter-half}
&&\epsfig{height=1.0in, file=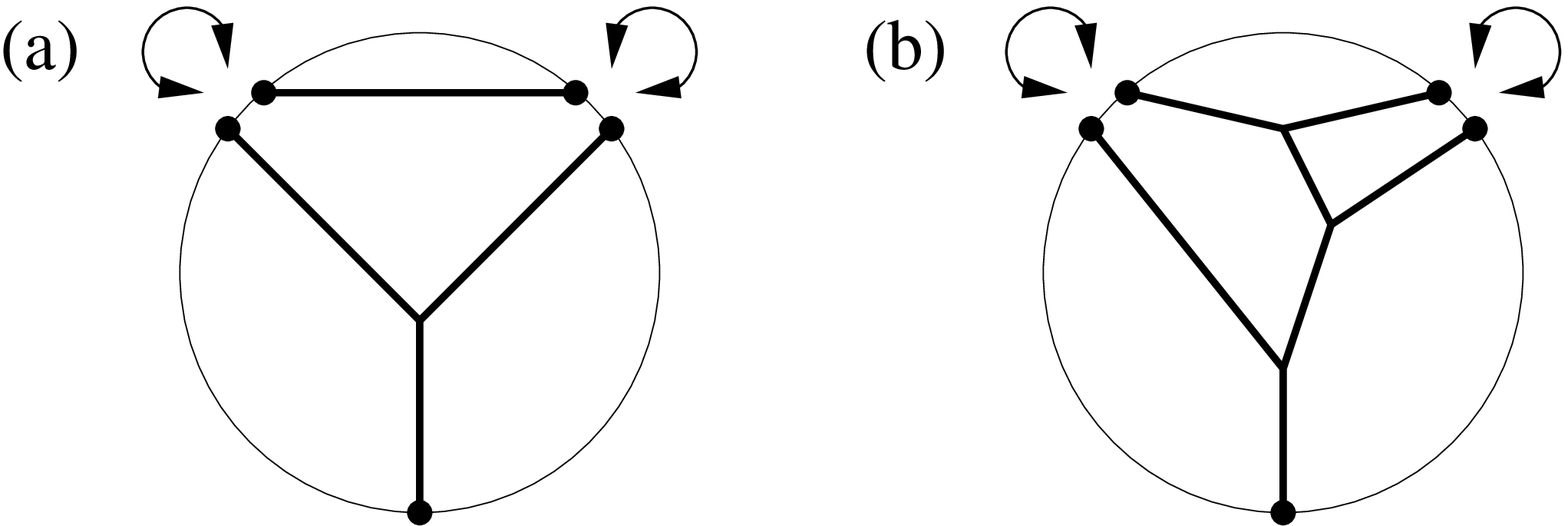, angle=0}
\hspace{2em}\phantom{\Bigg|}
\\
&&\hspace{4em}\mbox{
AdS description of 
equation (\ref{OOH:leading-born:normalized}). 
}
\nonumber
\end{eqnarray}

\subsection{
$\langle
\OO_{\tquarter} \OO'_{\tquarter} \OO''_{\tquarter} 
\rangle$ 
correlators
}

Similar arguments show that when all 
operators are \quarter-BPS, the normalized 
three-point functions are 
\begin{equation}
\label{OOO:leading-born:normalized}
{1\over \sqrt{N^{(2p+q)+(2r+s)+(2l+k)}}}
\langle 
\YY_{[p,q,p]} \YY_{[r,s,r]} \YY_{[l,k,l]} \rangle 
\sim 
\left\{
\matrix{
1
&
\mbox{(a) if all traces}\hfill
\cr &
\mbox{collapse pairwise}\hfill 
\cr
1/N^2 
&
\mbox{(b) if only one pair}\hfill
\cr &
\mbox{of traces collapses}\hfill
\cr 
1/N^4 
&
\mbox{(c) otherwise}\hfill
}
\right.
\end{equation}
and the corresponding AdS diagrams 
\begin{eqnarray}
\label {AdS:quarter-quarter-quarter}
&&\epsfig{height=1.1in, file=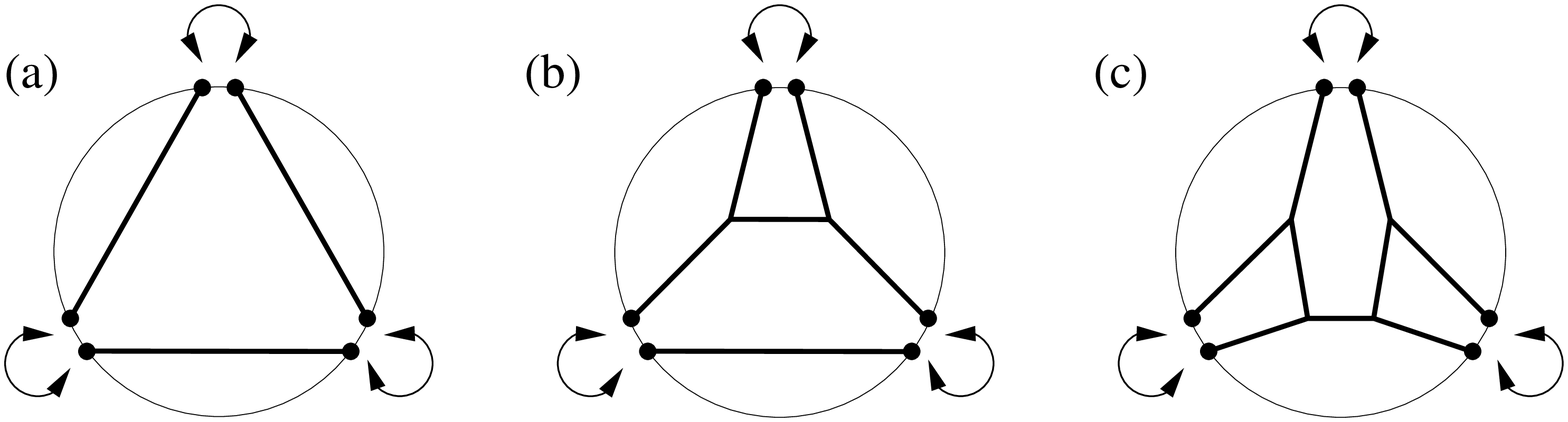, angle=0}
\hspace{1em}\phantom{\Bigg|}
\\
&&\hspace{7em}\mbox{
AdS description of 
equation (\ref{OOO:leading-born:normalized}). 
}
\nonumber
\end{eqnarray}
show the correct leading large $N$ behavior.

\subsection{
Detailed agreement between SYM and AdS
}

Unlike the \half-BPS calculations (e.g. \cite{LMRS}), 
this study does not provide a new independent check or application 
of the AdS/CFT correspondence. 
On the one hand, the definition of 
the predominantly double-trace \quarter-BPS 
operators in the SYM theory (in the large $N$ limit) 
is based on the OPE of \half-BPS primaries. 
On the other hand, AdS correlators of 
the duals of the \quarter-BPS operators 
(bound states of elementary SUGRA excitations)
are defined by the corresponding correlators of 
primary supergravity fields. 
Therefore, SYM correlators involving 
\quarter-BPS operators 
agree by construction 
with their SUGRA counterparts.

This is especially clear in the cases 
show in diagrams 
(\ref{AdS:half-half-quarter}a,b), 
(\ref{AdS:quarter-quarter-half}a), 
and (\ref{AdS:quarter-quarter-quarter}a). 
To leading order in $N$, these two- and three-point functions 
of \quarter-BPS scalar composite operators 
are expressed in terms of the previously studied 
two- and three-point functions of 
\half-BPS chiral primary operators.

\section{Conjectures} 

Let us summarize what has been done so far. 
First, \quarter-BPS primary operators were identified in 
Chapters \ref{chapter: BPS: 2pt} and \ref{chapter:systematic}: 
for general representations $[p,q,p]$ in the large $N$ limit; 
and for general $N$ in the case when $2p+q \le 7$. 
Second, three-point functions 
involving \half-BPS operators as well as 
several infinite families of \quarter-BPS operators 
were considered in this Chapter, 
also for arbitrary $N$. 
It was found that there are no $\OO(g^2)$ corrections 
to such correlators. 
Next, all three-point functions involving the \quarter-BPS primaries 
with $2p+q \le 7$ 
were 
computed for 
general $N$, and were shown to be protected at order $g^2$. 
In the large $N$ approximation, 
three point functions involving 
two \quarter-BPS primaries and one \half-BPS primary 
were shown to receive no $\OO(g^2)$ corrections, 
for general representations of the operators involved. 
Finally, we presented AdS considerations which reproduced 
many features of the CFT two- and three-point functions 
in the large $N$ limit.

Collecting the non-renormalization effects established above 
generates strong evidence for a number of natural conjectures, 
which we now state: 

(1) 
We conjecture that on the CFT side, 
for every $[p,q,p]$ representation of $SU(4)$ 
and arbitrary $N$, 
there are \quarter-BPS chiral primaries. 
Within each $[p,q,p]$, one of these operators is a linear 
combination of double and single trace scalar 
composites only; 
the other \quarter-BPS chiral primaries in $[p,q,p]$ 
also involve operators with higher numbers of traces. 

(2)
We speculate that two--point 
functions of \quarter-BPS operators, 
as well as three-point functions involving 
\half-BPS and \quarter-BPS operators, 
do not depend on the coupling $g^2$ of \NN=4 SYM. 
This non-renormalization also persist for all $N$, 
and is not just a large $N$ approximation. 

(3)
One of the group theory arguments 
of Section \ref{section:simplifications}, 
and the analysis of Section 
\ref{section:large N:3pt},
generalize straightforwardly to 
extremal correlators, 
i.e. $(n+1)$-point functions of the form 
$\langle \OO_0 (x_0) \OO_1(x_1) ... \OO_n(x_n) \rangle$ 
with $\Delta_0 = \Delta_1 + ... + \Delta_n$. 
So do the AdS considerations of Section \ref{section:SUGRA}. 
Therefore, we conjecture that arbitrary extremal correlators 
of \half- and \quarter-BPS chiral primaries are 
also protected. 


\section{Appendix}

\subsection{$[p,q,p] \otimes [r,s,r] \otimes [0,k,0]$ 
and BZ triangles}
\label{BZ-triangles}

Tensoring irreducible representations using Young tableaux 
can get quite tedious. 
Berenstein-Zelevinsky (BZ) triangles \cite{DiF} 
provide a powerful way to calculate the multiplicity of the 
scalar representation%
\footnote{
	It is conventional to choose 
	$\nu^*$ instead of $\nu$ for the third weight. 
	}
in $\lambda \otimes \mu \otimes \nu$.
We will discuss the construction for $SU(3)$ and $SU(4)$, 
the generalization to higher $SU(N)$ 
(but not to other Lie algebras, which is not currently known)
being straightforward.

For $SU(3)$, 
the triangles are constructed according to 
the following rules:
\begin{equation}
\label{BZ:su3}
\begin{tabular}{  c c c c c c c  }
         &          &          & $m_{13}$ &          &          &          \cr
         &          & $n_{12}$ &          & $l_{23}$ &          &          \cr
         & $m_{23}$ &          &          &          & $m_{12}$ &          \cr
$n_{13}$ &          & $l_{12}$ &          & $n_{23}$ &          & $l_{13}$ 
\end{tabular}
\end{equation}
where the nine non-negative integers 
$l_{ij}$, $m_{ij}$, $n_{ij}$ are related to the 
Dynkin labels 
$(\lambda_1,\lambda_2)$, $(\mu_1,\mu_2)$, $(\nu_1,\nu_2)$ 
of the highest weights of the three representations by 
\begin{equation}
\begin{tabular}{  c  c  c  }
$m_{13}+n_{12}=\lambda_1$ & $n_{13}+l_{12}=\mu_1$ & $l_{13}+m_{12}=\nu_1$ \cr
$m_{23}+n_{13}=\lambda_2$ & $n_{23}+l_{13}=\mu_2$ & $l_{23}+m_{13}=\nu_2$ 
\end{tabular}
\end{equation}
They must further satisfy the so-called hexagon conditions 
\begin{equation}
\label{hex:su3}
\begin{tabular}{  c  }
$n_{12}+m_{23}=m_{12}+n_{23}$ \cr
$l_{12}+m_{23}=m_{12}+l_{23}$ \cr
$l_{12}+n_{23}=n_{12}+l_{23}$ 
\end{tabular}
\end{equation}
This means that the length of opposite sides in the hexagon formed by 
$n_{12}$, $l_{23}$, $m_{12}$, $n_{23}$, $l_{12}$, and $m_{23}$ in 
(\ref{BZ:su3}) are equal, the length of a segment being the sum of its 
two vertices.

The number of such triangles gives the multiplicity ${\mathcal N}_{\lambda\mu\nu}$; 
if it is not possible to construct such a triangle, 
$\nu^*$ does not occur in the tensor product $\lambda \otimes \mu$. 

The integers in the BZ triangles have the following origin. 
Each pair of indices $ij$, $i<j$, on the labels of the triangle 
is related to a positive root of $SU(3)$. 
For $SU(N)$, positive roots can be written as 
$\e_i-\e_j$, $1 \le i \le j \le N$, 
in terms of orthonormal vectors $\e_i$ in ${\mathbf R}^N$. 

The triangle encodes three sums of positive roots:
\begin{equation}
\begin{tabular}{  c  }
$\mu + \nu - \lambda^* = \sum_{i<j} l_{ij} (\e_i-\e_j)$ \cr
$\nu + \lambda - \mu^* = \sum_{i<j} m_{ij} (\e_i-\e_j)$ \cr
$\lambda + \mu - \nu^* = \sum_{i<j} n_{ij} (\e_i-\e_j)$ 
\end{tabular}
\end{equation}
The hexagon relations (\ref{hex:su3}) can be seen as 
consistency conditions for these three expansions.

For $SU(4)$, the BZ triangles are defined in a similar way, 
in terms of 
\begin{equation}
\label{BZ:su4}
\begin{tabular}{  c c c c c c c c c c c  }
         &          &          &          &          & $m_{14}$ &          &          &          &          &          \cr
         &          &          &          & $n_{12}$ &          & $l_{34}$ &          &          &          &          \cr
         &          &          & $m_{24}$ &          &          &          & $m_{13}$ &          &          &          \cr
         &          & $n_{13}$ &          & $l_{23}$ &          & $n_{23}$ &          & $l_{24}$ &          &          \cr
         & $m_{34}$ &          &          &          & $m_{23}$ &          &          &          & $m_{12}$ &          \cr
$n_{14}$ &          & $l_{12}$ &          & $n_{24}$ &          & $l_{13}$ &          & $n_{34}$ &          & $l_{14}$ 
\end{tabular}
\end{equation}
eighteen non-negative integers, 
related to the Dynkin labels by 
\begin{equation}
\begin{tabular}{  c  c  c  }
$m_{14}+n_{12}=\lambda_1$ & $n_{14}+l_{12}=\mu_1$ & $l_{14}+m_{12}=\nu_1$ \cr
$m_{24}+n_{13}=\lambda_2$ & $n_{24}+l_{13}=\mu_2$ & $l_{24}+m_{13}=\nu_2$ \cr
$m_{34}+n_{14}=\lambda_3$ & $n_{34}+l_{14}=\mu_3$ & $l_{34}+m_{14}=\nu_3$ 
\end{tabular}
\end{equation}
Furthermore, an $SU(4)$ BZ triangle has three hexagons%
\footnote{
	The $SU(N)$ generalization is straightforward; 
	the BZ triangles are built out of three corner vertices
	and $\half (N-1)(N-2)$ hexagons. 
	}
\begin{equation}
\label{hex:su4}
\begin{tabular}{  c  c  c  }
$n_{12}+m_{24}=m_{13}+n_{23}$ & $n_{13}+l_{23}=l_{12}+n_{24}$ & $l_{24}+n_{23}=l_{13}+n_{34}$ \cr
$n_{12}+l_{34}=l_{23}+n_{23}$ & $n_{13}+m_{34}=n_{24}+m_{23}$ & $n_{23}+m_{23}=m_{12}+n_{34}$ \cr
$m_{24}+l_{23}=l_{34}+m_{13}$ & $m_{34}+l_{12}=l_{23}+m_{23}$ & $l_{13}+m_{23}=l_{24}+m_{12}$ 
\end{tabular}
\end{equation}

As an application, consider 
$\nu=[0,k,0] \subset [p,q,p] \otimes [r,s,r] = \lambda \otimes \mu$ 
of $SU(4)$; 
here all representations are self-conjugate. 
The restrictions on the $l_{ij}$, $m_{ij}$, $n_{ij}$ 
(these integers must all be non-negative) imply that 
the entries of the BZ triangle are actually 
\begin{eqnarray}
\label{BZ:ours}
m_{14} = l_{14} = m_{12} = l_{34} &=& 0 ,
\nonumber\\
n_{12} &=& p ,
\nonumber\\
n_{23} &=& n_{14} ,
\nonumber\\
n_{34} &=& r ,
\nonumber\\
l_{23} = m_{34} &=& p - n_{14} ,
\nonumber\\
l_{12} = m_{23} &=& r - n_{14} ,
\nonumber\\
l_{13} &=& \half (s + k - (2 p + q) + 2 n_{14}) , 
\nonumber\\
m_{24} &=& \half (q + k - (2 r + s) + 2 n_{14}) , 
\nonumber\\
n_{13} &=& \half (q - k + (2 r + s) - 2 n_{14}) , 
\nonumber\\
n_{24} &=& \half (s - k + (2 p + q) - 2 n_{14}) , 
\nonumber\\
m_{13} &=& \half ((2 p + q) + k - (2 r + s) ) , 
\nonumber\\
l_{24} &=& \half ((2 r + s) + k - (2 p + q) ) . 
\end{eqnarray}
All entries thus depend on a single parameter $n_{14}$ 
which is subject to restrictions 
$0 \le n_{14} \le p, r, \half(p+r-k)$; plus we get further constraints 
$k \ge | (2 p + q) - (2 r + s) |$, 
$p+q \ge r$, and $r+s \ge p$, etc.

Now, recall that $SO(6) \sim SU(4)$, and all our operators 
are in fact made of the scalars $\phi^I$ ($I=1, ... , 6$) 
which are in the 
fundamental $\mathbf 6$ of $SO(6)$. In terms of the Young diagrams 
for $SO(6)$, the representations involved are partitioned as 
\begin{eqnarray}
\mbox{
\setlength{\unitlength}{1em}
\begin{picture}(15,3)
\put(0,2){\framebox (3,1){$n_{14}$}}
\put(3,2){\framebox (5,1){$m_{34}$}}
\put(8,2){\framebox (3,1){$m_{24}$}}
\put(11,2){\framebox (4,1){$n_{13}$}}
\put(0,1){\framebox (3,1){$n_{14}$}}
\put(3,1){\framebox (5,1){$m_{34}$}}
\put(4,0){$p$}
\put(11,1){$q$}
\end{picture}}
\quad
\mbox{
\setlength{\unitlength}{1em}
\begin{picture}(15,3)
\put(0,2){\framebox (3,1){$n_{14}$}}
\put(3,2){\framebox (5,1){$m_{23}$}}
\put(8,2){\framebox (3,1){$l_{13}$}}
\put(11,2){\framebox (4,1){$n_{24}$}}
\put(0,1){\framebox (3,1){$n_{14}$}}
\put(3,1){\framebox (5,1){$m_{23}$}}
\put(4,0){$r$}
\put(11,1){$s$}
\end{picture}}
\nonumber\\
\mbox{
\setlength{\unitlength}{1em}
\begin{picture}(16,2.5)
\put(0,1){\framebox (5,1){$m_{34}$}}
\put(5,1){\framebox (3,1){$m_{24}$}}
\put(8,1){\framebox (5,1){$m_{23}$}}
\put(13,1){\framebox (3,1){$l_{13}$}}
\put(8,0){$k$}
\end{picture}}
\hspace{8em}
\end{eqnarray}

An especially convenient decomposition is when the $[0,k,0]$ state 
is made up of say only 1-s and $\bar2$-s. 
In which case, by symmetry in the vertices, there will be no 
contributions proportional to $\tilde C$ provided 
only two flavors are involved in the diagram. 
Unfortunately, this can be achieved only when $s+q \ge k$.

Alternatively, we can take a 
$[p,q,p]$ state made up of 
$n_{14} + m_{34} + m_{24}$ 1-s and 
$n_{14} + m_{34} + n_{13}$ 2-s; 
$[r,s,r]$ state made up of 
$n_{14} + m_{23} + l_{13}$ $\bar1$-s and 
$n_{14} + m_{23} + n_{24}$ $\bar2$-s; 
$[0,k,0]$ state made up of 
$m_{34} + m_{24}$ $\bar1$-s and 
$m_{23} + l_{13}$ 1-s, 
minus contractions. 
Unlike the previous decomposition, 
this one works for any 
$[0,k,0] \otimes [p,q,p] \otimes [r,s,r]$ 
containing the singlet.

\subsection{Partitioning a tableau into 2 flavors}
\label{partition-into-two}

It is often convenient to choose the operators 
to have only two distinct flavors. Here we shall
see that it can always be done. 

Consider an operator in the $[p,q,p]$ 
of $Gl(6)$ made of 
$n_1$ 1-s and $n_2$ 2-s, to be concrete. 
We have the following constraints: 
$p \le n_2 , n_1 \le p+q$. 
This state can be assigned 
an $SU(4)$ weight 
$w = n_1 (0,1,0) + n_2 (1,-1,1) = (n_2,n_1-n_2,n_2)$. 

Next, we project this onto the $SU(3) \times U(1)$; 
for example, we can choose $1 \to 1 + \bar 1$, $2 \to 2 + \bar 2$. 
Then $w$ contains terms with 
$b$ 1-s, $n_1 - b$ $\bar 1$-s, 
$c$ 2-s, $n_2 - c$ $\bar 2$-s; 
these 
have weights 
$w'_{b,c} 
= (n_2-n_1 + 2(b-c) , 2c-n_2 )^{2(b+c)-(n_1+n_2)}$. 

To make an irrep of $Gl(6)$ into one of $SO(6)$, 
we must subtract traces. Since traces 
have weight zero, contributions with $n$ contractions instead of 1
and $m$ instead of 2, 
are equivalent to $n_1'=n_1-2n$, $n_2'=n_2-2m$. 
They are projected onto 
$w'{}^{n,m}_{b,c} = 
(n_2-n_1 + 2(b+n)-2(c+m), 2(c+m)-n_2)^{2(b+n+c+m)-(n_1+n_2)}$. 
We see that for fixed $n_1$ and $n_2$, 
$w'{}^{n,m}_{b,c} = w'{}^{\tilde n,\tilde m}_{\tilde b,\tilde c}$ 
iff 
$b+n=\tilde b+\tilde n, c+m=\tilde c+\tilde m$. 

We are interested in having $b=n_1$, $c=n_2$, 
for example; then 
$w'_{n_1,n_2} = (n_1-n_2 , n_2 )^{n_1+n_2}$. 
In order for $w'{}^{n,m}_{b,c}$ to have the same weight we must have 
$n_1-n \ge b+n=n_1$, $n_2-m \ge c+m=n_2$, or $m=n=0$; 
likewise, traces also do not contribute to the projection 
onto $b=n_1$, $c=0$.

This means that $[p,q,p]$ states of $SO(6) \sim SU(4)$ 
which consist of $n_1$ of any $\phi_a$ and $n_2$ of any other 
$\phi_b$ minus various contractions,
project onto states in irreducible representations 
of $SU(3) \times U(1)$, ones  
containing $n_1$ $z_a$-s and $n_2$ $z_b$-s or 
$n_1$ $z_a$-s and $n_2$ $\bar z_b$-s, etc. 
{\textit {without having to subtract any traces}}.

\subsection{
Details of 
$\langle
\OO_{\tquarter} \OO_{\tquarter} \OO_{\tquarter} 
\rangle$
calculations}
\label{tedious details}

In Section \ref{section: all BPS} we needed to explicitly calculate 
several three-point functions 
$\langle
\OO_{[l,k,l]}(x) \OO_{[p,q,p]}(y) \OO_{[r,s,r]}(w) 
\rangle$
of \quarter-BPS operators. 
The flavor breakdown is discussed in Section \ref{section: all BPS}, 
see equations (\ref{QQQ:operators1}-\ref{QQQ:operators2}). 
For the five cases of (\ref{quarter-all: cases}), 
we discuss the combinatorial coefficients 
multiplying the Born diagram, 
as well as the ones in front of $\tilde B (x,w)$ and $\tilde B (w,y)$.

When operators $\OO_w$ are properly symmetrized, 
we can mark the $z_1$-s exchanged between $\OO_w$ and $\OO_x$ 
separately from the $z_1$-s exchanged between $\OO_w$ and $\OO_y$. 
As far as the combinatorial factors $\alpha_{\mathrm {free}}$, 
$\tilde \beta_{xw}$ and $\tilde \beta_{yw}$ 
are concerned, the difference is just a multiplicative factor. 
This would be equivalent to 
calculating the three-point functions with 
\begin{eqnarray}
\label{QQQ:operators1-new}
\OO_{[l,k,l]} (x) &\sim& [z_1^a z_2^b z_3^e] \\
\OO_{[p,q,p]} (y) &\sim& [\bar z_1^c \bar z_2^d \bar z_3^e] \\
\OO_{[r,s,r]} (w) &\sim& [\bar z_1^a z_1^c \bar z_2^b z_2^d] 
\label{QQQ:operators2-new}
\end{eqnarray}
rather than (\ref{QQQ:operators1}-\ref{QQQ:operators2})
instead,%
\footnote{
	As written in (\ref{QQQ:operators2-new}), $\OO_w$ is not even 
	a $[r,s,r]$ operator; we need to subtract $SO(6)$ traces. 
	But when calculating whether the three coefficients 
	$\alpha_{\mathrm {free}}$, 
	$\tilde \beta_{xw}$ and $\tilde \beta_{yw}$ 
	are zero or not, the answers are the same as if we had 
	done it properly.
	}
with the same 
$e \equiv \half [(2l+k) + (2p+q) - (2r+s)] \le l+k, p+q$, 
and integers $a$, $b$, $c$, $d$ partitioning 
$r+s = a+c$, $r = b+d$. 
This simplifies the calculations dramatically.

For operators in the [2,0,2] or [2,1,2] representations, 
we chose the \quarter-BPS operator from the beginning. 
In the other cases, several \quarter-BPS chiral primaries 
exist in each representation, so instead we choose the 
operators as $\OO_w = \sum_j C_w^j \OO_j$ for example 
(see 
Sections \ref{section:simplest}, 
\ref{6 and higher} 
and 
\ref{section: systematic: explicit} 
for the definitions). 
With $\OO_x$, $\OO_y$, $\OO_w$ as in 
(\ref{QQQ:operators1-new}-\ref{QQQ:operators2-new}), 
we list the representations and choices of flavors below:
\begin{eqnarray}
\label{202-202-222}
[2,0,2] \otimes [2,0,2] \otimes [2,2,2] : 
&&
\quad 
\langle 
[z_1 z_2^2 z_3]_x 
[\bar z_1 \bar z_2^2 \bar z_3]_y 
[z_1 z_2^2 \bar z_1 \bar z_2^2]_w 
\rangle 
\quad 
\\ 
\label{202-212-232}
[2,0,2] \otimes [2,1,2] \otimes [2,3,2] : 
&&
\quad 
\langle 
[z_1 z_2^2 z_3]_x 
[\bar z_1 \bar z_2^3 \bar z_3]_y 
[z_1 z_2^3 \bar z_1 \bar z_2^2]_w 
\rangle 
\quad 
\\ 
\label{202-212-313}
[2,0,2] \otimes [2,1,2] \otimes [3,1,3] : 
&&
\quad 
\langle 
[z_1 z_2^2 z_3]_x 
[\bar z_1^2 \bar z_2^2 \bar z_3]_y 
[z_1^2 z_2^2 \bar z_1 \bar z_2^2]_w 
\rangle 
\quad 
\\ 
\label{202-313-232}
[2,0,2] \otimes [3,1,3] \otimes [2,3,2] : 
&&
\quad 
\langle 
[z_1^2 z_3^2]_x 
[\bar z_1^2 \bar z_2^3 \bar z_3^2]_y 
[z_1^2 z_2^3 \bar z_1^2]_w 
\rangle 
\quad 
\\ 
\label{202-313-313}
[2,0,2] \otimes [3,1,3] \otimes [3,1,3] : 
&&
\quad 
\langle 
[z_1^2 z_3^2]_x 
[\bar z_1^2 \bar z_2^3 \bar z_3^2]_y 
[z_1^2 z_2^3 \bar z_1^2]_w 
\rangle 
\quad 
\end{eqnarray}
The coefficients 
$\alpha_{\mathrm {free}}$ and $\tilde \beta_{yw}$ 
take a long time to compute,
and are not particularly illuminating. 
For example, in the case of $[2,0,2] \otimes [3,1,3] \otimes [3,1,3]$ 
we find 
(after 2430 hrs 
of a {\textit {Mathematica}} computation on a Pentium-III with 1.4MHz speed)
\begin{eqnarray}
\alpha_{\mathrm {free}} &\!\!=\!\!& 
{(N^2-1) (N^2-4) \over 1036800 N^2} 
      (2332800 C^2_w C^2_y - 2332800 C^2_y C^4_w N - 
\\ \nonumber && \hspace{5em} 
       2332800 C^2_w C^4_y N + 10800 C^1_w C^1_y N^2 + 324000 C^2_w C^2_y N^2 - 
\\ \nonumber && \hspace{5em} 
       492480 C^2_y C^3_w N^2 - 492480 C^2_w C^3_y N^2 + 114048 C^3_w C^3_y N^2 + 
\\ \nonumber && \hspace{5em} 
       259200 C^1_y C^4_w N^2 + 259200 C^1_w C^4_y N^2 - 777600 C^4_w C^4_y N^2 - 
\\ \nonumber && \hspace{5em} 
       44100 C^1_y C^2_w N^3 - 44100 C^1_w C^2_y N^3 - 8280 C^1_y C^3_w N^3 - 
\\ \nonumber && \hspace{5em} 
       8280 C^1_w C^3_y N^3 + 950400 C^2_y C^4_w N^3 - 250560 C^3_y C^4_w N^3 + 
\\ \nonumber && \hspace{5em} 
       950400 C^2_w C^4_y N^3 - 250560 C^3_w C^4_y N^3 - 5875 C^1_w C^1_y N^4 - 
\\ \nonumber && \hspace{5em} 
       133200 C^2_w C^2_y N^4 + 90720 C^2_y C^3_w N^4 + 90720 C^2_w C^3_y N^4 - 
\\ \nonumber && \hspace{5em} 
       22752 C^3_w C^3_y N^4 - 52800 C^1_y C^4_w N^4 - 52800 C^1_w C^4_y N^4 + 
\\ \nonumber && \hspace{5em} 
       396000 C^4_w C^4_y N^4 + 900 C^1_y C^2_w N^5 + 900 C^1_w C^2_y N^5 + 
\\ \nonumber && \hspace{5em} 
       4920 C^1_y C^3_w N^5 + 4920 C^1_w C^3_y N^5 - 100800 C^2_y C^4_w N^5 + 
\\ \nonumber && \hspace{5em} 
       51840 C^3_y C^4_w N^5 - 100800 C^2_w C^4_y N^5 + 51840 C^3_w C^4_y N^5 + 
\\ \nonumber && \hspace{5em} 
       75 C^1_w C^1_y N^6 + 3600 C^2_w C^2_y N^6 - 2880 C^3_w C^3_y N^6 - 
\\ \nonumber && \hspace{5em} 
       50400 C^4_w C^4_y N^6) 
\\ \nonumber
\tilde \beta_{yw} &\!\!=\!\!& 
{(N^2-1)(N^2-4) \over 518400 }
      (172800 C^1_y C^2_w + 172800 C^1_w C^2_y - 
\\ \nonumber && \hspace{5em} 
       69120 C^1_y C^3_w - 
       69120 C^1_w C^3_y + 4147200 C^2_y C^4_w - 
\\ \nonumber && \hspace{5em} 
       1658880 C^3_y C^4_w + 4147200 C^2_w C^4_y - 1658880 C^3_w C^4_y + 
\\ \nonumber && \hspace{5em} 
       33450 C^1_w C^1_y N - 
       837000 C^2_w C^2_y N - 79920 C^2_y C^3_w N - 
\\ \nonumber && \hspace{5em} 
       79920 C^2_w C^3_y N + 197856 C^3_w C^3_y N + 457200 C^1_y C^4_w N + 
\\ \nonumber && \hspace{5em} 
       457200 C^1_w C^4_y N + 
       2678400 C^4_w C^4_y N - 107700 C^1_y C^2_w N^2 - 
\\ \nonumber && \hspace{5em} 
       107700 C^1_w C^2_y N^2 - 2640 C^1_y C^3_w N^2 - 2640 C^1_w C^3_y N^2 - 
\\ \nonumber && \hspace{5em} 
       910800 C^2_y C^4_w N^2 + 
       96480 C^3_y C^4_w N^2 - 910800 C^2_w C^4_y N^2 + 
\\ \nonumber && \hspace{5em} 
       96480 C^3_w C^4_y N^2 - 13175 C^1_w C^1_y N^3 + 
       18000 C^2_w C^2_y N^3 + 
\\ \nonumber && \hspace{5em} 
       83880 C^2_y C^3_w N^3 + 
       83880 C^2_w C^3_y N^3 - 43200 C^3_w C^3_y N^3 - 
\\ \nonumber && \hspace{5em} 
       100800 C^1_y C^4_w N^3 - 100800 C^1_w C^4_y N^3 - 
       597600 C^4_w C^4_y N^3 + 
\\ \nonumber && \hspace{5em} 
       1500 C^1_y C^2_w N^4 + 
       1500 C^1_w C^2_y N^4 + 9480 C^1_y C^3_w N^4 + 
\\ \nonumber && \hspace{5em} 
       9480 C^1_w C^3_y N^4 + 59760 C^3_y C^4_w N^4 + 59760 C^3_w C^4_y N^4 + 
\\ \nonumber && \hspace{5em} 
       125 C^1_w C^1_y N^5 - 5976 C^3_w C^3_y N^5)
\end{eqnarray}

In all cases (\ref{202-202-222}-\ref{202-313-313}), 
Born level correlators are nonzero for general 
$N$, and so are the order $g^2$ contributions for a 
random set of coefficients $C^j$. 
But when we set the $C^j$ to their proper values 
(to make $\OO_y$ and $\OO_w$ \quarter-BPS), 
we recover correlators which are nonvanishing 
($\alpha_{\mathrm {free}} \ne 0$) 
and protected at order $g^2$ 
($\tilde \beta_{xw} = \tilde \beta_{yw} = 0$).

%% file: nearpp-chapter.tex
\chapter{Strings in the near plane wave background and AdS/CFT}
\label{chapter:nearpp}


So far, we have been discussing the properties of 
supergravity modes, and the corresponding protected SYM operators. 
But we can do better than that. 
As it turns out, 
the GS superstring can be quantized exactly in the plane wave background 
\cite{Metsaev,Metsaev:2002re}, 
which can be viewed as the Penrose limit of the $AdS_5 \times S^5$ 
geometry \cite{BMN,Blau:2002dy}.
The limit involves scaling both the $AdS_5$ radius 
$R \rightarrow \infty$ and the R-charge $J \sim R^2$.
One considers 
states with finite plane wave light cone energy and momentum.
It was proposed by Berenstein, Maldacena and Nastase (BMN)
\cite{BMN} that such string states correspond
to single trace operators in the gauge theory with certain phases
inserted.
Remarkably, the parameter controlling perturbative expansion of scaling
dimensions of such operators is $\lambda' = g N / J^2$, which
can be made small to allow reliable gauge theory computations.
BMN were able to resum the diagrams weighted by powers of $\lambda'$
and show precise agreement between the scaling dimensions of SYM operators 
and the light cone energies of corresponding string states. 
This was further confirmed in \cite{Gross:2002su,Santambrogio:2002sb,Kristjansen:2002bb}.
The following development included studying string interactions both in 
the plane wave string theory and in the gauge theory
\cite{Kristjansen:2002bb,Spradlin:2002ar,Spradlin:2002rv,Klebanov:2002mp,Berenstein:2002sa,CFHMMPS,Kiem:2002xn,Huang:2002wf,Chu:2002pd,Lee:2002rm}. 

The plane wave limit is an improvement over 
being able to handle only supergravity states and protected operators. 
But we would still like 
to get closer to the full AdS string theory.
One way to gain insight is to do systematic perturbation 
theory around the plane wave limit, taking $1/R^2$ 
as a small parameter. 
This approach was tested in \cite{p1} on the
$AdS_3 \times S^3$ background with NS-NS flux.
String theory in this background is described by
an exactly solvable $SL(2) \times SU(2)$ WZNW model.
It was shown \cite{p1} that one can recover
the exact string spectrum at small coupling $g$ to the next to
leading order in $1/R^2$ expansion.

In this Chapter we determine
the leading order finite radius corrections to the string spectrum
in $AdS_5 \times S^5$. 
Specifically, we carry out systematic perturbation 
theory around the plane wave limit, taking $1/R^2$ 
as a small parameter. 
On the Yang Mills side, the corresponding calculation involves refining the 
definition of BMN operators and computing their scaling dimensions.
We work at 
small string coupling 
$g$, which corresponds
to computing only planar diagrams in the gauge theory.
Furthermore, we consider only the leading non-trivial term in 
the $\lambda'$ expansion. 
The calculation of scaling dimensions in SYM then reduces to
computing the matrix of two-point functions and its subsequent
diagonalization. 
We identify the gauge theory operator 
which corresponds to the light cone worldsheet Hamiltonian, 
and show that its matrix elements relevant for 
diagonalization agree with the string theory results.
Hence we conclude that to the accuracy we are working at,
the scaling dimensions of gauge theory operators 
agree with the spectrum of string states in $AdS_5 \times S^5$.%
\footnote{
Related issues were also addressed in \cite{Frolov:2002av,Schwarz,Tseytlin}. 
}

The Chapter is organized in the following way. 
In section \ref{section:worldsheet} 
we describe how to quantize the 
string in the background which includes 
the $\OO(1/R^2)$ corrections to the plane wave metric, 
and show how to compute the leading corrections to the 
spectrum of bosonic plane wave states.
In section \ref{section:SYM} we explain how the definition of BMN 
operators should be extended to include finite $J$ effects. 
There we also establish agreement 
between string and SYM results for a subset of 
matrix elements of the light cone Hamiltonian. 
In section \ref{section:discussion} we discuss our results
and mention possible future developments. 
In appendix \ref{section:gsw way} we present
an alternative technique, based on the formalism of \cite{GSW},
for computing $1/R^2$ corrections in string theory.
The results for  physical quantities are the same
as in section \ref{section:worldsheet}.
Appendix \ref{section:feynman rules} contains 
the tools we use in the SYM calculations. 
In appendix \ref{section:equalitygeneric} we generalize
the results of section \ref{section:SYM}.

\section{Corrections to the plane wave string spectrum}
\label{section:worldsheet}

In this section, we do perturbation theory 
on the worldsheet following the method described in \cite{p1}.
We start by outlining the procedure used in 
\cite{Metsaev,Metsaev:2002re,BMN} for deriving the 
leading order spectrum in the Penrose limit of $AdS_5 \times S^5$.
The $AdS_5 \times S^5$ metric is 
\begin{eqnarray}
\label{eq:exact metric}
ds^2 = R^2 
\left[
- dt^2 \cosh^2 \!\! \rho + d\rho^2 + \sinh^2 \!\! \rho \, d\Omega_3^2 
+ d\psi^2 \cos^2 \! \theta + d\theta^2 + \sin^2 \! \theta \, d\Omega_3'\!{}^2 
\right].
\end{eqnarray}
The Penrose limit of this geometry is obtained by zooming in 
on the neighborhood of a lightlike geodesic circling the equator of $S^5$. 
This is done by changing variables as 
\begin{eqnarray}
\label{eq:penrose variables:def}
X^+ = {1\over2} (t + \psi)
,\quad
X^- = {1\over2} (t - \psi) R^2 
,\quad
\rho = {r \over R}
,\quad
\theta = {y \over R},
\end{eqnarray}
and taking $R$ to be large, while keeping 
$|X^\pm|, r, y$ finite. 
At leading order in $1/R^2$, 
the $AdS_5 \times S^5$ metric (\ref{eq:exact metric}) reads 
\begin{eqnarray}
\label{eq:metric:leading order}
ds^2_0 = - 4 dX^- dX^+ - (r^2+y^2) dX^+ dX^+ + dr_i \, dr_i + dy_i \, dy_i.
\end{eqnarray}
Coordinates $y_i$ and $r_i$ parameterize two copies of $R^4$, 
but the $SO(8)$ symmetry of the metric (\ref{eq:metric:leading order}) 
is broken down to $SO(4) \times SO(4)$ by the
RR flux
\begin{eqnarray}
\label{eq:F-field:leading order}
F_{+1234} = F_{+5678} = \mbox{const}.
\end{eqnarray}

We would like to quantize type IIB superstring in the background 
(\ref{eq:metric:leading order}), (\ref{eq:F-field:leading order}). 
As was shown in \cite{Metsaev,Metsaev:2002re}, the way to do this is to 
look at the sigma-model part of the GS action, 
and use $\kappa$-symmetry in light-cone gauge 
to determine the rest of the worldsheet action. 
Bosons and fermions decouple for the plane wave background 
(\ref{eq:metric:leading order}) in light-cone gauge 
\cite{Metsaev,Metsaev:2002re}.
We will only be interested in the bosonic part 
of the full superstring action. 
The light cone gauge is specified by
\beqa
\label{eq:lc:polch:start}
  X^+&=&\tau, \\ \nonumber
  \p_\sigma \gss&=&0, \\ \nonumber
  {\mathrm {det}} \gamma_{\alpha \beta}&=&-1, 
\eeqa 
for bosonic fields; the worldsheet coordinates are 
$\tau \in (-\infty, \infty)$, $\sigma \in [0,l]$.
The worldsheet metric can be written as \cite{Polch}
\beq
  \gamma^{\alpha \beta}=\left(  \begin{array}{cc}
   -\gss(\tau) & \gst(\tau, \sigma)  \\
   \gst(\tau, \sigma) &  \gss^{-1}(\tau) (1-\gst^2(\tau, \sigma))
   \end{array} \right). 
\eeq

In this section we consider only the $y$ part of the theory.
The $r$ part can be included by noticing
that (\ref{eq:metric:leading order}) and (\ref{eq:F-field:leading order})
are invariant under $y \lra r$ while in the $\Or(1/R^2)$
correction to the plane wave metric $y$ and $r$ terms come
with opposite signs (see below).
This means that to restore the $r$ terms in the final result
one needs to copy the $y$ part, substitute $y \rightarrow r$ 
and flip the sign in front of the $\Or(1/R^2)$ terms.
This is confirmed in appendix \ref{section:gsw way}, where
explicit calculations are performed.

In the light cone gauge (\ref{eq:lc:polch:start}) 
the bosonic part of the Lagrangian is 
\beqa
\label{l00}
L_0 &=& {-}{1 \over 4 \pi} \int_0^l \left\{ \gss \bigg[
          4 {\dot X}^- {+}\sum_i ( y_i y_i{-} {\dot y}_i {\dot y}_i ) \bigg] 
\right.\nonumber\\&&\hspace{4em}\left.
{-} 2 \gamma_{\sigma\tau} 
\bigg[ 2 (X^-)' {-} \sum_i\dot y_i y_i' \bigg]
          {+}\gss^{-1} (1{-}\gamma_{\sigma\tau}^2) \sum_i y_i' y_i' \right\},
\hspace{3em}
\eeqa
where we used the leading order spacetime metric 
(\ref{eq:metric:leading order}). 
The equation of motion for the worldsheet metric (Virasoro constraints)
are
\beq
\label{ymi}
 (X^-)'={1 \over 2} \sum_i \dot y_i y_i'
, \qquad 
        {\dot X}^- = {1 \over 4} \sum_i
\left[ \dot y^i\dot y^i + y_i' y_i' - y_i y_i\right].
\eeq
One can use the equation of motion for 
$X^-$ and the leftover gauge freedom $\sigma \to \sigma + f(\tau)$ 
to set $\gamma_{\sigma\tau}=0$ in (\ref{l00}) \cite{Polch}. 
The equation of motion for the zero mode of $X^-$
implies that $\gss$ is related to the conserved light cone momentum 
$\PP_- = - i {\partial \over \partial X^-}$. 
Choosing the gauge 
\beq 
\label{gc}
l=2 \pi \eta,
\quad\mbox{where $\eta \equiv - {1\over2} \PP_-$},
\eeq
sets $\gss=1$ at the leading order in $1/R^2$.
The plane wave Hamiltonian that follows from  (\ref{l00}) can therefore be written as
\beq
\label{h0}
H_0={1 \over 4 \pi} \intl \sum_i \left[ 
           (2 \pi)^2 P^i_y P^i_y + y_i y_i  +y_i' y_i' \right].
\eeq 
where $P_y^i=\dot y_i/2 \pi$. 
The worldsheet theory of a light cone string is massive 
in the plane wave background.
The fields can be expressed in terms of eigenmodes
\beq
\label{xi}
  y_i={i \over \sqrt{2}} \sum_{n} {1 \over \sqrt{w_n}}
    \left[y^i_n-y^i_{n}{}^\dagger \right],
\eeq
where the  $\tau,\sigma$-dependent oscillators $y^i, y^i_n{}^\dagger$ are defined as 
\beq
  y^i_{n}=\alpha^i_{n} e^{-i w_n \tau -i n \sigma \over \eta},   \qquad
      y^i_{n}{}^\dagger=\alpha^i_{n}{}^\dagger e^{i w_n \tau +i n \sigma \over \eta}, \\ 
\eeq
and the frequencies are given by 
\beq
  w_n=\sqrt{\eta^2+n^2}.
\eeq
Substituting the field expansions into (\ref{h0}) diagonalizes
the plane wave Hamiltonian
\beqa
\label{h0o}
 H_0={1 \over \eta} \sum_{i,n} w_n N^i_{n},    
\eeqa
where 
$N^i_{n}= y^i_n{}^\dagger y^i_{n}$. 
The normal ordering constant cancels between bosons and fermions 
by virtue of spacetime supersymmetry, 
so we do not include it in (\ref{h0o}).
The leading terms in the expansion of $H_0$ in powers of $1/\eta^2$
are
\beq
\label{h0e}
   H_0=\sum_{i,n} N^i_n + {1 \over 2 \eta^2} \sum_{i,n} n^2 N^i_n+\Or\left({1 \over \eta^4}\right).
\eeq
In addition, we have the level matching condition 
\beq
\label{eq:level matching:w/s}
   \sum_{i,n} n N^i_n = 0.
\eeq

To compute $\OO(1/R^2)$ corrections 
to the string spectrum in the plane wave background, 
one would add the $\Or(1/R^2)$ correction $ds_1^2$ to 
the leading metric $ds_0^2$, write down the bosonic
part of the light cone Lagrangian, 
and then use $\kappa$-symmetry
to write the full GS action.
Subsequently the system can be quantized perturbatively in $1/R^2$. 
%
Expanding (\ref{eq:exact metric}) to next to leading order in $1/R^2$ we have
\begin{eqnarray}
\label{eq:metric:correction}
ds_1^2 = 
{1\over R^2} 
\left[
- 2 dX^- dX^+ (r^2-y^2) - {1\over3} (r^4-y^4) dX^+ dX^+ + 
{1\over3} (r^4 d\Omega_3{}^2 - y^4 d\Omega'_3{}^2) 
\right].
\hspace{-1.5em}\nonumber\\
\end{eqnarray}
%
The bosonic part of the $\Or(1/R^2)$ Lagrangian is therefore
quartic in the fields.
The leading form of the $\kappa$-symmetry then implies that
the fermionic part of the $\Or(1/R^2)$ GS action is at most 
bi-quadratic in bosons and fermions.
We are considering corrections to the spectrum of bosonic states, 
so the fermionic part of the action can only contribute
diagonal matrix elements of the type
\beq
\label{not}
   {1 \over R^2} \sum_{i,n} f(w_n) N^i_n,
\eeq
where $f(w_n)$ is some function.
Fixing the exact form of $f(w_n)$ in (\ref{not}) requires dealing with
the $\Or(1/R^2)$ fermionic part of the superstring  action.
This we have not bothered to do.
We also drop all terms that are due to the normal ordering
of bosonic operators in all subsequent calculations.

Using the identities
$dy_i dy_i = dy^2+y^2 d\Omega^2$ and $y dy = y_i dy_i$ 
we can write 
$y^4  d\Omega_3'{}^2 = y_i y_i dy_j dy_j - y_i y_j dy_i dy_j$
and deduce the correction to the leading order Lagrangian (\ref{l00})
\beq
\label{l1}
  L_1{=}{1 \over 4 \pi R^2} \intl \Bigg[ 
   {1 \over 3} \sum_i y_i^4
          {-} {1 \over 3} \sum_{i \neq j} \left[ y_i^2 ({\dot y}_j^2{-}y_j^2{-}(y_j')^2)
                {+}y_i y_j (y_i' y_j'{-}{\dot y}_i {\dot y}_j) \right]  
+{1 \over 2} y^2 {\dot X}^- \Bigg].
\eeq
Terms proportional to $\gst$ are higher order in $1/R^2$ 
and do not contribute to (\ref{l1}). 
As explained in \cite{p1}, for the purpose of computing the
leading corrections to the spectrum, the correction to the Hamiltonian
equals minus the correction to the Lagrangian.%
\footnote{
	One can convince oneself that this is the case by
	perturbing the Lagrangian, 
	calculating the canonically conjugate momenta, 
	and keeping only terms up to $\Or(1/R^2)$ in the Hamiltonian. 
	In \cite{p1} the zero mode of $X^-$ was treated separately, 
	but one can show that this is not necessary.
	} 
The correction to the plane wave Hamiltonian can
therefore be written as
\beqa
\label{h1:1 plus 2}
  H_1&=& {1 \over 4 \pi R^2} \intl \Bigg[ 
   -{1 \over 3} \sum_i y_i^4
          {+} {1 \over 3} \sum_{i \neq j} \bigg[ y_i^2 
  [ (2 \pi P^j_y)^2{-}y_j^2{-}(y_j')^2] \\ \nonumber && \quad
                {+}y_i y_j (y_i' y_j'{-}(2 \pi)^2 P^i_y P^j_y       ) \bigg]  
       - {1 \over 2} \sum_{i,j} y_i^2 [ 
                      (2 \pi P^j_y)^2+(y_j')^2-y_j^2 ] \Bigg],
\eeqa
where in rewriting the last term we used the Virasoro constraint 
[the second equation in (\ref{ymi})]. 

Next we expand (\ref{h1:1 plus 2}) in modes (\ref{xi}). 
We are interested in first order corrections 
to the energies, so we only need to compute matrix elements 
of $H_1$ between degenerate states. 
Plane wave string states are 
\beq
\label{pwstates}
 y_{n_1}^{i_1}{}^\dagger  \ldots  y_{n_k}^{i_k}{}^\dagger  \ldots |\eta \rangle.
\eeq
They 
are degenerate only if the two sets of worldsheet momenta 
$(n_1,\ldots n_k,\ldots)$ and $(n_1',\ldots n_k',\ldots)$ 
are permutations of one another. 
Thus the only relevant terms in $H_1$ 
are of 
the form $y_k y^\dagger_k y_l y^\dagger_l$. 
Diagonal contributions come from 
$y_k^i y_k^i{}^\dagger y^j_l y_l^j{}^\dagger$;
they add up to 
\beq
\label{eq:adrei:diag}
H_1^D = 
{1 \over 2 \eta R^2} 
\left( 
{1 \over 2} \sum_{i;n} {n^2 (N^i_n)^2 \over w_n^2}
-\sum_{i,j;m,n} {n^2 N^i_n N^j_m \over w_m w_n}
\right)
.
\eeq
The relevant off-diagonal terms are of the form
$y^i_m{}^\dagger y^i_n y^j_n{}^\dagger y^j_m$, $i \neq j$, $m \neq n$; 
and
$y^i_m{}^\dagger y^i_n{}^\dagger y^j_m y^j_n$, $i \neq j$.
These add up to 
\beq
\label{eq:adrei:off-diag}
  H_1^{OD}{=}{1 \over 2 \eta R^2 } \, \sum_{i \neq j;\,m \neq n}
  {n m \over w_n w_m}     
     ( y^i_m{}^\dagger y^i_n{}^\dagger y^j_m y^j_n  -y^i_m{}^\dagger y^i_n y^j_n{}^\dagger y^j_m)+
 {1 \over 4 \eta R^2 }  \sum_{i \neq j; n} {n^2 \over w_n^2}     
      y^i_n{}^\dagger y^i_n{}^\dagger y^j_n y^j_n.
\eeq     
Expanding (\ref{eq:adrei:diag}) and (\ref{eq:adrei:off-diag})
in powers of  $1/\eta$ we obtain
\beqa
\label{h1diag}
    H_1^D &=& 
{1 \over 2 \eta^3 R^2} \left({1 \over 2} \sum_{i;n} n^2 (N^i_n)^2 
        -\sum_{i,j;m,n} n^2 N^i_n N^j_m          \right) +\Or\left({1 \over \eta^5 R^2}\right)
\eeqa
and
\beqa
\label{h1offdiag}
  H_1^{OD}&=&{1 \over 2 \eta^3 R^2 } \, \sum_{i \neq j;\,m \neq n}
  n m      
     ( y^i_m{}^\dagger y^i_n{}^\dagger y^j_m y^j_n  {-}y^i_m{}^\dagger y^i_n y^j_n{}^\dagger y^j_m)
 \\ \nonumber && \quad
+{1 \over 4 \eta^3 R^2 }  \sum_{i \neq j; n} n^2 
      y^i_n{}^\dagger y^i_n{}^\dagger y^j_n y^j_n  
      {+}\Or\left({1 \over \eta^5 R^2}\right),
\eeqa  
respectively.
The leading $1/\eta$ term in $H_1$ is a sum of these two expressions.

An alternative derivation is given in appendix \ref{section:gsw way}, 
where more details are provided.


%
%
%
%

\section{Anomalous dimensions and AdS/CFT}
\label{section:SYM}

We now turn to 
the boundary $\NN=4$ Super Yang-Mills theory.
Our starting point will be the BMN operators \cite{BMN} which
correspond to plane wave states 
in the Penrose limit. 
One can still regard plane wave states as 
belonging to 
the Hilbert
space of the full $AdS_5 \times S^5$ theory, even though they are no longer
eigenstates of the full Hamiltonian.
As explained in the previous section, departing from the Penrose
limit corresponds to turning on perturbative corrections to the
plane wave Hamiltonian.
Eigenstates of the full Hamiltonian can 
be found using ordinary 
quantum-mechanical perturbation theory.

SYM operators which correspond to string eigenstates
must have definite conformal dimensions.
Such operators may be obtained from a complete set of operators 
by diagonalizing the matrix of their two-point functions.
This procedure is 
analogous to the
diagonalization of the string theory Hamiltonian.
We find that 
the spectra computed on both sides of the correspondence
match, and the operator defined by the
matrix of two-point functions is the SYM counterpart 
of the string Hamiltonian.

This section is organized as follows.
In section \ref{section:operators:pp} we define operators that 
correspond to plane wave 
states away from the strict Penrose limit.
In section \ref{section:anomalous dims} we show how the matrix of
two-point functions is related to the string Hamiltonian.
In section \ref{section:equalitysimple} 
we match the 
matrix elements
of the light cone Hamiltonian between the string and the gauge theory.
We analyze a simple case where all of the excited modes 
have distinct $SO(4)$ indices and none of them 
is excited more than once.
The most general case is treated in appendix \ref{section:equalitygeneric}.
Feynman rules are discussed in appendix \ref{section:feynman rules}.

\subsection{Operators}
\label{section:operators:pp}
The important assumption that we start with is that 
suitably refined BMN operators continue to correspond to
plane wave states, regarded as states in the Hilbert space
of $AdS_5 \times S^5$,  even away from the plane wave limit.
To define the right operators we will follow closely the
logic of BMN.
We start with the operator which corresponds to the light cone vacuum
\beq
\label{ovac}
   {1 \over \sqrt{\Omega} } \tr[ z^J] 
\quad \lra \quad  |\eta  \rangle,
\eeq
where $z={1\over\sqrt2} (\phi^5+i \phi^6)$ and $\Omega$
is a normalization constant 
(more about this below).
For the ground state (\ref{ovac}) there is a relation
$J=R^2 \eta$, but this gets modified by
$\Or(1/R^2)$ terms for excited states.

SYM operators which correspond to 
states with excited zero modes can be generated 
by acting on the light cone ground state (\ref{ovac}) 
with generators of the global symmetry group. 
The generators that we will be interested in are rotations in $ij$ plane,
denoted by $T_{ij}$ and their combinations 
$T_{i z}= {1\over \sqrt2} (T_{i5}+ i T_{i6})$
and 
$T_{i\bar z}={1\over \sqrt2} (T_{i5}- i T_{i6})$.
They act on the fields as
\beqa
  [T_{iz},z]&=&0, \qquad [T_{iz},{\bar z}]=\phi^i, \qquad 
[T_{iz}, \phi^j] = -z \, \delta_i^j, \\ \nonumber
  [T_{i\bar z},z]&=&\phi^i, \qquad  [T_{i\bar z},\bar z]=0, \qquad
  [T_{i\bar z}, \phi^j] = -{\bar z} \, \delta_i^j.
\eeqa
On the worldsheet we have a correspondence
\beq
   T_{iz} \lra  y_0^i, \quad  T_{i\bar z} \lra  y_0^i{}^\dagger.
\eeq
Consider as an example the operator corresponding to
the state $ y_0^i{}^\dagger y_0^j{}^\dagger |\eta \rangle, \, i \neq j$.
It is obtained by computing successive commutators of 
$T_{i\bar z}$ and  $T_{j\bar z}$ with (\ref{ovac}).
Either of these generators can turn any $z$ in the string of $z$-s into
$\phi^i$ or $\phi^j$ respectively.
The result is therefore the sum of 
$\tr[ z,\,\ldots \phi^i\, z\, \ldots \phi^j\, z \ldots]$
over all possible positions of inserted $\phi$'s:
\beq
\label{bps}
  {1 \over \sqrt{\Omega} } \left[ \sum_{a=0}^J \sum_{b=a}^J
    \tr[ z^a\, \phi^i\,z^{b-a}\, \phi^j\, z^{J-b}] +(i \lra j) \right]
\quad \lra \quad 
        y_0^i{}^\dagger y_0^j{}^\dagger |\eta \rangle.
\eeq
This formula has an obvious generalization for
higher number of $\phi$ insertions, as long as no label 
appears more than once.
If some of the $\phi$'s indices do coincide, 
$T_{i\bar z}$ can act on the same field. 
In this case, $z$ is first turned into $\phi^i$, 
and then into $-\bar z$.
For example, in the case of two $\phi$ insertions we have
\beq
\label{bps2}
  {1 \over \sqrt{\Omega} } \left( 2 \sum_{a=0}^J \sum_{b=a}^J
    \tr[ z^a\, \phi^i\,z^{b-a}\, \phi^i\, z^{J-b}] - \sum_{a=0}^{J+1}
               \tr[ z^a\, {\bar z}\,z^{J+1-a}] \right)
\quad \lra \quad 
        y_0^i{}^\dagger y_0^i{}^\dagger |\eta \rangle.
\eeq
To construct an operator with 
three $\phi$'s with the same index inserted,
one should act by  $T_{i\bar z}$ on both terms in (\ref{bps2})
to produce
\beq
\label{bps3}
  {1 \over \sqrt{\Omega} } \left( \sum
    \tr[ z \ldots \phi^i \ldots \phi^i \ldots \phi^i\ldots] - 3 \sum
               \tr[ z\ldots \phi^i\ldots {\bar z}\ldots] \right)
\quad \lra \quad 
        y_0^i{}^\dagger y_0^i{}^\dagger y_0^i{}^\dagger |\eta \rangle,
\eeq
where dots stand for a bunch of $z$'s and the sum is over all
possible positions of the insertions.
The second sum in (\ref{bps3}) has $J+1$ times
fewer terms than the first sum, and is subleading when it comes to
computing two-point functions.
Throughout this discussion we are interested in the subleading
corrections in $1/J \sim 1/\eta R^2$, and therefore 
we should keep this term.
If we act with $T_{i\bar z}$ one more time, 
a term 
$3 \sum \tr[ z\ldots {\bar z}\ldots {\bar z}\ldots]$ 
appears when $T_{i\bar z}$ hits the $\phi^i$ 
in the second sum in (\ref{bps3}).
This piece is $\Or(1/J^2)$ compared to the leading term, 
so we can drop it. 

In general, when an arbitrary number of zero modes
excited, the corresponding SYM operator is
\beqa
\label{genericbps}
 \Or&=& {\tilde \Or} - \Or_*, \\ 
 {\tilde \Or}
     &=& {1 \over \sqrt{\Omega} }  \sum  \tr[ z \ldots \phi^{i_1} \ldots \phi^{i_k} \ldots ], \\
 \Or_*&=& {1 \over \sqrt{\Omega} }
         \sum_{(p,q):  \,i_p=i_q}
         \tr[ z\ldots \phi^{i_1} \ldots \phi^{i_k} \ldots
               \check{\phi}^{i_p} \ldots  \check{\phi}^{i_q} 
                         \ldots {\bar z}\ldots],
\eeqa
where $\check{\phi}^{i_p}$ stands for ${\phi}^{i_p}$ being omitted from
the string of operators and the sum in $\Or_*$ runs over
all possible pairs of  $({\phi}^{i_p},{\phi}^{i_q}) $ with the same indices.
When writing 
(\ref{genericbps}), 
we omitted terms which appear when $T_{i\bar z}$ hits
the same field more than twice, as such are $\Or(1/J^2)$.
When all $\phi$'s inserted have different flavors, the
operator $\Or_*$ vanishes and we have $\Or= {\tilde \Or}$.

Next we turn to the construction of operators which correspond to 
general 
string states (\ref{pwstates}). 
Such operators must satisfy a few  necessary requirements.
First, if only the zero modes are excited,
they must reduce to the BPS operators described. 
Second, they must vanish unless the level matching condition 
\beq
\label{lm}
\sum_{i,n} n N^i_{n}=0,
\eeq
is satisfied.
Finally, our operators must reduce to the BMN operators 
as $J \rightarrow \infty$.

Suppose there is a total of $\N$ oscillators excited, 
\beq
   \N=\sum_i \N_i, \qquad \N_i=\sum_{n}  N^i_n.
\eeq
Due to the cyclicity of the trace,
\beq
\label{co}
  {1 \over \sqrt{\Omega} } 
    \tr[ z\ldots \phi^{i_1} \ldots \phi^{i_k} \ldots]
\eeq
is equivalent to $(J+\N)$ other terms in ${\tilde \Or}$
which are related to it by  cyclic permutations.
According to \cite{BMN}, at the leading order in the $1/J$ expansion, 
oscillators $y_n^i{}^\dagger$ correspond to 
insertions of $\phi^{i_k}$ with the phase $\exp({2 \pi i n_k a_k \over J})$,
where $a_k$ counts the number of $z$'s  to the left of this $\phi^{i_k}$.
One has to be be more careful when $1/J$ effects are taken into account.
In order for an operator to vanish when the level matching condition
is not satisfied, each sum over cyclically related terms in (\ref{co}) 
must vanish separately.
This happens when the phases assigned to the $\phi^{i_k}$ insertions are 
\beq
\label{qdef}
 q_{n_k}^{a_k} =  \exp\left( {2 \pi i n_k a_k \over J+\N}\right).
\eeq
Here $a_k$ counts {\textit all} operators 
appearing to the left of the  $\phi^{i_k}$ insertion, 
and not just the $z$-s.
Similar arguments can be made to fix the form of $\Or_*$.
Again, each $\phi^{i_k}$ insertion comes with a phase
given by (\ref{qdef}).
In order to satisfy the level matching condition we should
also assign a phase $q_{n_k+n_l}^{a_{\bar z}}$ to $\bar z$.

To summarize, we have a correspondence which relates
SYM operators and plane wave string states away from the strict
Penrose limit
\beq
\label{opdef}
  \Or= ({\tilde \Or} -O_* ) 
\quad\lra\quad 
   y_{n_1}^{i_1}{}^\dagger \ldots y_{n_k}^{i_k}{}^\dagger \ldots |\eta \rangle,
\eeq
where
\beqa
 {\tilde \Or}
     &=& {1 \over \sqrt{\Omega} } \sum
   \left( \prod_k q_{n_k}^{a_k} \right) \tr[ z \ldots \phi^{i_1} \ldots \phi^{i_k} \ldots ], \\
 \Or_*&=& {1 \over \sqrt{\Omega} }
         \sum_{(n_p, n_q):  \,i_p=i_q}
      \left( \prod_{k \neq p,q} q_{n_k}^{a_k} \right) q_{n_p+n_q}^{a_{\bar z}} \,
         \tr[ z\ldots \phi^{i_1} {\ldots} \phi^{i_k} {\ldots}
                \check{\phi}^{i_k} {\ldots}  \check{\phi}^{i_l} 
                         {\ldots} {\bar z}\ldots],
\nonumber\\
\eeqa
and the phases $q_{n_k}^{a_k}$ are given by (\ref{qdef}).

The normalization constant $\Omega$ will be chosen so that
the leading term in $1/J$ expansion of the
$\Or(g^0)$ two-point function is normalized to one.
This leading term is given by the interaction-free diagrams
\beq
\label{intfree}
\III{\phi^{i_1}_{n_1}}{\phi^{i_1}_{n_1}}{}{}{\phi^{i_k}_{n_k}}{\phi^{i_k}_{n_k}},
\eeq
where the subscript $n_k$ in $\phi^{i_k}_{n_k}$ means that 
the corresponding insertion of $\phi^{i_k}$ in the string of
operators comes with the phase $q_{n_k}^{a_k}$.
Expression 
(\ref{intfree}) contains only contractions
of the same $\phi^{i_k}_{n_k}$.
Interaction-free diagrams with contractions of  $\phi^{i_k}_{n_k}$
and  $\phi^{i_l}_{n_l}$ with $n_k \neq n_l$ are also allowed, as
long as $i_k=i_l$.
Such diagrams however are subleading in $1/J$.

From (\ref{intfree}) 
we 
infer
that 
\beq
  \Omega = c N^{J+\N} (J+\N) \Ot,
\eeq
where $c$ is an irrelevant numerical prefactor; 
$N^{J+\N}$ arises
from the number of color loops in (\ref{intfree}); 
and $(J+\N)$ takes 
care of the fact that performing a cyclic permutation 
in one of the operators entering the two-point function gives an equivalent diagram.
When no oscillators are excited more than once, there is no further choice
of contractions and $\Ot$ is equal to the number of ways $\N$ 
$\phi$'s can be distributed among $J$ $z$'s, $\Ot=\prod_{n=1}^\N (J+n)$.
When there are multiple excitations of the same mode, 
there can be $N^i_n!$ inequivalent permutations of the 
$\phi^{i}_{n}$ in either one of the operators.
This gives rise to $N^i_n!$ copies of the diagram (\ref{intfree}).
We conclude that in general, 
\beq
  \Ot= \prod_k N^{i_k}_{n_k}! \prod_{n=1}^\N (J+n).
\eeq


\subsection{Two-point functions and the light cone Hamiltonian}
\label{section:anomalous dims}

The light cone energy of a string state and its momentum are related
to the anomalous dimension $\Delta$ and $R$-charge $J$ of the corresponding 
operator as follows
\beqa
\label{eq:H-lc-p's way}
  H_{lc}&=&
- \PP_+ ~~ = i {\partial \over \partial X^+} = 
\Delta-J, \\
\label{etaj}
  \eta&=& 
- \half \PP_- = {i\over2} {\partial \over \partial X^-} = 
{\Delta + J \over 2 R^2}.
\label{eq:eta-p's way}
\eeqa 
One can find anomalous dimensions of the gauge theory operators 
by looking at two-point functions, and we are now going to explain 
this in detail.
We will only consider planar diagrams. 
This amounts to neglecting 
string amplitudes of genus one and higher. 
Furthermore, we will only look at the terms in $H_{lc}$ which
behave like
\beqa
\label{pws}
 && {1 \over \eta^{2a}} \sim \left( {R^2 \over J} \right)^{2a} = {(4 \pi g N)^{a} \over J^{2a}},
                 \\ \nonumber
 && {1 \over R^2 \eta^{2 a+1}} \sim {1 \over R^2} \left( {R^2 \over J} \right)^{2a+1} =
           {(4 \pi g N)^a \over J^{2a+1}}, 
\phantom{\Bigg|_\Bigg|}
\\ \nonumber 
           && \mbox{with $a = 0, 1$.}
\eeqa
On the string theory side, the first line in (\ref{pws}) corresponds to
the truncated expansion in powers of $1/\eta^2$ of the
plane wave Hamiltonian $H_0$.
The second line corresponds to the expansion of $H_1$.
Terms  in the two lines differ by a factor of $1/J$.
On the gauge theory side this factor arises when finite $J$ corrections
are taken into account, which leads to the modification of BMN operators,
explained in section \ref{section:operators:pp}.
The first perturbative (from the SYM point of view) correction to
the light cone energy in (\ref{pws}) corresponds to $a=1$,
which implies that $a=0$ term in the second line of (\ref{pws}) vanishes.
This is in complete accord with the expansion of $H_1$ in powers of $1/\eta$.

Consider a set of gauge theory operators $\Or_\alpha$ 
labeled by $\alpha=\{(i_k,n_k)\}$. 
We will be interested in the SYM operators 
which correspond to plane wave states 
with $\N$ worldsheet oscillators excited.
Their two-point functions can be arranged as
\beqa
\label{2g}
  \langle \Or_\alpha(x) {\bar \Or}_\beta(0) \rangle &=&
 \langle \Or_\alpha(x) {\bar \Or}_\beta(0) \rangle_{g^0}
 +\langle \Or_\alpha(x) {\bar \Or}_\beta(0) \rangle_{g^1} + \OO(g^2) \\ \nonumber &=&
             {1 \over |x|^{2(J+\N)}} \left[
         {\textbf T}_{\alpha\beta} - \F_{\alpha\beta} \log (\mu^2 x^2) 
+ \OO(g^2) 
\right].
\eeqa
Here, ${\textbf T}$ is a matrix of combinatorial factors 
which come from interaction-free diagrams in  
$\langle \Or_\alpha(x) {\bar \Or}_\beta(0) \rangle_{g^0}$, 
while ${\textbf F}$ captures the $\OO(g)$ effects of SYM interactions
in $\langle \Or_\alpha(x) {\bar \Or}_\beta(0) \rangle_{g^1}$. 
$\OO(g)$ contributions to the two point functions (\ref{2g}) 
come from diagrams of the type 
\beq
\label{4ptsample}
  \XII{}{}{}{}{}{}.
\eeq
In appendix \ref{section:feynman rules} we show that 
(\ref{4ptsample}) is equal to 
\beq
\label{eq:gamma value}
   \gamma \equiv -\beta \log \mu^2 x^2
          \equiv -{g N \over 2 \pi} \log \mu^2 x^2
\eeq
times a numerical factor 
determined by the fields which go into the 4-point vertex.

Operators $\OO_\alpha$ may not have well defined scaling dimensions 
at order $\OO(g)$; 
the same phenomenon occured in 
Chapters \ref{chapter: BPS: 2pt} and \ref{chapter:systematic}. 
To find pure operators and their anomalous dimensions, 
we need to transform to a basis of eigenstates of the dilatation
operator.
By a linear transformation, 
we should bring (\ref{2g}) to the form 
\beq
   {1 \over |x|^{2(J+\N)}} \left[{\textbf 1}-{\mathrm {diag}}[\{\lambda_\rho\}] 
	\log (\mu^2 x^2)\right],
\eeq
where the order $\OO(g)$ anomalous dimensions 
$\lambda_\rho$ are the eigenvalues of ${\textbf T}^{-1} \F$, 
and ${\textbf 1}$ is a unit matrix, 
see Section \ref{6 and higher}. 
The matrices in (\ref{2g}) have the form
\beqa
\label{eq:T definition}
  {\textbf T} &=& {\textbf 1}+ {1\over J}{\textbf T}^{(1)} 
+ \OO(1/J^2)
, \\
  \F      &=& \F^{(0)} +{1\over J} \F^{(1)}
+ \OO(1/J^2)
,
\label{ff}
\eeqa
since the operators $\OO_\alpha$ were chosen to be 
orthonormal at leading order, 
see the end of section \ref{section:operators:pp}. 
Hence, up to corrections that are higher order in $1/J$
\beq
\label{g2form}
 { {\textbf T}^{-1} \F }=  \F^{(0)}
     +{1 \over J} \left(  \F^{(1)} -  {\textbf T}^{(1)}  \F^{(0)} \right)
+ \OO(1/J^2)
.
\eeq

Finally, light cone energies of worldsheet states 
are related to the 
quantum numbers of operators in \NN=4 SYM 
as 
\beq
\label{lcesym}
\Delta-J=\N+\lambda_\rho
.
\eeq
In other words, $\N\,{\textbf 1}+{\textbf T}^{-1} \F$, 
plays the role of the light cone Hamiltonian.
In the next section we will show that 
$H_{lc}-\N=H_0+H_1-\N$ 
is identical to ${\textbf T}^{-1} \F$ computed in the
gauge theory 
[the $H_0$ and $H_1$ are given by
(\ref{h0e}), (\ref{h1diag}) and (\ref{h1offdiag})].
This means that to the accuracy we are working at, 
the spectrum of eigenstates of the light cone worldsheet Hamiltonian 
is the same as the spectrum of the dilatation operator 
in SYM.


\subsection{Equality of matrix elements} 
\label{section:equalitysimple}
Let us now show that ${\textbf T}^{-1} \F$ and $H_{lc}-\N$ indeed
have the same matrix elements that are relevant for the
diagonalization.
In this section we 
consider matrix elements between states 
with all modes having distinct SO(4) indices.
We also assume that no modes are excited
more than once, $N^i_n \le 1$.
In appendix C we generalize these results to 
matrix elements between arbitrary plane wave states. 

The 
relevant off-diagonal terms in  $H_{lc}$ 
are given by (\ref{h1offdiag}).
When sandwiched between 
\beq
\label{i1}
  |\X\rangle  =  y_{n_1}^{i_1}{}^\dagger  \ldots  y_m^i{}^\dagger  y_n^j{}^\dagger  |\eta \rangle,
 \qquad   m \neq n,
\eeq
and
\beq
\label{f1}
  |\X'\rangle  =  y_{n_1}^{i_1}{}^\dagger  \ldots  y_m^j{}^\dagger  y_n^i{}^\dagger  |\eta \rangle,
  \qquad  m\neq n,
\eeq
with $i \neq j$, the off-diagonal part of the Hamiltonian 
(\ref{h1offdiag}) 
gives rise to the following matrix element
\beq
\label{del1}
  \langle \X| H_1^{OD}  |\X' \rangle  =  
      -{1 \over J} \left({ R^2 \over J}\right)^2 m n \sqrt{ N^i_m N^j_n N^i_n{}' N^j_m{}'},
\eeq
where we expressed $\eta$ as 
\beq
    \eta={J \over R^2} \left( 1 + {1 \over 2 J} \sum_{i,m} N^i_m 
+ \OO(1/J^2)
\right).
\eeq
This follows from (\ref{eq:H-lc-p's way}) and (\ref{eq:eta-p's way}). 
The second term in the brackets gives an $\Or(1/J)$
correction when used in the leading order Hamiltonian 
(\ref{h0e}). 
We should also reinstate the normal ordering term (\ref{not}).
The SYM calculations will fix it to be ${1 \over R^2 \eta^3} \sum_{i,n} n^2 N^i_n$.
Combining these contributions, the diagonal matrix elements%
\footnote{
	Here and below $\Or$ stands for an arbitrary
	worldsheet state or SYM operator, 
	for example $\X$ or $\X'$. 
	Diagonal matrix elements are all given 
	by the same expression.} 
read 
\beqa
\label{hdiagx}
    \langle \Or| H_0{+}H_1^D{-}\N  |\Or \rangle  
&{=}& 
        {1 \over 2} \left({R^2 \over J} \right)^2 \sum_{i,n} n^2 N^i_n
\nonumber\\&&
         {+}{1 \over J} \left({R^2 \over J} \right)^2 
\bigg[
{-}
\!\!\!
\sum_{i,j,m,n} 
\!
n^2 N^i_m N^j_n 
{+}         
\sum_{i,n} {n^2  N^i_n (N^i_n{+}1) \over 4} 
\bigg].
\hspace{2.7em}
\eeqa
%
For the states considered in this subsection $N^i_n=1$ and 
off-diagonal elements of $H_1$ other than (\ref{del1}) vanish.

We will also denote the SYM operators corresponding to 
states (\ref{i1})-(\ref{f1}) by $\X$ and $\X'$. 
As explained in section \ref{section:operators:pp}, 
no $\tr[ z \ldots {\bar z} \ldots \phi^i \ldots]$ terms 
appear as long as all $i_k$ labels distinct. 
That is, $\X_*=\X_*'=0$, and $\X={\tilde \X}, \X'={\tilde \X'}$.
Contributions to ${\textbf T_{\X\X'}}$ and ${\textbf T}_{\Or\Or}$
come from the diagrams  similar to (\ref{intfree}), 
\beq
\label{intfree2}
\III{\phi^i_m}{\phi^i_{m'}}{\phi^j_n}{\phi^j_{n'}}
 {\phi^{i_k}_{n_k}}{\phi^{i_k}_{n_k}}.
\eeq
\\
The top and bottom rows in (\ref{intfree2}) 
correspond to the two SYM operators entering the
two-point function.
Summing the phases over positions of $\phi$'s 
we obtain
\beq
\label{2ptcs}
    {1 \over \Ot} \sum{}'  \prod_k r_k^{a_k} = \delta_{mm'}\delta_{nn'}-   
                  { \delta_{m+n,m'+n'} (1- \delta_{mm'}\delta_{nn'})\over J}
+\OO(1/J^2). 
\eeq
The prime on the sum in (\ref{2ptcs}) means that we count 
modulo cyclic permutations, 
and we defined 
\beq
 r_{n_k} \equiv q_{n_k} q_{n_k'}{}^*=
     \exp\left({ 2\pi i ( n_k-n'_k) \over J+\N}\right).
\eeq
Only $r_m$ and $r_n$ are different from one, so 
(\ref{2ptcs}) can be computed by making use of
the invariance under cyclic permutations and fixing $a_m=0$ (and so $r_m^{a_m}=1$). 
The $\Or(1/J)$ term in (\ref{2ptcs}) appears because the range
of $a_n$ is $[1,J+\N-1]$.
Contributions with more than two $r_{n_k} \ne 1$ 
are suppressed by at least $1/J^2$ compared to (\ref{2ptcs}), 
so we do not need to worry about them.
Comparing (\ref{2ptcs}) with 
(\ref{2g}) and (\ref{eq:T definition}), 
we arrive at
\beq
\label{tods}
  {\textbf T}^{(1)}_{\X\X}={\textbf T}^{(1)}_{\X'\X'}=0, \qquad {\textbf T}^{(1)}_{\X\X'}=-1.
\eeq

We now turn to the computation of $\F$.
Consider the diagrams that contribute both to 
$\langle \X(x) {\bar {\X'}}(0) \rangle_{g^1}$
and to the diagonal correlator $\langle \Or(x) {\bar \Or}(0) \rangle_{g^1}$. 
These are
\beq
\label{cr1}
 \IIX{}{}{}{}{\phi^{i_k}_{n_k}}{\bar z}{z}{\phi^{i_k}_{n_k}}+
\IIX{}{}{}{}{\phi^{i_k}_{n_k}}{\phi^{i_k}_{n_k}}{z}{\bar z} + (\phi^{i_k}_{n_k} \lra z, \phi^{i_k}_{n_k} \lra {\bar z}),
\eeq
and
\beq
\label{cr2}
\XIIm{\phi^{i_k}_{n_k}}{\phi^{i_k}_{n'_k}}{\phi^{i_l}_{n_l}}{\phi^{i_l}_{n'_l}}
           {z}{\bar z}{\phi^{i_k}_{n_k}}{\phi^{i_k}_{n_k}} +
\XIIm{\phi^{i_k}_{n_k}}{\phi^{i_l}_{n'_l}}{\phi^{i_l}_{n_l}}{\phi^{i_k}_{n'_k}}
               {z}{\bar z}{\phi^{i_k}_{n_k}}{\phi^{i_k}_{n_k}} + 
               (\phi^{i_k}_{n_k} \lra \phi^{i_l}_{n_l},\phi^{i_k}_{n_k'} \lra \phi^{i_l}_{n_l'} ).
\eeq
\\
The level matching condition gives 
\beq
\label{momc}
m+n=m'+n'.
\eeq
The diagrams in (\ref{cr2}) which contribute to 
${\textbf F}^{(1)}_{\X \X'}$
have $n_k=n'_l=m, n_l=n'_k=n$.
The contribution (\ref{cr1}) differs from the interaction-free
diagram (\ref{intfree2}) just by an overall factor
\beq
\label{dfactor}
   -\gamma \, ( q_{n'_k}{}^*+q_{n'_k}-2 ).
\eeq
Therefore, 
summing over possible configurations of fields 
gives (\ref{2ptcs}) times (\ref{dfactor}), 
for a particular $\phi^{i_k}_{n_k}$ participating in the 
interaction vertex in (\ref{cr1}). 
Since any one of the $\phi^{i_k}_{n_k}$ can 
be used in the interaction (\ref{cr1}), 
this must be further summed over $k$. 
We find 
\beq
\label{f0c}
   -\gamma \left[\delta_{mm'}\delta_{nn'}-
   { \delta_{m+n,m'+n'} (1- \delta_{mm'}\delta_{nn'})\over J} \right]
        \sum_k \, ( q_{n'_k}{}^*+q_{n'_k}-2 ).
\eeq
This expression overcounts certain diagrams 
which do not appear in (\ref{cr1}). 
More precisely, whenever two $\phi$'s 
are sitting next to each other 
the $\phi-\phi$ line cannot cross or touch a $z-\bar z$ line, 
either to the left or to the right. 
We will deal with such diagrams separately.

We can read off the $\Or(J^0)$ part of $\F$ from (\ref{f0c}) 
by using (\ref{2g}) and $\gamma=-\beta \log \mu^2 x^2$, 
\beq
\label{f0u}
   \F^{(0)}=-\beta \sum_k \, ( q_{n_k}{}^*+q_{n_k}-2 ).
\eeq
Expanding the $q$'s  in powers of  $1/J$ and taking the
leading term gives the result of BMN,
\beqa
\label{f0lo}
  \F^{(0)}&=& f^{(0)} {\textbf 1}, \\ 
  f^{(0)}&=&{2 \pi g N \over J^2} \sum_k (n_k)^2 = 
             {1 \over 2} \left( R^2 \over J \right)^2 \sum_n n^2 N^i_n.
\eeqa
To get $\Or(1/J)$ corrections to this result we have to
be more careful.
As explained above, in (\ref{f0c}) we overcounted 
the configuration of fields where 
two $\phi$'s appear next to one another 
in the top row of (\ref{cr1}), as in (\ref{cr2}): 
\beq
\label{nal}
(q_{n_k}^{a_k} q_{n_l}^{a_l} \ldots)  \tr[ \ldots \phi^{i_k} \phi^{j_l} \ldots]+(k\lra l).
\eeq
Now diagrams in (\ref{cr1}) where $\phi^{i_k}$ ($\phi^{j_l}$) interacts with 
$z-\bar z$ propagator to the right (left) are not allowed.
The value of such diagrams is
\beqa
\label{vdo}
  -\gamma ( q_{n'_k}^*+q_{n'_l}-2) q_{n_l} q_{n_l'}^* \,
     r_{n_k{+}n_l{-}n'_k{-}n'_l}^{a_k}  +(k\lra l)
=
\hspace{3em}
\nonumber\\
       -\gamma (q_{n_k}^*+q_{n_l}-2  q_{n_l} q_{n_l'}^*) \,
     r_{n_k{+}n_l{-}n'_k{-}n'_l}^{a_k}  +(k\lra l),
\eeqa
where we used (\ref{momc}).
Their contributions have to be substituted by the ones that appear
in (\ref{cr2}) instead. 
These are given by
\beq
\label{vdc}
   \gamma ( q_{n_l} q_{n_l'}^* - q_{n_l} q_{n'_k}^* ) r_{n_k{+}n_l{-}n'_k{-}n'_l}^{a_k}+(k\lra l).
\eeq
The difference of (\ref{vdc}) and (\ref{vdo}) is the same
for both diagonal ($n'_k=n_k, n'_l=n_l$) and off-diagonal 
($n'_k=n_l, n'_l=n_k$) cases, 
and equals
\beq
\label{add1}
   \gamma (q_{n_k}^*+q_{n_l}- q_{n_l} q_{n_k}^* -1 )  r_{n_k{+}n_l{-}n'_k{-}n'_l}^{a_k}
       +(k\lra l).
\eeq
This should be summed over $a_k$ and divided by the normalization constant $\Ot$.
Since the number of configurations with two $\phi$'s next to each other is 
down by $1/J$ compared to the total number of configurations, 
we pick up an overall factor of $1/J$. 
Configuration which have three and more $\phi$'s
next to each other are suppressed by even higher powers of $1/J$, 
and we can neglect them to the order we are working. 

The full result for 
$\langle \X(x) {\bar \X'}(0)\rangle_{g^1}$ is given by (\ref{f0c}), 
plus (\ref{add1}) with $n_k=m, n_l=n$.
Other terms in (\ref{cr2}) are $\Or(1/J^2)$ and are not important for us.
Since $m'=n, n'=m$ for an off-diagonal element, the
first term in (\ref{f0c}) vanishes, and we have
\beq
\label{fxod}
 \F_{\X\X'}^{(1)}
= -\beta   \left[ \sum_k \, ( q_{n'_k}{}^*{+}q_{n'_k}{-}2 )+
              (q_{m-n}^*{+}q_{m-n}{-}q_m^* {-} q_m {-}q_n^*{-}q_n {+}2 ) \right].
\eeq
To get the corresponding off-diagonal element of the light cone Hamiltonian, 
we should add 
$-[{\textbf T}^{(1)} \F^{(0)}]_{\X\X'}/J$
to $\F_{\X\X'}^{(1)}/J$, 
see (\ref{g2form}). 
According to (\ref{tods}) and (\ref{f0u}), such addition precisely
cancels the first term in (\ref{fxod}), 
and we find 
\beq
  [{\textbf T}^{-1} \F]_{\X\X'}= -{\beta   \over J}
                  (q_{m-n}^*{+}q_{m-n}{-}q_m^* {-} q_m {-}q_n^*{-}q_n {+}2 ).
\eeq       
Expanding the $q$'s in powers of $1/J$, taking the leading term and
substituting the value of $\beta$ we arrive at
\beq
\label{del1a}
   [{\textbf T}^{-1} \F]_{\X\X'}= -{1 \over J} \left({R^2 \over J}\right)^2 m n.
\eeq
This reproduces the string theory off-diagonal matrix element
(\ref{del1}), since for the states we are considering $N^i_n=1$.

Let us now compute the diagonal terms.
Now all of the diagrams in (\ref{cr2}) contribute, (\ref{add1})
should be summed over $k$ and added to (\ref{f0c}) with
$m'=m, n'=n$.
This gives
\beq
\label{diagmee}
 \langle \Or(x)\Or(0)\rangle_{g^1}=  -\gamma \sum_k \, ( q_{n_k}{}^*+q_{n_k}-2 )
                          +{\gamma \over 2 J} \sum_{k\neq l}
                          (q_{n_k}{+}q_{n_l}- q_{n_l} q_{n_k}^* -1 +c.c.).
\eeq       
Since ${\textbf T}_{\Or\Or}=1$, we have
\beq
\label{diagme}
   [{\textbf T}^{-1} \F]_{\Or\Or}=  -\beta \sum_k \, ( q_{n_k}{}^*+q_{n_k}-2 )
                          +{\beta \over 2 J} \sum_{k\neq l}
                          (q_{n_k}{+}q_{n_l}- q_{n_l-n_k} -1 +c.c.).
\eeq
The first term gives (\ref{f0lo}) at the leading order, 
however the definition (\ref{qdef}) of $q_{n_k}$ implies
that there is a $1/J$ correction to the leading term.
Expanding in powers of $1/J$ and keeping terms up to $\Or(1/J)$
one can write (\ref{diagme}) as
\beq
\label{tfdiag}
   [{\textbf T}^{-1} \F]_{\Or\Or}={1 \over 2} \left({R^2 \over J} \right)^2 \sum_n n^2  +
         {1 \over J} \left({R^2 \over J} \right)^2 \left(
         {-}\sum_{i,j,m,n} n^2 N^i_m N^j_n 
         {-}{1 \over 2} \sum_{k \neq l:\, i_k{\neq} i_l} n_k n_l  \right).
\eeq
Using the level matching condition 
(which now reads $\sum_k n_k=0$), 
we can write the last term in parenthesis as
\beq
  {-}{1 \over 2} \sum_{k \neq l:\, i_k{\neq} i_l} n_k n_l  =
     {1 \over 2} \sum_{k} n_k^2.
\eeq
Substituting this back into (\ref{tfdiag}) one can see that the
resulting expression is equal to the string theory result (\ref{hdiagx}).
In appendix \ref{section:equalitygeneric} we generalize
the results of this subsection to matrix elements between the
generic states.


\section{Summary and further developments}
\label{section:discussion}

It has been known for some time \cite{Metsaev,Metsaev:2002re} 
that type IIB string theory is solvable in the plane wave background,
which can be viewed as the Penrose limit of $AdS_5 \times S^5$.
BMN \cite{BMN} showed that the string spectrum in this background,
can be recovered from the boundary \NN=4 super Yang-Mills.
Motivated by these results, we analyzed the properties of
this correspondence when finite radius effects are included.
We found that to the leading order in $1/R^2$ and $\lambda'=g N/J^2$, 
the string theory spectrum matches the spectrum
of anomalous dimensions of (linear combinations of) BMN operators.
%
On the string side we have an interacting worldsheet 
theory, when the leading $\Or(1/R^2)$ corrections
to the plane wave metric are taken into account.
Leading corrections to the string spectrum can then be computed
with quantum mechanical perturbation theory.
On the SYM side, departing from the Penrose limit forces one to refine
the BMN operators, paying attention to $1/J$ corrections.
We nevertheless assume that these refined operators continue
to correspond to plane wave states even away from the Penrose limit.
Such operators however do not have definite scaling dimensions,
when $1/J$ corrections are included.
Finding the spectrum of scaling dimensions in SYM requires
one to compute the matrices of two-point functions
$\langle \Or_\alpha {\bar \Or}_\beta \rangle_{g^0} \sim {\textbf T}_{\alpha\beta}$
and 
$\langle \Or_\alpha {\bar \Or}_\beta \rangle_{g^1} \sim \F_{\alpha\beta}$.
Then, ${\textbf T}^{-1}\F$ 
is related to the light cone worldsheet Hamiltonian. 
We find matching between the matrix elements of these operators.

There is a number of questions raised by the results discussed here.
It would be interesting to see if the correspondence between 
the operators we define in section \ref{section:operators:pp} and
plane wave states is exact and holds for arbitrary values of AdS radius.
So far we matched the leading $1/R^2$, $\lambda'$ terms in
matrix elements of the light cone Hamiltonian.
We did not include the fermionic part of the superstring
in our analysis, which led to an undefined normal ordering
constant in diagonal matrix elements.
Incorporating fermions and extending the results of 
\cite{Metsaev,Metsaev:2002re}
to $\Or(1/R^2)$ corrections is an interesting open problem.
It would also be interesting to extend our analysis 
to higher powers of $\lambda'$.
This would require computing diagrams with multiple interactions,
but perhaps one may be able to come up with a resummation
technique similar to the one introduced in \cite{BMN}.
Extending our results to higher orders in $1/R^2$ seems 
more difficult technically, but might also deserve some interest.

Other possible extensions include studying backgrounds
that are more complicated than $AdS_5 \times S^5$.
Probing the strong coupling behavior of boundary
theories with fewer supersymmetries may be of
particular interest, but it remains to be seen how far one can go 
with this perturbative approach.



\section{Appendix}


\subsection{An alternative worldsheet discussion}
\label{section:gsw way}

In Section \ref{section:worldsheet} we discussed 
how to do the worldsheet calculations in the 
spirit of Polchinski \cite{Polch}. 
In this Appendix, we explain in detail 
how to fix the gauges using the method described 
in GSW \cite{GSW}. 
We find the same results for physical quantities 
as in Section \ref{section:worldsheet}.

\subsubsection{Penrose limit of $AdS_5 \times S^5$}
\label{section:leading order}

Before fixing any gauges, 
the bosonic part of the worldsheet action is 
\begin{eqnarray}
\label{eq:basic action:again}
S = - {1\over 4 \pi \alpha'} 
\int (d^2 \sigma) \sqrt{-\gamma} \gamma^{ab} G_{ab}
\end{eqnarray}
where 
the induced metric on the worldsheet is 
$G_{ab} \equiv \partial_a X^\mu \partial_b X^\nu G_{\mu\nu}$. 
Using reparametrization invariance and 
Weyl invariance, we can bring the worldsheet metric to the 
form 
\begin{eqnarray}
\label{eq:metric:gamma=eta}
\gamma_{ab} = \eta_{ab} = 
\left(\matrix{
-1 & 0 \cr 0 & 1
}\right)
\end{eqnarray}
in $(\tau,\sigma)$ coordinates. 
The leading order target space metric (\ref{eq:metric:leading order}) is 
\begin{eqnarray}
\label{eq:metric:leading order:again}
ds^2_0 = - 4 dX^- dX^+ - (r^2+y^2) dX^+ dX^+ + dX_I \, dX_I 
\end{eqnarray}
$I=1, ... , 8$. 
After fixing the worldsheet metric as in (\ref{eq:metric:gamma=eta}), 
the string action (\ref{eq:basic action:again}) becomes 
\begin{eqnarray}
\label{eq:action:leading}
S_0 &=& - {1\over 4 \pi \alpha'} 
\int (d\tau d\sigma) 
\left[
4 \dot X^- \dot X^+ + X^2 \dot X^+ \dot X^+ - \dot X_I \dot X_I 
\right.\nonumber\\&&\hspace{6em}\left.
- 4 (X^-)' (X^+)' - X^2 (X^+)' (X^+)' + X_I' X_I' 
\right]
\end{eqnarray}
%
The action (\ref{eq:action:leading}) is not completely gauge fixed. 
We still have the freedom to reparameterize 
the worldsheet coordinates holomorphically, 
\begin{eqnarray}
\label{eq:holomorphic:general}
\sigma^+ \to \tilde\sigma^+(\sigma^+)
,\quad
\sigma^- \to \tilde\sigma^-(\sigma^-)
\end{eqnarray}
where $\sigma^\pm = \tau \pm \sigma$ 
are the holomorphic and antiholomorphic 
worldsheet coordinates. 
Under 
(\ref{eq:holomorphic:general}), 
the new 
\begin{eqnarray}
\label{eq:holomorphic:new tau}
\tilde\tau = 
{1\over2} 
\left[
\tilde\sigma^+(\tau+\sigma) + \tilde\sigma^-(\tau-\sigma)
\right]
\end{eqnarray}
satisfies the free massless wave equation 
\begin{eqnarray}
\label{eq:holomorphic:new tau:wave eq}
\ddot{\tilde\tau} - \tilde\tau'' 
\equiv 
\left[
\partial_\tau^2 - \partial_\sigma^2 
\right] 
\tilde\tau 
= 0
\end{eqnarray}
$X^-$ enters the action (\ref{eq:action:leading}) linearly, 
so we can integrate it out, imposing its equation of motion 
as a constraint. 
This equation is 
$\ddot X^+ - (X^+)'' = 0$, 
and it has the form (\ref{eq:holomorphic:new tau:wave eq}). 
Hence we can choose the 
light-cone gauge 
\begin{eqnarray}
\label{eq:light-cone:X+}
X^+ = x^+ + p^+ \tau 
\end{eqnarray}
This exhausts all the gauge freedom in the problem. 
After integrating out $X^-$ and choosing the lightcone gauge, 
the action becomes 
\begin{eqnarray}
\label{eq:action:leading:lc}
S_0 &=& - {1\over 4 \pi \alpha'} 
\int (d\tau d\sigma) 
\left[
(p^+)^2 X^2 - \dot X_I \dot X_I + X_I' X_I' 
\right]
\end{eqnarray}
From this, 
we find the lightcone Hamiltonian to be 
\begin{eqnarray}
\label{eq:hamiltonian:leading}
H_0 
&=& {1\over 4 \pi \alpha'} \int_0^{2\pi} d\sigma 
\left[
(2 \pi \alpha')^2 P_I P_I + X_I' X_I' + (p^+)^2 X_I X_I 
\right]
\end{eqnarray}
where $P_I$ are the momenta conjugate to $X_I$. 
%
The Hamiltonian (\ref{eq:hamiltonian:leading}) is quadratic, 
and can be quantized exactly. 
Expand the $X_I$ and $P_I$ in modes as 
\begin{eqnarray}
\label{eq:oscillators}
X_I = 
\sum_{n=-\infty}^{+\infty} 
i \sqrt{\alpha' \over 2\varpi_n} 
\left[
a_n^I - a_n^I{}^\dagger
\right]
,\quad
2 \pi 
P_I = 
\sum_{n=-\infty}^{+\infty} 
\sqrt{\varpi_n \over 2 \alpha'} 
\left[
a_n^I - a_n^I{}^\dagger
\right]
\end{eqnarray}
where the frequencies are 
\begin{eqnarray}
\label{eq:frequencies}
\varpi_n = \sqrt{(p^+)^2 + n^2}
\end{eqnarray}
and the oscillators 
\begin{eqnarray}
\label{eq:oscillators:def}
a_n^I = \alpha_n^I e^{- i ( \varpi_n \tau - n \sigma )} 
,\quad
a_n^I{}^\dagger = 
\alpha_n^I{}^\dagger e^{+ i ( \varpi_n \tau - n \sigma )} 
\end{eqnarray}
close as 
$[ \alpha_m^I , \alpha_n^J{}^\dagger ] = \delta^{IJ} \delta_{mn}$. 
%
In terms of these oscillators, 
(\ref{eq:hamiltonian:leading}) reads 
\begin{eqnarray}
\label{eq:hamiltonian:leading:oscillators}
H_0 &=& 
\sum_{I=1}^8 \sum_{n=-\infty}^{+\infty}
\varpi_n 
\left[
N_n^I + {1\over2} 
\right] 
\end{eqnarray}
where the number operators are 
$N_n^I \equiv a_n^I{}^\dagger a_n^I$ 
(no sum on either $n$ or $I$). 
We will drop the normal ordering constants, 
since they cancel against the fermionic ones in the plane wave limit.


To compare space-time quantum numbers with worldsheet quantities, 
we look at the Noether charges associated with target space isometries. 
The relevant ones for us will be the energy 
$E = i \partial_t$, and the angular momentum $J = - i \partial_\psi$, 
where $t$ and $\psi$ are the global coordinates on $AdS$ 
used in (\ref{eq:exact metric}). 
In the dual CFT description, these correspond to the conformal dimension 
$\Delta = E$ and the $R$-charge $J$. 
We find 
\begin{eqnarray}
\label{eq:charges:general}
- i 
{\partial \over \partial X^\pm} 
\quad \lra \quad
\PP_\pm = 
\int_0^{2\pi} d\sigma P_\pm 
\label{eq:momenta:general}
,\quad\mbox{where } 
P_\mu = 
{
\delta S
\over 
\delta \dot X^\mu
}
= {1\over 2 \pi \alpha'} \, G_{\mu\nu}\dot X^\nu
\end{eqnarray}
are the momenta canonically conjugate to 
$X^\mu$. 
In the light-cone gauge, 
\begin{eqnarray}
\label{eq:charges:leading:lc:start}
\label{eq:charges-st:leading:lc:start}
{\Delta + J \over R^2} 
\quad \lra \quad 
i 
{
\partial
\over 
\partial X^-
}
~\lra~
- \PP_- &=& 
{2 p^+ \over \alpha' }
\\
\Delta - J 
\quad \lra \quad 
i 
{
\partial
\over 
\partial X^+
}
~\lra~
- \PP_+ &=& 
{1\over p^+} H_0 
= 
\sum_I \sum_n 
{\varpi_n \over p^+} 
N_n^I 
\label{eq:charges-st:leading:lc:end}
\label{eq:H=P+}
\label{eq:charges:leading:lc:end}
\end{eqnarray}
Given our gauge choice (\ref{eq:light-cone:X+}), 
$\PP_+$ and $H_0$ should differ by a factor of $-p^+$; 
the minus sign in (\ref{eq:H=P+}) comes about because 
$H = i \partial_t$,  
while 
$P = - i \partial_X$. 
The light-cone states 
\begin{eqnarray}
\label{eq:states:def:w/s:first}
| I_m , J_n , ... \rangle 
\equiv
a^I_m{}^\dagger a^J_n{}^\dagger \; ... \; 
|0, p^+ \rangle
\end{eqnarray}
have 
$p^+ = {J \over \sqrt{4 \pi g N}}$, 
with $R^4 = 4 \pi g N \alpha'^2$; 
$\Delta - J = 
[1 + {m^2 \over 2(p^+)^2}] 
+ [1 + {n^2 \over 2(p^+)^2}] 
+ \OO({1 \over(p^+)^4})$. 

Oscillators 
(\ref{eq:oscillators:def}) 
explicitly depend on time, so they are Heisenberg picture
operators. To go to the Schroedinger picture, we can just 
drop the time dependence and use the 
equations of motion which follow from 
the Hamiltonian
(\ref{eq:hamiltonian:leading}). 
These are 
\begin{eqnarray}
\label{eq:oscillators:schroedinger:def:start}
a_n^I &=& \alpha_n^I ~ e^{+ i n \sigma } 
,\quad
{d\over dt} a_n^I = - i \varpi_n \alpha_n^I ~ e^{+ i n \sigma } 
\\
\label{eq:oscillators:schroedinger:def:end}
a_n^I{}^\dagger &=& \alpha_n^I{}^\dagger e^{ - i n \sigma } 
,\quad
{d\over dt} a_n^I{}^\dagger 
= + i \varpi_n \alpha_n^I{}^\dagger ~ e^{- i n \sigma } 
\end{eqnarray}
It will be convenient to work with Heisenberg 
picture operators 
throughout, and convert the final expressions to 
the Schroedinger picture 
before doing 
perturbation theory.

\subsubsection{
Corrections to the Penrose limit of $AdS_5 \times S^5$} 
\label{section:corrections}

The $1/R^2$ correction to the space-time metric is 
given by ${1\over R^2} ds_1^2$ with 
\begin{eqnarray}
\label{eq:metric:correction:again}
ds_1^2 = 
- 2 dX^- dX^+ (r^2-y^2) - {1\over3} (r^4-y^4) dX^+ dX^+ + 
{1\over3} (r^4 d\Omega_3{}^2 - y^4 d\Omega'_3{}^2) 
\nonumber\\
\end{eqnarray}
%
Using the identities
$d r_i d r_i = dr^2 + r^2 d\Omega_3^2$ and 
$r d r = r_i d r_i$, 
we can write 
$r^4  d\Omega_3^2 = 
\left[
r_i r_i dr_j dr_j - r_i r_j dr_i dr_j
\right]$
and similarly for the $y$-s. 
This results in the contributions 
\begin{eqnarray}
\label{eq:X-ab}
X_{ab} \equiv {1\over 3} 
\left[
X_i X_i \; (\partial_a X_j) (\partial_b X_j)
-
X_i X_j \; (\partial_a X_i) (\partial_b X_j)
\right]
\end{eqnarray}
to the induced metric $G_{ab}$. 
$X$ can be either $r$ or $y$ in (\ref{eq:X-ab}), 
and the sums on the repeated $i$ and $j$ run from 1 to 4. 
The $i=j$ terms cancel in (\ref{eq:X-ab}).

After fixing the worldsheet metric as in (\ref{eq:metric:gamma=eta}), 
the bosonic part of the 
action 
becomes 
$S = S_0 + {1\over R^2} S_1$ with 
$S_0$ given in (\ref{eq:action:leading}), and 
\begin{eqnarray}
\label{eq:action:correction}
S_1 &\!=\!& - {1\over 4 \pi \alpha'} 
\int (d\tau d\sigma) 
\left\{
- 2 X^-
\left[ 
\partial_\tau (\dot X^+ ( r^2 - y^2 ) ) 
- 
\partial_\sigma ((X^+)' ( r^2 - y^2 ) ) 
\right] 
\right.\nonumber\\&&\hspace{3em}\left.
+ 
{1 \over 3} 
\left[ 
\dot X^+ \dot X^+ - (X^+)' (X^+)' 
\right]
(r^4 - y^4) 
- \left( r_{\tau\tau} - y_{\tau\tau} \right) 
+ \left( r_{\sigma\sigma} - y_{\sigma\sigma} \right) 
\right\}
\nonumber\\
\end{eqnarray}
(we integrated by parts so that 
derivatives of $X^-$ do not appear in $S_1$). 
Since the variable $X^-$ appears linearly in the action $S$, 
we can integrate it out, and impose its equation of motion 
as a constraint. 
Although this equation is no longer linear, 
it can be solved perturbatively in $1/R^2$. 
Writing 
\begin{eqnarray}
\label{eq:X+:breakdown}
X^+(\tau,\sigma) = 
X^+_0 
+ {1\over R^2} 
X^+_1 
\end{eqnarray}
where $X^+_{0,1}$ are both of order one, we get 
\begin{eqnarray}
\label{eq:eom:correction:linearized}
0 &=& 
\ddot{X}^+_0 
- 
(X^+_0)'' 
\nonumber\\&&
+ {1\over R^2} 
\left\{
\ddot{X}^+_1 
- (X^+_1)'' 
+ \partial_\tau 
\left[
\dot X^+_0 
\left( {r^2 - y^2 \over 2} \right) 
\right] 
- \partial_\sigma 
\left[
(X^+_0)' 
\left( {r^2 - y^2 \over 2} \right) 
\right] 
\right\} 
\nonumber\\
\end{eqnarray}
Since $X^+_0$ satisfies the 
free massless wave equation, we can take 
$X^+_0 = x^+ + p^+ \tau$. 
Thus the (modified) light-cone gauge choice is 
\begin{eqnarray}
\label{eq:lc gauge}
X^+(\tau,\sigma) = 
\left(
x^+ + p^+ \tau 
\right)
+ {1\over R^2} 
X^+_1 
\end{eqnarray}
To completely fix the gauge, we have to make sure that 
contributions of the form 
$x_1^+ + p_1^+ \tau$ and $e^{i n (\tau \pm \sigma)}$, 
are absent in the mode expansion of $X_1^+(\tau,\sigma)$. 
In terms of the original coordinate $X^+$ and the 
original $\tau$ and $\sigma$, this is a statement 
that 
\begin{eqnarray}
\label{eq:X1+:alternative def}
{1\over R^2} 
X_1^+(\tau,\sigma) \equiv X^+(\tau,\sigma) 
- {1\over 2\pi} \sum_{n, \pm} e^{i n (\tau \pm \sigma)} 
\int (d\sigma d\tau) X^+(\tau,\sigma) e^{- i n (\tau \pm \sigma)}
\end{eqnarray}

The leftover piece $X_1^+$ is not a new dynamical variable; 
rather, it depends on $r_i$ and $y_i$. 
It is defined to satisfy 
\begin{eqnarray}
\label{eq:eom:linearized}
\ddot{X}^+_1 
- (X^+_1)'' 
+ {1 \over 2} p^+ 
\partial_\tau 
\left( r^2 - y^2 \right) 
= 0
\end{eqnarray}
The $r^2$ and $y^2$ should be taken as their leading 
order versions (\ref{eq:oscillators}). 
Setting 
$r^i_n \equiv a^i_n$ and $y^i_n \equiv a^{i+4}_n$ 
in the mode expansions (\ref{eq:oscillators}), 
we find 
\begin{eqnarray}
\label{eq:X+ correction}
X^+_1 &=& 
{i p^+ \alpha' \over 2} 
\sum_{m \, , \; n} 
{ \varpi_n \over \sqrt{\varpi_m \varpi_n } }
{ 
\left[
( r_m^i r_n^i - r_m^i{}^\dagger r_n^i{}^\dagger ) - 
( y_m^i y_n^i - y_m^i{}^\dagger y_n^i{}^\dagger )
\right]
\over 
\left[(\varpi_m + \varpi_n)^2 - (m+n)^2 \right]
} 
\nonumber\\&+& 
{i p^+ \alpha' \over 2} 
\sum_{m \ne n} 
{ \varpi_n \over \sqrt{\varpi_m \varpi_n } }
{ 
\left[
( r_m^i r_n^i{}^\dagger - r_m^i{}^\dagger r_n^i ) - 
( y_m^i y_n^i{}^\dagger - y_m^i{}^\dagger y_n^i ) 
\right]
\over 
\left[(\varpi_m - \varpi_n)^2 - (m-n)^2 \right] 
} 
\hspace{3em}
\end{eqnarray}
Equation (\ref{eq:eom:linearized}) 
is solved in Heisenberg picture; 
the operator $X^+_1$ is determined in terms of 
the Heisenberg picture oscillators 
(\ref{eq:oscillators:def}). 
Since (\ref{eq:X+ correction}) 
contains no explicit time dependence, 
it can be interpreted as a Schroedinger picture expression 
(when the oscillators are taken to be in Schroedinger picture), 
and used in perturbative calculations of energies. 

The action in the modified lightcone gauge (\ref{eq:lc gauge}) reads 
\begin{eqnarray}
\label{eq:action:correction:lc:again}
S &=& 
- {1\over 4 \pi \alpha'} 
\int (d\tau d\sigma) 
\left\{
(p^+)^2 (r^2 + y^2) 
- \dot r_i \dot r_i 
- \dot y_i \dot y_i 
+ r_i' r_i' 
+ y_i' y_i' 
\phantom{{1\over 3}}
\right.\hspace{-3em}\nonumber\\&&\hspace{3em}\left.
+ {1\over R^2} 
\left[ 
\left( r_{\sigma\sigma} - y_{\sigma\sigma} \right) 
- \left( r_{\tau\tau} - y_{\tau\tau} \right) 
\phantom{{1\over 3}}
\right.\right.\nonumber\\&&\hspace{6em}\left.\left.
+ {1\over 3} (p^+)^2 ( r^4 - y^4 ) 
+ 2 p^+ \dot X_1^+ (r^2 + y^2) 
\right] 
\right\}
\hspace{-2.2em}
\nonumber\\
\end{eqnarray}
after integrating out $X^-$, i.e. 
after solving the constraint equation (\ref{eq:eom:correction:linearized}). 
As discussed in \cite{p1}, 
the first order correction to the Hamiltonian 
is minus the correction to the Lagrangian, 
$\delta H = - \delta L$. 
%
Hence 
the (modified-)lightcone Hamiltonian is 
\begin{eqnarray}
\label{eq:hamiltonian}
H &=& 
{1\over 4 \pi \alpha'} \int_0^{2\pi} \!\! d\sigma 
\left\{
\left[
(2 \pi \alpha' )^2
( P^r_i P^r_i + P^y_i P^y_i )
+
( r_i' r_i' + y_i' y_i' )
+ (p^+)^2 
( r^2 + y^2 )
\right]
\phantom{{1\over 3}}
\right.\hspace{-1em}\nonumber\\&&\left.\hspace{3em}
+ 
{1\over R^2} 
\left[
( r_{\sigma\sigma} - y_{\sigma\sigma} ) 
- ( r_{\tau\tau} - y_{\tau\tau} ) 
\phantom{{1\over 3}}
\right.\right.\nonumber\\&&\hspace{6em}\left.\left.
+ {1\over 3} (p^+)^2 
( r^4 - y^4 ) 
+ 2 p^+ \dot X^+_1 ( r^2 + y^2 ) 
\right]
\right\}
\hspace{-3em}
\nonumber\\
\end{eqnarray}
with $X_1^+$ given in (\ref{eq:X+ correction}). 
%

The conserved charges corresponding to 
${\Delta + J \over R^2}$ 
and 
$\Delta - J$
are 
\begin{eqnarray}
\label{eq:charges-st:correction:lc:start}
\label{eq:charges:correction:start}
-\PP_- &=& 
{2 p^+\over \alpha'} 
+ {4 p^+ \over 4 \pi \alpha'} \int_0^{2\pi} d\sigma \; 
{1\over R^2} \left( {\dot X_1^+ \over p^+} + {r^2 - y^2 \over 2} \right) 
\\
-\PP_+ &=& 
{1\over p^+} H
\label{eq:charges:correction:end}
\label{eq:charges-st:correction:lc:end}
\end{eqnarray}
%
In terms of the (Schroedinger picture) oscillators, 
\begin{eqnarray}
\label{eq:charges-st:correction:computed:start}
- \PP_- &=& 
{2 p^+\over \alpha'} 
\left\{
1 + {\alpha' \over 2 R^2} 
\left[ 
\sum_{i=1}^4 \sum_{n=-\infty}^{+\infty}
{1\over \varpi_n} (N_n^r{}^i - N_n^y{}^i)
\right] 
\right\}
\end{eqnarray}
Corrections of the form $a a$ and $a^\dagger a^\dagger$ 
precisely cancels 
between ${1\over p^+} \dot X_1^+$ and ${1\over 2}(r^2 - y^2)$ 
in (\ref{eq:charges-st:correction:lc:start}). 
For $p^+ \gg 1$, 
the worldsheet parameter $p^+$ is related to $J$ and $N$ as 
\begin{eqnarray}
\label{eq:ws p+ in terms of st J}
p^+ &=& 
{J \over\sqrt{ 4 \pi g N}} 
\left\{
1 + {1 \over J} 
\sum_{i=1}^4 \sum_{n=-\infty}^{+\infty}
\left[ N_n^y{}^i + {2 \pi g N n^2 \over J^2} N_n^r{}^i \right]
\right\}
\end{eqnarray}
to order $1/R^2$. 
Here we used $R^4 = 4 \pi g N \alpha'^2$, 
and wrote $(\Delta+J) = 2 J + (\Delta-J)$. 
In (\ref{eq:ws p+ in terms of st J})
the contributions of the $y$ and $r$ oscillators 
have rather different structure.

The Hamiltonian (\ref{eq:hamiltonian}) is relatively involved, 
so we analyze it in more detail. 
The leading order lightcone string states 
\begin{eqnarray}
\label{eq:states:def:w/s}
| a_p , b_q , ... \rangle 
= 
y^a_p{}^\dagger y^b_q{}^\dagger \; ... \; 
|0, p^+ \rangle
\end{eqnarray}
with worldsheet momenta 
$(p, q , ... )$ and $(p', q', ... )$ 
are degenerate only when the 
$(p, q, ... )$ and $(p', q', ... )$ 
are permutations of one another. 
Hence the only terms in $\delta H$ relevant 
for computing the first correction to 
the worldsheet energies, 
are the ones which permute the worldsheet momenta, 
namely 
$a_k a^\dagger_k a_k a^\dagger_k$ 
and 
$a_k a^\dagger_k a_l a^\dagger_l$.

Such terms in 
$\left[ (r_{\sigma\sigma}-y_{\sigma\sigma})
-(r_{\tau\tau}-y_{\tau\tau}) 
+ {1\over 3} (p^+)^2 (r^4-y^4) \right]$ 
combine as 
\begin{eqnarray}
\label{eq:terms from X-ss X-tt X4}
&&
2 \left({\alpha' \over 2} \right)^2 
\sum_k {(p^+)^2 \over \varpi_k^2} 
[ r_k^i r_k^i r_k^j{}^\dagger r_k^j{}^\dagger - 
y_k^i y_k^i y_k^j{}^\dagger y_k^j{}^\dagger]
\nonumber\\&+&
2 \left({\alpha' \over 2} \right)^2 
\sum_{k \ne l} {(p^+)^2 + \varpi_k \varpi_l - k l \over \varpi_k \varpi_l} 
[ r_k^i r_l^i r_l^j{}^\dagger r_k^j{}^\dagger - 
y_k^i y_l^i y_k^l{}^\dagger y_k^j{}^\dagger]
\nonumber\\&+&
2 \left({\alpha' \over 2} \right)^2 
\sum_{k \ne l} {(p^+)^2 - \varpi_k \varpi_l + k l \over \varpi_k \varpi_l} 
[ r_k^i r_l^j r_l^i{}^\dagger r_k^j{}^\dagger - 
y_k^i y_l^j y_l^i{}^\dagger y_k^j{}^\dagger]
\end{eqnarray}
and the term $2 p^+ \dot X_1^+ (r^2 + y^2)$ gives 
\begin{eqnarray}
\label{eq:terms from X-dot X-dot}
&-&
2 \left({\alpha' \over 2} \right)^2 
\sum_k 
[ r_k^i r_k^i r_k^j{}^\dagger r_k^j{}^\dagger - 
y_k^i y_k^i y_k^j{}^\dagger y_k^j{}^\dagger]
\nonumber\\&-&
2 \left({\alpha' \over 2} \right)^2 
\sum_{k \ne l} 
{2 (p^+)^2 (\varpi_k + \varpi_l)^2 
\over \varpi_k \varpi_l
[(\varpi_k + \varpi_l)^2 - (k+l)^2]} 
[ r_k^i r_l^i r_l^j{}^\dagger r_k^j{}^\dagger - 
y_k^i y_l^i y_k^l{}^\dagger y_k^j{}^\dagger]
\nonumber\\&-&
2 \left({\alpha' \over 2} \right)^2 
\sum_{k \ne l} 
{2 (p^+)^2 (\varpi_k - \varpi_l)^2 
\over \varpi_k \varpi_l
[(\varpi_k - \varpi_l)^2 - (k-l)^2]} 
[ r_k^i r_l^j r_l^i{}^\dagger r_k^j{}^\dagger - 
y_k^i y_l^j y_l^i{}^\dagger y_k^j{}^\dagger]
\hspace{2.5em}
\end{eqnarray}
Expressions 
(\ref{eq:terms from X-ss X-tt X4})-(\ref{eq:terms from X-dot X-dot}) 
appear in ${1 \over p^+} \delta H$ with an overall prefactor of 
\begin{eqnarray}
\label{eq:overall prefactor}
{1 \over 4 \pi \alpha'} \cdot 2 \pi \cdot {1 \over R^2} \cdot {1 \over p^+}
= {1 \over 2 \alpha' R^2 p^+} 
\end{eqnarray}
and we find 
\begin{eqnarray}
\label{eq:delta H}
{1\over p^+} \delta H &=&
{\alpha' \over 4 R^2 (p^+)^3} 
\sum_{i , j} 
\sum_{k}
{k^2 (p^+)^2 \over \varpi_k^2 }
(y^i_k y^i_k y^j_k{}^\dagger y^j_k{}^\dagger 
-r^i_k r^i_k r^j_k{}^\dagger r^j_k{}^\dagger) 
\nonumber\\&+&
{\alpha' \over 4 R^2 (p^+)^3} 
\sum_{i,j} 
\sum_{k \ne l}
{-2 k l (p^+)^2 \over \varpi_k \varpi_l }
(y^i_k y^i_l{}^\dagger y^j_l y^j_k{}^\dagger 
-r^i_k r^i_l{}^\dagger r^j_l r^j_k{}^\dagger) 
\nonumber\\&+&
{\alpha' \over 4 R^2 (p^+)^3} 
\sum_{i,j} 
\sum_{k \ne l} 
{2 k l (p^+)^2 \over \varpi_k \varpi_l }
(y^i_k y^j_k{}^\dagger y^i_l y^j_l{}^\dagger 
-r^i_k r^j_k{}^\dagger r^i_l r^j_l{}^\dagger) 
\nonumber\\&+&
... 
\end{eqnarray}
The ``...'' 
stands for terms not of the form 
$a a^\dagger a a^\dagger$, 
as well as terms with more than two distinct worldsheet momenta; 
we are also dropping corrections which are higher order 
in $1/R^2$ and $1/p^+$. 
The second and third lines of 
(\ref{eq:delta H}) cancel if $i = j$.

In deriving 
(\ref{eq:terms from X-ss X-tt X4})-(\ref{eq:terms from X-dot X-dot}), 
we have not been careful about the ordering of oscillators. 
This means that we may have overlooked some terms 
which involve commutators $[y^i_m , y^i_m{}^\dagger] = 1$. 
The only terms in (\ref{eq:delta H}) 
where this could happen come from the first line. 
This means we could be possibly neglecting 
\begin{eqnarray}
\label{eq:st:P+:correction:relevant:missing}
- \delta' \PP_+ 
= 
\left( {\alpha' \over 4 R^2 (p^+)^3} \right) 
\zeta \sum_{i} \sum_{k} {k^2 (p^+)^2 \over \varpi_k^2} 
(N_k^y{}^i - N_k^r{}^i) 
\end{eqnarray}
If we were to keep track of the ordering of oscillators, 
we would find $\zeta=1$. 
However, we have not analyzed the fermionic side, 
which can also produce similar terms. 

Finally, we compare the results of this Appendix with 
what we found in Section \ref{section:worldsheet}. 
We will only look at the $y$-oscillators. 
The difference between $p^+$ and $\eta$ is 
\begin{eqnarray}
\label{eq:p+ vs eta}
p^+ = 
\eta 
\left[ 
1 + {1\over 2 R^2} \sum_{i;n} {N^{yi}_n \over \varpi_n}
\right]
\end{eqnarray}
so the frequencies in the two approaches are related as 
\begin{eqnarray}
\label{eq:w/s frequencies}
\varpi_m = w_m 
\left[ 
1 + {\eta^2\over 2 R^2 w_m^2} \sum_{i;n} {N^{yi}_n \over w_n}
+ \OO(1/R^4)
\right]
\end{eqnarray}
Expressions 
(\ref{eq:delta H}) and (\ref{eq:st:P+:correction:relevant:missing}) 
then change trivially as 
$\varpi_n \to w_n$, $p^+ \to \eta$ at this 
order in $1/R^2$, while 
\begin{eqnarray}
\label{eq:H0:p+ vs eta}
{1\over p^+} \sum_{i;n} \varpi_n N^{yi}_n 
= 
{1\over \eta} \sum_{i;n} w_n N^{yi}_n 
- 
{1\over 2 R^2 \eta} \sum_{i,j;m,n} 
{n^2 N^{yi}_n N^{yj}_m \over w_m w_n} 
+ \OO(1/R^4)
\end{eqnarray}
Together, 
(\ref{eq:delta H}) and (\ref{eq:H0:p+ vs eta}) 
reproduce 
the sum of 
(\ref{h0o}), (\ref{eq:adrei:diag}) and (\ref{eq:adrei:off-diag}).


\subsection{\NNN=4 SYM}
\label{section:feynman rules}

Here, we give some details of the \NNN=4 SYM 
needed for the order $g_{\mathrm {Y\!M}}^0$ (tree) and $g_{\mathrm {Y\!M}}^2$ (one-loop level) 
calculations of Section 
(\ref{section:anomalous dims}). 
First we write down the \NNN=4 SYM action in terms 
of the fields we will be dealing with. 
When SUSY is broken down to \NNN=1, 
things much more cumbersome, 
so from the very beginning we use the \NNN=4 
Lagrangian \ref{L su3 form} from Chapter \ref{chapter: BPS: 2pt}, 
\begin{eqnarray}
\label{eq:lagrangian:su3 form}
\LL &=& \mbox{$1\over g_{\mathrm {Y\!M}}^2$ } \tr 
\left\{ 
- \mbox{$1\over 4$} F_{\mu\nu} F^{\mu\nu} 
+ i \lambda \sigma^\mu D_\mu \bar \lambda 
+ i \psi_j \sigma^\mu D_\mu \bar \psi^j 
+ D_\mu z_j D^\mu \bar z^j 
\right. \\ \nonumber &&\quad~~ \left. 
+ i \sqrt2 [ \lambda , \psi_j ] \bar z^j 
- \mbox{$i\over\sqrt2$} \e^{jkl} [ \psi_j ,  \psi_k ] z_l 
+ i \sqrt2 [ \bar \lambda , \bar \psi^j ] z_j 
- \mbox{$i\over\sqrt2$} \e_{jkl} [ \bar \psi^j , \bar \psi^k ] \bar z^l 
\right. \\ \nonumber &&\quad~~ \left. 
+ [ z_j , z_k ] [ \bar z^j , \bar z^k ] 
- \half [ z_j , \bar z^j ] [ z_k , \bar z^k ] 
\right\}
.
\end{eqnarray}
We leave the fields $z_1$, $\bar z^1$ as they are, 
and substitute 
\begin{equation}
\label{eq:z vs phi}
     z_j = \mbox{$1\over\sqrt{2}$} \left( \phi_j + i \phi_{j+3} \right) , 
\quad
\bar z^j = \mbox{$1\over\sqrt{2}$} \left( \phi_j - i \phi_{j+3} \right)  , 
\quad
\mbox{$j=2$ and 3.}
\end{equation}
The rest of the fields (gauge bosons and fermions) remain unchanged, 
and (\ref{eq:lagrangian:su3 form}) becomes 
\begin{eqnarray}
\label{eq:lagrangian:mixed form:def}
\LL = \LL_0 + \LL_1 + \LL_2 + \LL_{\mathrm {other}}
\end{eqnarray}
where
\begin{eqnarray}
\label{eq:lagrangian:mixed form:propagators}
\LL_0 &=& \mbox{$1\over g_{\mathrm {Y\!M}}^2$ } \tr 
\left\{ 
(\partial_\mu z_1) (\partial^\mu \bar z_1) 
+ 
\sum_k
\half 
(\partial_\mu \phi_k) (\partial^\mu \phi_k) 
\right\}
\end{eqnarray}
gives propagators for the scalars; 
\begin{eqnarray}
\label{eq:lagrangian:mixed form:order g}
\LL_1 &=& \mbox{$1\over g_{\mathrm {Y\!M}}^2$ } \tr 
\left\{ 
- i A^\mu [ z_1 , \partial_\mu \bar z_1 ] 
- i A^\mu [ \bar z_1 , \partial_\mu z_1 ] 
+ 
\sum_k
(- i A^\mu) [ \phi_k , \partial_\mu \phi_k ] 
\right. \nonumber \\ && \left. 
+ i \sqrt2 \; z_1 
\left( 
[ \bar \lambda , \bar \psi^1 ] - [ \psi_2,  \psi_3 ] 
\right) 
+ i \sqrt2 \; \bar z_1 
\left( 
[ \lambda , \psi_1 ] - [ \bar \psi^2,  \bar \psi^3 ] 
\right) 
\right. \nonumber \\ && \left. 
+ i \phi_2 
\left( 
[ \lambda , \psi_2 ] + [ \bar \lambda , \bar \psi^2 ] 
- [ \psi_3,  \psi_1 ] - [ \bar \psi^3,  \bar \psi^1 ] 
\right) 
\right. \nonumber \\ && \left. 
+ \phi_5 
\left( 
[ \lambda , \psi_2 ] - [ \bar \lambda , \bar \psi^2 ] 
+ [ \psi_3,  \psi_1 ] - [ \bar \psi^3,  \bar \psi^1 ] 
\right) 
\hspace{-3em}
\right. \nonumber \\ && \left. 
+ i \phi_3 
\left( 
[ \lambda , \psi_3 ] + [ \bar \lambda , \bar \psi^3 ] 
- [ \psi_1,  \psi_2 ] - [ \bar \psi^1,  \bar \psi^2 ] 
\right) 
\right. \nonumber \\ && \left. 
+ \phi_6 
\left( 
[ \lambda , \psi_3 ] - [ \bar \lambda , \bar \psi^3 ] 
+ [ \psi_1,  \psi_2 ] - [ \bar \psi^1,  \bar \psi^2 ] 
\right) 
\right\}
\hspace{-3em}
\end{eqnarray}
gives 3-field vertices; 
and 
\begin{eqnarray}
\label{eq:lagrangian:mixed form:order g^2}
\LL_2 &=& \mbox{$1\over g_{\mathrm {Y\!M}}^2$ } \tr 
\left\{ 
- [ A_\mu , z_1 ] [ A^\mu , \bar z_1 ]  
- \sum_k \half [ A_\mu , \phi_k ] [ A^\mu , \phi_k ]  
\right. \\ \nonumber &&\quad~~ \left. 
- \half [ z_1 , \bar z_1 ] [ z_1 , \bar z_1 ] 
+ \sum_k [ z_1 , \phi_k ] [ \bar z_1 , \phi_k ] 
+ \sum_{k > l} \half [ \phi_k , \phi_l ] [ \phi_k , \phi_l ] 
\right\}
\end{eqnarray}
contains 4-field interactions. 
Finally, 
\begin{eqnarray}
\label{eq:lagrangian:mixed form:other}
\LL_{\mathrm {other}} &=& \mbox{$1\over g_{\mathrm {Y\!M}}^2$ } \tr 
\left\{ 
- \mbox{$1\over 4$} F_{\mu\nu} F^{\mu\nu} 
+ i \lambda \sigma^\mu D_\mu \bar \lambda 
+ i \psi_j \sigma^\mu D_\mu \bar \psi^j 
\right\}
\end{eqnarray}
gives propagators for the gauge bosons and the fermions 
and their interactions with each other (at order $\OO(g_{\mathrm {Y\!M}}^2)$ 
these do not contribute to the diagrams we care about, and neither 
do the ghost terms). 
The Lagrangian (\ref{eq:lagrangian:mixed form:def}) 
has a leftover $SO(4)$ symmetry rotating the $\phi$-s.

\begin{figure}
{\begin{center}
\epsfig{width=4in, file=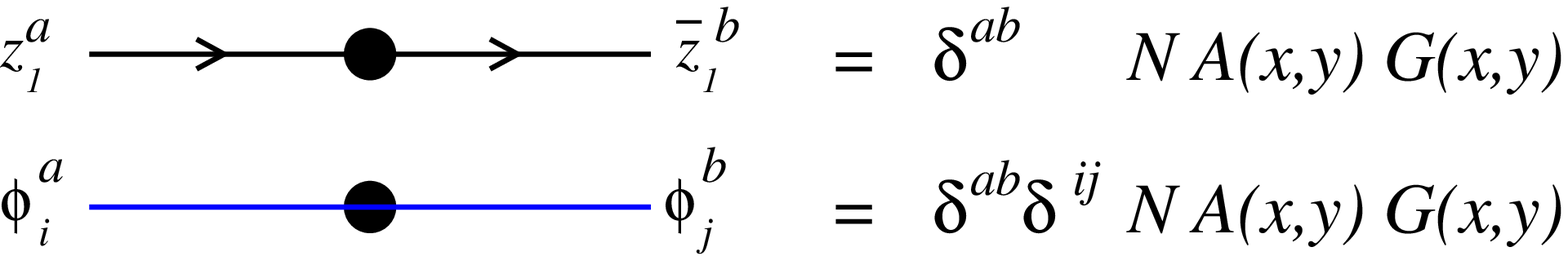, angle=0}\quad
\end{center}}
\caption{%
Order $g_{\mathrm {Y\!M}}^2$ corrections to scalar propagators 
consist of a gauge boson exchange and a fermion loop. 
\label {fig:propagator}
}%
\end {figure}

Feynman rules for the Lagrangian (\ref{eq:lagrangian:mixed form:def}) 
are somewhat awkward, but the tree and one-loop diagrams 
which involve only the scalars can be packaged in a convenient way. 
First, $\OO(g_{\mathrm {Y\!M}}^2)$ corrections to the 
scalar propagators are diagonal in color indices, 
see Figure \ref{fig:propagator}. 
Fermion loops cancel in $\langle \phi_2^a(x) \phi_5^b(y) \rangle_{g_{\mathrm {Y\!M}}^2}$ 
because of the way the signs work out in 
(\ref{eq:lagrangian:mixed form:order g}).

\begin{figure}
{\begin{center}
\epsfig{width=5.7in, file=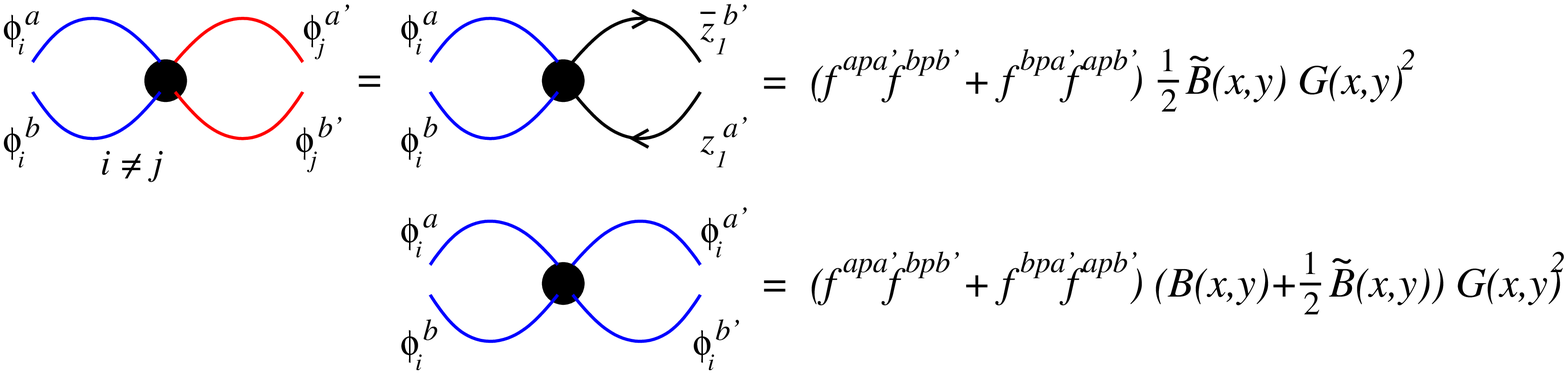, angle=0} 
\end{center}}
\caption{%
Order $g^2$ corrections to two point functions of operators of the form 
$\tr z ... \phi^1 z ... \phi^1$: 
four-field irreducible blocks.
These diagrams represent the net contribution 
of all contributing Feynman diagrams. 
\label {fig:four-point irreducible blocks:same}
}%
\end {figure}

Corrections to the 4-point irreducible blocks 
are more involved, but they can be related to the 
corresponding diagrams involving only $z$-s and $\bar z$-s. 
By comparing two-point functions of the protected 
operators in the [0,2,0] of $SU(4)$ 
written on the one hand in terms 
of $\phi$-fields, and on the other hand in terms of $z$-s and $\bar z$-s, 
we get the diagrams shown in 
Figure \ref{fig:four-point irreducible blocks}. 
Comparison of two-point functions of the Konishi scalar 
$\sum_{k=1}^6 \tr \phi^k \phi^k = \sum_{k=1}^3 \tr z^k \bar z^k$ 
produce the relations listed in 
Figure \ref{fig:four-point irreducible blocks:same}.

\begin{figure}
{\begin{center}
\epsfig{width=5.7in, file=diags2.eps, angle=0} 
\end{center}}
\caption{%
Order $g_{\mathrm {Y\!M}}^2$ corrections to two point functions of operators of the form 
$\tr z ... \phi^1 z ... \phi^2$: four-field irreducible blocks. 
When scalars $\phi^i$ are involved, 
the diagrams above represent the net contribution 
of all contributing Feynman diagrams, packaged in 
a way to mimic the \NN=1 component fields 
Feynman diagrams. 
(Thick lines would correspond to exchanges of auxiliary 
fields $F_i$ and $D$ in the \NN=1 formulation.) 
Diagrams with $z_2$ are given for comparison only. 
There are similar diagrams with 
one or both $z$-lines running in the opposite direction. 
\label {fig:four-point irreducible blocks}
}%
\end {figure}

The ``$D$-term'' contributions $A$ and $B$, 
and the four-field interaction ``$F$-term'' $\tilde B$ are 
defined by Figures \ref{fig:four-point irreducible blocks} 
and \ref{fig:four-point irreducible blocks:same}. 
As we saw in Section \ref{section:gauge-dependent} 
(also see \cite{CFHMMPS}), the 
$A$ and $B$ are not separately gauge invariant. 
These must appear as the gauge invariant combination $2A+B$, 
which vanishes 
in the \NNN=4 theory. So one only has to 
look at ``$F$-term'' contributions, which are all proportional to 
\begin{eqnarray}
\label{eq:b-tilde}
\gamma \equiv 
\half \tilde B (x,0) N = 
- {g_{\mathrm {Y\!M}}^2 N \over 4 \pi^2 } 
\log x^2 \mu^2 
\equiv 
- \beta 
\log x^2 \mu^2 
\end{eqnarray}
computed for example in Chapter \ref{chapter: BPS: 2pt} and in \cite{BMN}. 
In this Chapter, we are using the conventions of \cite{BMN}; 
in the Lagrangian (\ref{eq:lagrangian:su3 form}) we have 
$g_{\mathrm {Y\!M}}^2 = 2 \pi g$.

We only have to consider planar diagrams 
since we are interested in the leading large $N$ behavior. 
Put differently, 
\begin{eqnarray}
\label{eq:traces-leading}
\tr \! \left[ t^{a_1} ... t^{a_k} \right]
~
\tr \! \left[ t^{a_k} ... t^{a_1} \right] 
= \left( {N \over 2} \right)^k \left[ 1 + \OO(1/N^2) \right]
\end{eqnarray}
and $SU(N)$ traces of all other permutations of the generators 
(other than cyclic) are suppressed by $1/N^2$. 
To see this, one can use the ``trace merging formula''
\begin{equation}
\label{eq:merging traces:pp}
2 \left( \tr A t^c \right) \left( \tr B t^c \right) 
= 
\tr A B - 
\mbox{$1\over N$} \left( \tr A \right) \left( \tr B \right) 
\end{equation}
valid when $t^c$ are $SU(N)$ generators in the fundamental 
representation.

At one loop, 
all but the nearest neighbor interactions are suppressed. 
The relevant contributions in 
Figure \ref{fig:four-point irreducible blocks} 
have the form 
\begin{eqnarray}
\label{eq:one-loop combinatorial factor}
\XII{a}{a'}{b}{b'}{c_1}{c_J}
&=& 
\tr \! \left[ t^a t^b t^{c_1} ... t^{c_J} \right]
~
\tr \! \left[ t^{c_J} ... t^{c_1} t^{b'} t^{a'} \right] 
~
f^{abp} f^{a'b'p}
\nonumber\\
&=& \half \left( \half N \right)^{J-1} 
\tr \left( t^a t^b t^{b'} t^{a'} \right) f^{abp} f^{a'b'p}
\left[ 1 + \OO(1/N^2) \right]
\nonumber\\
&=& \half \left( \half N \right)^{J-1} 
\tr \left( t^a [t^p, t^a] t^{b'} [t^p , t^{b'}] \right) 
\left[ 1 + \OO(1/N^2) \right]
\nonumber\\
&=& \left( \half N \right)^{J+3}
\left[ 1 + \OO(1/N^2) \right]
\end{eqnarray}
The difference between the orderings $(ab)$ and $(ba)$ 
in (\ref{eq:one-loop combinatorial factor}) 
is a minus sign, 
\begin{eqnarray}
\label{eq:orderings}
\XII{a}{a'}{b}{b'}{c_1}{c_J}
= - 
\XII{b}{a'}{a}{b'}{c_1}{c_J}
\end{eqnarray}
Diagrams shown in the first two lines of 
Figure \ref{fig:four-point irreducible blocks:same}
have the form 
\begin{eqnarray}
\label{eq:one-loop combinatorial factor:blob}
\blobII{a}{a'}{b}{b'}{c_1}{c_J}
&=& 
\half [ f^{aa'p} f^{bb'p} + f^{ab'p} f^{ba'p} ] 
~ 
\tr \! \left[ t^a t^b t^{c_1} ... t^{c_J} \right]
~
\tr \! \left[ t^{c_J} ... t^{c_1} t^{b'} t^{a'} \right] 
\nonumber\\
&=& \half \left( \half N \right)^{J+3}
\left[ 1 + \OO(1/N^2) \right]
\end{eqnarray}
Only one of the two $f f$-terms contributes at this order; 
the other one is suppressed by at least $1/N^3$. 
The contribution (\ref{eq:one-loop combinatorial factor:blob}) 
is insensitive to $a \lra b$. 
The contributions (\ref{eq:one-loop combinatorial factor}) 
and (\ref{eq:one-loop combinatorial factor:blob}) 
come with a numerical prefactor of 
\begin{eqnarray}
\label{eq:prefactor:def}
{2 \over N} G(x,0)^{J+2} 
\gamma 
\end{eqnarray}
with $\gamma = - \beta \log x^2 \mu^2$ defined in (\ref{eq:b-tilde}).

To summarize, the tree level correlators are 
\begin{eqnarray}
\label{eq:tree diagrams}
\matrix{\epsfig{height=1in, file=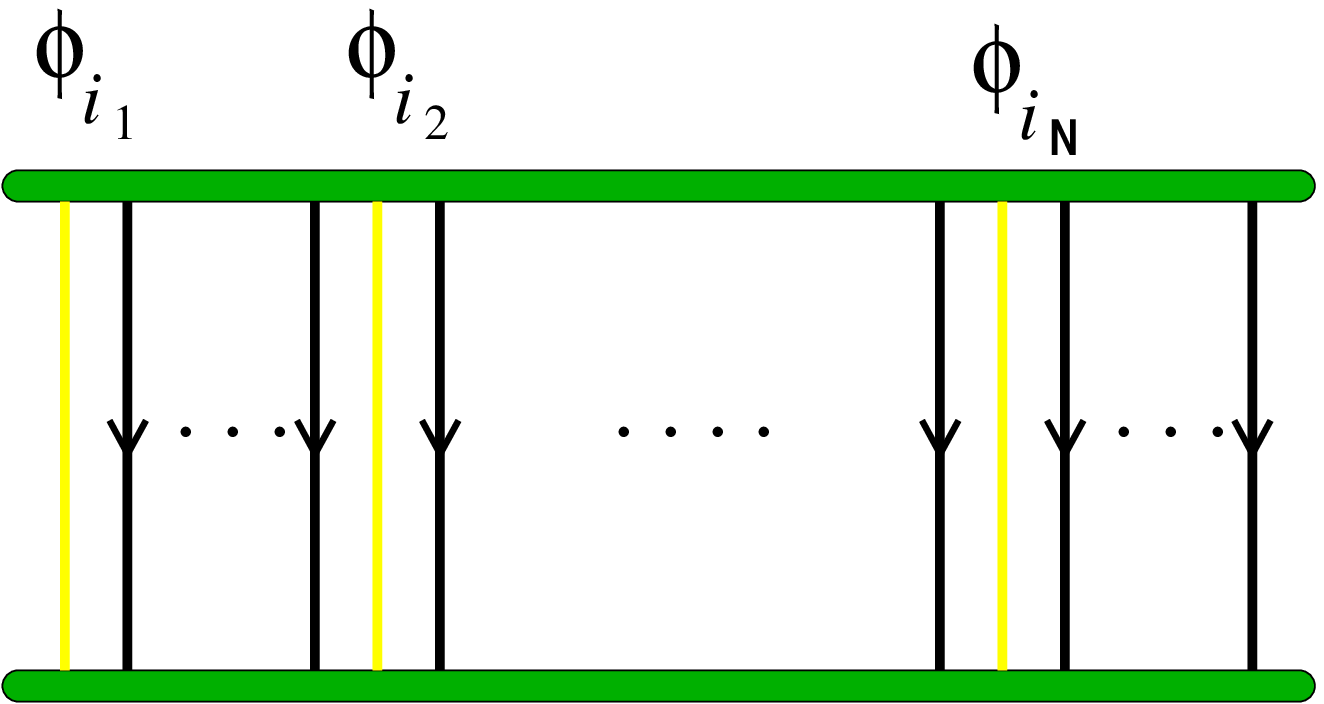}\quad}
= 
(\half G N)^{J+\N} 
\end{eqnarray}
and the relevant 
one-loop contributions 
can be schematically represented as 
\begin{eqnarray}
\label{eq:one-loop diagrams:one phi}
\matrix{{}\cr \epsfig{height=0.6in, file=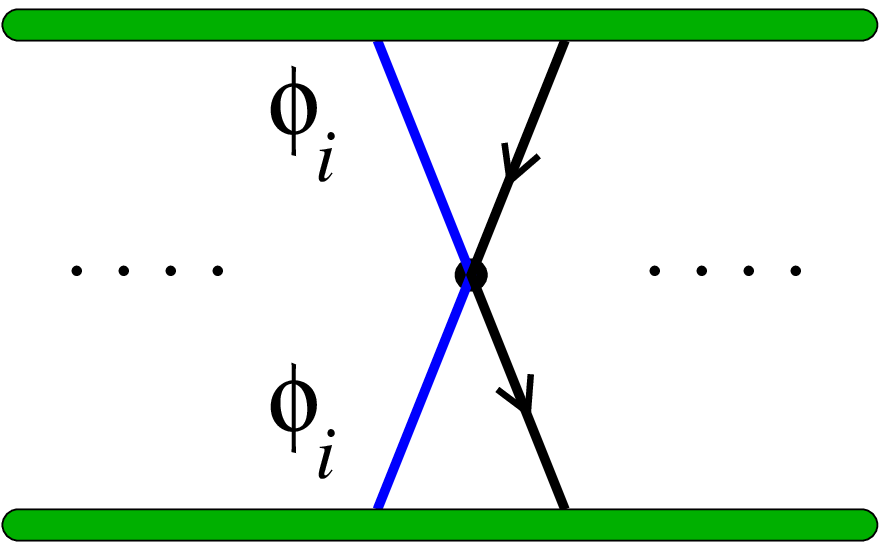}}
= - 
\matrix{{}\cr \epsfig{height=0.6in, file=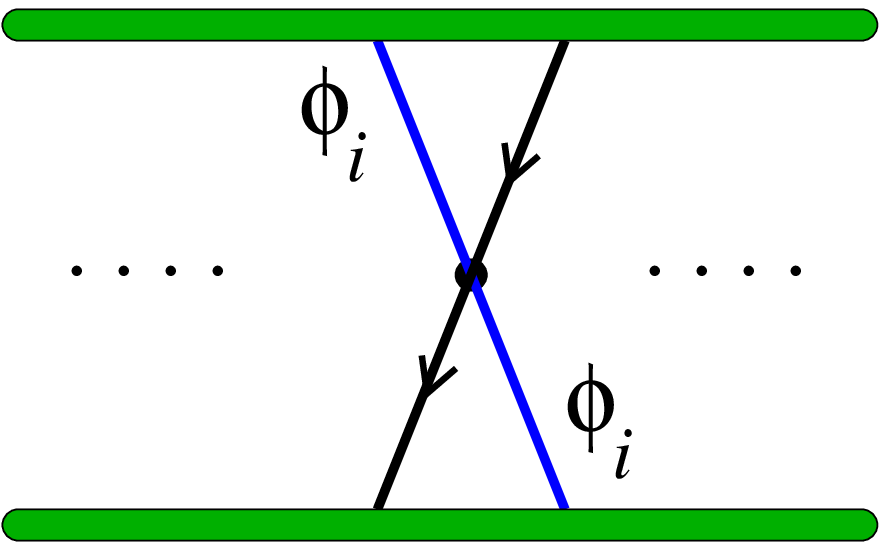}}
&=& 
\gamma \times (\half G N)^{J+\N} 
\end{eqnarray}
when only one $\phi$ is involved in the interaction, 
and  
\begin{eqnarray}
\label{eq:one-loop diagrams:two phi:distinct}
\matrix{{}\cr \epsfig{height=0.6in, file=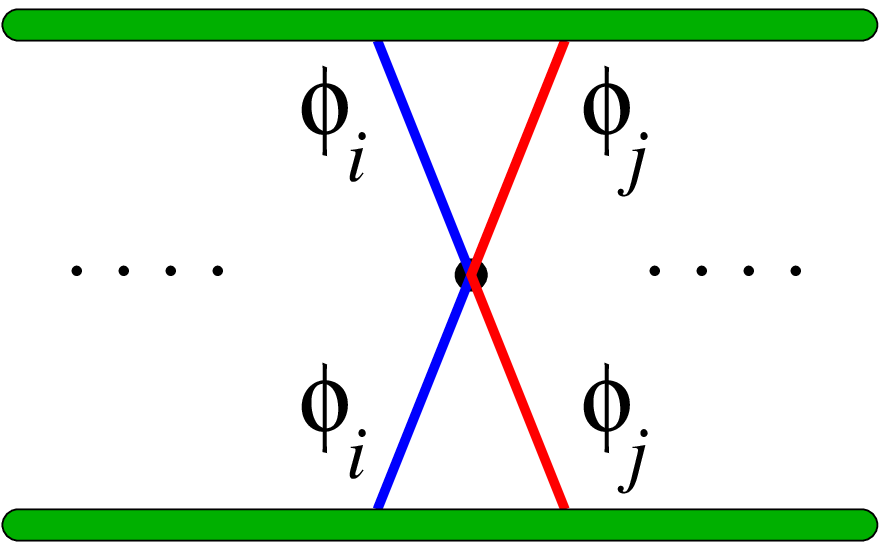}}
= - 
\matrix{{}\cr \epsfig{height=0.6in, file=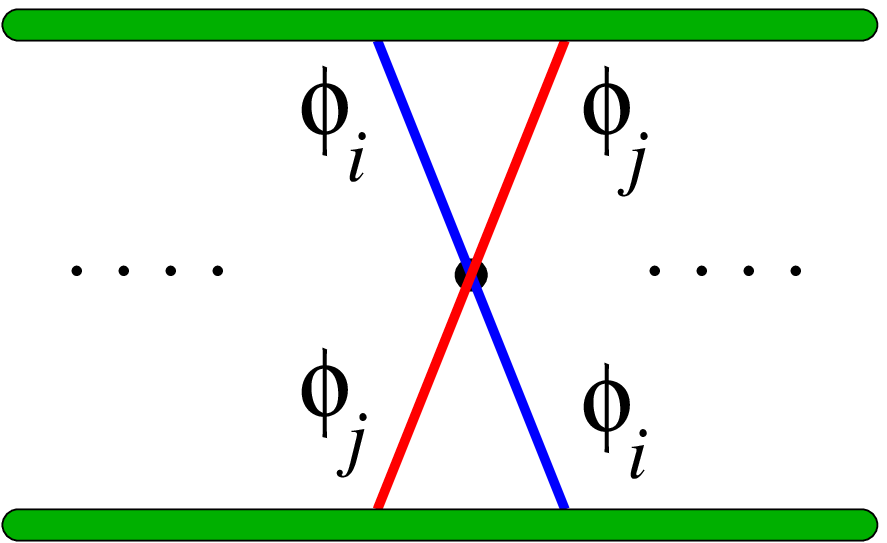}}
&=& 
\gamma \times (\half G N)^{J+\N} 
,\quad i \ne j
\end{eqnarray}
when two distinct $\phi$ within either trace interact. 
Furthermore, we have 
\begin{eqnarray}
\label{eq:one-loop diagrams:two phi:same}
\matrix{{}\cr \epsfig{height=0.6in, file=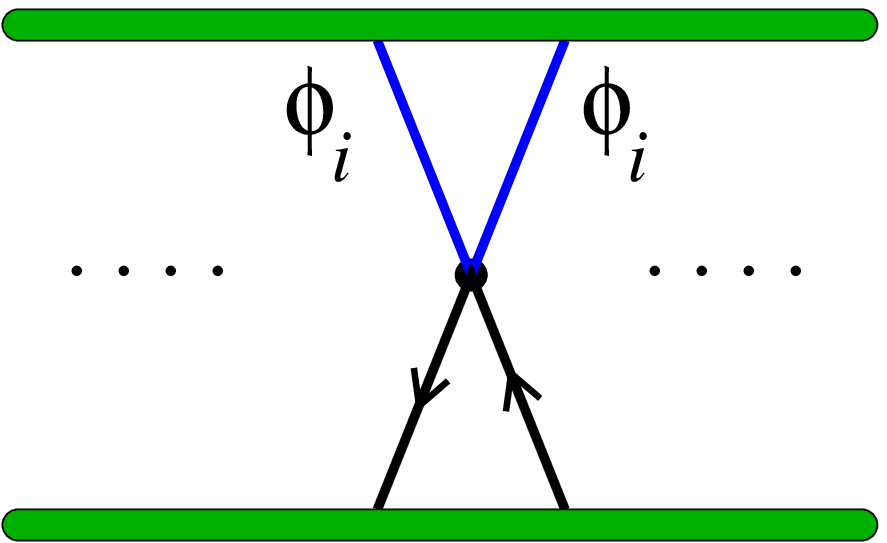}}
= 
\matrix{{}\cr \epsfig{height=0.6in, file=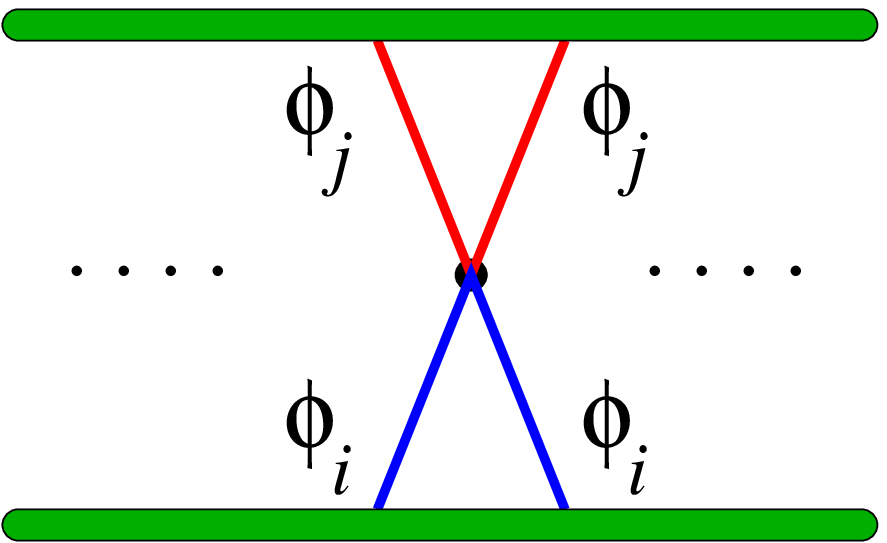}}
= 
\matrix{{}\cr \epsfig{height=0.6in, file=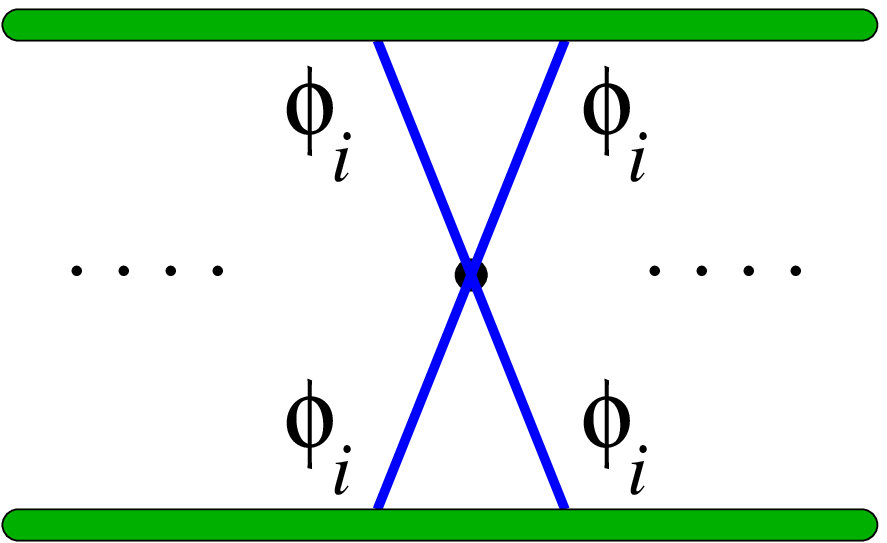}}
&=& 
{1\over2} \, \gamma \times (\half G N)^{J+\N} 
,\quad \!\! 
i \ne j
\nonumber\\
\end{eqnarray}
Finally, the diagrams which involve a $z z \bar z \bar z$ vertex 
can be read off from 
(\ref{eq:one-loop diagrams:two phi:distinct}) 
and (\ref{eq:one-loop diagrams:two phi:same}) 
by expanding the $z$ and $\bar z$ participating in the vertex 
in terms of the two remaining $\phi$'s, 
\begin{eqnarray}
\label{eq:one-loop diagrams:z-zbar}
\matrix{{}\cr \epsfig{height=0.6in, file=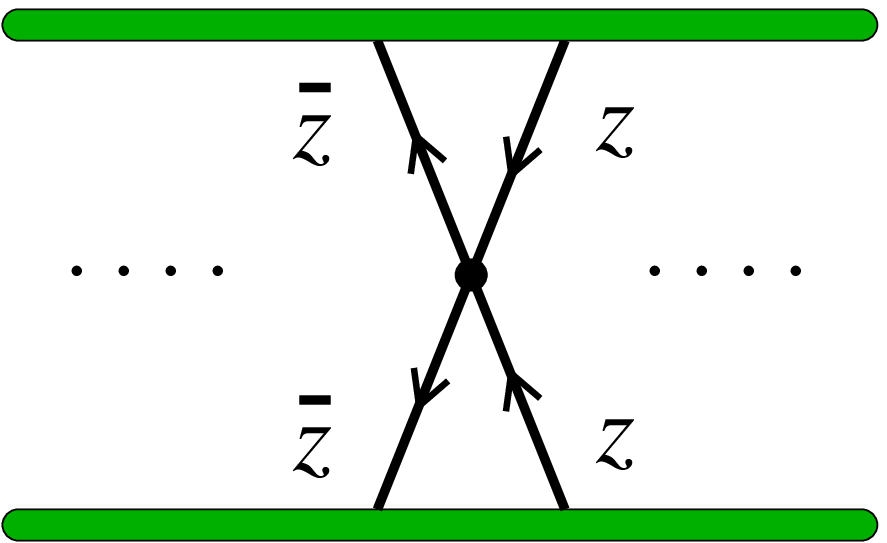}}
= 
- {1\over2} \, \gamma \times (\half G N)^{J+\N} 
, \hspace{2em}
\matrix{{}\cr \epsfig{height=0.6in, file=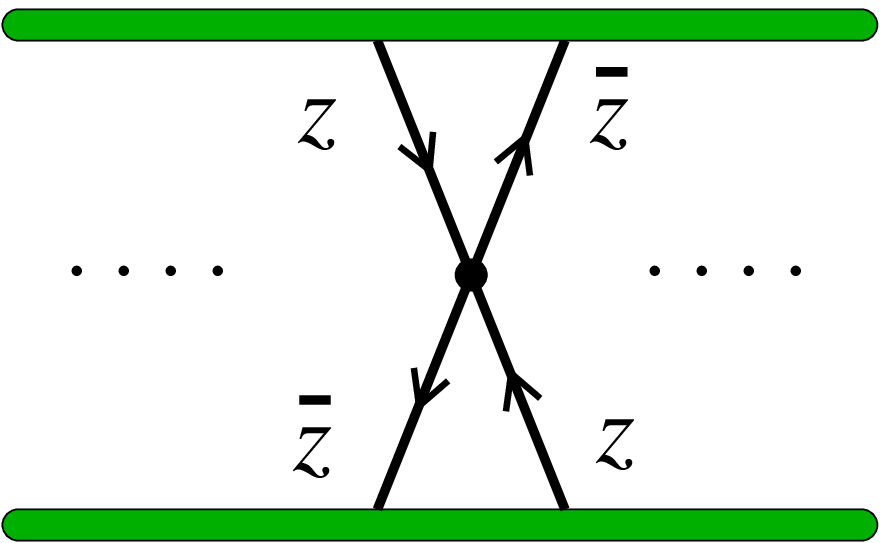}}
&=& 
{3\over2} \, \gamma \times (\half G N)^{J+\N} 
\nonumber\\
\end{eqnarray}
%
In the results 
(\ref{eq:tree diagrams})-(\ref{eq:one-loop diagrams:z-zbar}), 
we dropped terms suppressed by $1/N^2$.

\subsection{Equality of matrix elements: generic states}
\label{section:equalitygeneric}
In this appendix we will complete matching the matrix
elements of the light cone Hamiltonian between the two sides
of the AdS/CFT correspondence.
In section \ref{section:equalitysimple} we matched 
matrix elements for a subset of states.
There we considered states with all excited modes having distinct SO(4) indices $i_k$,
and no mode excited more than once.
We will now consider states with some $i_k$ being equal.
We initially restrict to the case with no modes excited
more than once, $N^i_n \le 1$, but will eventually generalize to most
general case.

In contrast with section \ref{section:equalitysimple}, $\Or_*$ no longer vanishes.
In addition to (\ref{del1}), we now have to consider off-diagonal elements between the states
\beq
\label{i2}
  |\Y\rangle
        =  y_{n_1}^{i_1}{}^\dagger  \ldots  y_m^i{}^\dagger  
        y_n^i{}^\dagger  |\eta \rangle, \qquad m \neq n,
\eeq
and
\beq
\label{f2}
  |\Y'\rangle
      =  y_{n_1}^{i_1}{}^\dagger  \ldots  
   y_m^j{}^\dagger  y_n^j{}^\dagger  |\eta \rangle, \qquad m \neq n,
\eeq
which are given by
\beq
\label{del2}
  \langle \Y|  H_1^{OD} |\Y' \rangle  =  
      {1 \over J} \left({ R^2 \over J}\right)^2 m n \sqrt{N^i_m N^i_n N^j_m{}' N^j_n{}'}.
\eeq
There is also an off-diagonal element given by (\ref{del1}),
but we have analyzed all diagrams contributing to it in 
section \ref{section:equalitysimple}.
Let us briefly explain why this is the case.
Consider $\Or(g^0)$ part of the contributing two-point function,
which we denote by $\langle \X {\bar \X}'\rangle_{g^0}$.
$\langle {\tilde \X} {\bar \X}_*'\rangle_{g^0}$ and
$\langle \X_*' {\bar {\tilde \X}} \rangle_{g^0}$
vanish, as there are no contributing interaction-free diagrams.
Although $\langle \X_* {\bar \X'}_*\rangle_{g^0}$  has 
nonvanishing terms, they are $\Or(1/J^2)$.
This is because $\X_*$ is itself $\Or(1/J)$ compared to ${\tilde \Or}$,
and an additional factor of $1/J$ will
appear because phases in $\X_*$ and $\X_*'$ do not
match exactly.
Similar conclusions can be made about $\Or(g)$ correlator 
$\langle {\tilde \X} {\bar \X'}\rangle_{g^1}$.

Let us compute the off-diagonal element (\ref{del2}) in the gauge theory.
The only contribution to $\langle \Y(x) {\bar \Y'}(0)\rangle_{g^0}$
comes from 
\beq
\label{intfree3}
 \langle \Y_*(x) {\bar \Y'}_*(0)\rangle_{g^0}=
{1 \over \Omega} \sum 
\III{{\check \phi}^i_m}{{\check \phi}^j_{m}}{{\check \phi}^i_n}{{\check \phi}^j_{n}}
 { \phi^{i_k}_{n_k}}{ \phi^{i_k}_{n_k}}={1 \over J},
\eeq
where the sum runs over all configurations of fields. 
No $\Or(g^0)$ diagrams appear in $\langle {\tilde \Y} {\bar \Y'}_* \rangle_{g^0}$,
$\langle \Y_* {\bar {\tilde \Y'}} \rangle_{g^0}$ and 
$\langle {\tilde \Y} {\bar {\tilde \Y'}} \rangle_{g^0}$.
Hence we have
\beq
\label{iijjod}
   {\textbf T}^{(1)}_{\Y\Y'}=1.
\eeq

Computation of $\F_{\Y\Y'}$ is more involved.
Possible contributions are
\beqa
\label{cr3}
\langle {\tilde \Y}(x) {\bar {\tilde \Y'}}(0) \rangle_{g^1}&=&
   {1 \over \Omega} {\sum} {\XIIm{\phi^i_m}{\phi^j_{m'}}{\phi^i_n}{\phi^j_{n'}}
           {z}{\bar z}{\phi^{i_k}_{n_k}}{\phi^{i_k}_{n_k}}} {+}
   (m {\lra} n) {+} (m' {\lra} n') {+}(m {\lra} n, m' {\lra} n') 
\nonumber\\ 
&=& { \gamma \over 2 J} (q_{n-m}+q_{n-m}^*+2),  
\phantom{\Bigg |}
\eeqa 
which holds both for $m'=n, n'=m$ (off-diagonal) and
$m'=m, n'=n$ (diagonal),
\beqa
 \label{cr4}
{-}\langle {\tilde \Y}(x) {\bar \Y'}_*(0) \rangle_{g^1}&=&
 {-}{1 \over \Omega} {\sum}  \XIIm{\phi^i_m}{z}{\phi^i_n}{\bar z}
           {z}{\bar z}{\phi^{i_k}_{n_k}}{\phi^{i_k}_{n_k}} {+}
 (m {\lra} n) {+} (z {\lra} {\bar z}) {+}(m {\lra} n, z {\lra} {\bar z})
\nonumber \\ 
&=& -{ \gamma \over 2 J} (q_m+q_m^*+q_n+q_n^*),
\phantom{\Bigg |}
\eeqa  
similar contribution from $ \langle \Y_*(x) {\bar \Y'}_*(0)\rangle_{g^1}$,
and
\beqa
\label{cr5}
\langle \Y_*(x) {\bar \Y'}_*(0) \rangle_{g^1}
&=&
{1 \over \Omega} {\sum}  
\IIX{z}{\bar z}{\;\;\bar z_{m{+}n}}{\;\;z_{m{+}n}}{\phi^{i_k}_{n_k}}{\bar z}{z}{\phi^{i_k}_{n_k}}
\nonumber \\ && \hspace{3em}
\phantom{\Bigg|}
{+}
(\phi^{i_k}_{n_k} {\lra} z){+}(\phi^{i_k}_{n_k} {\lra} {\bar z}) 
{+}(\phi^{i_k}_{n_k} {\lra} z,\phi^{i_k}_{n_k} {\lra} {\bar z})
\nonumber \\ 
&+& 
{1 \over \Omega} {\sum} \XIIm{z}{\bar z}{\bar z_{m{+}n}}{z_{m{+}n}}
           {z}{\bar z}{\phi^{i_k}_{n_k}}{\phi^{i_k}_{n_k}}
\nonumber \\ && \hspace{3em}
\phantom{\Bigg|}
{+}(z{\lra}{\bar z_{m{+}n}}){+}({\bar z}\lra z_{m{+}n}){+}
        (z{\lra}{\bar z_{m{+}n}},{\bar z}\lra z_{m{+}n})
\nonumber \\ 
&=& 
{\gamma \over J} \left( -
\!\!\!
\sum_{p:\, n_p \neq m,n} (q_{n_p}+q_{n_p}^*-2)
         -{1 \over 2} (q_{m+n}+q_{m+n}^*) + 3 \right)
\!\!
. 
\hspace{3.5em}
\eeqa
(The subscript in ${\bar z}_{n+m}$ stands for the phase
$q_{n+m}^{a_{\bar z}}$ which depends on the position of the $\bar z$ in the string
of operators.)
Combining (\ref{iijjod})--(\ref{cr5}) we have
\beq
\label{del2a}
  [{\textbf T}^{-1} \F]_{\Y\Y'}= -{\beta \over 2 J} (q_{m+n}-q_{m-n}+c.c.)=
             {1 \over J} \left({R^2 \over J}\right)^2 m n,
\eeq
which indeed agrees with (\ref{del2}), provided $N^i_n \le 1$.

Let us now turn to the diagonal matrix element.
Part of it was computed in section \ref{section:equalitysimple}
and is given by (\ref{tfdiag}). 
But now there are other contributions both to ${\textbf T}^{(1)}_{\Or\Or}$
and to $\F_{\Or\Or}$.
To update the former, we must take into account
\beq
\label{intfree4}
 \langle \Or_*(x) {\bar \Or}_*(0) \rangle_{g^0}= 
  {1 \over \Omega} \sum_{i,(m\neq n)} \sum
    \III{{\check \phi}^i_m}{{\check \phi}^i_{m}}{{\check \phi}^i_n}{{\check \phi}^i_{n}}
                    { \phi^{i_k}_{n_k}}{ \phi^{i_k}_{n_k}}=
     \sum_i {\N_i (\N_i-1) \over 2 J},
\eeq             
and
\beq
\label{intfree5}
 \delta \langle {\tilde \Or}(x) {\bar {\tilde \Or}}(0) \rangle_{g^0}= 
  {1 \over \Omega} \sum_{i,(m\neq n)} \sum
    \III{\phi^i_m}{\phi^i_{n}}{\phi^i_n}{\phi^i_m}
                    { \phi^{i_k}_{n_k}}{ \phi^{i_k}_{n_k}}= 
         -  \sum_i {\N_i (\N_i-1) \over  2 J},
\eeq
which cancels (\ref{intfree4}) to keep ${\textbf T}_{\Or\Or}=1$.
The $\Or(g^1)$ correlators related to (\ref{intfree4}) are
given by the sum of (\ref{cr5}) over pairs $(n_k\neq n_l):i_k=i_l$ 
with the substitution $m=n_k, n=n_l$: 
\beqa
\label{cr6}
\langle \Or_*(x) {\bar \Or}_*(0) \rangle_{g^1}
&=& 
{\gamma \over J}
\sum_{(n_k{\neq}n_l):i_k{=}i_l}
\left( 
3{-}\sum_{p{\neq}k,l} (q_{n_p}{+}q_{n_p}^*{-}2)
- {1 \over 2} ( q_{n_k{+}n_l}{+}q_{n_k{+}n_l}^* )
\right) 
\nonumber \\ 
&=&
{\gamma \over J} \Bigg[ -\left( \sum_i {\N_i (\N_i{-}1) \over 2} \right)
\sum_k (q_{n_k}{+}q_{n_k}^*{-}2) 
\nonumber \\ 
&&
\hspace{-1em}
\!+\!\!\!
\sum_{(n_k{\neq}n_l):i_k=i_l} \left\{ (q_{n_k}+q_{n_l}-2+c.c.)
                -{1 \over 2} (q_{n_k+n_l}+q_{n_k+n_l}^*) + 3 \right\} \Bigg].
\nonumber \\ 
\eeqa
The $\Or(g^1)$ counterpart of (\ref{intfree5}) is
\beqa
\label{cr7}
 \delta_1 \langle {\tilde \Or}(x) {\bar {\tilde \Or}}(0) \rangle_{g^1}
&=& 
      {\gamma \over J} \bigg[ \left( \sum_i {\N_i (\N_i{-}1) \over 2} \right) 
                \sum_k (q_{n_k}{+}q_{n_k}^*{-}2)
\bigg.\nonumber\\&&\bigg.
\hspace{2em}
+ 
\!\!\!
\sum_{(n_k{\neq}n_l):i_k{=}i_l} 
( q_{n_k} {+}q_{n_l}{-}2 q_{n_l{-}n_k}{+}c.c.) \bigg].
\quad
\eeqa
The first term in this expression is a value of the corresponding
interaction-free diagram times the sum of possible phases, while the
second term takes care of overcounted corrections (this technique
for computing $\Or(g^1)$ diagrams was explained in more detail
in section \ref{section:equalitysimple})
There is also a contribution which is a direct analog of (\ref{diagmee})
\beq
\label{cr8}
 \delta_2 \langle {\tilde \Or}(x) {\bar {\tilde \Or}}(0) \rangle_{g^1}
= 
  { \gamma \over J}  \sum_{(n_k{\neq}n_l):i_k{=}i_l} 
     \left[ (q_{n_k}+q_{n_l}-2+c.c.)+{1 \over 2}(q_{n_k-n_l}+q_{n_k-n_l}^*+2)  \right].
\eeq   
Finally, we should include the sum over pairs in (\ref{cr4}) and
the same term due to 
\beq
\label{cr9}
    -\langle \Y_*(x) {\bar{\tilde  \Y'}}(0){+} {\tilde \Y}(x) {\bar \Y'}_*(0)\rangle_{g^1}=
      - { \gamma \over  J} \sum_{(n_k{\neq}n_l):i_k{=}i_l} (q_{n_k}+q_{n_l}+c.c.),
\eeq
Combining (\ref{cr6})--(\ref{cr9}) we get
\beqa
\label{tfdiag2}
  \delta [{\textbf T}^{-1} \F]_{\Or\Or} &=& 
       -{\beta \over 2 J}  \sum_{(n_k{\neq}n_l):i_k{=}i_l} \left[
      3 q_{n_k{-}n_l}{+}q_{n_k{+}n_l}{-}4 q_{n_k}{-}4 q_{n_l}{+}4{+}c.c. \right] 
\hspace{3em}
\\ \nonumber
   &=&   {-}{1 \over J} \left( {R^2 \over J} \right)^2 \sum_{(n_k{\neq}n_l):i_k{=}i_l} n_k n_l,
\eeqa
which should be added to (\ref{tfdiag}).
In the case of $N^i_m \le 1$, (\ref{tfdiag2}) combined with the
last term in (\ref{tfdiag})
gives
\beq
\label{funnyc}
   -{1 \over J} \left( {R^2 \over J} \right)^2 \left[
       \sum_{(k,l):i_k{\neq}i_l} n_k n_l + \sum_{(n_k{\neq}n_l):i_k{=}i_l} n_k n_l \right]
    = {1 \over 2 J} \left( {R^2 \over J} \right)^2 \sum_k n_k^2,
\eeq
where we used the level matching condition.
Hence we again reproduce (\ref{hdiagx}).

Our last step will be generalization to the case of unconstrained $N^i_n$.
To see how (\ref{del1a}) is modified recall that all contributing
correlators should be divided by
\beq
\label{odf1}
   \sqrt{ N^i_m! N^j_n! N^i_n! N^j_m! N^i_m{}'! N^j_n{}'! N^i_n{}'! N^j_m{}'!\ldots}
\eeq
where $\ldots$ stands for other $N^{i_k}_{n_k}$ which will be cancelled
by the number of possible contractions, just as they are cancelled
in non-interacting diagrams to produce $T_{\Or\Or}=1+\Or(1/J)$.
On the other hand, the combinatorial factor that multiplies all the
correlators contributing to (\ref{del1a}) is
\beq
\label{odf2}
    N^i_m! N^j_n! N^i_n{}'! N^j_m{}'! \ldots
\eeq
The ratio of (\ref{odf2}) and (\ref{odf1}) is precisely
the factor $\sqrt{N^i_m N^j_n N^i_n{}' N^j_m{}'}$ which appears in (\ref{del1a}).
The combinatorial factor in (\ref{del2}) can be restored in the similar manner.

In addition to (\ref{del1}) and (\ref{del2}) we also need to consider
off-diagonal matrix elements between the states
\beq
\label{i3}
  |{\mathcal Z}\rangle
        =  y_{n_1}^{i_1}{}^\dagger  \ldots  y_n^i{}^\dagger  
        y_n^i{}^\dagger  |\eta \rangle, 
\eeq
and
\beq
\label{f3}
  |{\mathcal Z}'\rangle
      =  y_{n_1}^{i_1}{}^\dagger  \ldots  
   y_n^j{}^\dagger  y_n^j{}^\dagger  |\eta \rangle,
\eeq
which are given by
\beq
\label{del3}
  \langle {\mathcal Z}|  H_1^{OD} |{\mathcal Z}' \rangle  =  
      {1 \over 4 J} \left({ R^2 \over J}\right)^2  n^2 \sqrt{N^i_n (N^i_n-1) N^j_n{}'(N^j_n{}'-1)}.
\eeq
This can be computed similarly to (\ref{del2a}).
One should just multiply each term in (\ref{iijjod})--(\ref{cr5})
by 
\beq
\label{cf}
  {J^\N \over \sqrt{\Ot \, \Ot'}} {N^i_n (N^i_n-1) \over 2} {N^j_n{}' (N^j_n{}'-1) \over 2}
                           (N^i_n-2)!  \, (N^j_n{}'-2)!.
\eeq
The ingredients in (\ref{cf})  correspond to the normalization, the number of possible
choices of a pair out of  $N^i_n$ ($N^j_n{}'$) $\phi^i_n$'s ( $\phi^j_n$'s),
and the number of permutations of the leftover  $\phi^i_n$'s ( $\phi^j_n$'s).
Substituting $\Ot \approx \sqrt{J^\N N^i_n! N^j_n! \ldots}$ and
$\Ot' \approx \sqrt{ J^\N N^i_n{}'! N^j_n{}'! \ldots}$ into (\ref{cf}) one recovers
correct combinatorial factor in (\ref{del3}).

The expressions for diagonal matrix elements (\ref{tfdiag}) and (\ref{tfdiag2})
do not change when we allow $N^i_n>1$.
However (\ref{funnyc}) changes to
\beq
\label{funnyc2}
  {1 \over 2 J} \left( {R^2 \over J} \right)^2 \sum_{i,n} n^2 (N^i_n)^2.
\eeq
There is an additional contribution to the diagonal
matrix element, which is similar to (\ref{tfdiag2})
but with $n_l=n_k$.
To compute it, one has to follow the logic which led to 
(\ref{tfdiag2}) paying special attention to combinatorial factors.
We now have 
\beq
\label{intfree6}
 \delta \langle \Or_*(x) {\bar \Or}_*(0) \rangle_{g^0}= 
  {1 \over \Omega} \sum_{i,n} \sum
    \III{{\check \phi}^i_n}{{\check \phi}^i_{n}}{{\check \phi}^i_n}{{\check \phi}^i_n}
                    { \phi^{i_k}_{n_k}}{ \phi^{i_k}_{n_k}}=
     \sum_{i,n} {N^i_n (N^i_n-1) \over 4 J}
\eeq    
and
\beq
\label{intfree7}
 \delta \langle {\tilde \Or}(x) {\bar {\tilde \Or}}(0) \rangle_{g^0}= 0.
\eeq
since the diagram analogous to (\ref{intfree4}) with $m=n$ have been already
taken care of, and absorbed in the normalization constant.
The analog of (\ref{cr6}) is
\beqa
\label{cr10}
   \delta \langle \Or_*(x) {\bar \Or}_*(0) \rangle_{g^1}
&=& 
    {\gamma \over J}  \sum_{i,n} {N^i_n (N^i_n{-}1) \over 4} 
\nonumber\\&& \hspace{2em}\times
\bigg[
                                -\sum_k (q_{n_k}{+}q_{n_k}^*{-}2) 
    {+}2 (q_{n}{+}q_{n}^*{-}2)
          {-}{1 \over 2} (q_{2n}+q_{2n}^*) {+} 3  \bigg],
\nonumber\\&&
\eeqa
while the contribution similar to (\ref{cr7}) is absent.
The analog of (\ref{cr8}) is
\beq
\label{cr11}
  \delta_3\langle {\tilde \Or}(x) {\bar {\tilde \Or}}(0) \rangle_{g^1}{=} 
  { \gamma \over J} \sum_{i,n} {N^i_n (N^i_n-1) \over 4} 
     \left[ 4 (q_{n}+q_{n}^*-2)+2  \right].
\eeq   
Finally, there is an analog of (\ref{cr9}) given by
\beq
\label{cr12}
   - \delta \langle \Or_*(x) {\bar{\tilde  \Or'}}(0){+} {\tilde \Or}(x) {\bar \Or}_*(0)\rangle_{g^1} =
- { \gamma \over  J} \sum_{i,n} {N^i_n (N^i_n-1) \over 2} (q_{n}+q_{n}^*).
\eeq
Combining (\ref{intfree6})--(\ref{cr12}) we get the following contribution
to the diagonal matrix element from the $\phi^i_n/\phi^i_n$ interactions
\beq
    {\beta \over J} \sum_{i,n}{ N^i_n (N^i_n-1) \over 4} \left( 4 q_n - {q_{2 n} \over 2} +c.c.\right) 
   =  - {1 \over J} \left( {R^2 \over J} \right)^2 \sum_{i,n} {n^2 N^i_n (N^i_n-1) \over 4}.
\eeq
Adding this to (\ref{funnyc2}) and then  replacing the last term
in (\ref{tfdiag}) with the resulting expression we recover the string
theory result (\ref{hdiagx}).
This concludes the matching of matrix elements between the
string and the gauge theory.